\documentclass[universe,review,accept,moreauthors,pdftex,10pt,a4paper]{mdpi}
\firstpage{1} 

\articlenumber{7}
\doinum{10.3390/universe3010007}
\pubvolume{3}
\pubyear{2017}
\copyrightyear{2017}

\externaleditor{}
\history{}

\pdfoutput=1
 \usepackage{amssymb}
 \usepackage{amsmath}
  \usepackage{amsfonts}
  \usepackage{epsfig}
   \usepackage{soul}
   \usepackage{xspace}
   \usepackage{color}
   \usepackage{bm} 
\usepackage{booktabs} 
\usepackage{multirow}
\usepackage{soul} 
\usepackage{microtype}
\usepackage{mathcomp}
\makeatletter
\g@addto@macro{\UrlBreaks}{\UrlOrds}
\makeatother
\DeclareRobustCommand{\myurl}[1]{\texorpdfstring{\changeurlcolor{black}\url{#1}\changeurlcolor{blue}}{#1}}

\newcommand{\deta}{\mbox{$\Delta\eta$}}
\newcommand{\dphi}{\mbox{$\Delta\phi$}}
\newcommand{\vthree}{\ensuremath{v_3^2\{2\}}}
\newcommand{\pt}{\mbox{$p_T$}\xspace}
\newcommand{\pTtrig}{\mbox{$p^{\rm trig}_T$}\xspace}
\newcommand{\pTassoc}{\mbox{$p^{\rm assoc}_T$}\xspace}
\newcommand{\raa}{\mbox{$R_{AA}$}\xspace}
\newcommand{\rab}{\mbox{$R_{AB}$}\xspace}

\newcommand{\rpb}{\mbox{$R_{pPb}$}\xspace}
\newcommand{\Npart}{\mbox{$N_{\rm part}$}\xspace}
\newcommand{\Ncoll}{\mbox{$N_{\rm coll}$}\xspace}

\newcommand{\Nch}{\mbox{$N_{\rm ch}$}\xspace}
\newcommand{\Et}{\mbox{${\rm E}_T$}\xspace}

\def\mean#1{\left<#1\right>}
\newcommand{\sqs}{\mbox{$\sqrt{s}$}\xspace}
\newcommand{\sqsn}{\mbox{$\sqrt{s_{_{NN}}}$}\xspace}

\newcommand{\pp}{\mbox{$p$ $+$ $p$}\xspace}
\newcommand{\dau}{\mbox{$d$ $+$ Au}\xspace}

\newcommand{\pdau}{\mbox{$p(d)$ $+$ Au}\xspace}
\newcommand{\ppb}{\mbox{$p$ $+$ Pb}\xspace}
\newcommand{\pbe}{\mbox{$p$ $+$ Be}\xspace}
\newcommand{\pa}{\mbox{$p$ $+$ A}\xspace}
\newcommand{\auau}{\mbox{Au $+$ Au}\xspace}
\newcommand{\pbpb}{\mbox{Pb $+$ Pb}\xspace}
\newcommand{\cucu}{\mbox{Cu $+$ Cu}\xspace}
\newcommand{\cuau}{\mbox{Cu $+$ Au}\xspace}

\newcommand{\heau}{\mbox{$^{3}$He $+$ Au}\xspace}

\newcommand{\dNch}{\mbox{$dN_{\rm ch}/d\eta$}\xspace}

\newcommand{\ebj}{\mbox{$\varepsilon_{BJ}$}\xspace}
\newcommand{\PgU}{\mbox{$\Upsilon$}\xspace}
\newcommand{\PgUa}{\mbox{$\Upsilon\text{(1S)}$}\xspace}
\newcommand{\PgUb}{\mbox{$\Upsilon\text{(2S)}$}\xspace}
\newcommand{\PgUc}{\mbox{$\Upsilon\text{(3S)}$}\xspace}

\newcommand{\sNN}{$\sqrt {{s_{\rm NN}}}$}
\newcommand{\MV}{$\sigma^{2}/M$}
\newcommand{\Ss}{$S\sigma$}
\newcommand{\KV}{$\kappa\sigma^{2}$}

\Title{Phenomenological Review on Quark--Gluon Plasma: Concepts vs. Observations}

\Author{Roman Pasechnik $^{1}$,* and Michal \v{S}umbera $^{2}$}
\AuthorNames{Roman Pasechnik, Michal \v{S}umbera}

\address{%
$^{1}$ \quad Department of Astronomy and Theoretical Physics, Lund University, SE-223 62 Lund, Sweden\\
$^{2}$ \quad Nuclear Physics Institute ASCR
250 68 \v{R}ez/Prague, Czech Republic; sumbera@ujf.cas.cz}

\corres{Correspondence: Roman.Pasechnik@thep.lu.se}

\abstract{In this review, we present an up-to-date phenomenological summary of research developments in the physics of the Quark--Gluon Plasma (QGP). 
A short historical perspective and theoretical motivation for this rapidly developing field of contemporary particle physics is provided. In addition, 
we introduce and discuss the role of the quantum chromodynamics (QCD) ground state, non-perturbative and lattice QCD results on the QGP properties, as well as 
the transport models used to make a connection between theory and experiment. The experimental part presents the selected results 
on bulk observables, hard and penetrating probes obtained in the ultra-relativistic heavy-ion experiments carried out at the Brookhaven National Laboratory Relativistic Heavy Ion Collider (BNL RHIC) and CERN Super Proton Synchrotron (SPS) and 
Large Hadron Collider (LHC) accelerators. We also give a brief overview of new developments related to the ongoing searches of 
~the QCD critical point and 
to the collectivity in small \mbox{(\pp and \pa) systems.}}



\begin{document}

\section{Introduction}

Quark--gluon plasma (QGP) is a new state of nuclear matter existing at extremely high temperatures and densities when composite states
called hadrons (protons, neutrons, pions, etc.) lose their identity and dissolve into a soup of their constituents---quarks and gluons. The existence of 
this novel phase of matter was proposed in the mid-seventies \cite{Collins:1974ky, Cabibbo:1975ig}, just ten years after the birth of the {\it Quark Model} 
of hadrons \cite{GellMann:1964nj, Zweig:1981pd}, and two years after it was realised that the candidate non-Abelian field theory of inter-quark forces---{\it quantum chromodynamics} (QCD) \cite{Fritzsch:1973pi}---predicts their weakening at short distances, the so-called {\it asymptotic freedom} 
\cite{Gross:1973id, Politzer:1973fx}. 

Contrary to atoms and molecules which can be ionized to reveal their constituents, quarks and gluons are never found free, but are confined inside 
the hadrons. This situation is quite similar to decomposing the magnet into two when trying to isolate its north pole from its south pole. 
Even deeper goes the analogy \cite{Nambu:1976ay} between field lines confining quarks inside the hadrons and the magnetic field in the 
~vicinity of a~superconductor which expels the magnetic flux lines ({\it Meissner effect}). If two magnetic poles are surrounded by a superconducting medium, 
the field is confined into a thin tube. A~hadronic string with quark and antiquark sitting at its end points has a similar one-dimensional field confined 
not by a superconducting medium, but by the vacuum\footnote{The string picture of hadrons also makes it possible to explain the transition from 
a hadronic to a QGP state of matter at finite temperature $T$ as a phase transition from the ordered state to the disordered state (see Ref.~\cite{Yagi:2005yb}, p.~44). 
With $T$ approaching the critical temperature of deconfinement $T_{\rm c}$, the effective string tension decreases, and $q$ and $\bar{q}$ attached 
to its end points lose their correlation.}.

Experimental attack on producing~QGP under laboratory conditions started in the world-leading particle physics facilities at CERN (Conseil Europ{\'e}en pour la Recherche Nucl{\'e}aire, Geneva, Switzerland) and Brookhaven National Laboratory (BNL, New York, NY,~USA) in the late 1980s~\cite{Schmidt:1992ge, Stock:2008ru, Schukraft:2015dna}. In the year 2000, after finishing the main part of its heavy-ion program 
at the Super Proton Synchrotron (SPS) accelerator, CERN announced circumstantial evidence for the creation of a new state of matter in \pbpb collisions \cite{Heinz:2000bk}. The real discovery 
of QGP took place in 2005, when four international collaborations studying \auau collisions at the Relativistic Heavy Ion Collider (RHIC) at BNL announced the results of 
their first five years of measurements \cite{Arsene:2004fa, Back:2004je, Adams:2005dq, Adcox:2004mh}.  Surprisingly, the properties of the new state of matter 
\cite{Gyulassy:2004zy} differed markedly from predictions made over many years before its discovery\footnote{For a representative collection of papers tracing 
the development of theoretical ideas on the QCD deconfining phase transition before nineties
, see Ref.~\cite{Rafelski:2003zz}.}.

This review aims at giving a short historic introduction into the vast research field of QGP physics and the underlined phenomenological aspects with a comprehensive 
list of corresponding references. Effects of the hot/dense medium such as the nuclear suppression, initial-state interactions, in-medium energy loss, color screening 
and saturation are important for a proper understanding of the collective phenomena in heavy-ion collisions, and are included in the scope of this review. Besides, we have 
qualitatively overviewed and confronted with existing observations such fundamental theoretical concepts as the QCD vacuum and 
phase diagram, equation of state (EoS) of deconfined QCD matter, initial-state effects, collectivity, flow, hydrodynamic properties of the QGP, and associated 
electromagnetic effects.

The paper is organized as follows. Section~\ref{sec:matter} presents a short history of the theoretical understanding of extreme states matter. 
The phase diagram of QCD is discussed in Section~\ref{sec:phases}. 
Section~\ref{sec:experiment} gives a historical perspective on experiments operating with with collisions of heavy ions. Section~\ref{sec:signatures} 
describes the basic signatures of QGP production, while current developments in QGP research are provided in Section~\ref{sec:developments}. The concluding
remarks are given in Section~\ref{sec:conclusions}. For further reading on fundamental concepts and the latest studies of QGP dynamics, we recommend several 
textbooks~\cite{Csernai:1994xw, Letessier:2002gp,Yagi:2005yb,Vogt:2007zz, Florkowski:2010zz, Rak:2013yta} and reviews \cite{BraunMunzinger:2007zz,
Shuryak:2008eq,Sarkar:2010zza,Florkowski:2014yza,Roland:2014jsa,Nouicer:2015jrf} published in recent years.
 
\section{Matter under Extreme Conditions}
\label{sec:matter}

The properties of matter under extreme conditions at high values of state variables have always attracted the curiosity of scientists, owing to the possibility 
of advancing to new domains of the phase diagram and producing the exotic states of matter in laboratory \cite{Zeldovich:1966, Fortov:2011}. 
The first attempts to discuss the properties of matter at densities well above the normal nuclear density $\rho_0$ = 2 $\times$ 10$^{14}$ g$\cdot$cm$^{-3}$ 
\mbox{($\approx$0.16 GeV$\cdot$fm$^{-3}$)} date back to the seminal Oppenheimer--Volkoff paper from 1939 \cite{Oppenheimer:1939ne}. A study of the gravitational 
stability of a new phase of neutron matter suggested a few years earlier by Landau~\cite{Landau1932} led them to carry out their computations 
to several tens of $\rho_0$ before smoothly extrapolating the results to the black-hole singularity. In 1962, when discussing relativistic limitations of the equation 
of state (EoS) of matter made of baryons interacting via a massive vector field, Zeldovich used the density twenty times exceeding $\rho_0$ \cite{Zeldovich:1962}. 
In 1976, the same value of density was shown by Baym and Chin \cite{Baym:1976yu} to be energetically favourable for the neutron matter--quark phase transition. 

In cosmology, a very dense matter with $\rho~\propto$~10$^6$~g$\cdot$cm$^{-3}$ ($\approx$1 eV$\cdot$fm$^{-3}$) was first studied in 1946 by Gamow 
\cite{Gamow:1946eb} when discussing the relative abundances of elements in the universe. The key discovery of the cosmic microwave background radiation by 
Penzias and Wilson in 1965 \cite{Penzias:1965wn} not only provided a strong basis for the hot universe scenario which was used by Gamow, but also  motivated Sakharov~\cite{Sakharov:1966fva} to push it {ad extremum}. Considering the properties of hot matter at densities when gravitational interaction 
between photons becomes significant, he established that the absolute maximum of the temperature of any substance in equilibrium with radiation is 
on the order of Planck temperature $T_P =\sqrt{\frac{\hbar c^5}{G k^2}}~\propto$~10$^{32}$ K ($\approx$10$^{22}$ MeV). Unfortunately, the theoretical 
apparatus of that period was completely inadequate to deal with the thermal history of the universe from $T_P$ downwards, but even at temperatures 
twenty orders of magnitude lower \cite{Huang:1970iq}. 

The problem was due to two successful but mutually conflicting contemporary models of hadrons: the Bootstrap model \cite{Hagedorn:1965st, Rafelski:2016hnq} 
based on the hypothesis that all hadrons are composite of one another, and---at that time not fully developed composite model of hadrons---the Quark model \cite{GellMann:1964nj, Zweig:1981pd}. The Bootstrap model predicted that after reaching some limiting value of temperature---the so-called {\it Hagedorn temperature} ($T_H$ = 170--180 MeV) 
that can be estimated from the spectrum of hadronic masses~\cite{Rafelski:2016hnq, Majumder:2010ik}---the subsequent heating of strongly interacting matter will lead 
to the creation of more and more massive hadron species, but not to an increase of its temperature. The quark model, on the other hand, predicted relic cosmological quarks  \cite{Zeldovich:1967rg} roaming free through our universe. The leftover quarks were predicted to be as abundant as gold atoms \cite{Zeldovich:1967}. 

The conflict was finally resolved in 1973 when Gross, Wilzek \cite{Gross:1973id}, and Politzer \cite{Politzer:1973fx} discovered the asymptotic freedom in non-Abelian 
gauge theories of elementary particle interactions. Shortly after the idea of asymptotic freedom was introduced, two groups---Collins and Perry \cite{Collins:1974ky}, and 
Cabibbo and Parisi \cite{Cabibbo:1975ig}---independently realized its fascinating consequence for the properties of hot and dense matter. The first group argued 
that since the interaction between quarks weakens as quarks get closer at sufficiently high density, these quarks are no longer confined inside the hadrons and 
become free. The superdense matter at densities higher than the nuclear one consists of a quark soup. The other group re-interpreted the existence of the Hagedorn 
limiting temperature $T_H$ as a signal of a second-order phase transition between the hadronic and quark--gluon phases of matter. 

The discovery of asymptotic freedom also paved the way to our current understanding of the evolution of the early universe. The commonly-accepted scenario 
of the subsequent cooling of the universe assumes a series of first- or second-order phase transitions associated with the various spontaneous symmetry-breakings 
of the basic non-Abelian gauge fields  \cite{Linde:1978px, Bailin:2004zd, Boyanovsky:2006bf}. The Standard Model (SM) of elementary particles predicts two 
such transitions \cite{Boyanovsky:2006bf}. One taking place at temperatures of a few hundred GeV is responsible for the spontaneous breaking of 
the electroweak (EW) symmetry, providing masses to elementary particles. It is also related to the EW baryon-number-violating processes, which had 
a major influence on the observed baryon-asymmetry of the universe \cite{Trodden:1998ym}. The lattice simulations have shown that the EW transition 
in the SM is an analytic crossover \cite{Kajantie:1996mn}.

The second---the transition of QGP to hadronic matter---happens at $T<200$ MeV, and is related to the spontaneous breaking of the chiral symmetry of the non-Abelian 
theory of strong interactions, which is based on the SU(3)$_{\text c}$ color group: the QCD. The nature of this phase transition affects to a great extent our understanding 
of the evolution of the early universe \cite{Boyanovsky:2006bf}. For instance, in a strong first-order phase transition, the QGP supercools before bubbles of hadron 
gas are formed. Since the hadronic phase is the initial condition for nucleosynthesis, the inhomogeneities in this phase could have a strong effect on the nucleosynthesis 
epoch \cite{Boyanovsky:2006bf}. Here, the lattice non-perturbative QCD calculations developed since the late 1970s \cite{Susskind:1979up} (for a recent review, see 
Ref.~\cite{Petreczky:2012rq}) can be of great help. Knowing that the typical baryon chemical potentials $\mu_B$ are much smaller than the typical hadron 
masses ($\mu_B \approx 45$ MeV at \sqsn = 200 GeV \cite{Adams:2005dq} and negligible in the early universe), we can use the lattice QCD calculations 
performed at $\mu_B=0$. The results \cite{Aoki:2006we} not only confirm the previous finding \cite{Susskind:1979up} that confinement of quarks into hadrons 
is strictly a low-temperature phenomenon, but provide strong evidence that the QCD transition is also a crossover, and thus the above mentioned scenarios---and 
many others---are ruled out. The same conclusion was made in Ref.~\cite{Bhattacharya:2014ara}, where the first lattice analysis of the QCD phase transition in a model 
with chiral quarks having physical masses was performed. Numerical simulations on the lattice also indicate that at vanishing $\mu_B \approx 0$ MeV, the two phase 
transitions which are possible in the QCD---de-confining and chiral symmetry restoring---occur at essentially the same~point \cite{Cheng:2008}.

The situation at large $\mu_B$ and $T$ is more complicated (see the left panel of Figure~\ref{fig:phase_diagram}). Here, the wealth of novel QCD phases is predicted to exist \cite{Rischke:2004}, including the so-called quarkyonic phase \cite{McLerran:2007}.  At $T \approx 0$ MeV and $\mu_B \geq 1 $ GeV, a variety of color superconducting 
phases occur \cite{Rischke:2004, Buballa:2005}. Somewhere on the phase boundary at $\mu_{B} \approx 400$ MeV, a critical point separating the first- and second-order 
phase transitions is predicted \cite{Rischke:2004}. The search for this point is now underway at RHIC \cite{Aggarwal:2010cw}, and some of the recent results will be discussed 
in Section~\ref{sec:developments}.

\section{Phases of QCD Matter}
\label{sec:phases}

Although the quark matter was mentioned as early as 1970 by Itoh \cite{Itoh:1970} in the context of neutron stars, the term ``hadronic plasma'' was first introduced 
in 1977 by Shuryak \cite{Shuryak:1977ut} to describe a new state of matter existing at temperatures above 1 GeV. This makes a good analogy with a classical 
gaseous plasma case, when electrically neutral gas at high enough temperatures turns into a statistical system of mobile charged particles \cite{Ichimaru:1982zz}. While 
in the latter case their interactions obey the U(1)$_{\text em}$ gauge symmetry of Quantum Electrodynamics (QED), in the QCD case, the interactions between plasma 
constituents is driven by their SU(3)$_{\text c}$ color charges. For this reason, the SU(3)$_{\text c}$ plasma is now called the quark--gluon plasma (QGP). For an exhaustive 
collection of papers tracing the development of theoretical ideas on the topic of QGP up to 1990, see Ref.~\cite{Rafelski:2003zz}. For a summary of later developments, 
see recent reviews \cite{Shuryak:2008eq, Braun-Munzinger:2015hba}.  

Let us note that, contrary to the first oversimplified expectations \cite{Rafelski:2003zz}, strongly interacting multi-particle systems feature numerous emergent phenomena 
that are difficult to predict from the underlying QCD theory, just like in condensed matter and atomic systems where the interactions are controlled by the QED theory. 
In addition to the hot QGP phase, several additional phases of QCD matter were predicted to exist \cite{BraunMunzinger:2008tz, Fukushima:2010bq}. In particular, 
the long-range attraction between quarks in the color anti-triplet ($\bf \bar 3$) channel was predicted to lead to the color superconductivity (CSC) with 
the condensation of $^{1}S_0$ Cooper pairs \cite{Bailin:1983bm}. The analysis of CSC two-flavor deconfined quark matter at moderate densities 
\cite{Buballa:2002wy} has revealed quite spectacular properties of this novel phase of matter, such as the spontaneous breakdown of rotation invariance 
manifested in the form of the quasi-fermion dispersion law. At high baryon density, an interesting symmetry breaking pattern $\text{SU}(3)_{\text c}\times 
\text{SU}(3)_{\text R} \times \text{SU}(3)_{\text L} \times \text{U}(1)_{\text B} \rightarrow \text{SU}(3)_{\text c+L+R} \times \text{Z}(2)$ leading to 
the formation of quark Cooper pairs was found in QCD with three massless quark flavours (i.e.,~$m_u$ = $m_d$ = $m_s$ = $0$) \cite{Alford:1998mk, Fukushima:2010bq}. 
This breaking of color and flavor symmetries down to the diagonal subgroup SU(3)$_{\text c+L+R} $ implies a simultaneous rotation of color and flavor degrees of freedom, 
called the color--flavor locking (CFL). Let us note that the CSC and CFL phases of deconfined QCD matter might play an important role when studying the EoS of neutron 
stars \cite{Becker:2008}. Another interesting phase is the matter--pion condensate studied by Migdal \cite{Migdal:1978az}.
\begin{figure}[H]
\centering
\includegraphics[width=.46\textwidth]{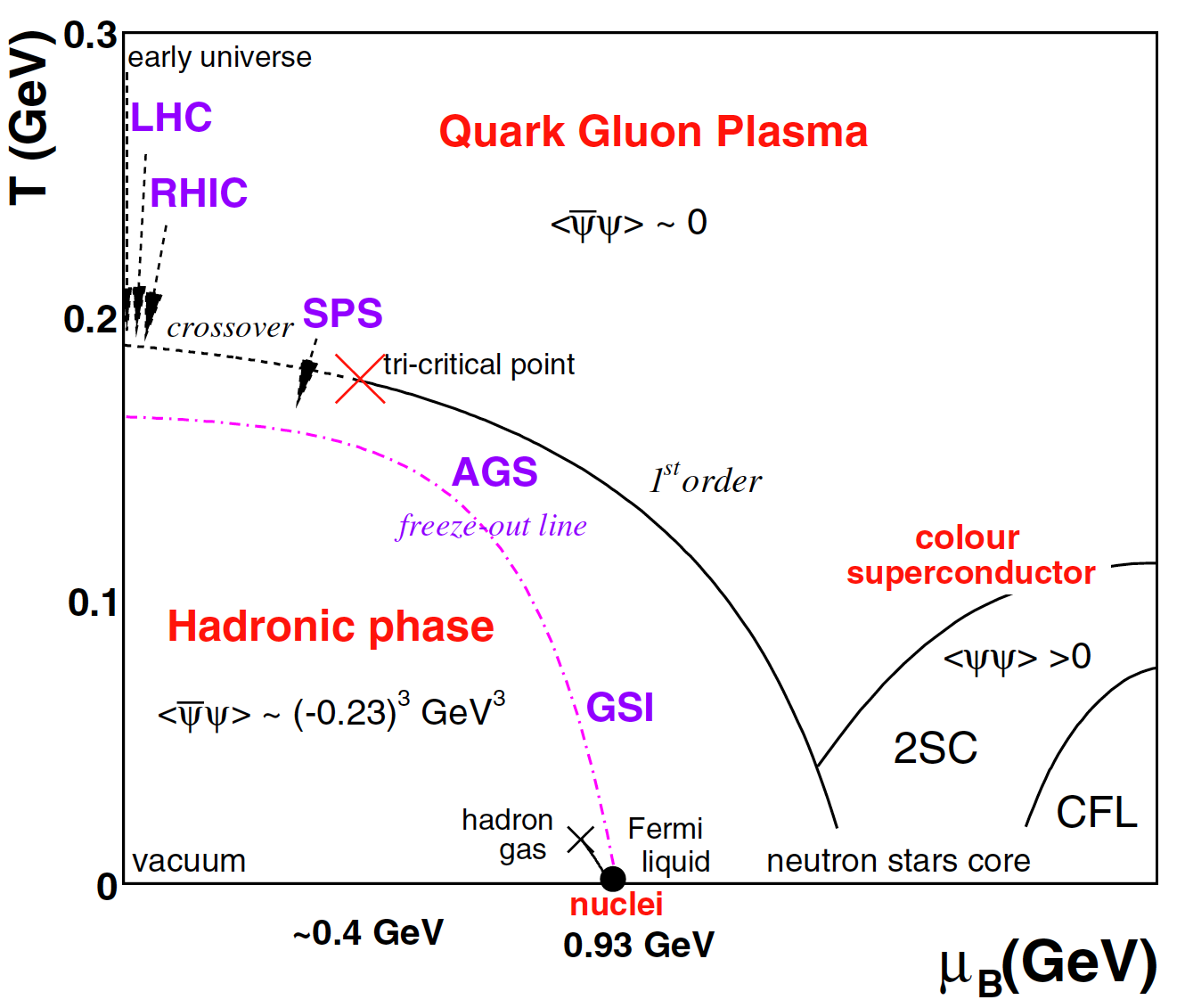}
\quad
\includegraphics[width=.48\textwidth]{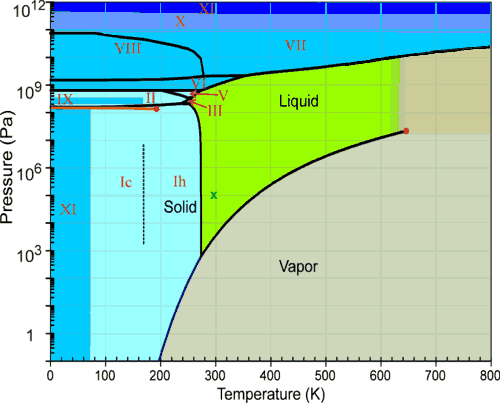}
\caption{\textbf{Left}: The schematic phase diagram of quantum chromodynamics (QCD) in terms of $T$ and $\mu_B$ state variables adapted from 
Ref.~\cite{D'Enterria:2007xr}; \textbf{Right}: The phase diagram of water illustrates a similar complexity, and 
is taken from Ref.~\cite{Chaplin}. AGS: Alternating Gradient Synchrotron; CFL: color--flavor locking; LHC: Large Hadron Collider; RHIC: Relativistic Heavy Ion Collider; SPS: Super Proton Synchrotron.}
\label{fig:phase_diagram}
\end{figure} 

Current knowledge on the QCD phase diagram is summarized on the left panel of Figure~\ref{fig:phase_diagram}. The~arrows indicate the expected crossing 
through the deconfinement transition during the expansion phase in heavy-ion collisions at different accelerators. The red and black full circles denote 
the critical endpoints of the chiral and nuclear liquid--gas phase transitions, respectively. The (dashed) freeze-out curve indicates where hadro-chemical 
equilibrium is attained at the final stage of the collision. The~nuclear matter ground-state at $T = 0$ and $\mu_{\rm B}$ = 0.93 GeV, and the approximate 
position of the QCD critical point at $\mu_{\rm B}$ $\sim$ 0.4 GeV are also indicated. The dashed line is the chiral pseudo-critical line associated with 
the crossover transition at low temperatures. Comparing this diagram to the phase diagram of water shown on the right panel, one notices that (at least 
theoretically) the complexity of the former approaches the latter.


\subsection{The Role of the QCD Ground State}
\label{sec:QCDvac}

The quantum ground state of QCD plays an immensely important role in both particle physics and cosmology. In particular, the quark--gluon condensate
is responsible for spontaneous chiral symmetry breaking, color confinement, and hadron mass generation (for a comprehensive review 
on the QCD vacuum, see Refs.~\cite{Shifman:1978bx,Shifman:1978bx-2,Schafer:1996wv,Diakonov:2002fq,Diakonov:2009jq} and references therein). It determines 
 the properties (and possibly the generation mechanism) of the quark--gluon plasma, and dynamics of the phase transitions and hadronization.
The latter phenomena are the most critical QCD phenomena taking place beyond the Perturbation Theory (PT), and thus are very difficult to explore 
by means of the well-known approaches. This strongly motivates further even deeper studies in this direction.

Let us start with the classical Yang--Mills (YM) gauge theory in the ${\rm SU}({\rm N}_{\rm c})$ (${\rm N}_{\rm c}=3$ for QCD) 
determined by the gauge-invariant Lagrangian
\begin{equation}
\mathcal{L}_{\rm cl}=-\frac{1}{4}F^a_{\mu\nu}F_a^{\mu\nu}\,, \label{L}
\end{equation}
where
\begin{equation*}
F^a_{\mu\nu}=\partial_\mu A^a_\nu - \partial_\nu A^a_\mu +
g_s\,e^{abc} A^b_\mu A^c_\nu
\end{equation*}
is the gluon field stress tensor with ${\rm SU}({\rm N}_{\rm c})$ adjoint $a,b,c=1,\dots\, N_c^2-1$ and Lorentz 
$\mu,\nu=0,1,2,3$ indices, and with the strong coupling constant $g_s$. The generating functional of such 
a classical theory is given by the Euclidean functional integral
\begin{eqnarray} \label{FuncI}
Z\propto \int [DA]\, e^{-S_{\rm cl}[A] + \int J^a_\mu A^a_\mu d^4x} \,, \qquad
S_{\rm cl}[A]=\int \mathcal{L}_{\rm cl} d^4x \,,
\end{eqnarray}
which is dominated by minima of the classical action $S_{\rm cl}[A]$ corresponding to the classical vacuum state 
with $F^a_{\mu\nu}=0$ unaltered by quantum corrections. The field excitations about the classical YM vacuum 
are referred to as instantons \cite{Belavin,Vainshtein:1981wh}.

In fact,  the classical YM equations of motion corresponding to Equation~(\ref{L}) are form-noninvariant with 
respect to small quantum fluctuations which break the conformal invariance of the gauge theory~\cite{Savvidy}---the effect known as {\it the conformal} (\emph{or trace}) \emph{anomaly}. Indeed, there is no threshold for the vacuum polarisation 
of a massless quantum gluon field by its classical component such that the solutions of the classical YM equations 
are unstable w.r.t the radiative corrections and cannot be used in physical applications. 
The conformal anomaly in QCD has notable implications; for example, in Cosmology, leading to an appearance 
of the Lorentz-invariant negative-valued contribution to the cosmological constant, 
\begin{eqnarray}
\epsilon^{\rm QCD}&=&\frac{\beta(g_s^2)}{8}\, \langle0|:F^a_{\mu\nu} F_a^{\mu\nu}:|0\rangle + 
\frac14 \sum_{q=u,d,s} \langle0 | :m_q\bar qq: | 0\rangle \simeq -(5\pm 1)\times 10^{-3}~ \text{GeV}^4\,,
 \label{Lambda-QCD}
\end{eqnarray}
where the one-loop expression for the QCD $\beta$-function $\beta=-b\alpha_s/(4\pi)$, $b=b_{\rm eff}=9$ accounting 
for three light flavours $u,\,d,\,s$ (for pure gluodynamics, $b=b_g=11$) is typically used. Besides the wrong 
~sign, the QCD 
vacuum density $\epsilon^{\rm QCD}$ is over forty orders of magnitude larger in absolute value than the positive cosmological 
constant observed in astrophysical measurements,
\begin{equation}
\epsilon_{\rm CC}>0\,, \qquad \Big| \frac{\epsilon_{\rm CC}}{\epsilon^{\rm QCD}}\Big| \simeq 10^{-44} \,,
\end{equation}

The nonperturbative QCD vacuum effect is expected to be dynamically cancelled at macroscopically large distances 
in the course of cosmological expansion (see Ref.~\cite{Pasechnik:2016sbh} and references therein). A dynamical mechanism of such 
a cancellation of vacua terms is yet unknown (for the existing scenarios discussed in the literature, see Refs.~\cite{Pasechnik:2016sbh,
Pasechnik:2013poa,Pasechnik:2013sga,Pasechnik:2016twe}.

Consider a consistent effective Lagrangian formulation of the YM theory incorporating the conformal 
anomaly. In the corresponding variational technique, the strong coupling $g_s$ is treated 
as an operator depending on operators of quantum gluon fields by means of the RG~equations in the operator form.
Namely, the gauge field operator $\mathcal{A}_\mu^a$ is considered as a variational variable which---together with 
the corresponding stress tensor operator---are related to those in the standard normalisation as follows
\begin{eqnarray}
\mathcal{A}_\mu^a\equiv g_s A_\mu^a \,, \qquad  \mathcal{F}^a_{\mu\nu}\equiv g_s F^a_{\mu\nu}=
\partial_\mu\mathcal{A}^a_\nu-\partial_\nu\mathcal{A}^a_\mu
+f^{abc}\mathcal{A}_\mu^b\mathcal{A}_\nu^c \,. \label{AF}
\end{eqnarray}

The effective action and Lagrangian operators of the quantum gauge theory 
is given in terms of the gauge-invariant operator of the least dimension $J$ by \cite{Pagels}
\begin{equation}
S_{\rm eff}[\mathcal{A}] = \int \mathcal{L}_{\rm eff} d^4x \,, \qquad \mathcal{L}_{\rm
eff}=-\frac{J}{4g_s^2(J)}\,, \qquad J=\mathcal{F}^2\equiv \mathcal{F}^a_{\mu\nu}\mathcal{F}_a^{\mu\nu}=2g_{s,*}^2(B^2 - E^2)\,, 
\label{Lrg}
\end{equation}
respectively, whose variation w.r.t $\mathcal{A}_\mu^a$ leads to the energy--momentum tensor of the gauge theory
\begin{equation}
\displaystyle
T_{\mu}^{\nu,{\rm g}}=\frac{1}{g_s^2}\Big[1-\frac12\beta(g_s^2)\Big]
\Big(-\mathcal{F}^a_{\mu\lambda}\mathcal{F}_a^{\nu\lambda} - \frac{1}{4}\delta_{\mu}^{\nu}\,J\Big) -
\frac{\delta_{\mu}^{\nu}\beta(g_s^2)}{8g_s^2}\,J \,, \qquad g_s^2 = g_s^2(J) \,.
\label{S-tem}
\end{equation}

In Equation~(\ref{Lrg}), as a normalisation point, one can choose, for example, the strong coupling in the minimum of the effective action 
$g_{s,*}^2=g_s^2(J=J^*)$. One distinguishes chromomagnetic $\langle B^2 \rangle > \langle E^2 \rangle$ 
and chromoelectric $\langle B^2 \rangle < \langle E^2 \rangle$ condensates, such that one or both of the corresponding ground-state 
solutions (minima of the effective action) should be stable in order to contribute to the physical QCD~vacuum. 

The strong coupling dependence on $J$ is determined by the RG evolution equation
\begin{equation}
\displaystyle 2J\frac{dg_s^2}{dJ}=g_s^2\,\beta(g_s^2) \,, \qquad g_s^2=g_s^2(J) \,,
\label{rg}
\end{equation}

The effective action (\ref{Lrg}) can be considered as an effective classical model \cite{Pagels} which possesses well-known properties 
of the full quantum theory such as (i) local gauge invariance; (ii) RG evolution and asymptotic freedom; (iii) correct quantum vacuum 
configurations; and (iv) trace anomaly.

In Ref.~\cite{Pasechnik:2016twe}, it was noticed that the effective YM equation of motion in an expanding universe 
with the conformal metric $g_{\mu\nu}=a^2\mathrm{diag} (1,\,-1,\,-1,\,-1)$ ($g\equiv {\rm det} (g_{\mu\nu})$)
\begin{eqnarray} 
\label{YMeq}
\left(\frac{\delta^{ab}}{\sqrt{-g}}\partial_\nu\sqrt{-g}-f^{abc}\mathcal{A}_\nu^c\right)
\left[\frac{\mathcal{F}_b^{\mu\nu}}{g_s^2\,\sqrt{-g}}\,
\Big(1-\frac12\beta\big(g_s^2\big)\Big)\right]=0 
\end{eqnarray}
has a partial nonperturbative solution $\beta(g_{s,*}^2)=2$, where $g_{s,*}^2=g_s^2(J^*)$ is the solution of the RG Equation (\ref{rg}) 
evaluated in the minimum of the effective action (\ref{Lrg}) $J^*=\langle J \rangle$. The corresponding value of the ground state density
\begin{eqnarray}
T^{0,{\rm g}}_{0}=-\frac{J^*}{4g_{s,*}^2} \equiv \mathcal{L}_{\rm eff}\Big|_{J=J^*}
\end{eqnarray}
indicates that the QCD vacuum $\epsilon^{\rm QCD}<0$ indeed has a chromomagnetic nature $\langle B^2 \rangle > \langle E^2 \rangle$ for 
$g_{s,*}^2>0$ in the deeply non-perturbative domain. This means that the corresponding solution is stable; namely, any small perturbation around 
the vacuum state effectively vanishes at the typical QCD time scale $\Delta t \sim 1/\Lambda_{\rm QCD}$.

In asymptotically free gauge theories like QCD, the quantum vacuum configurations are controlled by the strong coupling regime. 
Performing an analysis in Euclidean spacetime, in Ref.~\cite{Pagels} it was shown that the vacuum value of the gauge invariant 
$\langle J \rangle$ in a strongly-coupled quantum gauge theory does not vanish as it does in the classical gauge theory, and 
the corresponding functional integral is not dominated by the minima of the classical action (\ref{FuncI}). 
Moreover, it was shown that there are no instanton solutions to the effective action (\ref{Lrg}), such that the ground state of 
the quantum YM theory does not contain the classical instanton configurations. Instead, the quantum vacuum can be understood
as a state with ferromagnetic properties which undergoes spontaneous magnetisation, providing a consistent description of the 
nonperturbative QCD vacuum and confinement alternative to the conventional instanton model.

How to understand the smallness of the observed cosmological constant within the effective QCD action approach? One way
elaborated in Ref.~\cite{Pasechnik:2016twe} is to assume that such a compensation happens due to the presence of an additional QCD-like 
dynamics---Mirror QCD---with a confinement scale \mbox{$\Lambda_{\rm mQCD}\gg \Lambda_{\rm QCD}$}. The corresponding nonperturbative
Mirror QCD vacuum contribution may have an opposite sign to that in QCD, and can therefore compensate the QCD one at a certain
time scale in the course of cosmological expansion. Another interesting possibility explored in Ref.~\cite{Pasechnik:2013poa} is to assume that the QCD vacuum 
is degenerate itself, and at a given time scale consists of two opposite-sign (quantum-topological and quantum-wave) contributions. Then,
as soon as such a compensation occurs, the observable small cosmological constant can be generated by means of weak gravitational 
interactions in the QCD vacuum. Both possibilities, however, require a fine tuning of vacuum parameters in order to provide an exact
compensation of bare (zeroth-order in gravitational interactions) QCD contributions to the ground state density.

Can one avoid such a major fine tuning problem? The stable ground-state solution $J^*>0$ can actually be both chromomagnetic 
(when $g_{s,*}^2>0$, $\langle B^2 \rangle > \langle E^2 \rangle$, and $\epsilon^{-}<0$) and chromoelectric (corresponding 
to $g_{s,*}^2<0$, $\langle B^2 \rangle < \langle E^2 \rangle$, and $\epsilon^{+}>0$). Indeed, the standard argument in favor of the 
positive definiteness of $g^2_s$ is given in a classical YM theory, where
\begin{eqnarray}
\mathcal{F}^2 \propto -\frac{\partial}{\partial t}A_i^a\frac{\partial}{\partial t}A_i^a
\end{eqnarray}
in Minkowski space, such that $g^2_s<0$ would lead to infinitely fast growth of the field $A_i^a$ 
and action $S_{\rm cl}=\int \mathcal{L}_{\rm cl}d^4x$ would not have a minimum. In the quantum case, however, $g^2_s$
is a function of $J$, and can take negative values as long as the effective action $S_{\rm eff}$ has a minimum for $g^2_s<0$.
Besides, in close vicinity of the ground-state solution $\beta(g_{s,*}^2) = 2$, the corresponding solution of the RG Equation~(\ref{rg}) takes a linear behaviour
\begin{eqnarray}
\frac{d\ln g_s^2}{d\ln J}~\approx~1 \,, \qquad g_s^2~=~\pm |g_{s,*}^2| \frac{J}{J^*} \,, \qquad J,\,~J^* >0 \,,
\end{eqnarray}

Adopting that the stable attractor solution $J\to J^* >0$ is realised at macroscopically large time scales, 
the net QCD ground-state density would then asymptotically vanish \cite{Pasechnik:2016sbh}:
\begin{eqnarray}
\epsilon^{\pm} \to \pm \frac{J^*}{4|g_{s,*}^2|} \,, \qquad \epsilon^{-}(T) + \epsilon^{+}(T) \to 0 \,, \qquad T\ll T_{\rm QCD}=
\Lambda_{\rm QCD}\sim 100\;{\rm MeV} \,,
\end{eqnarray}
if both contributions coexist in the QCD vacuum, thus canceling each other beyond the confinement radius or after the QCD phase 
transition epoch in the cosmological history of the universe. The latter is an important example of conformal anomalies' cancelation
in the classical limit of a YM theory without any fine tuning. In the deconfined (QGP) phase (i.e., at temperatures $T\gtrsim T_{\rm QCD}$), 
the chromoelectric contribution $\epsilon^{+}(T)$ should quickly vanish such that $\epsilon^{\rm QCD}\simeq \epsilon^{-}(T=T_{\rm QCD})$, 
providing a consistency with hadron physics phenomenology. This effect becomes plausible as long as $\epsilon^{+}(T)$ is attributed to the
ground state of hadronic degrees of freedom, which indeed becomes relevant only as soon as QGP is hadronised \cite{Pasechnik:2013poa}.

\begin{figure}[H]
\centering 
 \includegraphics[width=.49\textwidth, height=0.4\textwidth]{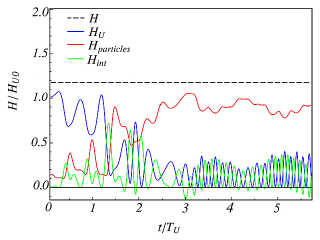}
\caption{Real-time evolution of the energy density components of the interacting Yang--Mills (YM) system---homogeneous gluon 
condensate $H_U$ and inhomogeneous waves $H_{particles}$ versus the interactions' contribution $H_{int}$ and the total 
Hamiltonian $H=H_U+H_{particles}+H_{int}$. Reproduced from Ref.~\cite{Prokhorov:2013xba}.}
\label{fig:swap}
\end{figure}

What is the possible role of the QCD vacuum in the generation of QGP? As was demonstrated in Ref.~\cite{Prokhorov:2013xba} within a semi-classical 
analysis, an interacting YM system of homogeneous gluon condensate and inhomogeneous wave modes evolves in real time in such a way that 
the amplitude of waves parametrically grows at the expense of decaying gluon condensate. The corresponding effect of the energy ``swap'' from 
the condensate to the gluon plasma waves is illustrated in Figure~\ref{fig:swap}. Together with the growth of the plasma wave amplitudes,
a vacuum average of their bi-linear products do not vanish and grow as well, inducing the positive-valued component $\epsilon^{+}(T)$ of 
the ground-state density. This~effect---if it holds in the full quantum formulation---can then be the basis for the QGP generation and reheating 
mechanism, both in heavy ion collisions and in Cosmology. A similar mechanism may be responsible for reheating of the cosmological plasma 
and particle production in the end of Cosmic Inflation due to decay of the dominant spatially-homogeneous chromoelectric condensate  (inflaton) 
driving the inflationary epoch.

Can such parameteric growth of plasma modes due to a decay of the gluon condensate be studied in particle (e.g., heavy-ion) collisions?
Can this effect be detectable, and if so, be used as a tool for the investigation of the dynamical evolution of the quantum ground state in QCD?
Answers to these and other related questions are big unknowns, and very little has been done so far in this direction. An interesting insight 
into the problem of QCD ground state can be offered by the low-$p_\perp$ ($<$200 MeV) spectra of pions measured at the LHC, 
which show up to $\sim$30\%--50\% enhancement compared to the hydrodynamic models (see Refs.~\cite{Abelev:2013vea, Abelev:2013pqa, 
Adam:2015pbc}). A possible interpretation could be found in the framework of a hypothesis about hadronization and freeze-out processes in 
chemical non-equilibrium~\cite{Petran:2013lja}. Among the possible reasons for the non-equilibrium dynamics are the QGP supercooling 
\cite{Csorgo:1994dd, Shuryak:2014zxa} and gluon condensation \cite{Blaizot:2011xf} phenomena. A particularly interesting possibility 
has been proposed in Refs.~\cite{Begun:2013nga, Begun:2014rsa, Begun:2015ifa}, where it was shown that the Bose--Einstein pion condensate at 
the level of 5\% can account for the missing low-$p_\perp$ charged pion yields coming from a coherent source in \pbpb collisions (\sqs=~2.76~TeV) 
at various centralities. Moreover, if there is such a condensate, there must be large fluctuations of pions, which should be seen starting 
from the fourth moment of the multiplicity distribution \cite{Begun:2016cva}. Further studies of the non-equilibrium QCD dynamics accounting for 
the ground state are certainly required from both theoretical and experimental standpoints.


\subsection{Strongly Interacting Quark--Gluon Plasma}

The surprising fact that the deconfined matter found at RHIC \cite{Arsene:2004fa, Back:2004je, Adams:2005dq, Adcox:2004mh} does not behave as 
a gas of almost-free quarks and gluons, but as a strongly interacting liquid \cite{Shuryak:2008eq, Gyulassy:2004zy} was anticipated by only a few~\cite{Plumer:1985tq, Rischke:1992uv, Bannur:1995np}. The fact that QGP close to the critical temperature $T_c$ is a strongly interacting system 
was used in Ref.~\cite{Thoma:2004sp, Bannur:2005yx} to exploit its analogy with strongly-coupled classical non-relativistic plasmas~\cite{Ichimaru:1982zz} in order to understand experimental observations and to interpret the lattice QCD~results. 

By definition, plasma is a state of matter in which charged particles interact via long-range (massless) gauge fields \cite{Shuryak:2008eq}. 
This distinguishes it from neutral gases, liquids, or solids in which the inter-particle interaction is of short range. So, plasmas themselves can be gases, 
liquids, or solids, depending on the value of the plasma parameter $\Gamma$, which is the ratio of interaction energy to kinetic energy of the particles 
forming the plasma \cite{Ichimaru:1982zz}.

A non-relativistic electromagnetic plasma is called strongly-coupled if the interaction energy (Coulomb energy) between the particles is larger than 
the thermal energy of the plasma particles; i.e., if the Coulomb coupling parameter $\Gamma_{\rm EM}$ = $q^2$/($aT^2$) > 1, where $q$ is 
the particle charge, $a$ is the interparticle distance, and $T$ is the plasma temperature (in the system of units where $\hbar$ = $c$ = $k_B$ = 1).  
Let~us note that the strongly-coupled classical electromagnetic plasmas are not exotic objects at all \cite{Ichimaru:1982zz}. For example, table salt 
(NaCl) can be considered as a crystalline plasma made of permanently charged ions Na$^+$ and Cl$^-$ \cite{Shuryak:2008eq}. At $T\approx 10^{3}$ K 
(still too small to ionize non-valence electrons), it transforms into a molten salt, which is a liquid plasma with $\Gamma \approx 60$. An estimate 
of the plasma parameter for QGP was considered in Ref.~\cite{Thoma:2004sp} where it was found that $\Gamma_{\rm QGP}$ = 2C$\alpha_{\rm s}$/($aT^2$), 
where---depending on the type of plasma---$C$ = 4/3 or $C$ = 3 is the Casimir invariant for quarks or gluons and $a$ is the interparton distance $a$ = 0.5 fm. 
For QGP at temperature only slightly above the critical de-confining temperature (i.e., $T = 200$ MeV), the corresponding coupling constant $\alpha_{\rm s}$ = 
0.3--0.5, and $\Gamma_{\rm QGP}$ = 1.5--6, the plasma can be considered as a strongly interacting one.

The strongly interacting plasmas $\Gamma \geq 1$ are also a special case of strongly correlated systems where {\it correlated behavior} means 
a deviation from the trivial ideal gas behavior \cite{Bonitz:2010}.
Prominent properties of all strongly correlated systems (see Figure~\ref{fig:plasmas}) 
can be quantified by a few dimensionless parameters: the coupling parameter $\Gamma$, the degeneracy parameter $\chi=n\lambda_{th}^3$,  and 
the Brueckner parameter $r_s=a/a_B$, where $n$ is the number density of the particles, $\lambda_{th}=\sqrt{2\pi/(m T)}$ is the thermal de Broglie 
wavelength, $a$ is the average interparticle distance, $m$ is the particle mass, and $a_B=1/(m e^2)$ is the Bohr radius. 

Strongly interacting plasmas that can be studied in laboratory are ultracold atomic Fermi gases~\cite{Giorgini:2008zz}; in particular, 
strongly-coupled {$^6$Li atoms \cite{Johnson:2010zz, Cao:2010wa}. A distinctive property of these plasmas is that---similarly to the strongly-
coupled QGP (sQGP)---their shear viscosity-to-entropy density ratio $\eta/s$ (see Sec.~\ref{hydrodynamics}
for definition) characterizing how close 
the fluid is to a perfect liquid \cite{Kovtun:2004de} is effectively negligible \cite{Shuryak:2008eq, Schafer:2009dj, Cao:2010wa}. Cold atomic gases 
are produced in optical or magneto-optical traps containing typically $10^5--10^6$ atoms \cite{Thomas:2010zz}. The hydrodynamic 
behaviour is observed when the trapping potential is modified, or if the local density or energy density is modified using laser beams~\cite{Cao:2010wa}. In this way, the scattering length $a$ (and hence the interaction strength between the atoms) can be made almost 
infinite~\cite{Shuryak:2008eq}. This is also the case of data point $^6$Li ($a=\infty$) shown in Table~\ref{tab_eta}, where the thermodynamical 
parameters for several other substances of interest are summarized. For H$_2$O and $^4$He, two points are displayed. First are the data 
at atmospheric pressure and temperatures just below the boiling point and the $\lambda$ transition, 
respectively. These data points roughly correspond to the minimum of $\eta/n$ at atmospheric pressure. Second are the data near the critical point, 
which roughly corresponds to the global minimum of $\eta/s$. 

\begin{figure}[H]
\centering
\includegraphics[width=.8\textwidth]{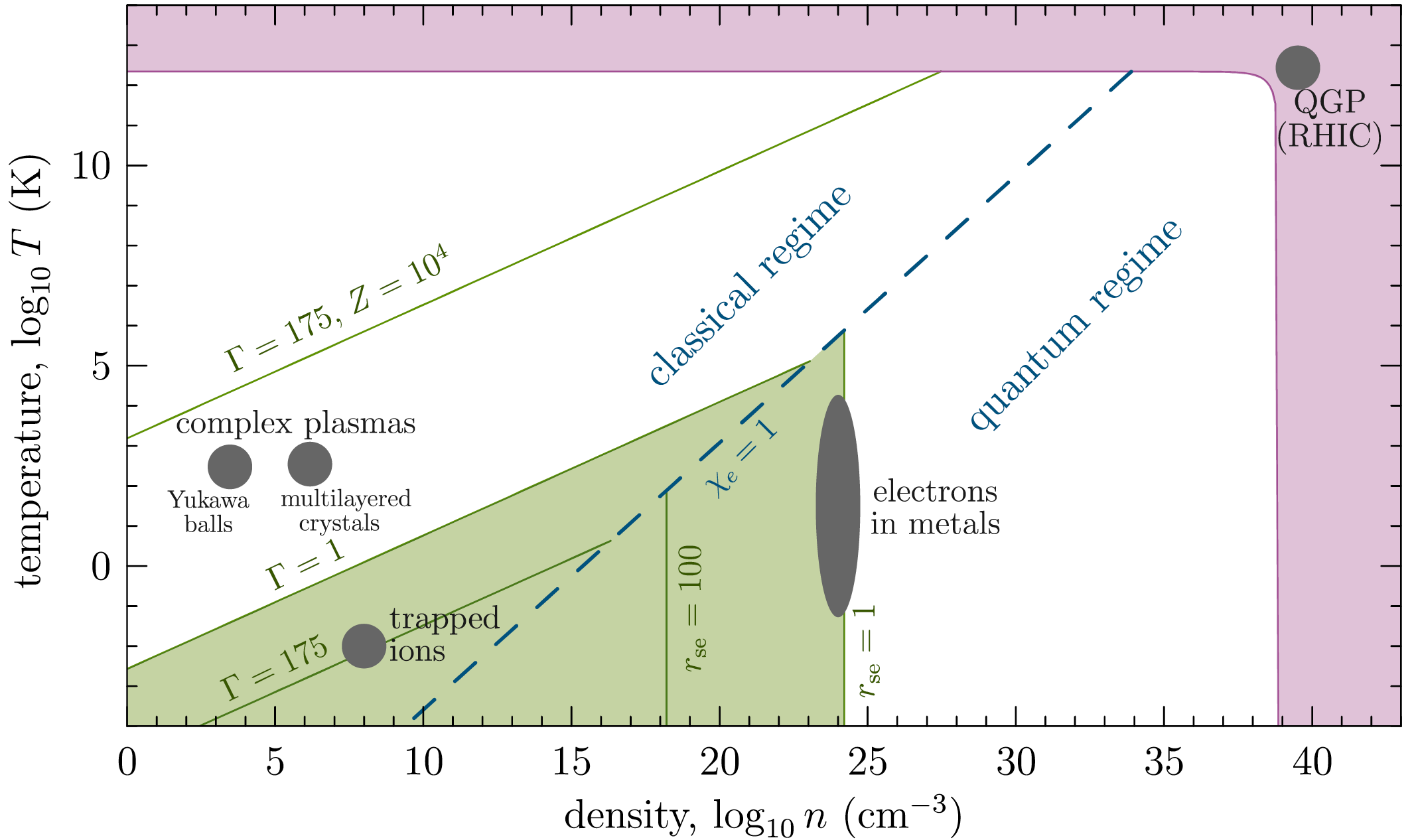}
\caption{Examples of strongly correlated systems in thermodynamic equilibrium include complex plasmas, trapped ions, and the quark--gluon plasma (QGP)
extending along the outer (\textbf{pink}) area, while dot shows
~the conditions at RHIC. The figure has been reproduced from Ref.~\cite{Bonitz:2010}.}
\label{fig:plasmas}
\end{figure}   

\begin{table}[hbt]
\begin{center}
\begin{tabular}{cccccc} 
\toprule
\textbf{Fluid} & \boldmath{$p$} \textbf{[Pa]} & \boldmath{$T$} \textbf{[K]} & \boldmath{$\eta$} \boldmath{[\textbf{Pa}$\cdot$\textbf{s}]} & 
   \boldmath{$\eta/n$}   \boldmath{$[\hbar]$} & 
   \boldmath{$\eta/s$}  \boldmath{$[\hbar/k_B]$} \\ \midrule
H$_2$O   &  0.1$\times 10^6$   & 370    & $2.9\times 10^{-4}$ 
         &   85               & 8.2     \\
$^4$He   &  0.1$\times 10^6$   & 2.0    & $1.2\times 10^{-6}$ 
         &  0.5               & 1.9   \\
H$_2$O   & 22.6$\times 10^6$   & 650    & $6.0\times 10^{-5}$ 
         & 32                 & 2.0  \\
$^4$He   & 0.22$\times 10^6$   & 5.1    & $1.7\times 10^{-6}$ 
         & 1.7                & 0.7  \\
$^6$Li ($a=\infty$)   
         &  12$\times 10^{-9}$ & 23$\times 10^{-6}$ & $\leq$1.7 $\times$ $10^{-15}$  
         &  $\leq$1          & $\leq$0.5  \\
QGP      &  88$\times 10^{33}$ &  2$\times 10^{12}$ & $\leq$5 $\times$ $10^{11}$  
         &                    & $\leq$0.4 \\ 
\bottomrule
\end{tabular}
\end{center}
\vskip -.4cm \caption{\label{tab_eta}
The viscosity $\eta$, the viscosity over density $\eta/n$ ratio, and the viscosity over
entropy density $\eta/s$ ratio for several fluids at particular values of pressure $p$ 
and temperature $T$ (from Ref.~\cite{Schafer:2009dj}). }
\end{table}


\subsection{QCD at High Temperatures and Vanishing Chemical Potentials}
\label{subsec:muzero}

The grand canonical partition function in ${\rm SU}({\rm N}_{\rm c})$ gauge theory (such as QCD) with
${\rm N}_{\rm f}$ fermion flavours having a common chemical potential $\mu$ reads
\begin{eqnarray}
Z(T,\mu)=\int DU~ e^{-S_g(T)}~ \prod_{f=1}^{N_f} {\rm Det}~ M(m_f, \mu ,T) \,,
\end{eqnarray}
where $M$ is the Dirac operator, $S_g$ is the gauge part of the QCD action which depends on temperature $T$ 
through boundary conditions. In the Hamiltonian formulation,
\begin{eqnarray}
Z(T,\mu)={\rm Tr}~{\rm exp} \Big[ -\frac{\hat{H}}{T} - \frac{\mu \hat{N}}{T} \Big] \,,
\end{eqnarray}
where $\hat{N}$ is the number operator, and $\hat{H}$ is the Hamiltonian. It is needless to mention that the analytic 
properties of the free energy $F(T,\mu)$
\begin{eqnarray}
F(T,\mu)=-T \, {\rm ln} Z(T,\mu)
\end{eqnarray}
as a function of general complex $\mu$ are known to be useful for studying the phase structure of 
QCD on the lattice \cite{Gupta:2003ji}.

Let us consider QCD thermodynamics at high temperatures and zero chemical potentials---the region relevant for the LHC and also partly 
 for RHIC---by recalling how the basic bulk thermodynamic observables can be obtained from the grand canonical partition function with vanishing 
quark chemical potentials, $Z(T,V)\equiv  Z(T,\mu)|_{\mu\to 0}$ \cite{Cheng:2008}. The grand canonical potential, $\Omega(T,V)$, normalized 
in such a way that it vanishes at zero temperature,
\begin{eqnarray}
\displaystyle
\Omega(T,V)= T~ {\rm ln} Z(T,V) -\Omega_0\,,  \qquad  \Omega_0 =\lim_{T\rightarrow 0} T~ {\rm ln} Z(T,V) ~,
\label{gcpotential}
\end{eqnarray}
can be used to obtain the thermal part of the pressure ($p$) and energy density ($\epsilon$) 
\begin{eqnarray}
\displaystyle
p  =  \frac{1}{V} \Omega(T,V)\,, \qquad
\epsilon = \frac{T^2}{V} \frac{\partial \Omega(T,V) / T}{\partial T}~~,~~ 
\label{freeenergy}
\end{eqnarray}
both vanishing at small temperature, by construction. Using these relations, one can express the difference between 
$\epsilon$ and $3p$; i.e.,~the thermal contribution to the trace of the energy--momentum tensor $\Theta^{\mu\mu} (T)$ 
(also called the trace anomaly or the interaction measure) in terms of a derivative of the pressure with respect to temperature:
\begin{eqnarray}
\displaystyle
\frac{\Theta^{\mu\mu} (T)}{T^4} \equiv \frac{\epsilon - 3p }{T^4}  =  
T \frac{\partial}{\partial T} (p/T^4) ~~.
\label{delta}
\end{eqnarray}

In fact, it is $\Theta^{\mu\mu} (T)$ which is the basic thermodynamic quantity conveniently calculated on the lattice as 
the total derivative of $\ln Z$ with respect to the lattice spacing $a$ \cite{Bazavov:2014pvz}:
\begin{equation}
\Theta^{\mu\mu}=\epsilon -3p = -\frac{T}{V}\frac{d\ln Z}{d \ln a}\; .
\label{e3p_dlnZ}
\end{equation}

Before moving to the results of lattice calculations, it is useful, for comparison, to recall a description of the strongly interacting matter below deconfinement 
temperature T$_{\text c}$. Here, all thermodynamic quantities are expected to be well-described by the hadron resonance gas (HRG) model consisting of 
non-interacting hadrons as proposed by Hagedorn in the mid 1960s \cite{Hagedorn:1965st} (see also Ref.~\cite{Rafelski:2016hnq}). The trace anomaly in 
the HRG model is given by 
\begin{eqnarray}
\displaystyle
\left( \frac{\epsilon - 3p}{T^4}\right)^{HRG} =\sum_{m_i\le m_{max}}
\frac{d_i}{2\pi^2}
\sum_{k=1}^\infty
\frac{(-\eta_i)^{k+1}}{k}
\left( \frac{m_i}{T}\right)^3 K_1\left(\frac{km_i}{T}\right) \; ,
\label{e3plow}
\end{eqnarray}
where $K_1(\tfrac{km_i}{T})$ is a modified Bessel function, the different particle species of mass $m_i$ have degeneracy factors $d_i$ and 
$\eta_i = -1 (+1)$ for bosons (fermions), and the sum runs over all known hadrons up to the resonance mass of $m_{max}=2.5$~GeV.

The results on the temperature dependence of the trace anomaly and suitably normalized pressure, energy density, and entropy density from lattice calculations, 
together with the HRG predictions, are shown on the left and right panels of Figure~\ref{fig:cont}, respectively. The vertical band in the right panel marks the crossover 
region, $T_c=(154\pm 9)$~MeV. The horizontal line at $95 \pi^2/60$$\approx$15.6 corresponds to the ideal Stefan--Boltzmann (SB) gas limit for the energy 
density of relativistic massless gas consisting of $N_f=3$ quark flavours and gluons with $N_c=3$ colors having altogether $g$ degrees of freedom:
\begin{equation}
\frac{3p_{SB}}{T^4}=\frac{\epsilon_{SB}}{T^4}=g\frac{\pi^2}{30}~~,~~ \qquad g=2(N^2_c -1) + \frac{7}{2}N_cN_f=\frac{95}{2}~~.
\label{dof}
\end{equation}

The fact that even at $T \propto$ 400 MeV, the pressure, energy density, and entropy density of the QGP are far from their ideal gas values indicates 
substantial remaining interactions among the quarks and gluons in the deconfined phase. It is interesting to compare (at least qualitatively) this behaviour with 
the gaseous two-component plasma of particles with charge $q=\pm ze$. The pressure normalized to that of the ideal gas can be deduced from the standard 
textbook formula (e.g., Ref.~\cite{Landau:1980mil})
\begin{equation}
\frac{p}{p_{id}}= 1 - \frac{\sqrt{\pi}}{3}\sqrt{\frac{ n q^6}{T^3}}\; ,
\label{wplasma}
\end{equation}
which is valid for $n \ll T^3/q^6$. The non-ideal behaviour of gaseous two-component plasma thus increases very rapidly 
~with charge $q$ of the plasma 
particles, but much more slowly 
~with their density $n$,  and decreases rather quickly with the plasma temperature $T$.

\begin{figure}[H]
\includegraphics[width=7.5cm]{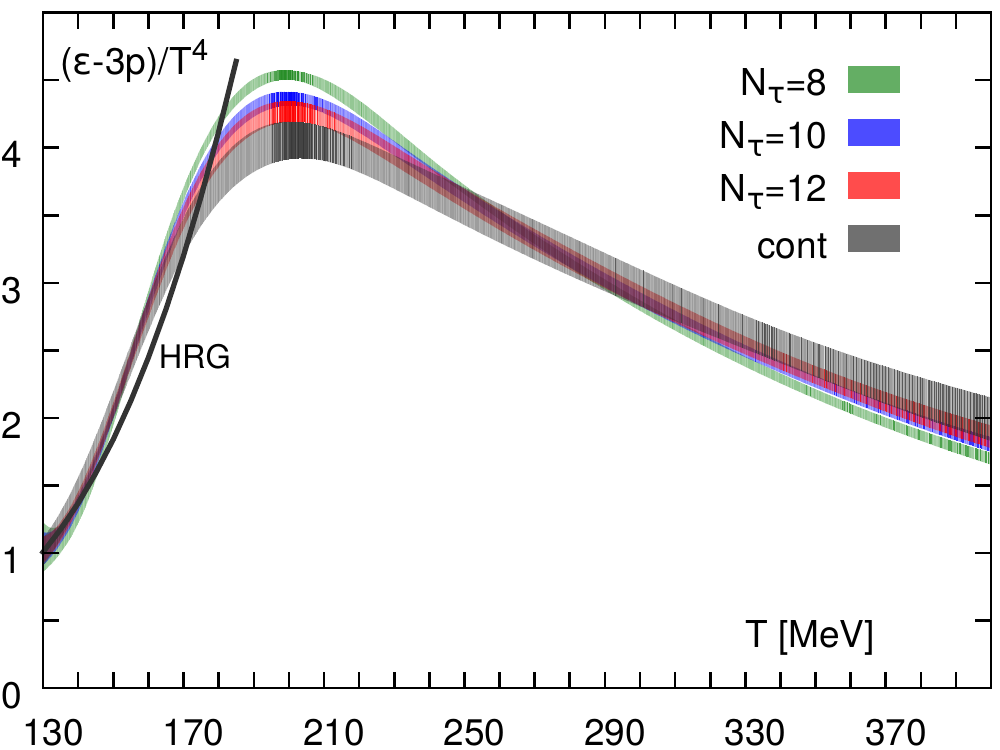}~~~
\includegraphics[width=7.5cm]{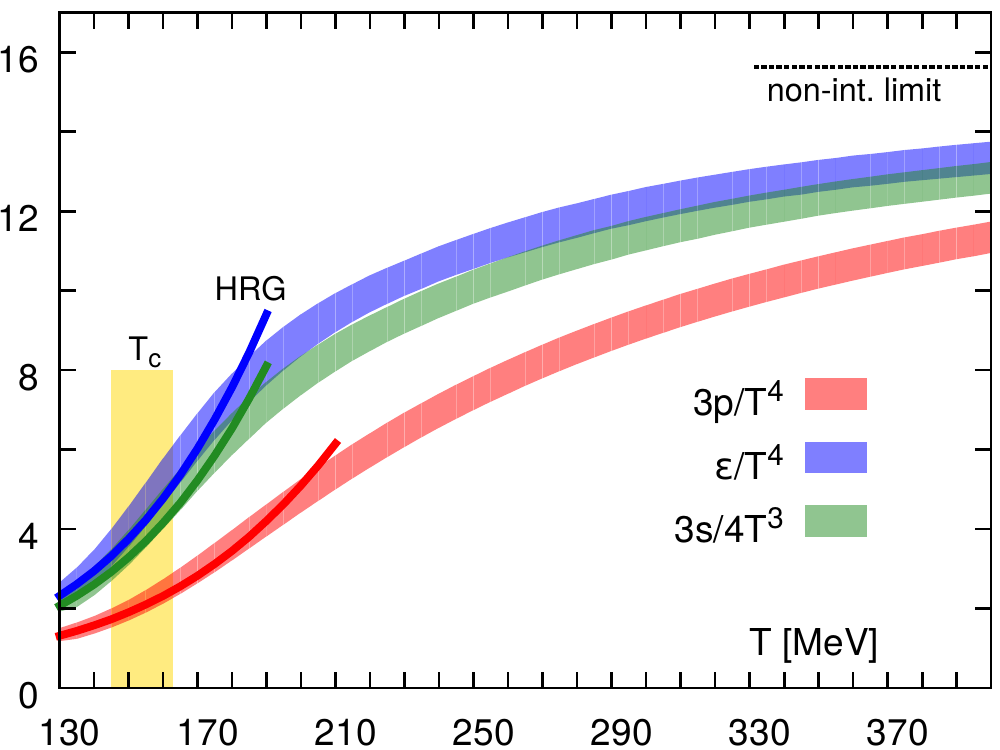}
\caption{\textbf{Left}: The continuum-extrapolated trace anomaly for several values of the lattice spacing $aT=1/N_\tau$ and its continuum extrapolation; 
\textbf{Right}: The continuum-extrapolated values of suitably normalized pressure, energy density, and entropy density as functions of the temperature. 
Both figures have been reproduced from Ref.~\cite{Bazavov:2014pvz}. The darker lines in both figures show the corresponding predictions 
of the hadron resonance gas (HRG) model.}
\label{fig:cont}
\end{figure}

\subsubsection{Softest Point in the EoS of Deconfined QCD Matter}
\label{softestpoint}

An important property of the QGP phase transition is the presence of a local minimum in the ratio of 
pressure to energy density $p/\epsilon$ as a function $\epsilon$ \cite{Shuryak:1986nk, Hung:1994eq}.  
The possible existence of this {\it softest point} in the QCD EoS is distinguishable by a very small sound velocity 
of the deconfined medium, and has thus been suggested as a signal of the first-order phase transition.
This also becomes evident in second-order derivatives of the QCD partition function with respect to temperature. The speed of sound, $c_s$, 
is related to the inverse of the specific heat, $C_V= {\rm d}\epsilon /{\rm d}T$,
\begin{equation}
c_s^2 = \frac{\partial p}{\partial \epsilon}  = 
\frac{{\rm d} p/{\rm d} T}{{\rm d} \epsilon/{\rm d} T} 
= \frac{s}{C_V} \; , \qquad
\frac{C_V}{T^3}=\left. \frac{\partial \epsilon}{\partial T}\right|_V \equiv \left( 
4 \frac{\epsilon}{T^4} + T 
\left. \frac{{\partial} (\epsilon/T^4)}{\partial T}\right|_V \right)
\, .
\label{CV}
\end{equation}

The quantity $T {{\rm d} (\epsilon/T^4)}/{{\rm d} T}$ can be calculated directly from the trace anomaly and its derivative with respect to temperature,
\begin{equation}
T \frac{{\rm d} \epsilon/T^4}{{\rm d} T} = 3 \frac{\Theta^{\mu\mu}}{T^4}
+T \frac{{\rm d} \Theta^{\mu\mu}/T^4}{{\rm d} T} \; .
\label{dedT}
\end{equation}

In Figure~\ref{fig:cs2} (left panel), we show the speed of sound as a function of temperature. The softest point of the EoS predicted in Ref.~\cite{Hung:1994eq} 
at $T\simeq (145-150)$~MeV  (i.e., at the minimum of the speed of sound) lies on the low temperature side of the crossover region.  At this point, 
the speed of sound is only slightly below the corresponding HRG value. Furthermore, the value $c_s^2\simeq 0.15$ is roughly half-way between zero, 
the value expected at a second-order phase transition with a divergent specific heat, and the value for an ideal massless gas, $c_s^2=1/3$ 
\cite{Bazavov:2014pvz}. At the high temperature end, $T$~$\sim$~350~MeV, it~reaches within $10\%$ of the ideal gas value.
\begin{figure}[H]
\includegraphics[trim={ 0 0.1cm 0 0cm},clip,width=.465\textwidth]{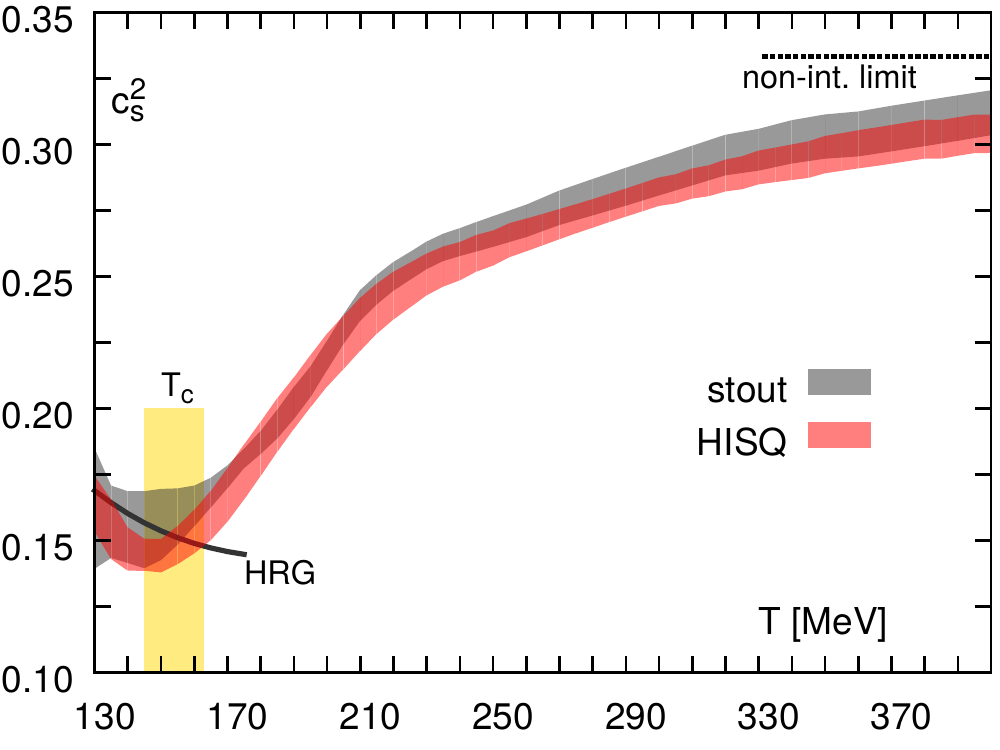} \qquad
\includegraphics[width=.475\textwidth, height=.345\textwidth]{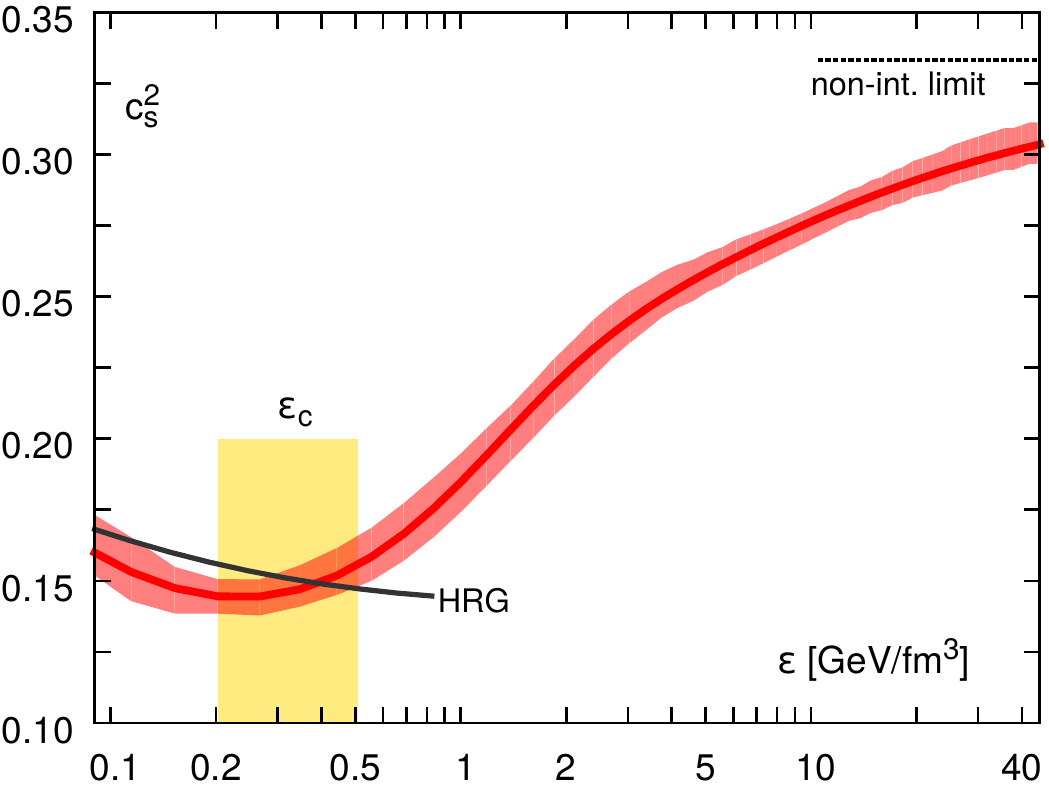}
\caption{The speed of sound squared $c^2_s $ from lattice QCD and the HRG model versus (\textbf{Left}) temperature $T$ and  (\textbf{Right}) energy density $\epsilon$.
The figure has been reproduced from Ref.~\cite{Bazavov:2014pvz}. The vertical yellow bands mark the location of the crossover region $T_c=(154\pm 9)$~MeV 
and the corresponding range in energy density, $\epsilon_c=(0.18-0.5)~{\rm GeV/fm}^3$, respectively.}
\label{fig:cs2}
\end{figure}

The softest point of the EoS is of interest for the phenomenology of heavy ion collisions, as it characterizes the temperature and energy density range in which the expansion 
and cooling of matter slows down. The system spends a longer time in this temperature range, and one expects to observe characteristic signatures from this regime. 
The quantity $c_s^2$ as a function of the energy density is shown in Figure~\ref{fig:cs2} (left). At the softest point, the energy density is only slightly above that of 
normal nuclear matter, $\rho_0=160~{\rm MeV}/{\rm fm}^3$. In the crossover region, $T_c=(154 \pm 9)$~MeV, the energy density varies from 
$180~{\rm MeV}/{\rm fm}^3$ at the lower edge to $500~{\rm MeV}/{\rm fm}^3$ at the upper edge---slightly above the energy density inside the proton 
$\epsilon_{\rm proton}=~450~{\rm MeV}/{\rm fm}^3$. 

Thus, t}he QCD crossover region~starts at or close to the softest point of the EoS, and the entire crossover region corresponds to relatively small values of 
the energy density, (1.2--3.1)$\epsilon_{\rm nuclear}$. This~value is about a factor of four smaller than that of an ideal quark--gluon gas in this temperature~range.  

\subsubsection{Testing the Properties of the Medium with Infinitely Heavy Static Test Charges}
\label{melting}

An important property of the QGP medium is the color screening: the range of interaction between heavy quarks becomes inversely proportional to the temperature. 
This effect also forms the basis of the most common description of dynamics of quarkonia (mesons consisting of heavy $Q\bar{Q}$), produced in heavy ion 
collisions---the potential between the heavy quarks $c \bar{c}$ or $b \bar{b}$ becomes screened by deconfined quarks and gluons, and the heavy quarks 
separate from each other, leading to a suppression of quarkonia yields \cite{Matsui:1986dk}.

On the lattice, this phenomenon is studied using (infinitely) heavy static test charges \cite{Brambilla:2010cs, Borsanyi:2015yka}. The color screening effect
is estimated from the spatial correlation function $G(r,T)$ of a static quark and anti-quark, which propagate in Euclidean time from $\tau = r = 0$ 
to $\tau = r = 1/T$, where $T$ is the temperature. The free energy of static quark pair $Q\bar{Q}$ is then calculated as the logarithm of the correlator 
$F(r, T) = -T \ln G(r, T)$ \cite{Bazavov:2009us, Brambilla:2010cs}. In the zero temperature limit, the singlet free energy coincides with the zero 
temperature potential calculated on the lattice \cite{Bazavov:2011nk}. However, as argued in Ref.~\cite{Borsanyi:2015yka}, using the free energies 
instead of potentials is preferable, since the latter are not gauge invariant. On the other hand, the gauge-invariant static quark--antiquark pair free energy 
is a non-perturbatively well-defined quantity that carries information about the deconfinement properties of the QGP. 

\begin{figure}[H]
\centering
\begin{minipage}[b]{0.43\linewidth}
\includegraphics[width=1.02\linewidth]{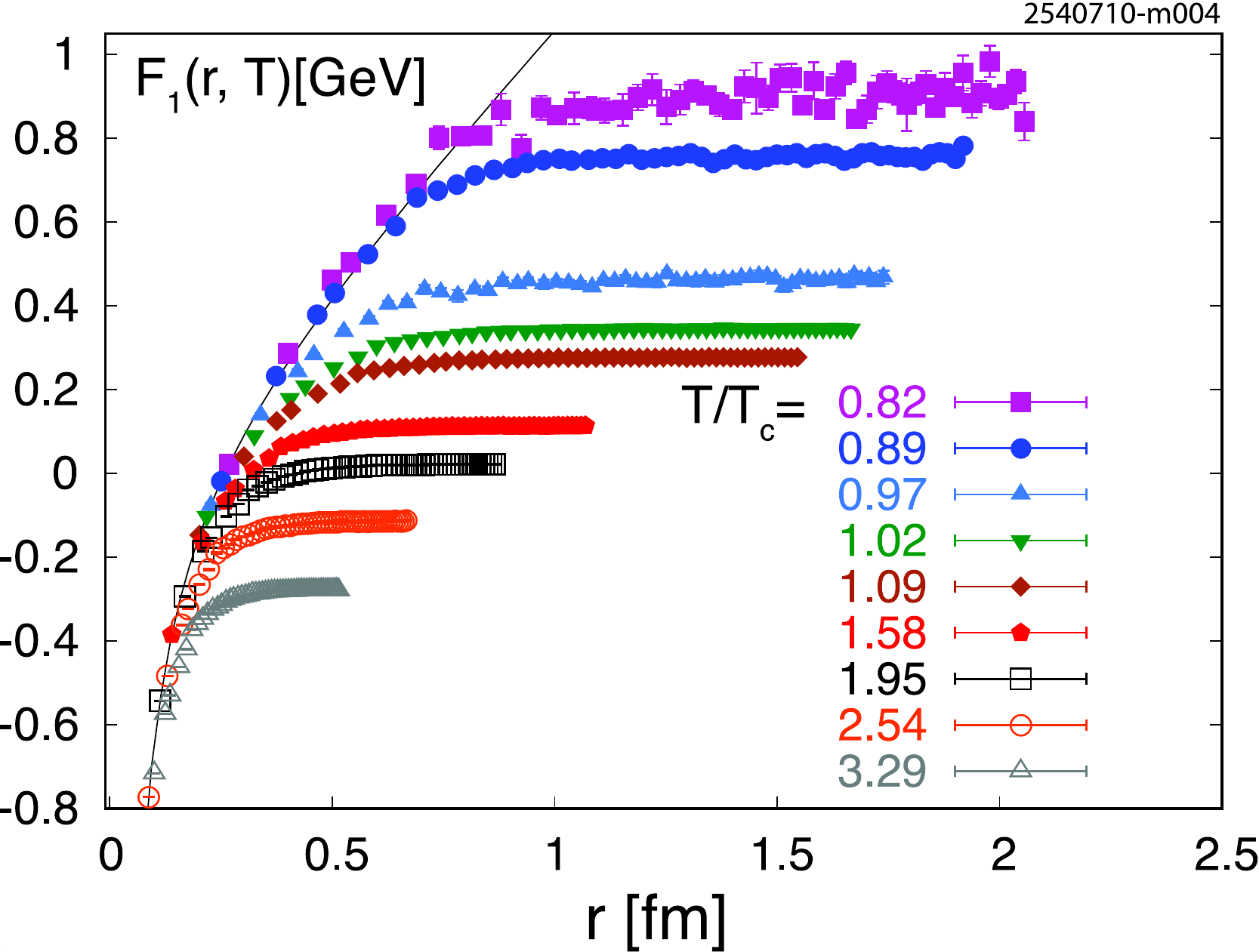}
\includegraphics[trim={ 1.8cm 0 0 2.cm},clip,width=1.15\linewidth]{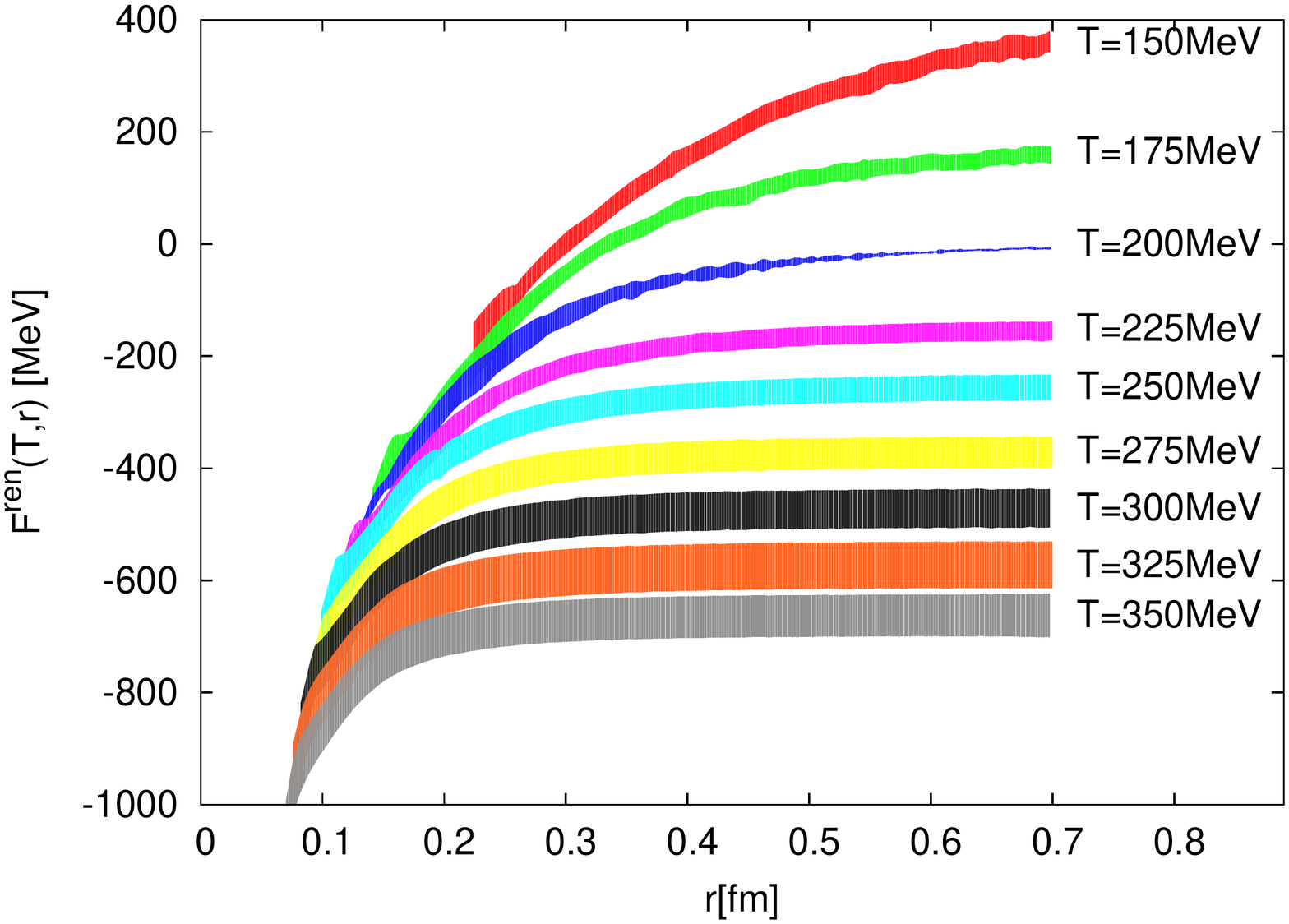}
\end{minipage}
\begin{minipage}[b]{0.46\linewidth}
\includegraphics[trim={ 0 0 0 0cm},clip,width=\linewidth]{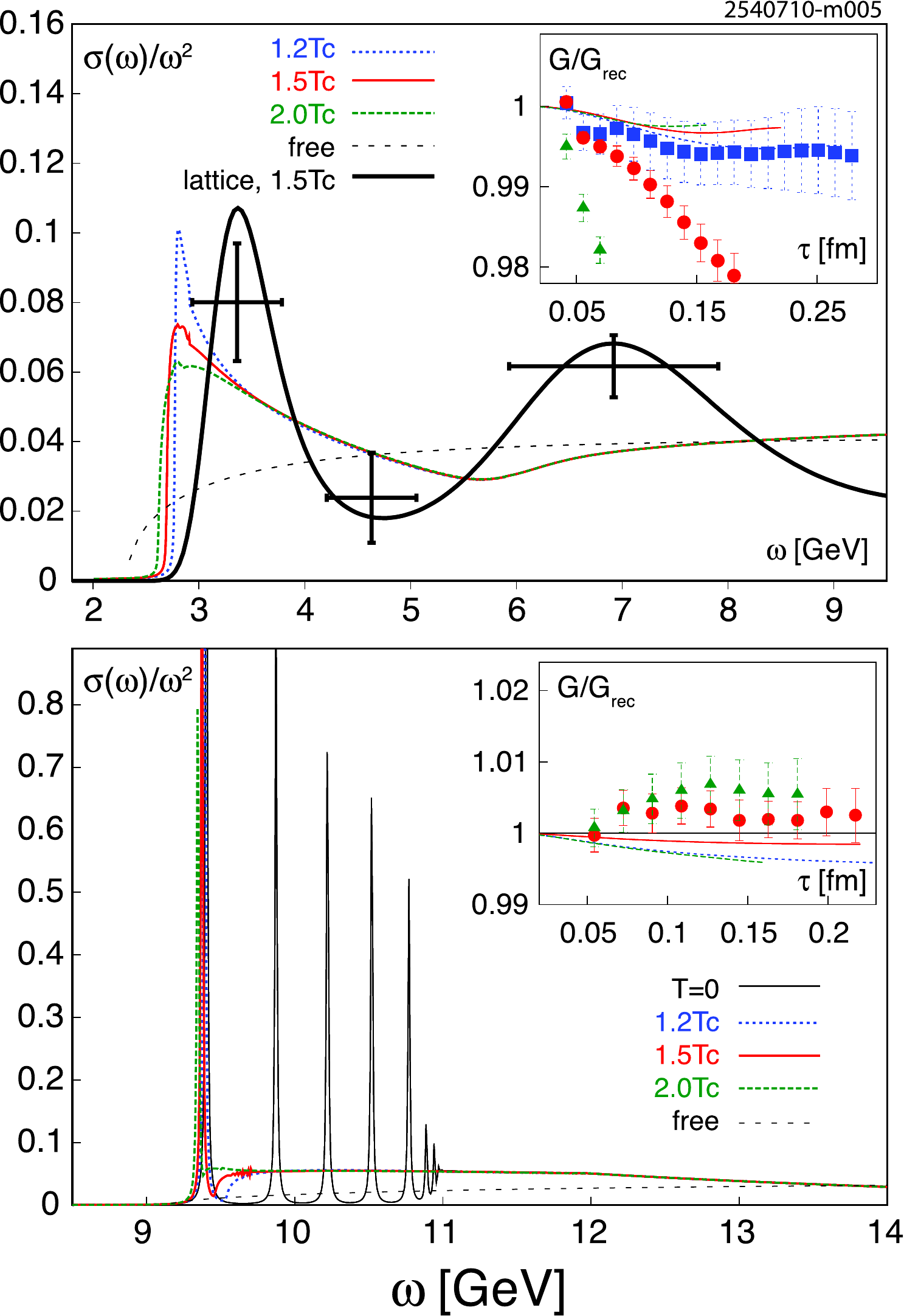}
\end{minipage}
\caption{\textbf{Left}: Heavy-quark-singlet free energy versus quark separation calculated in 2 $+$ 1 flavor QCD for  (\textbf{top}) different values 
of $T/T_c$ \cite{Brambilla:2010cs} and (\textbf{bottom}) continuum-extrapolated values of the static $Q\bar{Q}$ free energy for different temperatures 
\cite{Borsanyi:2015yka}. The solid black line on the right top plot is a parameterisation of the zero temperature potential;
 ~\textbf{Right}: The S-wave 
(\textbf{upper}) charmonium  and (\textbf{lower}) bottomonium spectral functions calculated in potential models. Insets: correlators compared to lattice data. 
The~dotted curves are the free spectral functions. Reproduced from Ref.~\cite{Brambilla:2010cs}.
}
\label{fig:spectral}
\end{figure}

The heavy-quark free energies for different temperatures $T$ of the medium from two different lattice calculations are presented in Figure~\ref{fig:spectral} (left). 
The solid black line on the upper plot is a parameterisation of the zero temperature potential. One can see that with increasing temperature the free energy---and 
hence, the spatial correlations between $Q$ and $\bar{Q}$---get more and more diluted.

The correlation functions in time variable are related to the spectral functions $\sigma(\omega,T)$ by 
\begin{equation}
 G(\tau, T) = \int_{0}^{\infty} d\omega\; \sigma(\omega, T) \frac{\cosh(\omega(\tau - 1/(2T)))}{\sinh(\omega/2T)} \,.
\label{spectral_function}
\end{equation}

While a stable $Q\bar{Q}$ quarkonium state in the vacuum contributes  a $\delta$-function-like peak at the value of its mass $m_H$ to the spectral function, 
in the medium, it gives a quasi-particle-like smeared peak with the width being the thermal width. As  the temperature increases, the width increases, and 
at sufficiently high temperatures, the contribution from the meson state in the spectral function becomes sufficiently broad so that it is no longer meaningful 
to speak of it as a well-defined state (Figure~\ref{fig:spectral}, right). The effect is more prominent for the lighter mesons like charmonia, consisting of $c\bar{c}$ 
pairs, and much weaker for the bottomonia---$b\bar{b}$ mesons.


\subsection{QCD at High Temperatures and Non-Zero Chemical Potentials}
\label{subsec:munonzero}

As direct lattice QCD calculations at non-zero $\mu_B$ are not yet possible, one has to analyze the EoS using Taylor expansion in 
quark chemical potentials $\mu_u$, $\mu_d$, and $\mu_s$: \cite{Petreczky:2012rq, Ding:2015ona} 
\begin{eqnarray}
\frac{p}{T^4}&=&\frac{1}{VT^3}\ln Z(T,\mu_u,\mu_d,\mu_s)=\sum_{ijk} \frac{1}{i! j! k!} \chi_{ijk}^{uds} 
\left(\frac{\mu_u}{T}\right)^i \left(\frac{\mu_d}{T}\right)^i \left(\frac{\mu_s}{T}\right)^j \\
\chi_{ijk}^{uds}&=&\frac{\partial^{\,i+j+k}p/T^4}{\partial(\mu_{u}/T)^{i}\partial(\mu_{d}/T)^{j}\partial(\mu_{s}/T)^{k}}~~,
\end{eqnarray}
where $\mu_u$, $\mu_d$, and $\mu_s$ are related to the chemical potentials corresponding to the baryon number $B$, electric charge $Q$, 
and strangeness $S$ of hadrons, as follows:
\begin{eqnarray}
\mu_u=\tfrac13\mu_B+\tfrac23\mu_Q \,, \quad
\mu_d=\tfrac13\mu_B-\tfrac13\mu_Q \,, \quad
\mu_s=\tfrac13\mu_B-\tfrac13\mu_Q-\mu_S\,.
\label{chem}
\end{eqnarray}

The EoS at non-zero $\mu_{B,Q,S}$ can thus be obtained from the coefficients $\chi_{ijk}^{BQS}$ of the Taylor expansion in hadronic chemical 
potentials expressed via $\chi_{ijk}^{uds}$ \cite{Petreczky:2012rq}. Here, we report the result \cite{Ding:2015ona} for \mbox{$\mu_Q$ = $\mu_S$ = 0}, 
which sufficiently illustrates the relative importance of higher-order corrections in different temperature and $\mu_B$ regions. The Taylor series for 
the pressure is given by
\begin{equation}
\frac{p(T,\mu_B)-p(T,0)}{T^4} =
\frac{1}{2} \chi_2^B(T) \left(\frac{\mu_B}{T} \right)^2 \left[ 1 +
\frac{1}{12}\frac{\chi_4^B(T)}{\chi_2^B(T)}  
\left(\frac{\mu_B}{T} \right)^2 \right] + {\cal O}(\mu_B^6) \;.
\label{PTaylor}
\end{equation}

\begin{figure}[H]
\includegraphics[scale=0.6]{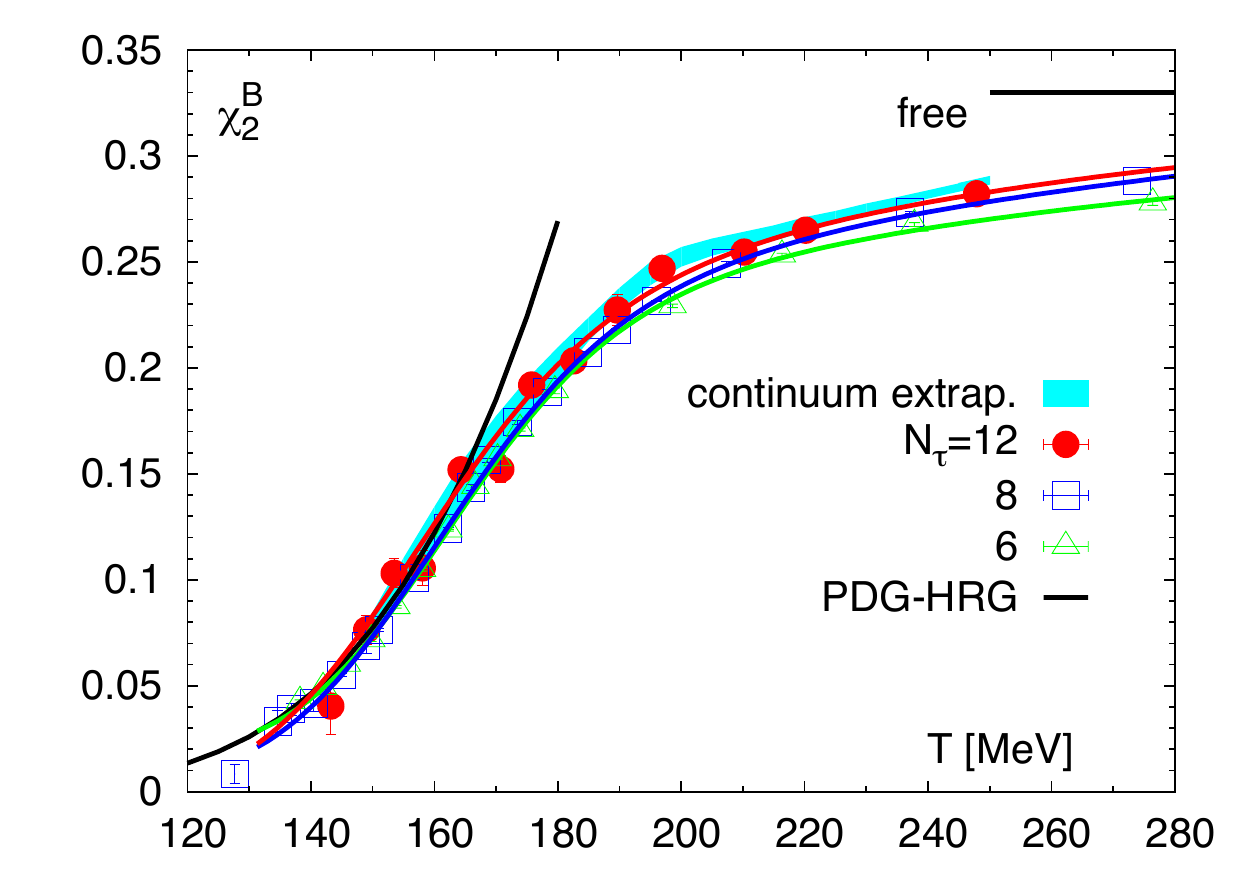} \quad
\includegraphics[scale=0.6]{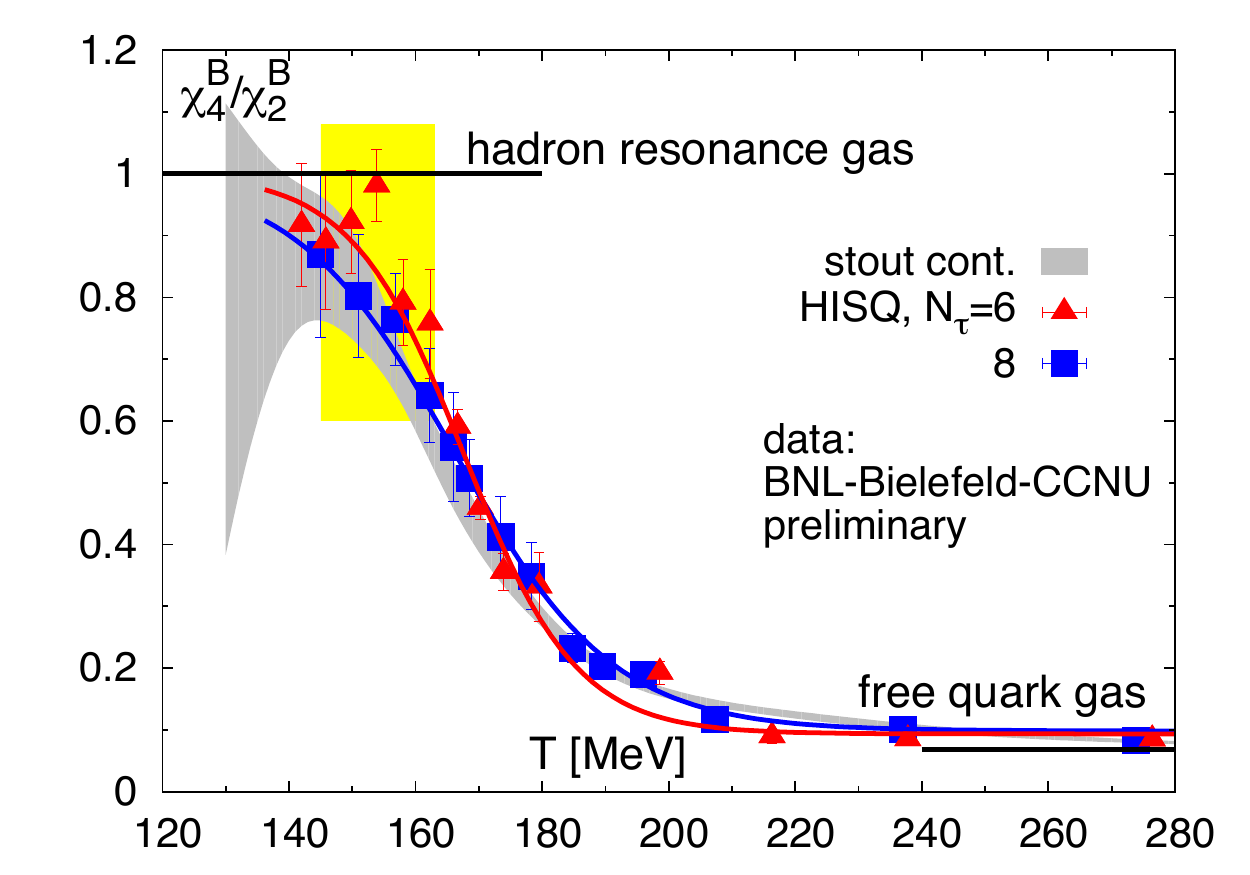}
\caption{The expansion coefficients of the pressure at the non-zero baryon chemical potential adopted from Ref.~\cite{Ding:2015ona}. 
\textbf{Left}: The leading-order correction; \textbf{Right}: The relative contribution of the next-to-leading order l{correction.} BNL: Brookhaven National Laboratory.}
\label{fig:Bcumulants}
\end{figure}

The leading-order correction to the pressure at non-vanishing $\mu_B$ is proportional to the quadratic fluctuations of the net baryon number. The next-to-leading
order corrections are proportional to the quartic fluctuations. In Figure~\ref{fig:Bcumulants}, we show $\chi_2^B(T)$ (left) and $\chi_2^B(T)/\chi_4^B(T)$ (right). 
With increasing temperature, the ${\cal O}(\mu_B^4)$ correction rapidly loses importance relative to the leading ${\cal O}(\mu_B^2)$ term. Moreover, the results for 
the $\mu_B$-dependent contribution to the total pressure evaluated for different values of $\mu_B/T$ \cite{Ding:2015ona} suggest that the EoS given by 
Equation~(\ref{PTaylor}) works well for all values of the chemical potential below $\mu_B/T$ = 2, corresponding to the region of nuclear collisions at energies 
\mbox{$\sqrt{s_{NN}}\ge 20$~GeV.}

Let us note that $\chi_{ijk}^{BQS}$ are interesting in their own right, as they are related to the fluctuations and correlations of conserved charges. 
The latter are sensitive to the underlying degrees of freedom, which could be hadronic or partonic, and so they are used as sensitive probes of deconfinement. 
While the off-diagonal expansion coefficients are related to correlations among conserved charges (e.g., $\chi_{11}^{XY}=\frac{1}{VT^3}\langle N_XN_Y\rangle$), 
the diagonal ones describe their second- and higher order fluctuations
\begin{equation} \hspace{-.2cm}
\chi_2^X = \frac{1}{VT^3}\langle N_X^2\rangle\,,  \qquad \chi_4^X = \frac{1}{VT^3}\left(\langle N_X^4\rangle - 
3 \langle N_X^2\rangle^2\right)\,, \quad
\rm{etc} \,.
\label{fluc}
\end{equation}

\section{Study of Hot and Dense Nuclear Matter Using Nuclear Collisions}
\label{sec:experiment}

\subsection{Heavy Ion Accelerators}

The basic hopes and goals associated with investigations of very hot and dense nuclear matter in laboratory were first formulated in mid-seventies 
\cite{Chapline:1974zf, Scheid:1974zz,Sobel:1975bq}. It was the experience with astrophysical objects like supernovae and neutron stars, and with 
thermonuclear ignition which led the authors to the idea that the nuclear matter shock compression \cite{Zeldovich:1966} of about five-fold normal 
nuclear density should be accomplished in violent head-on collisions of heavy nuclei \cite{Stock:2008ru}. The goal was to find out the response of 
the nuclear medium under compression by pressure resisting that compression; i.e., to study the nuclear matter EoS. The original question was: 
is such a bulk nuclear matter EoS accessible within the dynamics of relativistic heavy ion collisions? \cite{Nagamiya:1982kn, Stock:1985xe}. 
The prospect of observing a phase transition in highly compressed nuclear matter \cite{Lee:1974ma} was lurking behind.

The interest in collisions of high-energy nuclei as a possible route to a new state of nuclear matter was substantially strengthened with the arrival of QCD 
as the microscopic theory of strong interactions. The particle physics community began to adapt existing high-energy proton 
accelerators to provide heavy-ion nuclear beams in the mid-1970s. The Berkeley Bevalac and {JINR}
~Synchrophasotron started to accelerate nuclei to kinetic energies 
from few hundreds of MeV to several GeV \mbox{per nucleon \cite{Stock:2008ru, Stock:1985xe}}. By the mid-1980s, the first ultra-relativistic nuclear beams 
became available. Silicon and gold ions were accelerated to 10 GeV/nucleon at Brookhaven's Alternating Gradient Synchrotron (AGS)~\cite{Schmidt:1992ge}. 
The first nuclear collisions took place at CERN in the early 1980s when alpha particles were accelerated to {\it center-of-mass energy per nucleon--nucleon pair} 
$\sqrt{s_{NN}}=64$~GeV at the {ISR}
~collider. The new era of research began at CERN in fall 1986, when oxygen---and later on in~the summer of~1990, sulphur ions---were injected into the SPS and accelerated up to \mbox{an energy of 200~GeV/nucleon} \mbox{(\sqsn = 19.6 GeV)} \cite{Schmidt:1992ge, Stock:2008ru, Schukraft:2015dna}. 
However, the genuine heavy ion program only started in 1994, after the CERN accelerator complex was upgraded with a new lead ion source which 
was linked to pre-existing interconnected accelerators, the Proton Synchrotron (PS) and the SPS. Seven large experiments involved (NA44, NA45/CERES, 
NA49, NA50, NA52, WA97/NA57, and WA98)  studied different aspects of \pbpb and Pb+Au collisions at \sqsn = 17.3 GeV and \sqsn = 8.6 GeV~\cite{Stock:2008ru, Schukraft:2015dna}.

In the meantime, at the Brookhaven National Laboratory (BNL), the Relativistic Heavy Ion Collider (RHIC) \cite{Harrison:2002es} rose up from the ashes 
of the {ISABELLE/CBA} $\bar{p}p$ collider project abandoned in 1983 by particle physicists. In 1984, the first proposal for a dedicated nucleus--nucleus machine 
accelerating gold nuclei up to \sqsn = 200 GeV was submitted. Funding to proceed with the construction was received in 1991, and on 12 June 2000, 
the first \auau collisions at \sqsn = 130 GeV were recorded by the BRAHMS, PHENIX, PHOBOS and STAR experiments \cite{Arsene:2004fa, Back:2004je, 
Adams:2005dq, Adcox:2004mh}.

The idea of the Large Hadron Collider (LHC) \cite{Evans:2008zzb} dates even further back, to the early 1980s. Although CERN's Large Electron Positron 
Collider (LEP, which ran from 1989 to 2000) was not yet built, scientists considered re-using the 27-kilometer LEP ring for an even more powerful  $pp$ machine 
running at the highest possible collision energies (\sqs = 14 TeV) and intensities. The ion option (\sqsn~=~5.4~TeV per nucleon--nucleon pair for \pbpb collisions) 
was considered since the beginning. The LHC was approved in December 1994, and its official inauguration took place on 21 October 2008.  The first proton--proton 
collisions occurred on 23 November 2009, and the first Pb $+$ Pb collisions on 7 November~2010. The ALICE, ATLAS, and CMS experiments are currently 
involved in the heavy-ion program at the LHC \cite{BraunMunzinger:2007zz, Roland:2014jsa}.

\begin{figure}[H]
\centering
\includegraphics[trim={ 0 1.8cm 0 0cm},clip,width=0.45\textwidth, height=6.85cm]{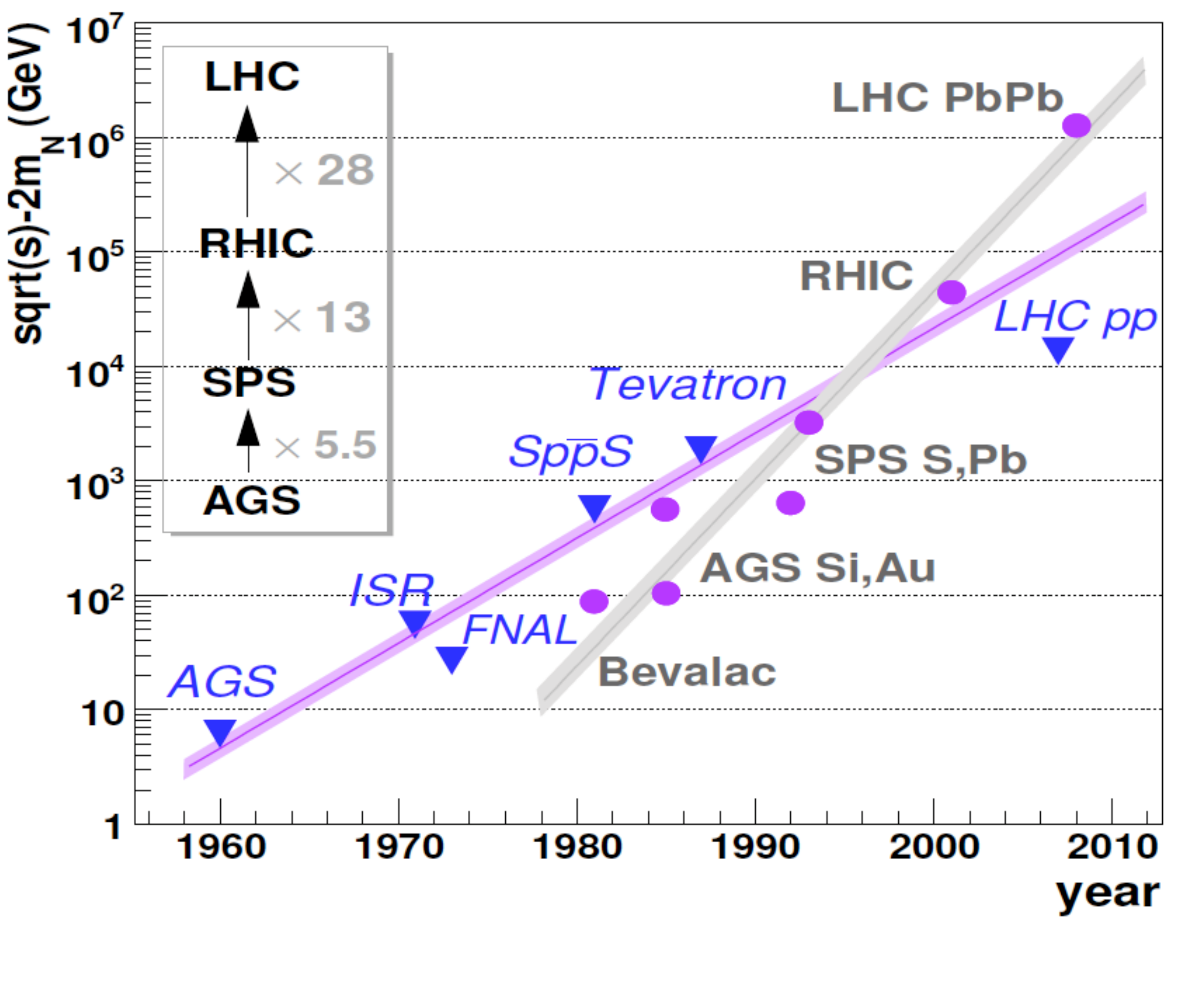}\quad
\includegraphics[trim={ 0 0 0 0},clip,width=0.5\textwidth]{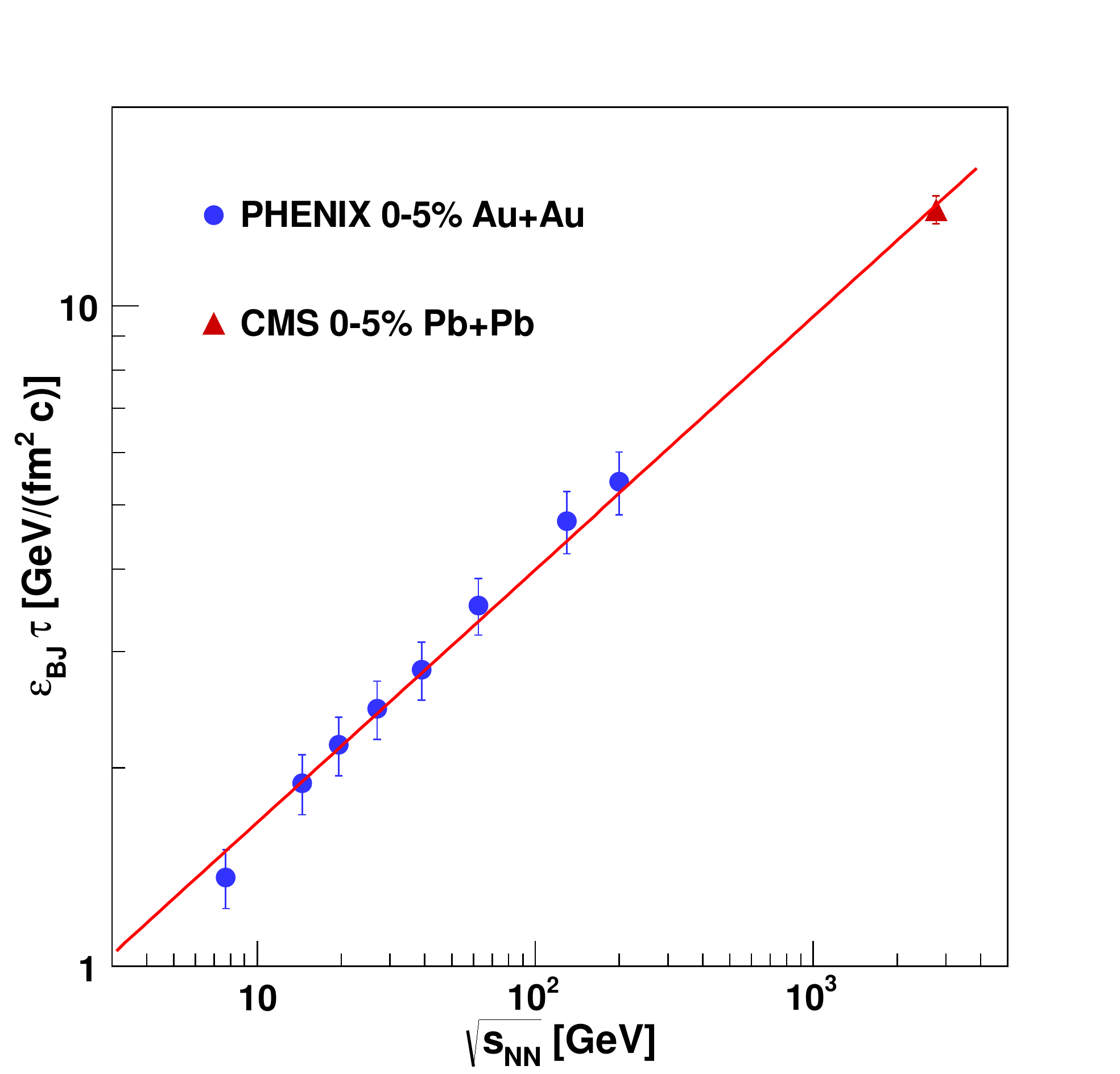}
\caption{\textbf{Left}: Available center-of-mass energy \sqs -2m$_{\rm N}$ versus time for (anti)proton (blue triangles) and ion (magenta circles) 
accelerators, adapted from Ref.~\cite{D'Enterria:2007xr}; \textbf{Right}: The Bjorken estimate of the initial energy density \ebj  (Equation~\ref{ebj}) 
multiplied by $\tau$ calculated from the data on transverse energy distributions in 5$\%$ most central \auau \cite{Adare:2015bua} and \pbpb 
\cite{Chatrchyan:2012mb} collisions as a function of {c.m.s.}
~energy per one nucleon--nucleon pair \sqsn. The red line corresponds to a power law fit. 
From Ref.~\cite{Mitchell:2016fqp}.}
\label{fig:accelerators}
\end{figure}

Many years ago, American accelerator physicist M. Stanley Livingston noted that advances in accelerator technology increase the energy records achieved by 
new machines by a factor of ten every six years. This trend is illustrated on the left panel of Figure~\ref{fig:accelerators}, summarising the worldwide advances 
in high-energy accelerators in the period of 1960--2008. One can see that the increase in the energy is even faster for the ion accelerators than for the proton 
accelerators. However, even in the most central nucleus--nucleus collisions, not all of the
~energy could be converted into a thermalised form of energy needed for 
the phase transition from hadronic into a QGP state of matter to occur. The experimentally accessible quantity measuring this transformation is the density of 
transverse energy \Et per unit of pseudorapidity $\eta$. The latter can be used to estimate the initial energy density using the Bjorken {formula}~\cite{Bjorken:1982qr}:
 \begin{equation}
  \epsilon_{BJ} = \frac{1}{S_\perp \tau}\frac{d\Et}{d\eta}\Big \arrowvert_{\eta=0}\;\; ,
  \label{ebj}
\end{equation}
where $S_\perp$ is the transverse overlap area of the nuclei and $\tau$  is the time scale for the evolution of the initial non-equilibrium state of matter into a 
(locally) thermalized system. The dependence of $\epsilon_{BJ} \tau$ on \sqsn for most central \auau and \pbpb collisions is presented on the right panel of 
Figure~\ref{fig:accelerators}. Even for a rather pessimistic value of the eqilibration time $\tau$ = 1 fm/{\it c} \cite{Adcox:2004mh}, the achieved energy density 
increases from 1.4 to 14 GeV/fm$^3$. It is thus not only much higher than than the normal nuclear density, but 3--30 times bigger than the energy density 
inside the nucleon $\epsilon_{\rm N}$ = $0.45~{\rm GeV}/{\rm fm}^3$, and is definitely higher than the 1 GeV/fm$^3$ scale required for the QCD deconfining 
transition from the lattice calculations that were discussed in Section~\ref{subsec:muzero}.\vspace{-12pt}


\subsection{Heavy Ion Collisions as a Source of Strong Electromagnetic Fields}

The important quantity determining the lifetime of heavy ion beams inside the accelerator tube (and hence their potential to produce an adequate number 
of nuclear collisions) is the loss of ions in the bunch. Its decay rate $\lambda_T$ is given by the formula:
 \begin{equation}
\lambda_T=-\frac{1}{N}\frac{dN}{dt}=\frac{N_{IR}L\sigma_T}{kN} \;\; ,
  \label{declum}
\end{equation}
in which $k$ is the number of bunches, $N$ is the number of particles per bunch, $N_{IR}$ is the number of interaction regions, $\sigma_T$ is the total 
cross-section, and $L$ is the available luminosity:
\begin{equation}
L=f\gamma \frac{N^2 k}{4\pi \epsilon \beta^*}F  \;\; ,
  \label{luminosity}
\end{equation}
which---in addition to $k$ and $N$---also depends on the revolution frequency $f$, the Lorentz factor of the beam $\gamma $, the emittance $\epsilon$, the beta function 
at the collision point $\beta^*$,  and the geometric luminosity reduction factor $F$ due to the crossing angle at the interaction point \cite{Evans:2008zzb}. 
Since the event rate for a certain process is given by its cross-section times the luminosity $N_{evt}=L\sigma_{evt}$, the possibility of studying rare phenomena 
depends on the maximum luminosity accessible. At the same time, the rate of background processes (which in general have large cross-sections) will increase $L$, 
 at some point reaching the maximum event rate that the experiment can handle. Secondary beams created by these background processes can limit the collider 
heavy ion luminosity, since they have a different charge-to-mass ratio than the primary beam, and can be lost in cryogenically cooled magnets.

Quite surprisingly, at RHIC and the LHC, these background processes are not part of the strong nuclear interaction cross-section $\sigma_R$  determined primarily by 
the nuclear geometry \mbox{$\sigma_R\approx \pi r_0^2 (A_I^{1/3}+A_{II}^{1/3} )^2$}, but are solely accounted for by the coherent action of all electric charges in 
colliding nuclei. The cross-section of electromagnetic processes---primarily due to the creation of $e^+e^-$ pairs with subsequent $e^-$ atomic shell capture and 
electro-magnetic dissociation---becomes important for ions with $Z > 30$. It is as large as hundreds of barns; i.e., about 30 (60) times larger than $\sigma_R$ for 
\auau collisions at RHIC (\pbpb collisions at the LHC) \cite{Fischer:2013uwj}.

A classical description of the electromagnetic action of a fast-moving charged particle on another one based on the equivalence between the perturbative action of its field 
and the flux of electromagnetic radiation dates back to Fermi \cite{Fermi:1924tc}, Weizs\"{a}cker \cite{vonWeizsacker:1934nji}, and Williams \cite{Williams:1934ad}. 
This equivalence is true as far as the effects caused by different spectral components add up incoherently (i.e.,~a perturbation caused by the fields is small enough).
The solution for the time-dependent electromagnetic fields mutually seen by the two incident ions can be found, for example, in the textbook on ``Classical Electrodynamics''~\cite{Jackson:1998nia}. The longitudinal ($\parallel$) and transversal ($\perp$) field components induced by a heavy ion $I$ passing a target $II$ at distance $b$ and 
with velocity $\beta$ are given by the following formulas:
\begin{equation}
E_{\parallel}(t) = \frac{-Z_I e \gamma \beta t}{(b^2 + \gamma^2 \beta^2 t^2)^{3/2}} \,,  \quad
\vec{E}_{\perp}(t) = \frac{Z_I e \gamma \vec{b}}{(b^2 + \gamma^2 \beta^2 t^2)^{3/2}} \,,  \quad 
B_{\parallel}(t)=0 \,,  \quad  \vec{B}_{\perp}(t) = \vec{\beta} \times \vec{E}_{\perp}(t) \,.
\label{emgfields}
\end{equation}

Let us note that for $\gamma \gg 1$, these fields act on a very short time scale of order $\Delta t \propto b/\gamma$. During this time, fields $\vec{E}_{\perp}(t)$ and 
$\vec{B}_{\perp}(t)$ are {\it equivalent} to a linearly polarized pulse of radiation incident on a target in the beam direction. Thus, according to the equivalent photon method, 
the strong and rapidly time-varying field of the point charge $Z_I$ is seen by a passing charge as a flux of virtual (nearly real) photons with intensity
\begin{equation}
I(\omega, b)=\frac{1}{4\pi} \arrowvert \vec{E}(\omega) \times \vec{B}(\omega) \arrowvert  \approx  \frac{1}{2\pi} \arrowvert 
E_{\perp}(\omega)\arrowvert ^2 \sim Z_I^2  \;\; ,
\end{equation}
where $\vec{E}(\omega)$, $\vec{B}(\omega) $, and $E_{\perp}(\omega)$ are the Fourier components of the fields $\vec{E}$, $\vec{B}, $ and $E_{\perp}$. 
The energy spectrum of these photons falls as $\propto 1/E_{\gamma}$, up to a maximum energy $E^{max}_{\gamma} = \gamma/ b_{min}$. The interaction 
between the colliding nuclei becomes dominantly electromagnetic for impact parameters $b$ exceeding the size of the radii of colliding nuclei 
$b > b_{min}= R_I + R_{II} = \sqrt{\sigma_R/\pi}$.

Interactions between ultra-relativistic nuclei taking place at $b > b_{min}$ are called the ultra-peripheral collisions. By taking advantage of the photon fields carried by 
relativistic nuclei, they are used  to study photoproduction and two-photon physics at hadron colliders. This field of ultra-relativistic heavy ion collisions is sometimes 
called {\it ``non-QGP'' physics}, and is thus outside the scope of this article. We refer the interested reader to reviews \cite{Baur:2001jj, Bertulani:2005ru, Baltz:2007kq, 
Klein:2015qna} where more detailed information on these aspects can be~found.


\subsubsection{Quark--Gluon Plasma in a Strong Magnetic Field}

More important from the point of view of QGP physics are the strong magnetic fields accompanying ultra-relativistic heavy ion collisions \cite{Tuchin:2013ie}. 
Consider the collision of two identical nuclei of radius $R$ with electric charge $Ze$, and use the Biot--Savart law to estimate the magnitude of the perpendicular 
magnetic field they create in the center-of-mass frame 
\begin{equation}
B_{\perp}\sim \gamma \, Ze\, \frac{b}{R^3} \;\; .
\label{maxfield}
\end{equation}

Here, $\gamma=\sqsn/2m_N$ is the Lorentz factor. At RHIC, heavy ions are collided at $\sqsn= 200$~GeV per nucleon, hence $\gamma=100$. 
Using $Z=79$ for Gold and $b\sim R_A\approx 7$~fm, we estimate $eB\approx m_\pi^2\sim 10^{18}$~G. At the LHC at \sqsn= 5.02~ TeV and $Z=82$, 
this value is even 30 times larger. To appreciate how strong this field is, compare it with the magnetic field of a neutron star $10^{10}$--$10^{13}$~G \cite{Becker:2008}, 
or that of its slowly rotating magnetic variant, the magnetar, $10^{15}$~G \cite{Kouveliotou:2003tb}. It is very likely the strongest magnetic field in nature, though 
existing only for a minute period of time.

Calculation with the realistic distribution of protons in a nucleus shows that magnetic field rapidly decreases as a power of time, and after the first 3 fm/{\it c} drops 
from its maximal value (\ref{maxfield}) by more than three orders of magnitude \cite{Kharzeev:2007jp}. However, different estimates to be discussed in the next 
sections indicate that a strongly interacting thermalised medium is formed as early as 0.5~fm/{\it c}. Therefore, a more realistic calculation going beyond the above 
field in the vacuum calculation has to include the response of the medium determined by its electrical conductivity. It has been found by lattice calculations \cite{Ding:2010ga} 
that the gluon contribution to the electrical conductivity of static quark--gluon plasma is 
\begin{equation}
\sigma= (5.8\pm 2.9)~\frac{T}{T_c} ~\text{MeV}\,.
\label{qgpconductivity}
\end{equation} 

This result was confirmed and further extended by more elaborate lattice simulations with 2 $+$ 1 dynamical flavors for temperatures T = (120--350)  MeV 
\cite{Amato:2013naa,Aarts:2014nba}. The calculations have shown that $\sigma T$  already starts to deviate from zero for $T < T_c$ (i.e.,~in the confined phase), 
and increases towards the QGP value (\ref{qgpconductivity}) and further on. The non-zero electrical conductivity in the QGP and (probably also) in the hadronic phase 
when taken at its face value would inevitably lead to a substantially prolonged lifetime of the magnetic field inside the medium, and might thus even influence 
the hadron decay widths \cite{Filip:2015mca}.

A plethora of novel non-dissipative transport phenomena related to the interplay of quantum anomalies with the magnetic field and vorticity in systems with chiral fermions, 
including the QGP, is reviewed in Ref.~\cite{Kharzeev:2015znc}. The most direct effect of magnetic field $\vec{B}$ on the QGP is the induction of electric currents carried by 
the charged quarks and antiquarks in the plasma, and later, by the charged hadrons. In Ref.~\cite{Gursoy:2014aka}, it was suggested that it may leave its imprint on 
the azimuthal distributions and correlations of the produced charged hadrons. Charged particles moving along the magnetic field direction $y$ are not influenced by 
the magnetic Lorentz force, while those moving in the $xz$-plane (i.e.,~in the reaction plane) are affected the most. The result is an azimuthally anisotropic flow of 
expanding plasma in the $xy$-plane, even when the initial plasma geometry is completely spherically symmetric.

Another effect is related to the chiral symmetry restoration. In such a state, within a localized region of space-time, gluon fields can generate nontrivial topological charge 
configurations that lead to parity violation in strong interactions \cite{Lee:1974ma}. In ultra-relativistic heavy-ion collisions, interactions between quarks and these gluonic 
states can lead to an imbalance in left- and right-handed quarks which violates parity symmetry \cite{Kharzeev:1998kz}. The presence of a strong magnetic field induced 
by the spectator protons transforms this chirality imbalance into an electromagnetic current perpendicular to the reaction plane. This interesting phenomenon stemming 
from the interplay of chirality, magnetic field, and the chiral anomaly is called the Chiral Magnetic Effect (CME) \cite{Kharzeev:2015znc}.

Several manifestations of the phenomena related to strong magnetic fields produced in \auau or \pbpb collisions have been reported by RHIC \cite{Abelev:2009ac, 
Adamczyk:2014mzf, Adamczyk:2015eqo} and the LHC \cite{Abelev:2012pa} experiments. Some doubts on the prevailing interpretation were cast by the recent observation 
of charge-dependent azimuthal correlations also in \ppb collisions at the LHC \cite{Khachatryan:2016got}. Moreover, in the presence of elliptic flow (for its definition, see 
Ref.~\ref{bulk}), practically all conventional two-particle correlations like the local charge conservation \cite{Schlichting:2010qia} may contribute to the reaction-plane-dependent correlation function used to quantify the CME \cite{Skokov:2016yrj}. Obviously, more investigations are needed. The program of varying the magnetic field 
by a controlled amount while keeping all else fixed by using nuclear isobars (pairs of nuclei with the same mass number $A$ but different charge $Z$) is now under 
consideration at RHIC. The most attractive isobars are Zr$+$Zr and Ru$+$Ru, or some other combinations having charge differences of four like Sn$^{124}$/Xe$^{124}$, 
Te$^{130}$/Ba$^{130}$, and Xe$^{136}$/Ce$^{136}$ \cite{Skokov:2016yrj}.

An interesting suggestion addressing the simultaneous effects of the huge vorticity of nearly-perfect fluid and the strong magnetic field generated in non-central heavy-ion 
collisions was made in Ref.~\cite{Becattini:2016gvu}. The authors suggest the measurement of a global polarization of the final hadrons in order to estimate the thermal 
vorticity due to the large orbital momentum of colliding nuclei as well as the electromagnetic field developed in the plasma stage of the collision.

\subsection{Transport Models}

\begin{figure}[H]
\centering
\begin{tabular}{cc}
\includegraphics[scale=0.33]{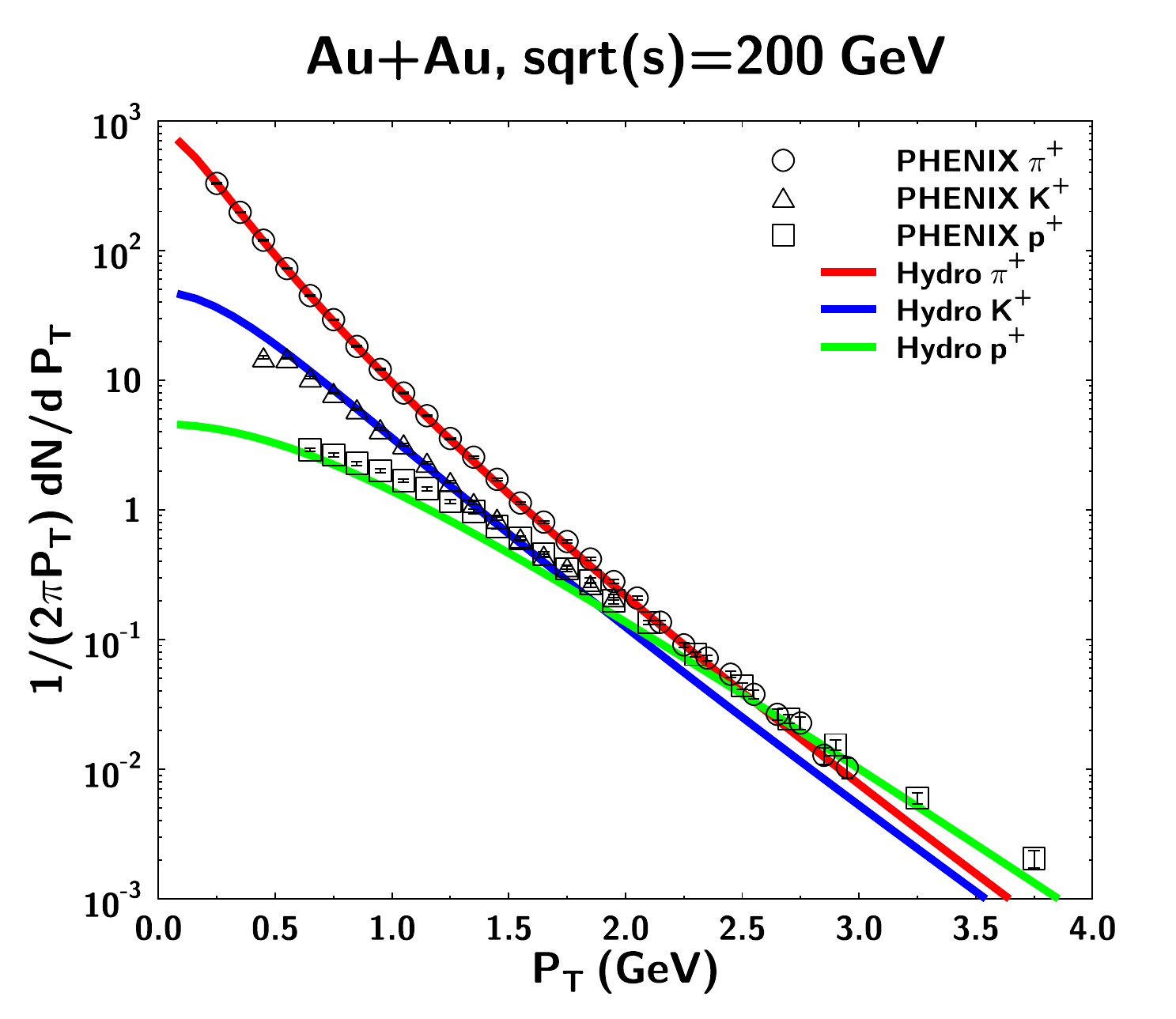}  
&\includegraphics[scale=0.3]{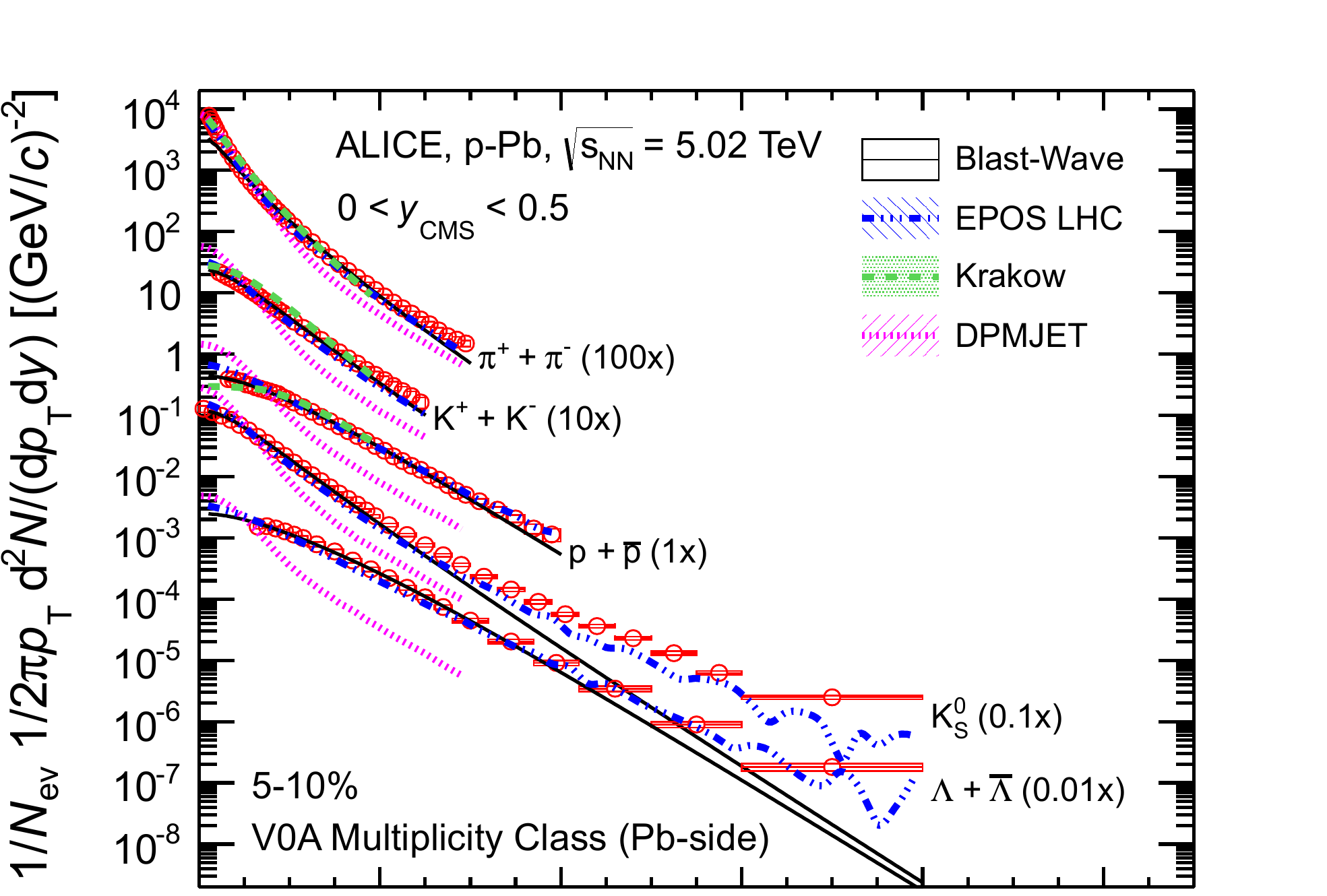}\\  
\includegraphics[scale=0.33]{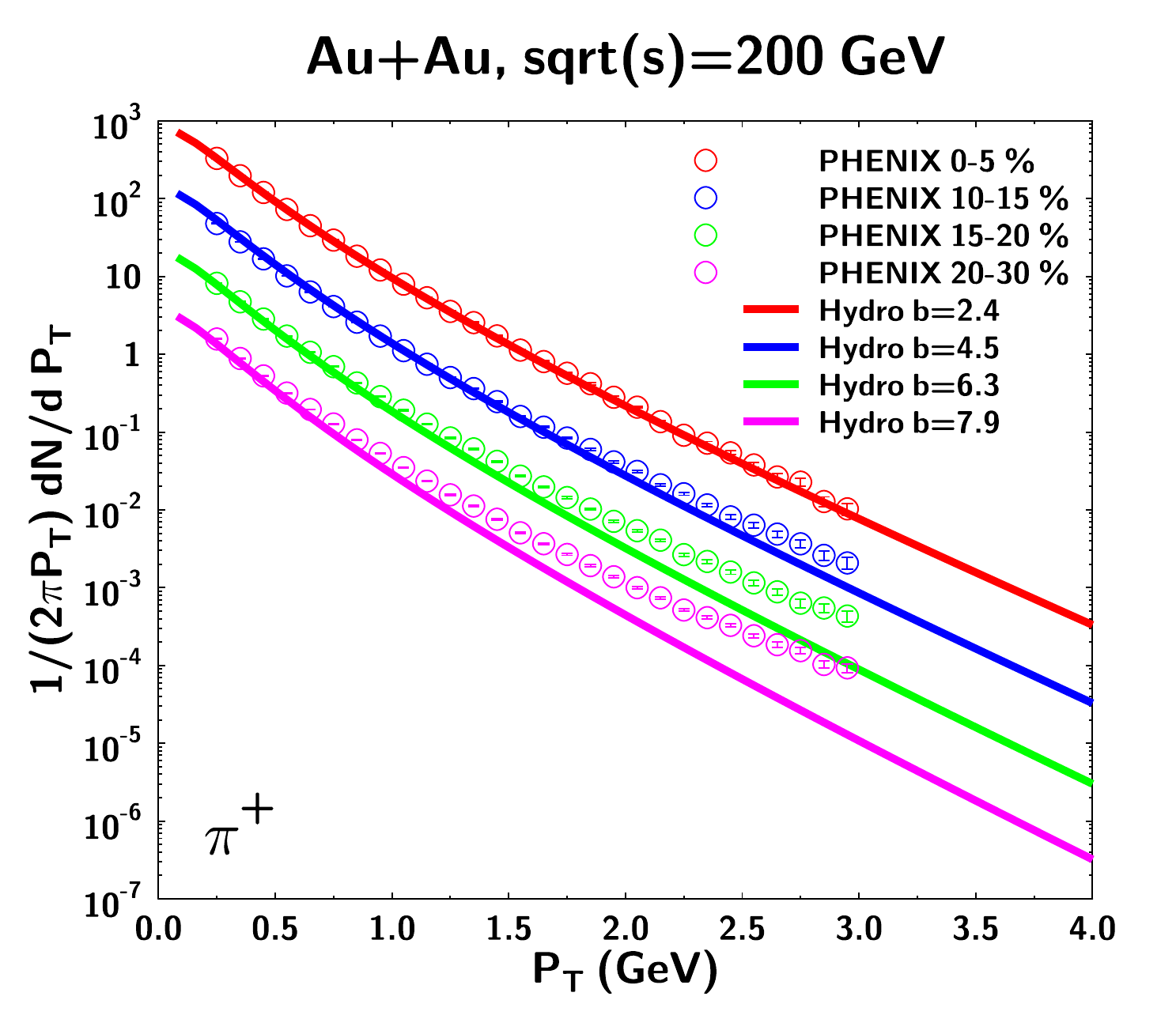}
&\includegraphics[scale=0.3]{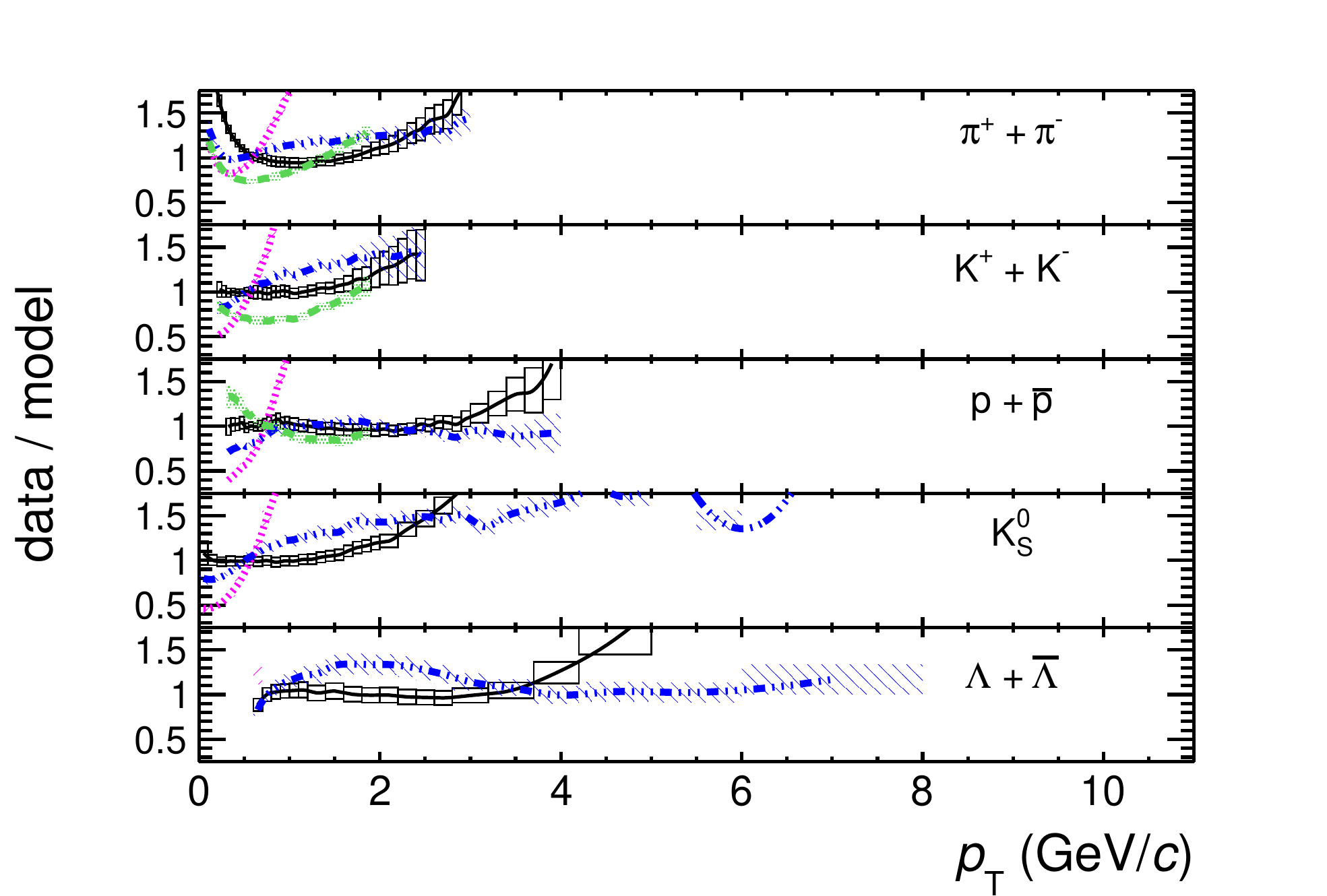} \\ 
\end{tabular}
\caption{\textbf{Left}: Transverse momentum spectra of pions, kaons, and protons emitted into the incident lead nucleus hemisphere ($0 < y_{cms} < 0.5$) 
in 5--10\% most central \ppb  collisions at \sqsn= 5.02~TeV. Data (\textbf{red} circles) are compared to two microscopic models: {EPOS LHC}
~\cite{Pierog:2013ria}, 
{DPMJET} \cite{Roesler:2000he}, to the hydrodynamics calculation \cite{Bozek:2012qs}, and to the Blast-Wave fit with formula Equation~(\ref{blastwave}). 
Reproduced from Ref.~\cite{Abelev:2013haa}; \textbf{Right}: Comparison of the experimental  \pt-spectra of $\pi^+$, $K^+$, and $p^+$ from \auau 
collisions at  (\textbf{top})  $\sqsn= 200$ GeV and (\textbf{bottom})
~(for the case of $\pi^+$) of their centrality dependence \cite{Adler:2003cb} with the hydrodynamical 
model calculations. Reproduced from Ref.~\cite{Nonaka:2006yn}.
}
\label{fig:models}
\end{figure}

One of the main tasks of the theory is to link experimental observables to different phases and manifestations of the QCD
matter. To achieve this goal, a detailed understanding of the dynamics of heavy-ion reactions is essential. This is facilitated
by transport theory, which helps to interpret or predict the quantitative features of heavy-ion reactions. It is particularly
well-suited for a non-equilibrium situation, finite size effects, non-homogeneity, N-body phase space, particle/resonance
production and freeze-out, as well as for collective dynamics.~Microscopic
\cite{Aichelin:1991xy, Werner:1993uh, Gyulassy:1994ew, Bass:1998ca, Humanic:1998ji, Roesler:2000he, Lin:2004en, Xu:2016hmp, Pierog:2013ria}, 
~macroscopic (hydrodynamical) \cite{Kolb:2000sd, Bozek:2012qs, Akamatsu:2013wyk, deSouza:2015ena}, or hybrid \cite{Lokhtin:2008xi, Hirano:2012kj, Shen:2014vra} 
transport models attempt to describe the full time evolution from the initial state of a heavy-ion reaction up to the freeze-out of 
all initial and produced particles after the reaction. This is illustrated in Figure~\ref{fig:models}, where a comparison of the data from 
heavy-ion collisions to the microscopic and hydrodynamical models is~presented.


The hadronic cascade models (some with mean-field interactions) have succeeded in reproducing the gross and many-detailed features of
the nuclear reactions measured at SIS, AGS, and SPS~\mbox{\cite{Bass:1998ca, Aichelin:1991xy, Werner:1993uh, Humanic:1998ji}}. 
They have become indispensable for experimentalists who wish to identify interesting features in their data or to make predictions 
to plan new experiments. The main strength of the models based on superposition of $pp$ collisions, relativistic geometry, and final-state 
hadronic rescattering is not that it gives a precise agreement with experiment for individual observables in particular kinematic regions, 
but in its ability to give an overall qualitative description of a range of observables in a wide kinematic region. The price to be paid for this 
simplicity is to assume that either hadrons or hadron-like objects can exist at the earliest stage of the heavy-ion collision just after the two 
nuclei pass through each other; i.e.,~that the hadronization time in the frame of the particle is short and insensitive to the environment 
in which it finds itself. The general success of these models at lower energies can nonetheless easily lead to misconceptions at higher energies. 
The main concern is the relevance of these models at high particle densities, which are so characteristic for collisions of heavy systems. Here, 
all the models based on hadronic dynamics are fundamentally inconsistent~\cite{Muller:1999ys}. Studying the size of
~the fraction of the energy 
contained in known hadrons and that  temporarily stored in  more elusive objects (such as pre-hadronized strings), it was found \cite{Bass:1998ca} 
that up to a time of 8 fm/{\it c}, most of the energy density resides in strings and other high-mass continuum states that have not fully decayed. 
The physical properties of these objects are poorly known, even when they occur in isolation \cite{Kittel:2005}---not to speak about their interactions 
(or even their existence) in a dense environment. The application of these models to the early phase of collision of two ultra-relativistic heavy nuclei 
is therefore ill-founded \cite{Muller:1999ys}.

A complementarity between the microscopic and macroscopic descriptions becomes obvious for the case of strongly-interacting plasmas. 
The fact that neither Boltzmann equation nor cascades can be used for liquids stems from the fact that particles are strongly correlated 
with several neighbours at all times. The very idea of ``scattering'' and cross-section involves particles coming from and going to infinity: 
it is appropriate for dilute gases, but not for condensed matter where interparticle distances do not exceed the range of the forces 
at any time \cite{Shuryak:2008eq}.

The idea to use the laws of ideal hydrodynamics to describe the expansion of the strongly interacting matter formed in high energy
hadronic collisions was first formulated by Landau in~1953~\cite{Landau:1953}. Later on, Bjorken \cite{Bjorken:1982qr} 
discovered a simple scaling solution that provides a natural starting point for more elaborate solutions in the ultra-relativistic domain. 
The phenomenological success of the Landau model was  a big challenge for high energy physics for decades \cite{Weiner:2005gp}. 
First,~because hydrodynamics is a classical theory, and second because
~it assumes local equilibrium. Both of these assumptions imply many degrees of freedom, and it is by no means clear that the highly excited but still small systems produced in violent nuclear 
collisions satisfy the criteria justifying treatment in terms of a macroscopic theory \cite{Heinz:2009xj}. Therefore, the Landau model (and
other statistical models of strong interactions) were considered up to the mid-seventies as exotic approaches, lying outside of mainstream 
physics \cite{Weiner:2005gp}. Then, the authors of Refs.~\cite{Scheid:1974zz, Sobel:1975bq} realized that the exploitation of hydrodynamics 
in an interpretation of data is the only chance of proving  the existence of a new state of matter in laboratory. This is a trivial corollary 
of the well-known fact that a state of matter is defined by its EoS, and there is no other way to get information about the EoS than by 
using hydrodynamics~\cite{Stoecker:1986ci, Heinz:2009xj, Weiner:2005gp}.

\subsubsection{Elements of Relativistic Hydrodynamics}
\label{hydrodynamics}

Let us now briefly recall some of the basic results from relativistic hydrodynamics~\mbox{\cite{Csernai:1994xw, Yagi:2005yb, Vogt:2007zz, Hirano:2008hy, Gale:2013da, Jaiswal:2016hex} }
upon which the contemporary models are based. The basic hydrodynamical equations describe the energy--momentum and the current conservation
\begin{equation}
\partial_\mu T^{\mu\nu} = 0\,, \quad \partial_\mu j_i^\mu = 0\, ,
\label{conservationT}
\end{equation}
where $j_i^\mu, i=B, S, Q$ is the conserved current.  Both quantities can be decomposed into time-like and space-like components using natural 
projection operators, the local flow four-velocity $u^\mu$,  and the second-rank tensor perpendicular to it $\Delta^{\mu\nu} = g^{\mu\nu} - u^\mu u^\nu$:
\begin{eqnarray}
T^{\mu\nu} &=& \epsilon u^\mu u^\nu - p\Delta^{\mu\nu} + W^\mu u^\nu + W^\nu u^\mu + \pi^{\mu\nu}\,,\label{decompositionT}\\
j_i^\mu    &=& n_i u^\mu + V_i^\mu\,,\label{decompositionN}
\end{eqnarray}
where $\epsilon=u_\mu T^{\mu\nu}u_\nu$ is the energy density, $p=p_s+\Pi  = -\frac13\Delta_{\mu\nu}T^{\mu\nu}$ is the hydrostatic + bulk\ pressure, 
$W^\mu=\Delta^{\mu}_{\ \alpha}T^{\alpha\beta}u_\beta$ is the energy\ (or\ heat)\ current, $n_i=u_\mu j^{\mu}_i$ is the charge\ density, 
$V_i^\mu = \Delta^{\mu\ }_{\ \nu} j_i^\nu$ is the charge\ current, and $\pi^{\mu\nu} = \left< T^{\mu\nu}\right> $ is the shear\ stress\ tensor. 
The angular brackets in the definition of the shear stress tensor $\pi^{\mu\nu}$ stand for the following operation:
\begin{equation}
\left<A^{\mu\nu}\right> = \left[\frac12(\Delta^{\mu\ }_{\ \alpha}\Delta^{\nu\ }_{\ \beta}+\Delta^{\mu\ }_{\ \beta}\Delta^{\nu\ }_{\ \alpha})-\frac13\Delta^{\mu\nu}\Delta_{\alpha\beta}\right]A^{\alpha\beta}\,.
\end{equation}

To further simplify our discussion, we restrict ourselves in the following to only the one conserved charge, and denote the corresponding baryon current 
as $ j^\mu = j_B^\mu $. The various terms appearing in the decompositions (\ref{decompositionT}) and (\ref{decompositionN}) can then be grouped 
into ideal and dissipative parts
\begin{eqnarray}
T^{\mu\nu} &=& T_{id}^{\mu\nu} + T_{dis}^{\mu\nu} = \left[\epsilon u^\mu u^\nu - p_s\Delta^{\mu\nu}\right]_{id} +\left[-\Pi\Delta^{\mu\nu}+W^\mu u^\nu+W^\nu u^\mu + \pi^{\mu\nu}\right]_{dis}\\
j^\mu &=& j_{id}^\mu + N_{dis}^\mu \; =  \left[nu^\mu \right]_{id} + \left[V^\mu \right]_{dis}.
\end{eqnarray}

Neglecting the dissipative parts, the energy--momentum conservation and the current conservation (\ref{conservationT}) define {\it ideal hydrodynamics}. 
In this case (and for a single conserved charge), a solution of the hydrodynamical Equation (\ref{conservationT}) for a given initial condition describes 
the space-time evolution of the six variables---three state variables $\epsilon(x)$, $p(x)$, $n(x)$, and three space components of the flow velocity $u^\mu$. 
However, since (\ref{conservationT}) constitute only five independent equations, the sixth equation relating $p$ and $\epsilon$---the EoS---has to be added by hand 
to solve them. For this, one can either use the relativistic non-interacting massless gas EoS or its generalization to the case of a non-zero interacting measure 
$\Theta^{\mu\mu} (T)= \epsilon - 3p$. In addition to many different phenomenological parameterizations of $\Theta^{\mu\mu}$, one can exploit the relation~
(\ref{e3p_dlnZ}) to obtain the EoS directly from the lattice QCD simulations. Examples of this approach were given in Section~\ref{softestpoint} (see particularly 
Equations~\ref{CV} and \ref{dedT}), and are illustrated in Figure~\ref{fig:cs2}.

Two definitions of flow can be found in the literature \cite{Csernai:1994xw, Yagi:2005yb, Hirano:2008hy}; one related to the flow of energy (Landau) 
\cite{Landau:1953} reads
\begin{equation}
\label{eq:landau}
u_L^\mu=\frac{T^{\mu\ }_{\ \,\nu}u_L^{\nu}}{\sqrt{u_L^\alpha T_{\alpha\ }^{\ \beta}T_{\beta\gamma}u_L^{\gamma}}}=\frac{1}{e}T^{\mu\ }_{\ \,\nu}u_L^{\nu}\,,
\end{equation}
while the other relating to the flow of conserved charge (Eckart) \cite{Eckart:1940te} as follows:
\begin{equation}
u_E^\mu=\frac{j^\mu}{\sqrt{j_\nu j^\nu}}\,.
\end{equation}

Let us note that $W^\mu = 0$ ($V^\mu = 0$) in the Landau (Eckart) frame. In the case of vanishing dissipative currents, both definitions represent a common flow. 
In ultra-relativistic heavy-ion collisions, the Landau definition is more suitable when describing the evolution of matter in the region with a small or zero baryon number 
deposition (i.e.,~when $j = j_B = 0$) like the mid-rapidity region at the LHC and at the top RHIC energy; see Figure~\ref{fig:phase_diagram}. In this case, the heat conduction 
effects can be neglected.

In order to solve the hydrodynamic equations with the dissipative terms, it is customary to introduce the following two phenomenological definitions (so-called 
\textit{constitutive equations}) for the shear stress tensor $\pi^{\mu\nu}$ and the bulk pressure $\Pi$ \cite{Hirano:2008hy},
\begin{eqnarray}
\pi^{\mu\nu} = 2\eta\left<\nabla^{\mu}u^\nu\right>\;, \quad
\Pi = -\zeta \partial_\mu u^\mu = -\zeta \nabla_\mu u^\mu\,, \label{bp}
\end{eqnarray}
where the coefficients $\eta$ and $\zeta$ are called the \textit {shear viscosity} and \textit{bulk viscosity}, respectively. 

For the boost-invariant Bjorken flow \cite{Bjorken:1982qr} which is also called the one-dimensional Hubble flow, since velocity 
in the $z$ direction, $v_z$, is proportional to $z$,
\begin{eqnarray}
u^\mu_{\mathrm{BJ}}=\frac{x^\mu}{\tau} = \frac{t}{\tau}\left(1,0,0,\frac{z}{t}\right)\,,
\end{eqnarray}
where $\tau=\sqrt{t^2-z^2}$ is the proper time, one obtains the following equation of motion \cite{Hirano:2008hy}:
\begin{equation}
\frac{d\epsilon}{d\tau}=-\frac{\epsilon+p_s}{\tau}\left(1-\frac{4}{3\tau T}\frac{\eta}{s}-\frac{1}{\tau T}\frac{\zeta}{s}\right)\,.
\label{be1}
\end{equation}

Neglecting the last two terms in Equation~(\ref{be1}), one obtains the famous Bjorken solution of ideal hydrodynamics \cite{Bjorken:1982qr}. 
The last two terms on the right-hand side in Equation~(\ref{be1}) describe a compression of the energy density due to viscous corrections. The first one is due 
to the shear viscosity in compressible fluids, while the second comes from the bulk viscosity. Two dimensionless coefficients in the viscous 
correction, $\eta/s$ and $\zeta/s$, reflect the intrinsic properties of the fluids; see Table~\ref{tab_eta} and Figure~\ref{fig_v1d} (left panel). 
The value $\eta/s=1/4\pi$ has been obtained in the framework of $\mathcal{N}=4$ SUSY
~Yang--Mills theory~\cite{Kovtun:2004de}. The conformal 
nature of this theory gives $\zeta/s=0$ automatically. Moreover, \mbox{$\eta/s=\mathcal{O}(0.1-1)$} for gluonic matter is obtained from the lattice calculations 
of pure $SU(3)$ gauge theory \cite{Nakamura:2004sy}, while the bulk viscosity $\zeta$ has a prominent peak around $T_c$ resulting from 
the trace anomaly of QCD \cite{Karsch:2007jc}.

Hydrodynamics provides an effective description of a system that is in local thermal equilibrium, and can be derived from the underlying kinetic description 
through Taylor expansion of the entropy four-current $S^\mu$ = $s u^\mu$ in gradients of the local thermodynamic variables. In zeroth order in gradients, one 
obtains ideal fluid dynamics. Then, the higher orders describe dissipative effects due to irreversible thermodynamic processes such as the frictional energy dissipation 
between the fluid elements that are in relative motion, or heat exchange of the fluid element with its surroundings on its way to approach thermal 
equilibrium with the whole fluid. The relativistic Navier--Stokes description given above in Equation~(\ref{bp})---which accounts only for terms that are \textit{linear} 
in velocity gradient---leads, unfortunately, to severe problems.  Qualitatively, it can be seen from Equation~(\ref{bp}) observing what happens if one of 
the thermodynamic forces $\langle\nabla^{\mu}u^\nu\rangle$ or $\nabla_\mu u^\mu$ is suddenly switched off/on \cite{Muronga:2003ta}. In this case, 
the corresponding thermodynamic flux $\pi^{\mu\nu}$ or $\Pi$ which is a purely local function of the velocity gradient also instantaneously vanishes/appears. 
The linear proportionality between dissipative fluxes and forces causes an instantaneous (acausal) influence on the dissipative currents, leading to numerical 
instabilities \cite{Hiscock:1983zz}.

Any numerical implementation of relativistic dissipative fluid dynamics thus requires the inclusion of terms that are second order in gradients \cite{Gale:2013da}. 
The resulting equations for the dissipative fluxes $\pi^{\mu\nu}$  and $\Pi$  are relaxation-type equations \cite{Jaiswal:2016hex} with microscopic relaxation times 
$\tau_{\pi}\equiv 2\eta \alpha$ and $\tau_{\Pi}\equiv \zeta \beta$, which encode the time delay between the appearance of thermodynamic gradients that drive 
the system out of local equilibrium and the associated build-up of dissipative flows in response to these gradients, thereby restoring causality \cite{Jaiswal:2016hex,
Heinz:2013th}. Since the relaxation times must be positive, the Taylor expansion coefficients $\alpha$ and $\beta$ must all be larger than zero. Accounting for 
non-zero relaxation times at all stages of the evolution constrains departures from local equilibrium, thereby both stabilizing the theory and improving its quantitative precision.

\subsubsection{Blast Wave Parametrization}

Interpretation of the results of hydrodynamical calculations or of the experimental data in terms of the collective flow of matter \cite{Abelev:2008ab, Abelev:2013vea} 
is greatly facilitated by the use of the analytical so-called {\it Blast~Wave} (BW) parametrization \cite{Siemens:1978pb, Schnedermann:1993ws, Retiere:2003kf}. 
Within the boost-invariant scenario of Bjorken \cite{Bjorken:1982qr}, and for the full azimuthal symmetry which is valid in central collisions of two nuclei, 
the velocity field of expanding matter is given by 
\begin{equation}
u^\mu(\rho,\eta)
= (\cosh\rho~\cosh\eta, ~\vec e_r \sinh\rho, ~\cosh\rho~\sinh\eta ) \;,
\end{equation}
where $\rho = \tanh^{-1} \beta_T$ and $\eta$ are transverse and longitudinal rapidities, respectively, and $\vec e_r$ is the unit vector in the transverse plane. 
The transverse velocity distribution $\beta_T(r)$ of the thermalized matter in the region $0\le r \le R$ is described by a self-similar profile
\begin{equation}
\label{betar}
\beta_T( r ) = \beta_s \left( {r\over R} \right)^k  \quad,
\end{equation}
where $\beta_s$ is the surface velocity, and parameter $k$ is usually given the value $k$ = $2$ to resemble the solutions of hydrodynamics \cite{Schnedermann:1993ws}. 
The spectrum of locally thermalized matter is constructed as a superposition of the individual thermal components \cite{Cooper:1974mv}:
\begin{equation}
\label{Ed3ndp3CooperFrye}
E{{d^3N}\over{d^3p}} = 
                                 {{g}\over{(2\pi)^3}} \int e^{-(u^\nu p_\nu -\mu )/T_{kin}}
				p^\lambda d\sigma_\lambda	\;,
\end{equation}
where $\sigma$ is the hypersurface defining a borderline between the hydrodynamical behaviour and free-streaming particles (the so-called {\it freeze-out hypersurface}), 
and $T_{kin}$ is the {\it temperature of the kinetic freeze-out}. Boosting each component with the transverse rapidity $\rho = \tanh^{-1} \beta_T$, one obtains 
the transverse momentum spectra of particles from the collective radial flow of expanding matter:
\begin{equation}
{{dN}\over{\pt \,d\pt}} \propto
	\int_0^R r\,dr \, m_T 	I_0\Big(\frac{p_T \sinh\rho}{T}\Big)
				K_1\Big(\frac{m_T \cosh\rho}{T}\Big)\; ,
\label{blastwave}
\end{equation}
\noindent  where $m_T=\sqrt{(m^2+\pt^2)}$, $I_0(x)$, and $K_1(x)$ are the Bessel functions.
 
Formulas for the case of non-central collisions when the transverse shape (\ref{betar}) is controlled not by one but two parameters, $R_x$ and $R_y$, 
can be found in Ref.~\cite{Retiere:2003kf}. In full generality, there are eight parameters describing the blast wave parametrization: $T, \rho_0, \rho_2, R_x, R_y, a_s, \tau_0$, 
and $\Delta \tau$. Here, $T$ is the temperature, $\rho_0$ and $\rho_2$ describe the strength of the zero- and second-order oscillation of the transverse rapidity, 
the parameter $a_s$ corresponds to a surface diffuseness of the emission source, and $\tau_0$ and $\Delta \tau$ are the mean and width of a Gaussian longitudinal 
proper time $\tau = \sqrt{t^2-z^2}$ freeze-out distribution.

\begin{figure}[H]
\centering
\includegraphics[width=.475\textwidth]{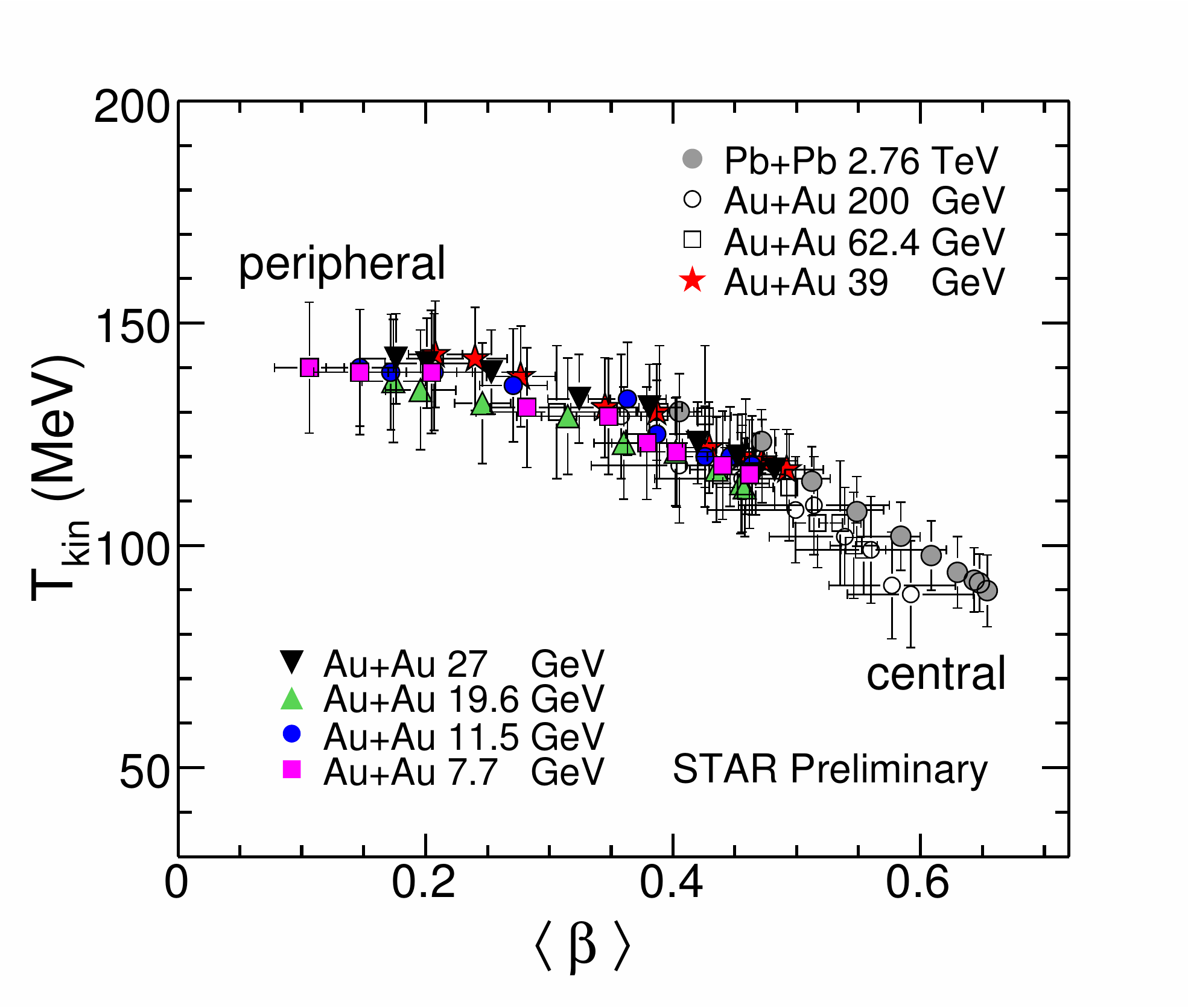} 
\includegraphics[width=.475\textwidth, height=0.4\textwidth]{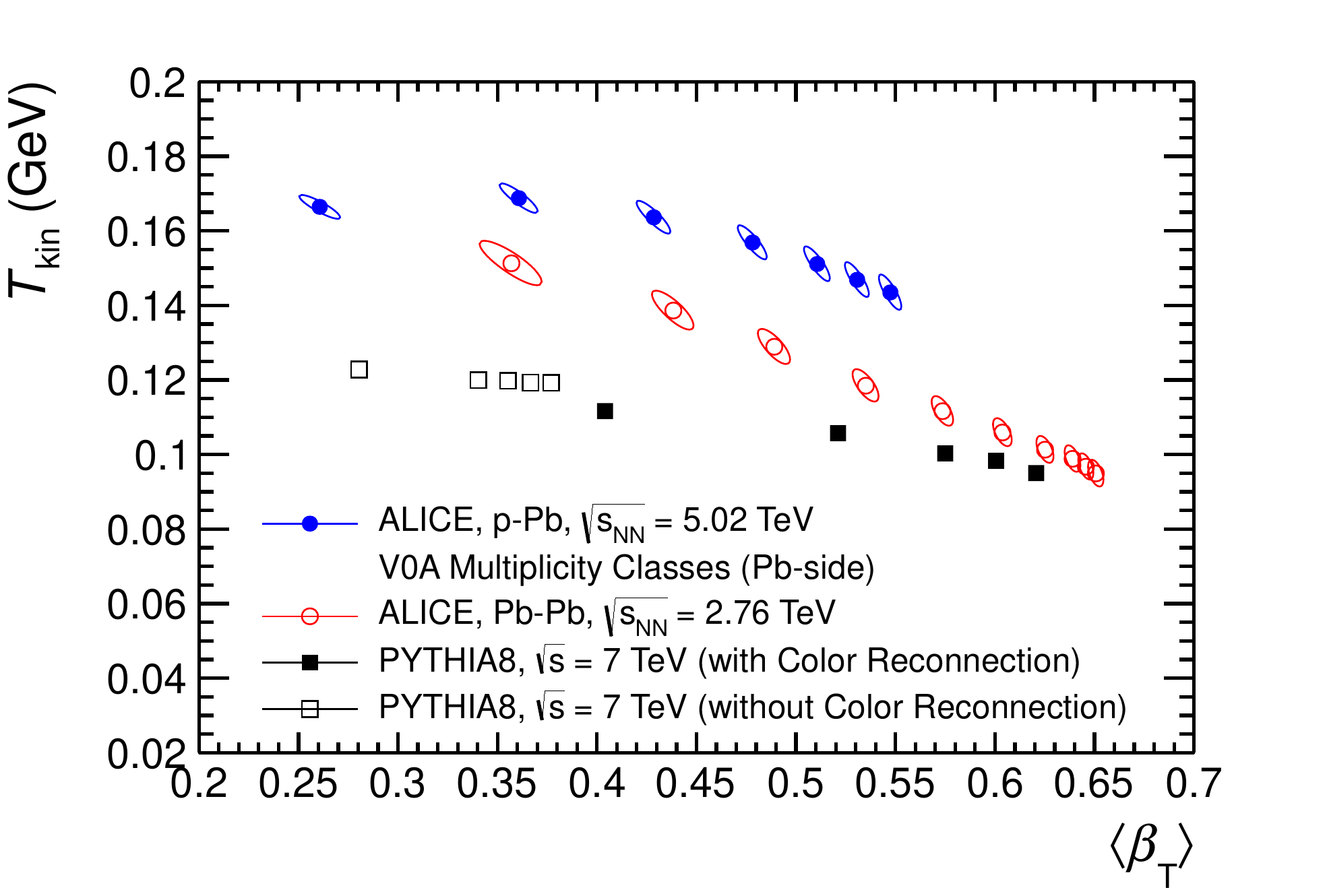}
\caption{\textbf{Left}: Variation of the blast wave parameters $T_{kin}$ and $\left<\beta \right>$ obtained from the fits to the spectra of pions, kaons, protons, and 
their anti-particles produced in \auau and \pbpb collisions at energies  \sqsn = 7.7 GeV $-$ 2.76 TeV and different collision centralities. The centrality increases 
from left to right for a given energy. Reproduced from Ref.~\cite{Kumar:2014tca}; \textbf{Right}: The same, but for \pbpb at \sqsn ~=~2.76~TeV and \ppb at \sqsn = 5.02 TeV.  MC~simulations of \pp collisions at \sqs = 7 TeV using the PYTHIA8 event generator \cite{Sjostrand:2007gs} with and without color reconnection are shown 
as open and filled squares, respectively. Reproduced from Ref.~\cite{Abelev:2013haa}.
}
\label{fig:freeze-out}
\end{figure}

In Figures~\ref{fig:models} and \ref{fig:freeze-out}, the examples of the BW fit analysis are presented. As can be seen from Figure~\ref{fig:freeze-out}, the kinetic freeze-out 
temperature $T_{kin}$ that determines the shape of the \pt-spectra of particles is strongly anti-correlated with the radial flow velocity $\left<\beta \right>$: 
the higher  $T_{kin}$ is, the lower $\left<\beta \right>$ is, and vice versa. Nevertheless, the radial flow reveals itself as a shoulder structure at small transverse 
momenta in the \pt-spectra of $\Lambda$s, protons, and kaons; see Figure~\ref{fig:models}. For the pions, there is almost no sensitivity to distinguish between 
the two cases---a reduction of temperature is almost compensated by the radial flow. 

%
%


\subsection{Initial State Description of Nuclear and Hadronic Interactions}
\label{initial}

An indispensable part of the full description of the experimental data from heavy-ion collisions comes not only from the understanding of its dynamics starting 
from the moment of thermalization, but also at earlier times. In particular, the question of where the observed (local) thermalization of deconfined matter comes 
from is still quite open \cite{Baier:2000sb, Xu:2004mz, Kurkela:2011ti, Romatschke:2016hle}. The importance of event-by-event initial state fluctuations on 
anisotropic collective flow and other final state observables is also worth mentioning \cite{Broniowski:2008vp, Petersen:2010zt, Heinz:2011mh, Noronha-Hostler:2016eow}. 
Since these topics currently remain a significant source of uncertainty in predicting the final state observables,  in the next two paragraphs we will provide 
two alternative ways of describing the initial state of the collision.

\begin{figure}[H]

\centering
\includegraphics[width=120mm]{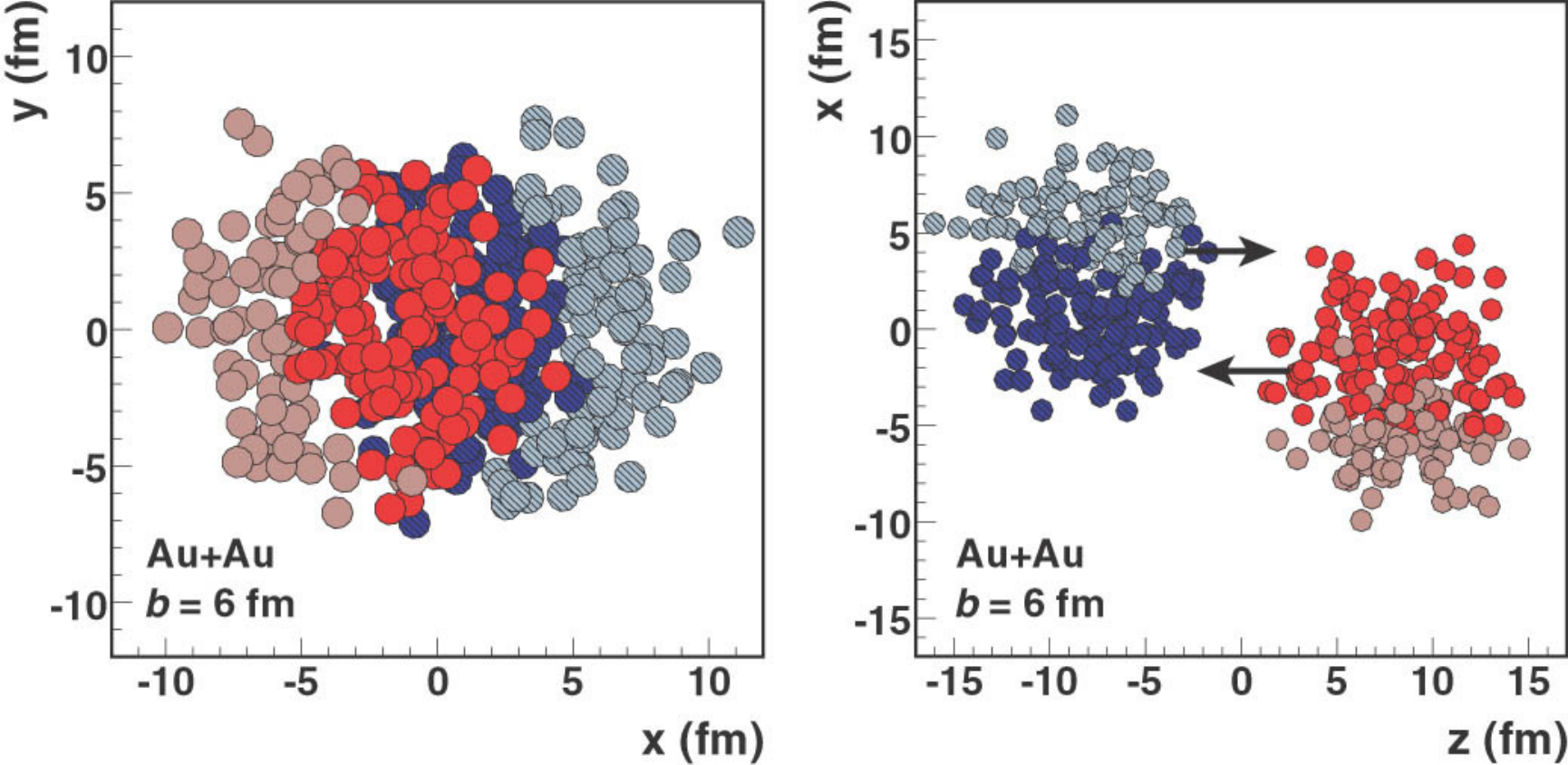}
\caption{
A Glauber Monte-Carlo event (\auau at \sqsn = 200~GeV) viewed in the transverse plane (\textbf{left} panel) and along the beam axis (\textbf{right} panel). 
The nucleons are drawn with diameter $\sqrt{\sigma^{in}_{NN}/\pi}$. Darker disks represent the participants, and lighter disks represent the spectators. 
Reproduced from Ref.~\cite{Miller:2007ri}.
\label{fig:glauber_mc_event}
}
\end{figure}

In high-energy nucleus--nucleus (A $+$ B) interactions, the de Broglie wavelength of the nucleons (N) of the incoming nucleus is much smaller than 
the inter-nucleon distances inside the partner nucleus. To each incoming nucleon, the positions of the nucleons within the partner nucleus appear to be frozen. 
After a single elementary NN (elastic or inelastic) collision, both participating nucleons acquire a transverse momentum, which in the majority of cases is very small 
compared to their longitudinal one, and so the longitudinal momenta before and after the collision are very close to each other $p_z \approx p_{z'}$. 
High incident energies and small scattering angles mean that the scattering is dominated by a large orbital momentum $\ell$, and so it is convenient 
to replace the partial-wave expansion of the scattering amplitude by an impact parameter $b$ = (1 $+$ $\ell$)/p representation. The A $+$ B collision can thus be 
described using a semi-classical approach due to {Glauber} \cite{Glauber:1955qq, Czyz:1969jg, Glauber:1970jm, Glauber:2006gd}, 
which treats the nuclear 
collision as multiple NN interactions \cite{Joachain:1973yv, Franco:1978yw}. The nucleons which have suffered at least one NN collision are called the {\it participants}, and
those who have avoided it are called the {\it spectators} (see Figure~\ref{fig:glauber_mc_event}). The total number of spectators and participants thus adds up 
to N$_{\rm{spec}}$+\Npart = A $+$ B. On the other hand, the total number of collisions suffered by all participants fulfils inequality \Ncoll $\ge$ \Npart/2.

\subsubsection{Glauber Model}
\label{Glauber}

Using the nuclear mass number density $\int dz d^2{\mathbf s} ~\rho_{A,B}(z,{\mathbf s} )$ = A,B and the inelastic NN cross-section $\sigma^{in}_{NN}$, 
we can express \Npart and \Ncoll analytically \cite{Yagi:2005yb, Vogt:2007zz, Miller:2007ri, Florkowski:2014yza}:
\begin{eqnarray}
\displaystyle
\Npart(b)&=&\int d^2{\mathbf s} ~T_A({\mathbf s})\Big(1-e^{-\sigma^{in}_{NN}T_B({\mathbf s})}\Big)
+ \int d^2{\mathbf s} ~ T_B({\mathbf {s-b}})\Big(1-e^{-\sigma^{in}_{NN}T_A({\mathbf s})}\Big) \\
\Ncoll(b)&=&\int d^2{\mathbf s} ~ \sigma^{in}_{NN} T_A({\mathbf s})T_B({\mathbf {b-s}}) \equiv \sigma^{in}_{NN}T_{AB}(\mathbf b)\;,
\label{Glauber}
\end{eqnarray}
where $T_{A}(\mathbf b)$ and $T_{AB}(\mathbf b)$ are the nuclear thickness and the nuclear overlap functions, respectively:
\begin{eqnarray}
\displaystyle
T_{A}(\mathbf b)\equiv \int dz \rho_{A}(z, {\mathbf s})\;, \quad  T_{AB}(\mathbf b) \equiv \int d^2{\mathbf s}~ T_A({\mathbf s})T_B({\mathbf {b-s}}) \,.
\label{thickness}
\end{eqnarray}

\begin{figure}[H]
\centering
\includegraphics[trim={ 0 .43cm 0 .4cm}, clip, width=.5\textwidth, height=.4\textwidth]{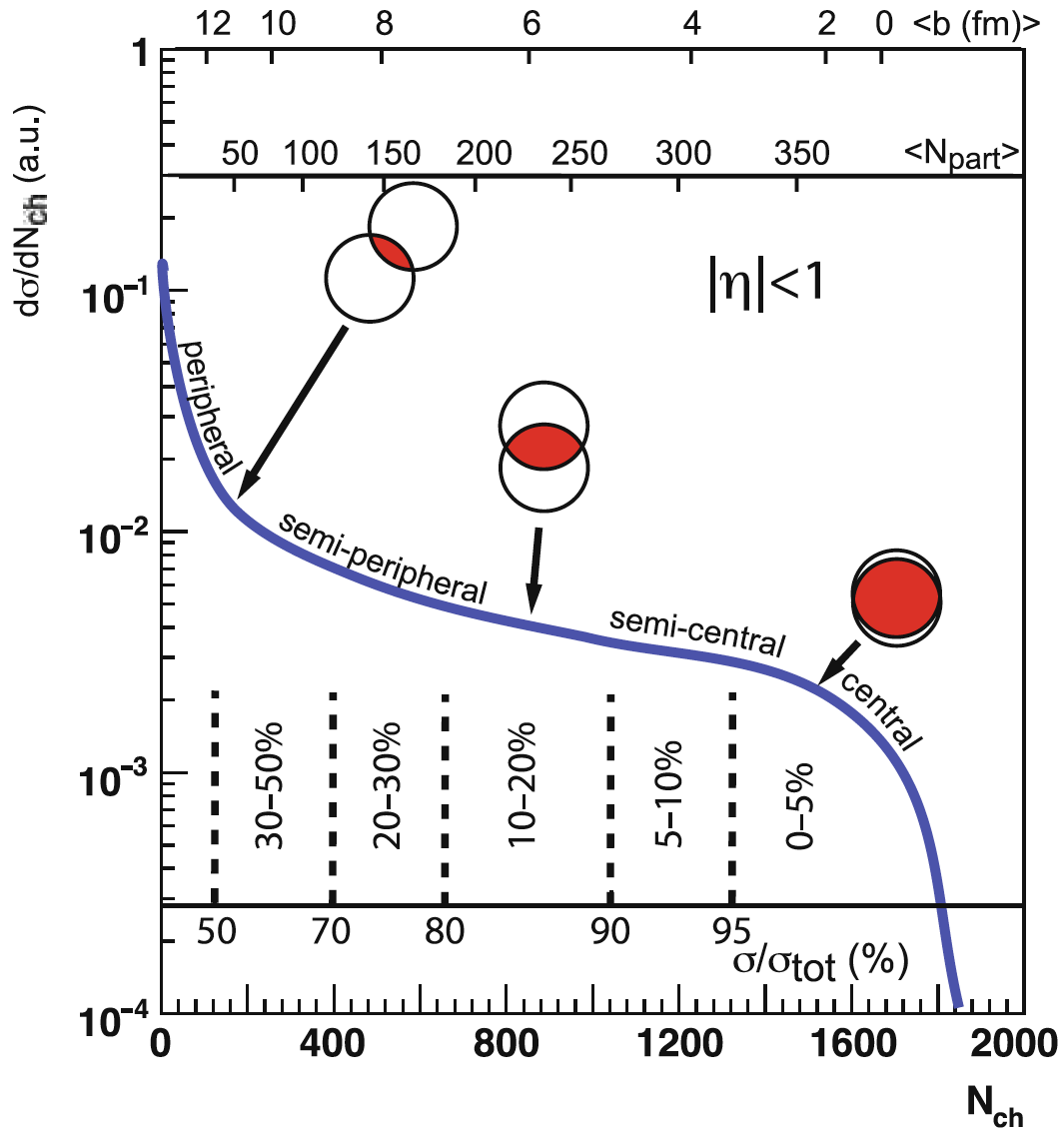} 
\quad\includegraphics[width=.435\textwidth, height=.4\textwidth]{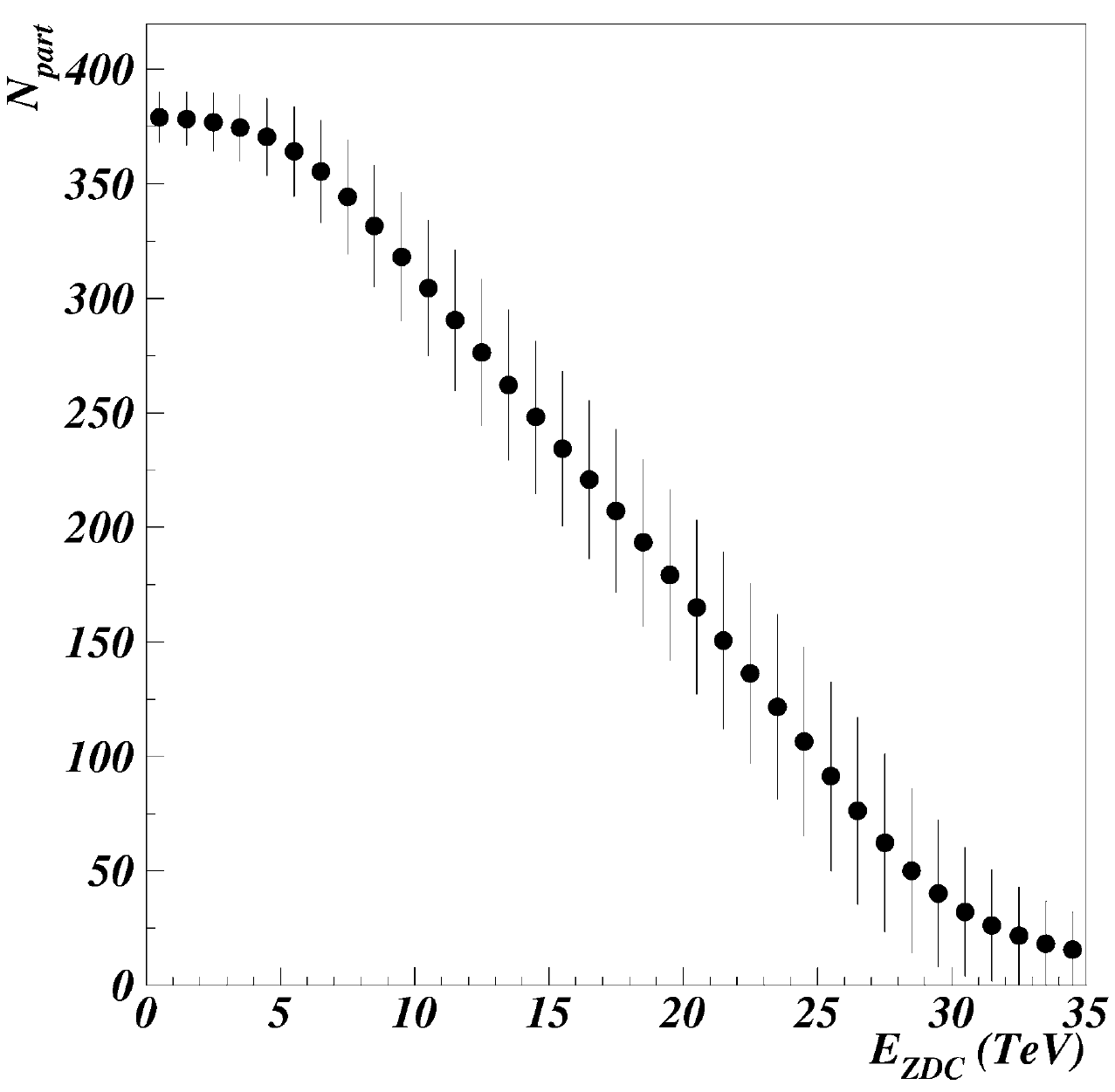}
\caption{\textbf{Left}: A cartoon showing the centrality definition from the final-state charged particle multiplicity \Nch and its
correlation with the average impact parameter $\langle b\rangle$ and the mean number of nucleons participating in the collision <\Npart>. 
Reproduced from Ref.~\cite{Sarkar:2010zza}; \textbf{Right}: The number of participants \Npart as a function $E_{ZDC}$ in \pbpb collision 
at incident momentum per nucleon 158  GeV/{\it c} (\sqsn= 17.3 GeV) calculated in the Glauber approach. Reproduced from Ref.~\cite{Abreu:2001kd}. 
The~error bars represent the r.m.s.
~of the \Npart distribution at fixed $E_{ZDC}$.}
\label{centrality1}
\end{figure}

\begin{figure}[H]\vspace{-24pt}
\centering
\includegraphics[trim={ 0 0 1.9cm 0}, clip, width=.45\textwidth, height=.435\textwidth]{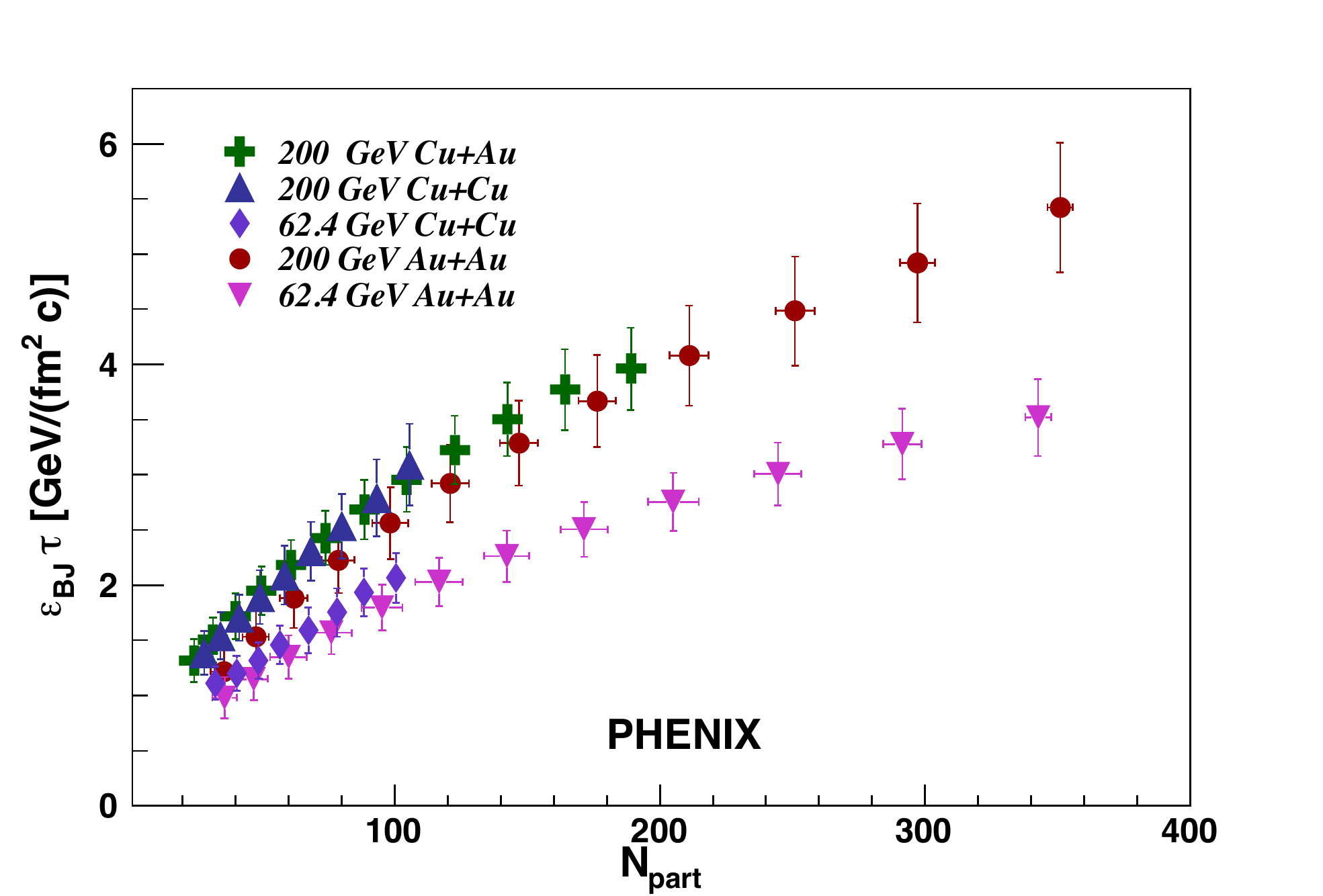} \quad \quad
\includegraphics[trim={ .3cm 0 0 0}, clip, width=.45\textwidth, height=.39\textwidth]{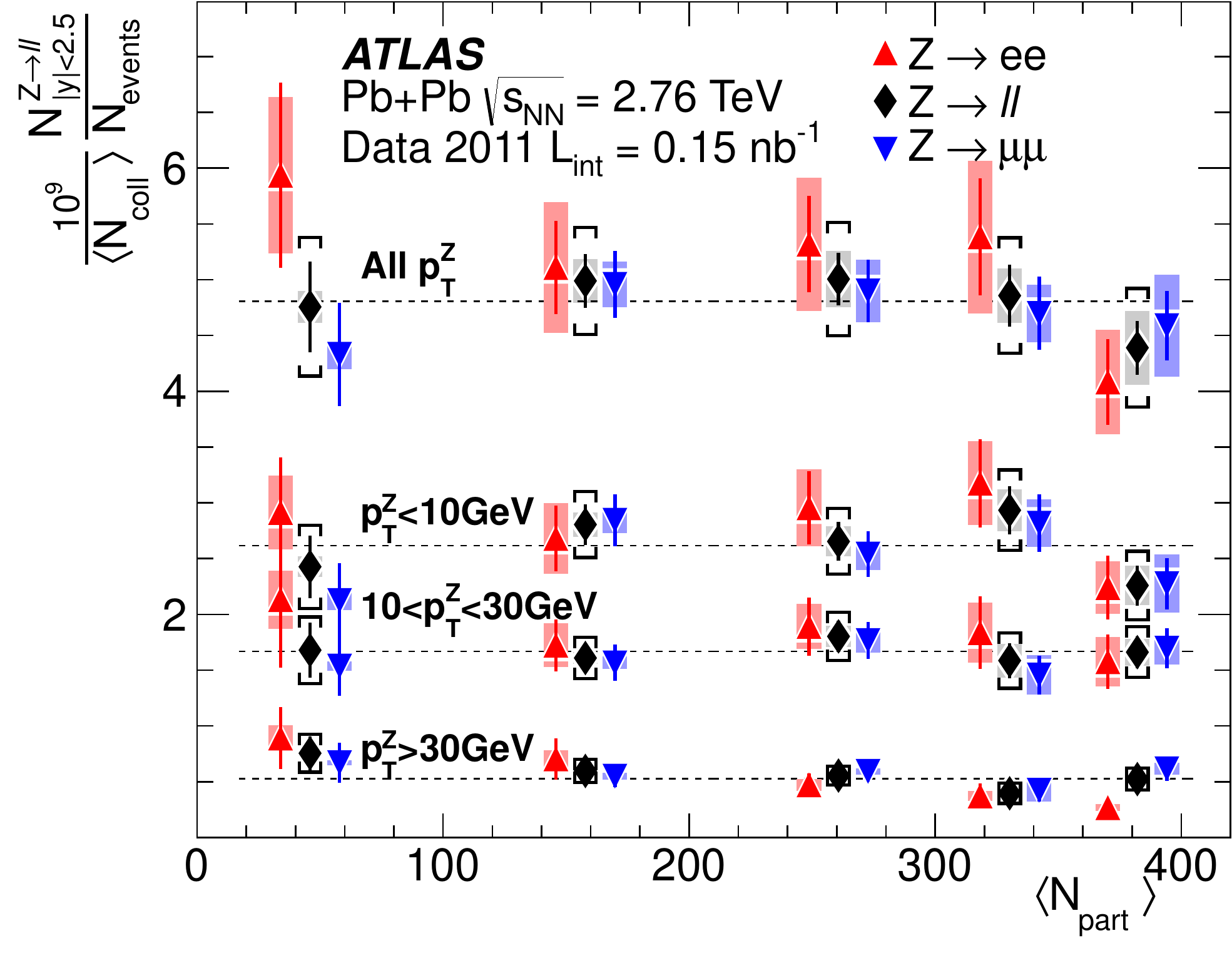} 
\caption{\textbf{Left}: The Bjorken estimate of the initial energy density \ebj  (Equation~\ref{ebj}) multiplied by the thermalization time $\tau$ as 
a function of \Npart for different colliding systems (\auau, \cuau, \cucu) at two energies (\sqsn = 62.4 and 200 GeV). Reproduced from 
Ref.~\cite{Adare:2015bua}; \textbf{Right}: The dependence of $Z$ boson yields divided by average number of collisions <\Ncoll> on the average number 
of participants <\Npart> for \pbpb collisions at \sqsn = 2.76 TeV. Reproduced from Ref.~\cite{Aad:2012ew}.}
\label{centrality2}
\end{figure}

The Glauber model calculations are also often carried out via {Monte Carlo} \cite{Miller:2007ri, Broniowski:2007nz, Loizides:2014vua}. 
Nucleons inside the colliding nuclei 
are distributed randomly according to a nuclear density profile. At a given impact parameter, $\mathbf b$, the impact parameter $\mathbf s$ of all pairs of nucleons 
is calculated. Interaction occurs when $\pi s^2$ < $\sigma^{in}_{NN}$;  see Figure~\ref{fig:glauber_mc_event}. The calculated \Npart(b) and \Ncoll(b) are then used 
to make a contact with the measured bulk observables, like the multiplicity of charged particles \Nch measured by the tracking detectors at midrapidity (Figure~\ref{centrality1}, left panel), or the energy left by the spectator nucleons in the Zero Degree Calorimeter $E_{ZDC}$ (Figure~\ref{centrality1}, right panel). The left panel also explains how 
different bins in multiplicity of charged particles \Nch can be transformed into the bins in collision centrality. The left panel of Figure~\ref{centrality2} illustrates that even 
for quite different nuclear systems (\auau, \cuau, \cucu), the number of participants \Npart selects the collisions with almost the same energy density \ebj. 
The right panel of Figure~\ref{centrality2} shows that centrality dependence of the yields of EW-interacting particles (like direct photons, $W^\pm$, or $Z$-bosons) 
is completely determined by the \Ncoll. 

A generalisation of the Glauber model beyond its original non-relativistic potential description of the scattering process was first formulated by 
Gribov \cite{Gribov:1968fc, Gribov:1968jf}, who used the effective field theory to describe multiple interactions proceeding via the Pomeron exchange. 
The interference terms appearing naturally in this treatment of multiple scattering automatically assure  the unitarity of the theory \cite{Abramovsky:1973fm}. 
With the advent of the QCD, the parton-based multiple scattering models became popular \cite{Braun:1988pk, Drescher:2000ha, Ostapchenko:2010vb, 
Pasechnik:2010zs, Pasechnik:2010cm}. Such models allow the treatment of A $+$ B and NN collisions on more equal footing. Their participants could thus be not only 
nucleons, but also the pQCD
~partons, the valence quarks \cite{Braun:1988pk, Nemchik:2008xy}, or any effective sub-nucleon degrees of freedom \cite{Loizides:2016djv}.


\subsubsection{QCD Scattering in the Dipole Picture: Initial-State Energy Loss and Shadowing} 
\label{dipole}

The color dipole formalism \cite{nik,zkl} provides phenomenological means for the universal treatment of inclusive and diffractive \pp, \pa, and A$+$A 
collisions at high energies. In reactions involving a hard scale such as the Deep Inelastic Scattering (DIS), Drell--Yan (DY), heavy quark production, etc., 
this formalism effectively accounts for the higher-order QCD corrections  and enables the quantification of the initial state interaction, saturation, 
gluon shadowing, and nuclear coherence effects~\cite{nik,nik_dif,k95,bhq97,kst99,krt01,npz-1,npz-2}. At small $x$, the eigenstates of interaction correspond 
to color dipoles with a definite transverse separation, the main ingredients of the dipole picture. Provided that the initial projectile state can be considered 
as a collection of dipoles, any production process in the target rest frame is then viewed as a superposition of universal dipole-target partial amplitudes 
$f_{\rm el}({\bf b},{\bf r};x)$ at different dipole separations ${\bf r}$, impact parameters ${\bf b}$, and Bjorken $x$. Integrating the amplitude over ${\bf b}$,
\begin{eqnarray}
 \sigma_{\bar qq} = \int d^2b\,2{\rm Im} f_{\rm el}({\bf b},{\bf r};x)\,,
 \label{Sig-dip}
\end{eqnarray}
one arrives at the universal dipole cross-section $\sigma_{\bar qq}=\sigma_{\bar qq}(r,x)$, which is determined phenomenologically. For example, 
the DIS process in the target rest frame is viewed as a scattering of the ``frozen'' $q\bar q$ dipole of size 
$r$$\sim$$1/Q$, originating as a fluctuation of the virtual photon $\gamma^* \to q\bar q$ off the target nucleon. The DY pair production is considered 
as a bremsstrahlung of massive $\gamma^*$~(and~$Z^0$~boson) off a projectile quark before after~and the quark scatters off the target \cite{k95}.
The projectile $q\bar q$ dipole probes the dense gluonic field in the target at high energies when nonlinear (e.g., saturation) effects due to multiple scatterings 
can become relevant. The latter, however, cannot be reliably predicted due to unknown soft and non-perturbative QCD dynamics. 

In the limit of point-like color-neutral dipole $r\to 0$, the corresponding dipole cross-section vanishes quadratically $\sigma_{\bar qq}\propto r^2$ due 
to color screening, which is known as the color transparency~\mbox{\cite{zkl,BBGG,BM}}. A naive saturated shape of the dipole cross-section known as 
the Golec-Biernat--W\"usthoff ansatz originally fitted to the DIS data \cite{GolecBiernat:1998js}
\begin{equation}
\label{gbw}
\sigma_{q\bar{q}}\propto 1 - e^{-\frac{r^2 Q_s^2(x)}{4} } \,,  \qquad Q_s^2(x) \propto \left( \frac{x_0}{x} \right)^\lambda{,}
\end{equation}
with the energy-dependent saturation scale $Q_s$ denoting a characteristic boundary between the linear and non-linear QCD regimes. Such a simple
ansatz has proven to successfully capture the major features of a vast scope of observables in \pp and \pa, while the {DGLAP}~evolution of the
gluon density in the target affects the saturation scale at large scales $\mu^2$ such that a QCD-corrected ansatz $Q_s^2\propto \alpha_s(\mu^2)\,xg(x,\mu^2)$ 
is often used in practice \cite{Bartels:2002cj,kmw-1,kmw-2,ipsatnewfit}.

The saturation effects are amplified in nuclear collisions, such that the nuclear saturation scale $Q_s^A$ is expected to be enhanced w.r.t. the nucleon one by 
a factor of $A^{1/3}$. Then, the dipole--nucleus cross-section $\sigma_{q\bar{q}}^A$ can be represented in terms of the forward partial dipole--nucleus amplitude 
$f_{\rm el}^A$ analogically to Equation~(\ref{Sig-dip}). Evolution of this amplitude in rapidity $Y=\ln(1/x)$ is predicted, for example, in the Color Glass Condensate formalism 
(see below) {;} while no first-principle calculation capable of predicting the ${\bf b}$-dependence exists, only phenomenological fits inspired by saturation physics 
are implemented (aside from what is mentioned above, see Refs.~\cite{iim,kkt,dhj,buw,Soyez2007,Kowalski:2003hm,amirs}).

In heavy-ion collisions at high energies, due to the absence of final state interactions, the DY pair production process on nuclear targets in the dipole picture is often 
considered to be an excellent probe accessing the impact parameter dependence of the Initial State Interaction (ISI) effects as well as nuclear shadowing and nuclear 
broadening---the crucial information that cannot be derived within the parton model (e.g., ~\cite{Basso:2015pba,Basso:2016ulb,Goncalves:2016qku} and references therein). 
The ISI effects emerging due to multiple rescattering of projectile partons in a medium prior to a hard scattering are only relevant close to kinematic boundaries due 
to energy conservation significantly suppressing the nuclear cross-sections~\cite{kopeliovich-isi-1,kopeliovich-isi-2}; e.g., when the Feynman variable 
$x_L\equiv x_F=2 p_L/\sqrt{s} \to 1$ and $x_T=2 p_T/\sqrt{s} \to 1$. In~order to account for the ISI-induced energy loss in the Glauber approximation, 
one sums up over initial state interactions in a \pa collision at a given impact parameter $b$, leading to a nuclear ISI-improved projectile quark {PDF}
~\cite{kopeliovich-isi-1,kopeliovich-isi-2}:
\begin{equation}
\label{eq-ISI}
q_{f}(x,Q^2) \Rightarrow q_{f}^A(x,Q^2,{\bf b}) = C_v \, q_{f}(x,Q^2)\,
\frac{e^{-\xi \sigma_{\rm eff}T_A({\bf b})}-e^{-\sigma_{\rm eff}T_A({\bf b})}}
{(1-\xi)(1-e^{-\sigma_{\rm eff}T_A({\bf b})})} \,, \qquad \xi = \sqrt{x_L^2 + x_T^2} \,,
\end{equation}
where $T_A({\bf b})$ is the nuclear thickness function defined in Equation~(\ref{thickness}), $C_v$ is fixed by the Gottfried sum rule, and $\sigma_{\rm eff}=20$~mb 
is the effective cross-section controlling the multiple ISI. The ISI-induced energy loss can induce a significant suppression at large $M_{l\bar{l}}$, $p_T$, and 
forward rapidities. In fact, it has been noticed earlier in $\pi^0$ production in central $dAu$ collisions \cite{phenix-isi-dAu}, and in direct photon production 
in central $AuAu$ collisions \cite{phenix-isi-AuAu-1,phenix-isi-AuAu-2} at midrapidity and large transverse momenta (where no shadowing is expected) by 
the PHENIX Collaboration.

In the course of propagation, a projectile parton experiences multiple rescatterings in the color medium of a target nucleus. These cause a notable suppression of 
the gluon radiation and energy loss by the parton---a QCD analog of the Landau--Pomeranchuk--Migdal (LPM) effect in QED \cite{Landau:1953um,Landau:1953um-2,Migdal:1956tc} 
(for~more details, see Ref.~\cite{LPM-review} and references therein). In heavy ion collisions, the LPM effect was studied in Refs.~\cite{LPM-ion,LPM-ion-1,LPM-ion-2}.

In the nuclear DY reaction, for example, the shadowing is controlled by the coherence length corresponding to the mean lifetime of $\gamma^*$-quark fluctuations
\begin{equation}
\label{eq-cl}
 l_c = \frac{1}{x_2 m_N} \frac{(M_{l\bar{l}}^2 + p_T^2)(1-\alpha)}{\alpha(1-\alpha) M_{l\bar{l}}^2 + \alpha^2 m_f^2 + p_T^2} \,,
\end{equation}
given in terms of the quark mass $m_f$, nucleon mass $m_N$, dilepton invariant mass $M_{l\bar{l}}$, its transverse momentum $p_T$, and light-cone momentum 
fractions of dilepton $\alpha$ and the target gluon $x_2=x_1-x_F$. In Ref.~\cite{Basso:2016ulb}, it was noticed that the long coherence length (LCL) compared to the nuclear 
radius $R_A$ (namely, $l_c\gg R_A$) is practically useful in kinematic domains of RHIC and the LHC experiments. 

The quark shadowing in the LCL limit is automatically accounted for by scattering of the leading Fock components in the dipole formula which represents a 
process-dependent convolution of the light-cone wave function for a given Fock state with a superposition of partial dipole amplitudes. In~\pa collisions, the latter are given by 
the eikonal Glauber--Gribov form \cite{zkl}:
\begin{eqnarray}
{\rm Im} f_{\rm el}^A({\bf b},{\bf r};x) =
1-\exp\left(-\frac{1}{2}\,T_A({\bf b})\,\sigma_{q\bar{q}}({\bf r},x)\right) \,,
\label{Na}
\end{eqnarray}
which resums high-energy elastic scatterings of the dipole in a nucleus. Such an eikonalisation is justified in the LCL regime where the transverse separation of partons 
in the projectile multiparton Fock state is ``frozen'' in the course of propagation of the dipoles through the nuclear matter such that it becomes an eigenvalue of 
the scattering operator. 

At large scales and at small collision energies ($\sqrt{s} \lesssim 200$ GeV), one recovers $l_c \lesssim 1$ fm such that the LCL approximation fails and an alternative description 
is necessary. In Refs.~\cite{Zakh,Zakh-1,Zakh-2}, a universal path-integral approach to multiple dipole interactions with a nuclear target which can be used for any value of $l_c$ 
has been developed. It is also known as the Green function approach, and consistently incorporates the color transparency and quantum coherence effects responsible 
for nuclear shadowing. Recently, the Green functions technique accounting for the nuclear attenuation of the colorless dipole in the medium has been employed in 
Ref.~\cite{Goncalves:2016qku} in analysis of the nuclear coherence effects in DY reaction in \pa collisions. In Ref.~\cite{BDMPS-2}, it was demonstrated that the Green function technique is equivalent to the diagrammatic formulation developed by the Baier--Dokshitzer--Mueller--Peigne--Schiff Collaboration (e.g., 
Refs.~\cite{BDMPS-1,BDMPS-1-1,BDMPS-1-2}). In Ref.~\cite{Nemchik:2014gka}, it was understood that the basic reason for the observed suppression of high-$p_T$ 
hadron production in heavy ion collisions is the attenuation of early produced colorless dipoles (``pre-hadrons'') propagating through a dense absorptive matter, but is not 
an energy loss because of a much shorter production length than was typically assumed.

In the case of DY, the lowest Fock state $|q\gamma^*\rangle$ scattering is not enough in the LHC kinematics domain, where the higher
Fock components containing gluons (e.g., $|q\gamma^*\,g\rangle$, $|q\gamma^*\,gg\rangle$ etc.) also become noticeable, causing an additional suppression known as
gluon shadowing (GS). These components are heavier than $|q\gamma^*\rangle$, and thus have a shorter coherence length. In the high energy limit, the additional
suppression factor $R_G$ induced by the GS corrections has been computed in Ref.~\cite{kopeliovich-gs,kopeliovich-gs-1,kopeliovich-gs-2} using the Green function 
technique, going beyond the LCL limit as
\begin{eqnarray}
R_G(x,Q^2,{\bf b}) \equiv \frac{x g_A(x,Q^2,{\bf b})}{A\cdot x g_p(x,Q^2)} 
\approx 1 - \frac{\Delta\sigma}{\sigma_{tot}^{\gamma^*A}} \,,
\end{eqnarray}
in terms of the inelastic correction $\Delta\sigma$ to the total nuclear DY cross-section $\sigma_{tot}^{\gamma^*\,A}$, related to the formation
of a next-to-leading $|q\bar{q}\,g\rangle$ Fock component. The GS is a leading-twist phenomenon effective at small $x_2\ll 1$ such that $R_G \to 0$ very slowly 
at $Q^2\to \infty$.


\subsubsection{Color Glass Condensate} 
\label{CGC}

In addition to the sQGP matter, another instance where quarks and gluons cannot be treated as independent degrees of freedom is the case of parton coherence. 
A generalization of pQCD to hard collisions of small-$x$ ($x\ll1$) partons (called also a {\it semi-hard regime}) was first discussed by Gribov, Levin, and Ryskin 
\cite{Gribov:1984tu}. The basic failure of the standard DGLAP approach \cite{Gribov:1972ri, Dokshitzer:1977sg, Altarelli:1977zs} is that it predicts too-fast 
~increase 
of small-$x$ parton density with the scale $Q^2$. Consequently, the growth of hadronic cross-sections proceeds at rate which would sooner or later violate unitarity. 
The proposed solution---parton recombination and saturation---is at variance with the standard assumption that the partons themselves can be considered as 
independent free particles. The parameter determining the probability of parton--parton recombination is the ratio of the parton--parton cross-section to the square 
of the average distance between partons. The fact that the cross-section of such a semi-hard process (which now complies with the unitarity) increases rapidly 
with incident energy gives rise to expectations that (at least asymptotically) bulk particle production in hadron--hadron collisions can be described via 
pQCD \cite{Gyulassy:2004zy}. 

The modern implementation of the above ideas is the Color Glass Condensate (CGC) formalism~\cite{McLerran:1993ka, Kharzeev:2004if, JalilianMarian:2005jf, 
Gelis:2010nm}---a natural generalization of pQCD to dense partonic systems. When applied to heavy nuclei, it predicts strong color fields in the initial stage 
of the collision. The strength of the fields is due to the condensation of low-$x$ gluons into single macroscopic (i.e., classical) field state called 
the CGC. Since the characteristic scale of the parton saturation grows as $Q_{s} \propto A^{1/3}$ \cite{Gyulassy:2004zy, JalilianMarian:2005jf}, 
it is enhanced on nuclear targets. According to the CGC-motivated phenomenology, the saturation phenomena are expected to show up (if not already) 
in \pdau collisions at RHIC and , for sure, 
~in nuclear collisions at the LHC. For example, due to the gluon saturation, the growth of the inelastic nucleon--nucleon 
cross-section $\sigma^{in}_{NN}$ with increasing collision energy \sqs may result in a broadening of the nucleon density distribution in position space. 
This in turn leads to a natural smoothing of the initial energy density distribution in the transverse plane of the matter created near midrapidity in 
heavy-ion collisions \cite{Heinz:2011mh}. 

The CGC is described by an effective field theory that separates two kinds of degrees of freedom---fast frozen color sources and slow dynamical 
color fields. The basic evolution equation of such an effective field theory is an RG equation known as the Jalilian--Marian--Iancu--McLerran--Weigert--Leonidov--Kovner (JIMWLK) 
equation \cite{JalilianMarian:1997jx,JalilianMarian:1997gr,Kovner:2000pt,Weigert:2000gi,Iancu:2000hn,Ferreiro:2001qy}, which reflects the independence of 
physical quantities with respect to variations of the cutoff separating these degrees of freedom (for more details, see Ref.~\cite{Gelis:2010nm} and references therein).

A supporting argument for the CGC as a possible state of QCD matter comes from successful analysis of HERA~data in terms of geometrical scaling
\cite{GolecBiernat:1999qd}. The geometrical scaling is the statement that the total  $\gamma^{*}p$ cross-section depending a priori on two independent variables---the photon 
virtuality $Q^{2}$ and the Bjorken variable $x$---is a function of a single variable $\tau=Q^{2}/Q^{2}_{s}$, where the so-called saturation scale $Q^{2}_{s}$ 
depends nontrivially on $x$, with dimensions given by a fixed reference scale $Q^{2}_{0}$. However, calculations of Ref.~\cite{Caola:2008xr} show that 
the standard linear leading-order DGLAP perturbative evolution is able to explain the geometric scaling. The situation with CGC applicability at current energies 
is thus unsettled (see also Refs.~\cite{Basso:2015pba, Basso:2016ulb}). The experimental data from RHIC and the LHC, as well as exploitation of non-CGC-based models \cite{Nemchik:2008xy} are needed to resolve this problem.

There are two popular representative models of the initial state which are based on the CGC---the KLN
~model \cite{Kharzeev:2001yq} and 
the IP Glasma
~{model} \cite{Schenke:2012wb}. 
A Monte Carlo implementation of KLN CGC initial state \cite{Drescher:2006ca, Heinz:2011mh, Hirano:2012kj} 
is based on the number distribution of gluons produced in the transverse plane given by the $k_T$-factorisation formula \cite{Kharzeev:2001yq}:
\begin{eqnarray}
  \frac{dN_g}{d^2 r_{\perp}dy} & = &\kappa
   \frac{4N_c}{N_c^2-1}
    \int     \frac{d^2p_\perp}{p^2_\perp}
      \int \frac{d^2k_\perp}{4} \;\alpha_s(Q^2)\nonumber  \\
       & \times & \phi_A(x_1,({\mathbf p}_\perp +{\mathbf k}_\perp)^2 /4)\;
       \phi_B(x_2,({\mathbf p}_\perp{-}{\mathbf k}_\perp)^2/4)~.
      \label{eq:ktfac}
\end{eqnarray}

Here, $p_\perp$ and $y$ denote the transverse momentum and rapidity of the produced gluons, and $x_{1,2} = p_\perp\exp(\pm y)$/\sqsn are 
the light-cone momentum fractions of the colliding gluons. The~running coupling $\alpha_s(Q^2)$ is evaluated at the scale 
$Q^2$ = $\max( ({\mathbf p}_\perp -{\mathbf k}_\perp)^2/4,({\mathbf p}_\perp +{\mathbf k}_\perp)^2/4)$. The gluon distribution 
function is given by
\begin{equation}
\label{eq:uninteg}
  \phi_{A}(x,k_\perp^2;{\mathbf r}_\perp)\sim
    \frac{1}{\alpha_s(Q^2_{s,A})}\frac{Q_{s,A}^2}
       {{\rm max}(Q_{s,A}^2,k_\perp^2)}~.
\end{equation}

An overall normalisation factor $\kappa$ is chosen to fit the multiplicity data in most central \auau collisions at RHIC. In the MC-KLN model \cite{Kharzeev:2001yq}, 
the saturation momentum is parameterized by assuming that the saturation momentum squared is 2 GeV$^2$ at $x=0.01$ in \auau collisions at $b=0$ fm 
at RHIC, where $\rho_\text{part}=3.06$ fm$^{-2}$; i.e.,
\begin{equation}
Q_{s,A}^2 (x; {\mathbf r}_\perp)  =  2\ \text{GeV}^2
\frac{\rho_{A}({\mathbf r}_\perp)}{1.53\ \text{fm}^{-2}}
\left(\frac{0.01}{x}\right)^{\lambda} \ .
\label{eq:qs2}
\end{equation}

Here, $\lambda$ is a free parameter which is expected to be in the range of $0.2<\lambda<0.3$ from the global analysis of $e+p$ scattering for 
$x<0.01$~\cite{GolecBiernat:1998js, GolecBiernat:1999qd}. 

\begin{figure}[H]
\centering
\begin{minipage}[t]{5 cm}
\includegraphics[scale=0.18,angle=270]{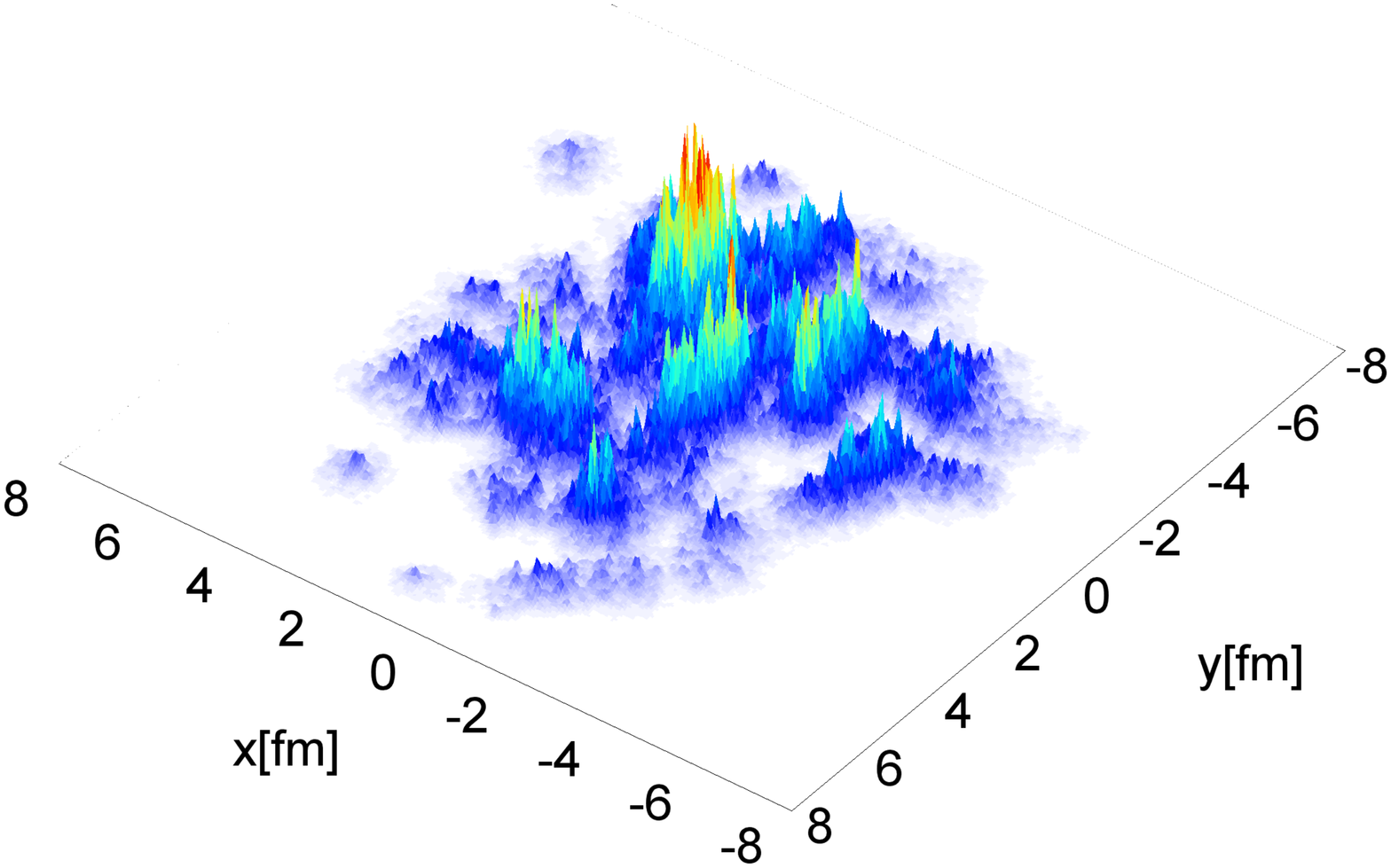}
\end{minipage}
\begin{minipage}[t]{5 cm}
\includegraphics[scale=0.18,angle=270]{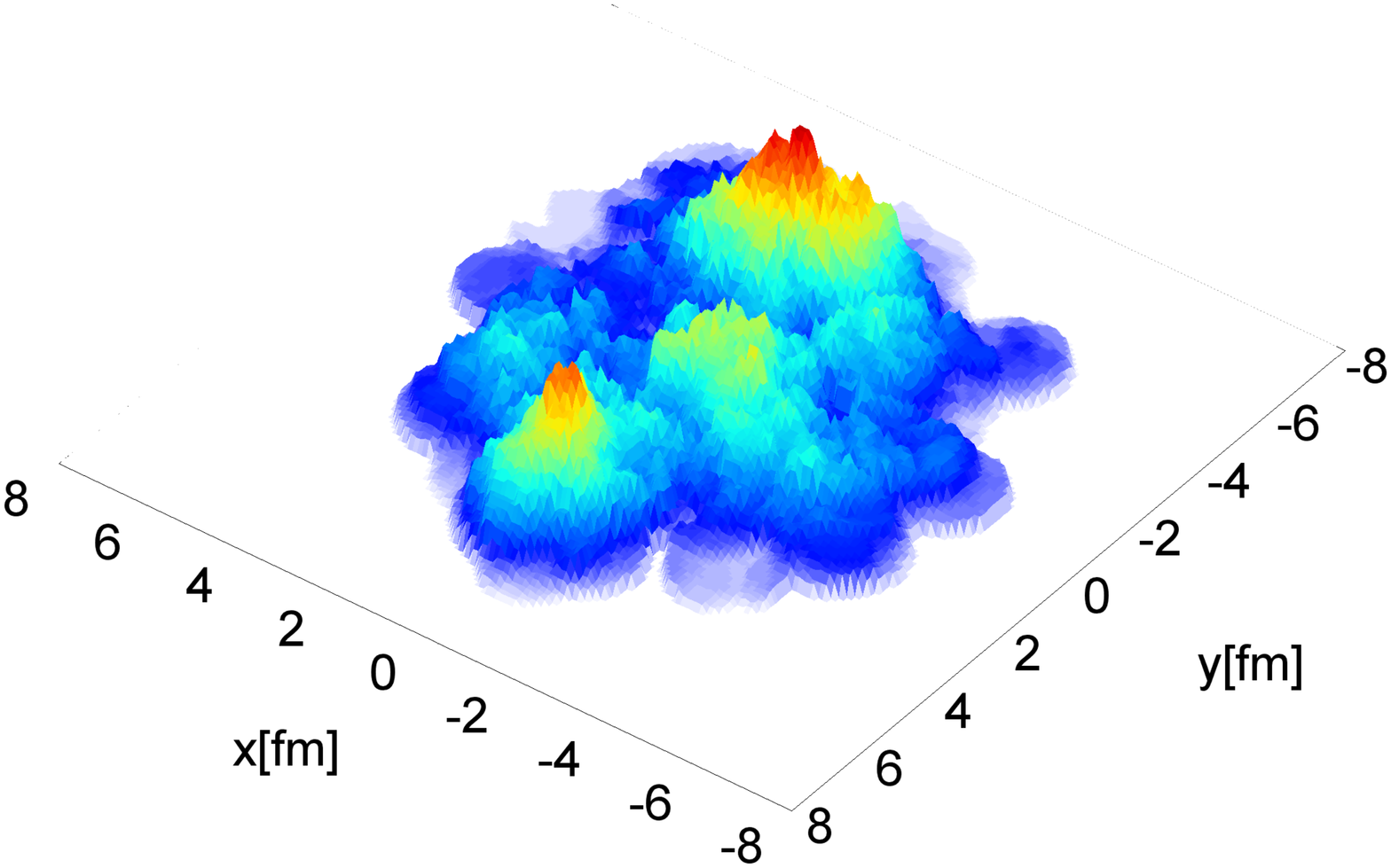}
\end{minipage}
\begin{minipage}[t]{5 cm}
\includegraphics[scale=0.18,angle=270]{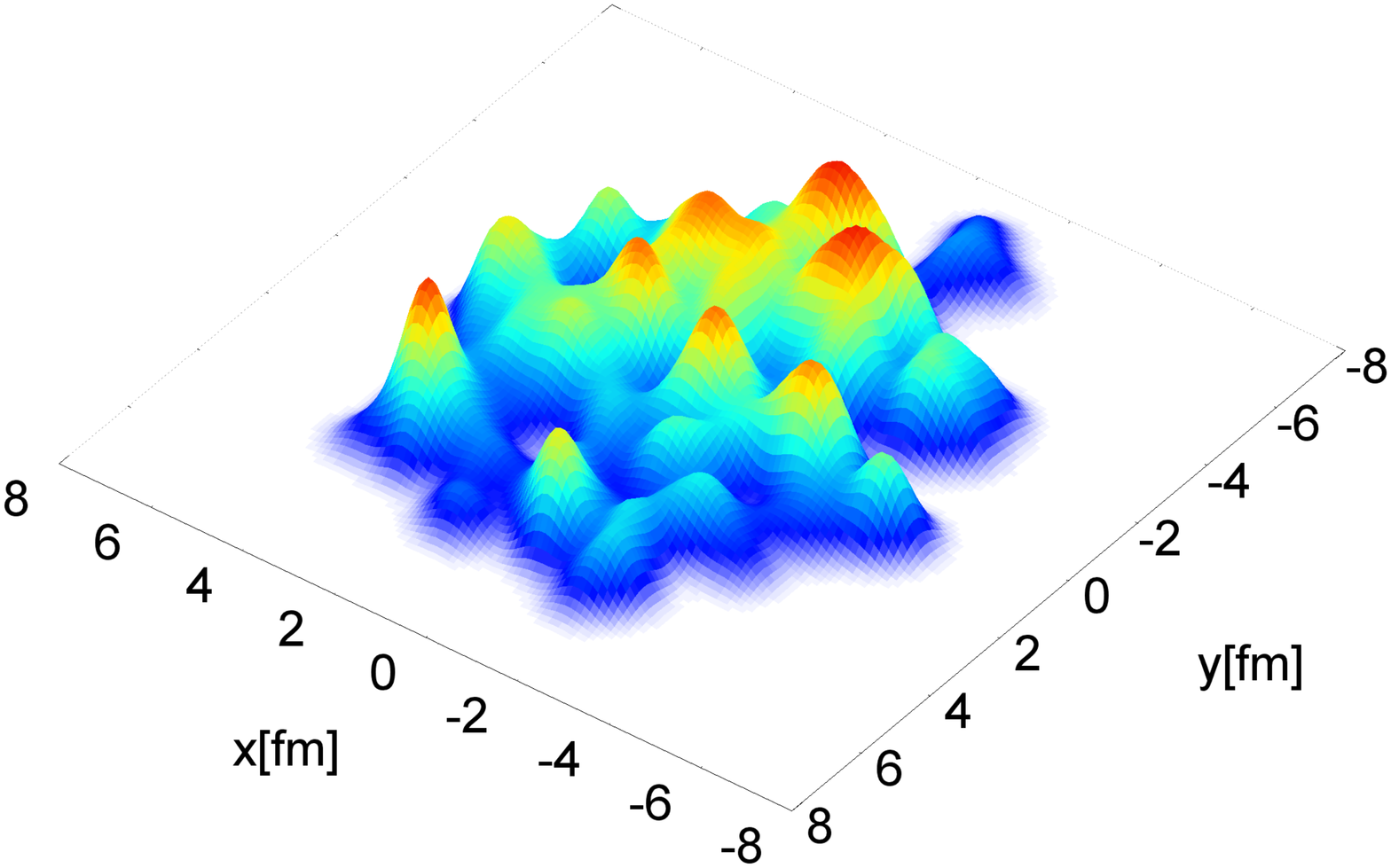}
\end{minipage}
\caption{Examples of the initial energy density distribution from (\textbf{left}) the IP-Glasma model at $\tau = 0$ fm, (\textbf{middle}) the MC-KLN model, 
and  (\textbf{right}) the MC-Glauber model. Reproduced from Ref.~\cite{Schenke:2012wb}.}
\label{fig:IP-Glasma}
\end{figure}

The IP-Glasma model \cite{Schenke:2012wb} solves the classical Yang--Mills equations in which initial charge distributions of two colliding nuclei are sampled 
from a Gaussian distribution with the impact parameter and Bjorken $x$-dependent color charge distributions.
A parameterization of $x$ and impact parameter dependence of the saturation scale is taken from the IP-Sat (Impact Parameter Saturation) 
model \cite{Bartels:2002cj, Kowalski:2003hm}. Fluctuations in the IP-Glasma model have a length scale on the order of the inverse of 
the saturation scale $Q_{s}^{-1}({\mathbf x}_{\perp})$ $\sim$ 0.1-- 0.2 fm. A comparison of the initial energy density distribution among 
the IP-Glasma, MC-KLN, and MC-Glauber models is shown in Figure~\ref{fig:IP-Glasma}.

\begin{figure}[H]
\begin{center}
\hskip -.5cm \includegraphics[width=.5\linewidth]{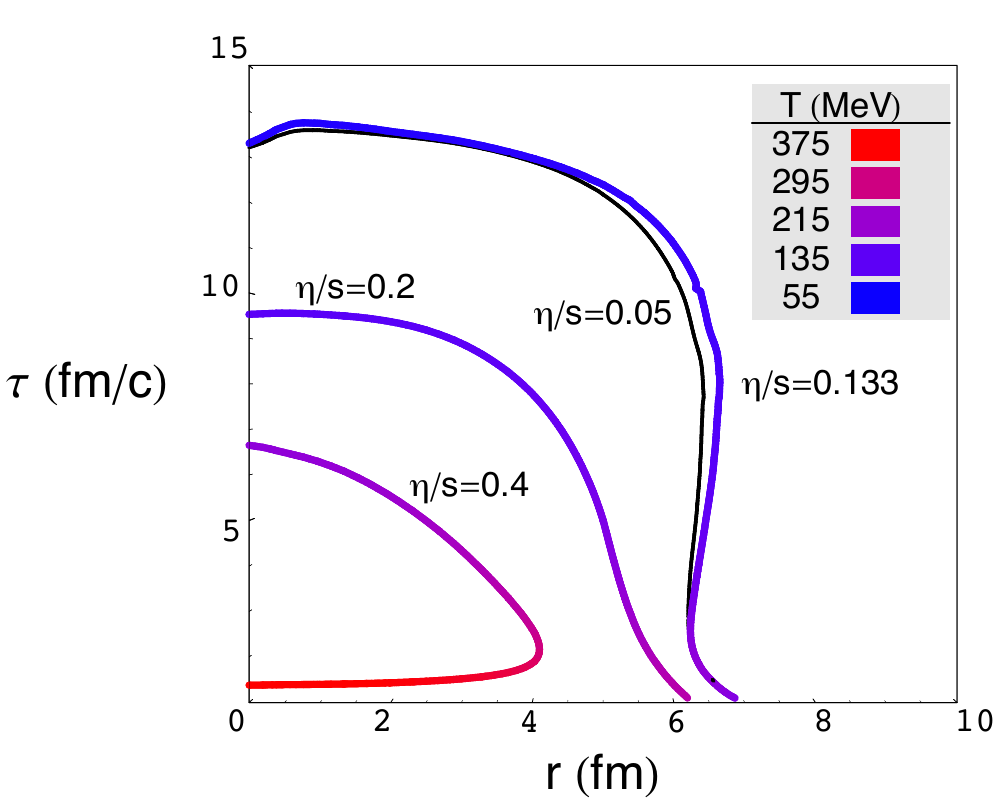}
\includegraphics[width=.52\linewidth, height=.395\linewidth]{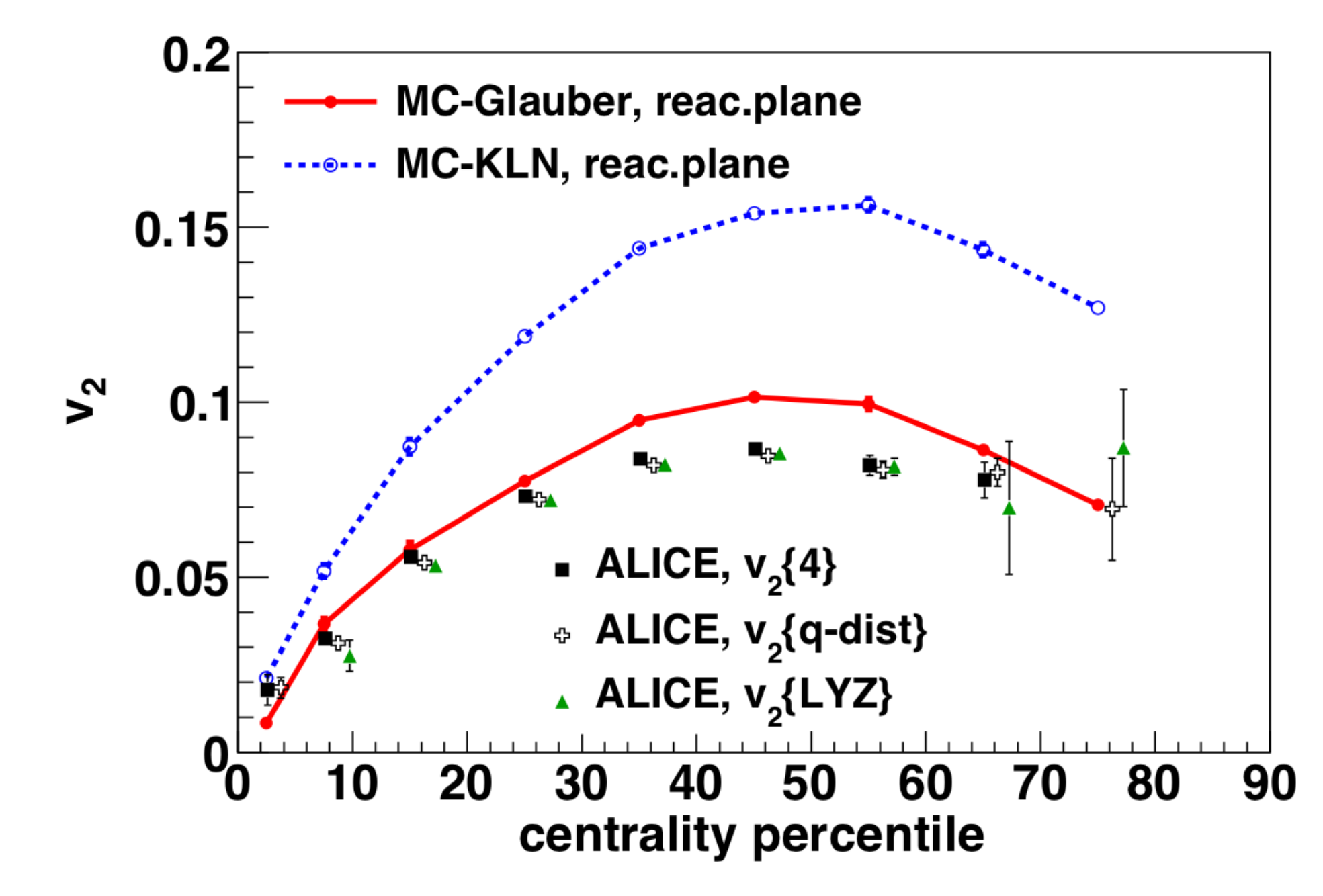}
\end{center}
\vspace*{-.6cm} \caption{\textbf{Left}: Location~of freeze-out surfaces for central \auau collisions \cite{Schafer:2009dj}; 
\textbf{Right}: The centrality dependence of the elliptic flow of charged hadrons from Pb--Pb collisions at \sqsn~=~ 2.76~TeV~\cite{Hirano:2010je}. 
The ALICE data are from Ref.~\cite{Aamodt:2010pa}.}
\label{fig_v1d}
\end{figure}

Figure~\ref{fig_v1d} provides two examples of transport model calculations. The left panel shows the location of freeze-out surfaces for 
central \auau collisions at several fixed values of the shear viscosity-to-entropy density ratio $\eta/s$ obtained from a numerical solution 
of viscosious
~hydrodynamics~\cite{Schafer:2009dj}. The~shading corresponds to the freeze-out temperature. The freeze-out occurs when 
the viscous terms become large compared to the ideal terms. Note that hydrodynamics breaks down not only at late, but also at early times 
(see the curve $\eta/s$ = 0.4 in Figure~\ref{fig_v1d}). The right panel displays the centrality dependence of the elliptic flow coefficient $v_{2}$ (Equation~\ref{vN}) 
for two models for the initial density in the transverse plane---one is motivated by the parton saturation (CGC), and the other exploits nucleons only (Glauber). 
The calculations \cite{Hirano:2010je} were done within a hybrid model, where the expansion of the QGP starting at $\tau_{0}$ = 0.6 fm/c is described by 
ideal hydrodynamics with a state-of-the-art lattice QCD EoS, and the subsequent evolution of hadronic matter below switching temperature $T_{sw}$ = 155 MeV 
is described using a hadronic cascade model. This nicely illustrates the strength of hydrodynamics---either the viscosity of QGP from RHIC to the LHC increases, 
or the CGC initial condition is ruled out \cite{Hirano:2010je}.


\subsubsection{Other Initial State Models} 
\label{Oism}
For completeness, let us here briefly list some other approaches to initial state description used in the analysis of ultra-relativistic heavy ion collisions. 
Fluctuating initial conditions given by a multi-phase transport (AMPT) model \cite{Lin:2004en} were applied in an event-by-event partonic transport plus 
hydrodynamics hybrid approach \cite{Pang:2012he, Bhalerao:2015iya, Xu:2016hmp} to study collective flow. In Refs.~\cite{Paatelainen:2013eea,Niemi:2015qia}, NLO
~pQCD together with saturation-like suppression of low-energy partons were used to calculate the initial energy densities and formation times 
which have been used further on in 3D hydro calculations of space-time evolution of the QCD matter with dissipative fluid dynamics, event by event. 
A new initial conditions model for high-energy \pp, \pa, and A+A collisions that deposit entropy proportional to the generalized mean of nuclear overlap 
density was introduced in Ref.~\cite{Moreland:2014oya}. The model assumes that N one-on-one nucleon collisions produce the same amount of entropy 
as a single N-on-N collision.

\section{Experimental Signatures of Deconfined QCD Matter}
\label{sec:signatures}

\begin{figure}[H]
\centering
\includegraphics[width=0.9\textwidth]{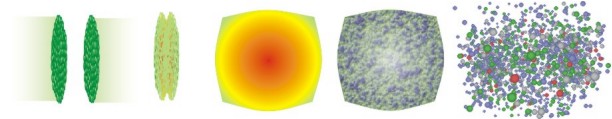}
\caption{Cartoon of a collision of two ultra-relativistic nuclei. 
\textbf{Left} to \textbf{Right}: the two nuclei approach, collide, form a hot and dense equilibrated system, 
QGP expands and hadronizes, and finally hadrons rescatter and freeze out. 
The figure is taken from Ref.~\cite{Bass_time_evolution}.}
\label{fig:evolution}
\end{figure}

Evolution of the high energy nucleus--nucleus collision is schematically depicted in Figure~\ref{fig:evolution}. Two~Lorentz-contracted pancakes 
of nuclear matter collide, thermalize, and form a deconfined QGP medium which expands, cools down, and hadronises to final state hadrons. 
Experimentally we do not observe each stage separately, but only through the time-integrated final state quantities---the momentum spectra of hadrons, 
photons, or leptons, particle multiplicities, energy flow, {etc.}  Nevertheless, some time ordering of different processes giving rise to the final state 
observables exists. At very early collision times when colliding matter thermalizes, the entropy is produced which later---after (almost) isotropic expansion---
transforms into particle multiplicities \cite{Landau:1953, Broniowski:2008vp, Florkowski:2014yza}. Early collision times also favour production of 
high \pt {partons} \cite{Roland:2014jsa, Bielcikova:2016lgh} 
or heavy quarks ({\it c, b}) \cite{Andronic:2015wma}. The formation of QGP reveals 
itself in many ways, including radiation of low momentum direct or virtual photons serving as a thermometers, enhanced production of hadrons 
containing strange ({\it s}) quarks~\cite{Rafelski:1982pu, Koch:1986ud, Letessier:2002gp}, and melting of $c\bar{c}$ or $b\bar{b}$ mesons 
\cite{Matsui:1986dk, Brambilla:2010cs, Andronic:2015wma} called quarkonia. The subsequent rapid expansion of deconfined matter having 
more than ten times the degrees of freedom than the hadronic matter (see~Equation~\ref{dof})---and therefore also much higher internal pressure---
produces a strong radial flow which leaves its imprint on the spectra of final state particles and their yields \cite{Teaney:2000cw, Andronic:2008gu, 
Heinz:2009xj, Broniowski:2008vp, Florkowski:2014yza}.  

In the following, we present several examples of observables related to different stages of dynamics of nucleus--nucleus collisions at high-energies.

\subsection{Bulk Observables} 
\label{bulk}

\begin{figure}[H]
\centering
\includegraphics[width=0.45\textwidth]{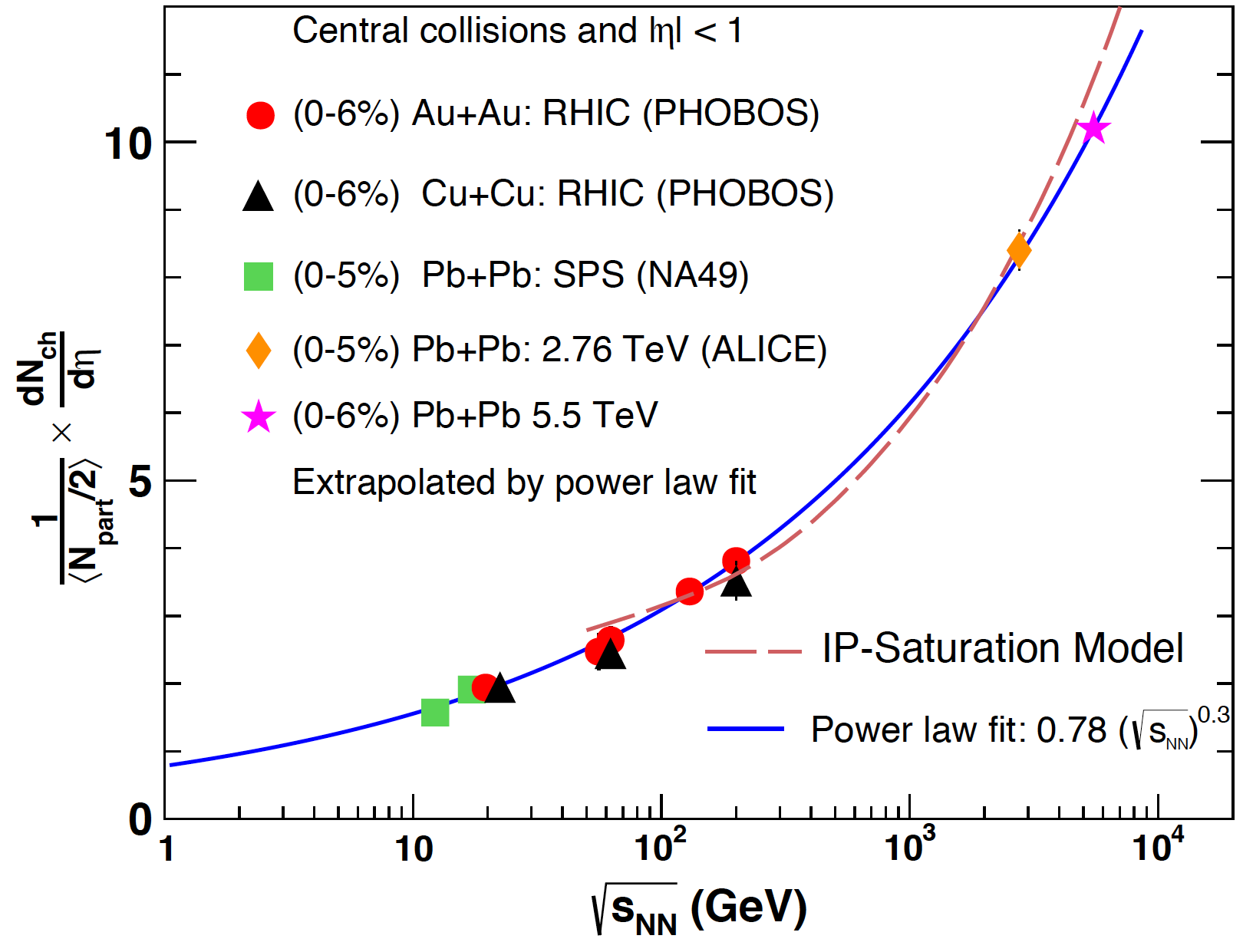} \qquad
\includegraphics[width=0.45\textwidth, height=.35\textwidth]{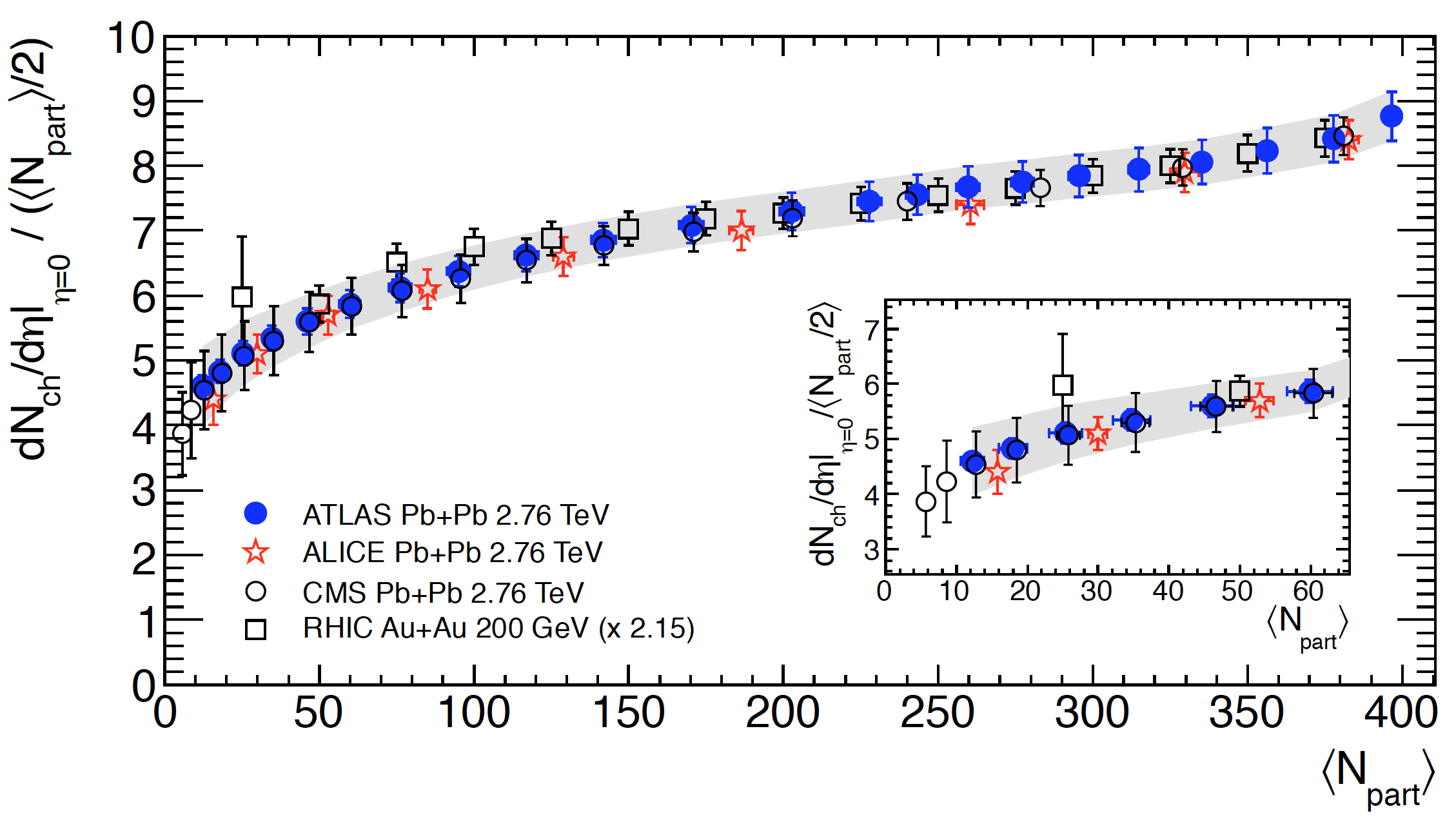}
\caption{\textbf{Left}: The charged particle rapidity density per participaiting pair, \dNch/(0.5\Npart), in central \auau, \cucu, and \pbpb collisions from SPS to 
LHC energies. The star denotes an extrapolation to \pbpb collisions at 5.5 TeV. The IP
-saturation model calculation \cite{Tribedy:2011aa} is illustrated by a dashed curve. Adapted from Ref.~\cite{Nouicer:2015jrf}; \textbf{Right}: \dNch/(0.5\Npart) vs. $\left<\Npart\right>$  in \pbpb and \auau collisions at the LHC 
and RHIC, respectively. The RHIC data are multiplied by 2.15. The~inset shows the $\left<\Npart\right>$ < 60 region in more detail. Reproduced from 
Ref.~\cite{ATLAS:2011ag}.}
\label{fig:MultMid}
\end{figure}

Traditionally, the very first measurements of heavy-ion collisions at a new energy regime comprise the charged-particle density at midrapidity 
{\dNch}$\big |_{\eta=0}$, also including its centrality dependence. Its collision-energy dependence for the 5\% (6\%) most central heavy-ion collisions---normalized per participant pair (i.e., $\left<\Npart\right>$/2)---is presented in Figure~\ref{fig:MultMid} (left panel). The right panel of Figure~\ref{fig:MultMid} 
shows that the normalized charged-particle density is rising with centrality, which means that the particle multiplicity at mid-rapidity increases faster 
than \Npart, presumably due to the contribution of hard processes to the particle production \cite{Roland:2014jsa}. However, this increase is very similar 
to that observed at the top RHIC energy.

\begin{figure}[H]
\centering
\includegraphics[width=.79\textwidth, height=0.3\textwidth]{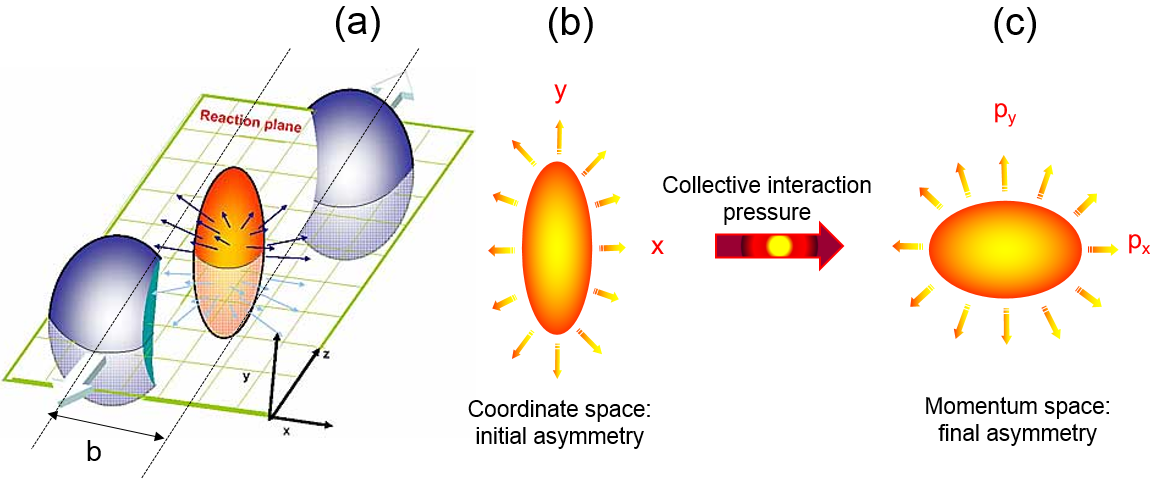} 
\caption{(\textbf{a}) A non-central collision of two nuclei leads to an almond-shaped interaction volume;
(\textbf{b}) This initial spatial anisotropy with respect to the reaction plane  translates via pressure gradients 
into (\textbf{c}) a momentum anisotropy of the produced particles. Reproduced from Ref.~\cite{Nouicer:2015jrf}.}
\label{v2_schema}
\end{figure}

One of the most celebrated predictions of the collective behaviour of matter created in non-central collisions of ultra-relativistic nuclei concerns its evolution 
in the transverse plane, which results from the pressure gradients due to spatial anisotropy of the initial density profile \cite{Ollitrault:1992bk, Heinz:2009xj} 
(see Figure~\ref{v2_schema}). The azimuthal anisotropy is usually quantified by the Fourier coefficients~\cite{Poskanzer:1998yz}:
\begin{equation}
v_n = \mean{\cos[ n (\phi-\Psi_n)]},
\label{vN}
\end{equation}
where $\phi$ is the azimuthal angle of the particle, $\Psi_n$ is the angle of the initial state spatial plane of symmetry, and $n$ is the order of the harmonic. 
In a non-central heavy ion collision, the beam axis and the impact parameter define the reaction plane azimuth $\Psi_{\rm RP}$. For a smooth matter 
distribution in the colliding nuclei, the plane of symmetry is the reaction plane $\Psi_n = \Psi_{\rm RP} $, and the odd Fourier coefficients are zero 
by symmetry. However, due to fluctuations in the matter distribution (including contributions from fluctuations in the positions of the participating 
nucleons in the nuclei---see Figure~\ref{fig:glauber_mc_event}), the plane of symmetry fluctuates event-by-event around the reaction plane. 
This plane of symmetry is determined by the participating nucleons, and is therefore called the participant plane $\Psi_{\rm PP}$~\cite{Manly:2005zy}.  
Since the planes of symmetry $\Psi_n$ are not known experimentally, the anisotropic flow coefficients are estimated from measured correlations 
between the observed particles \cite{Aamodt:2010pa, ALICE:2011ab}.

In the following, we shall restrict ourselves to the properties of the Fourier coefficients $v_n$ with $n =~2$ and $n = 3$, which provide the dominant contributions 
to the observed azimuthal {\it elliptic} and {\it triangular} asymmetry, respectively. The sensitivity of $v_2$ to initial condition is illustrated on Figure~\ref{fig_v1d} (right panel), 
where the centrality dependence of the elliptic flow in \pbpb collisions at \sqsn=2.76 TeV is shown. For more details on the corresponding initial state models, 
see Section~\ref{CGC}.

\begin{figure}[H]
\centering
\includegraphics[width=.5\textwidth, height=0.36\textwidth]{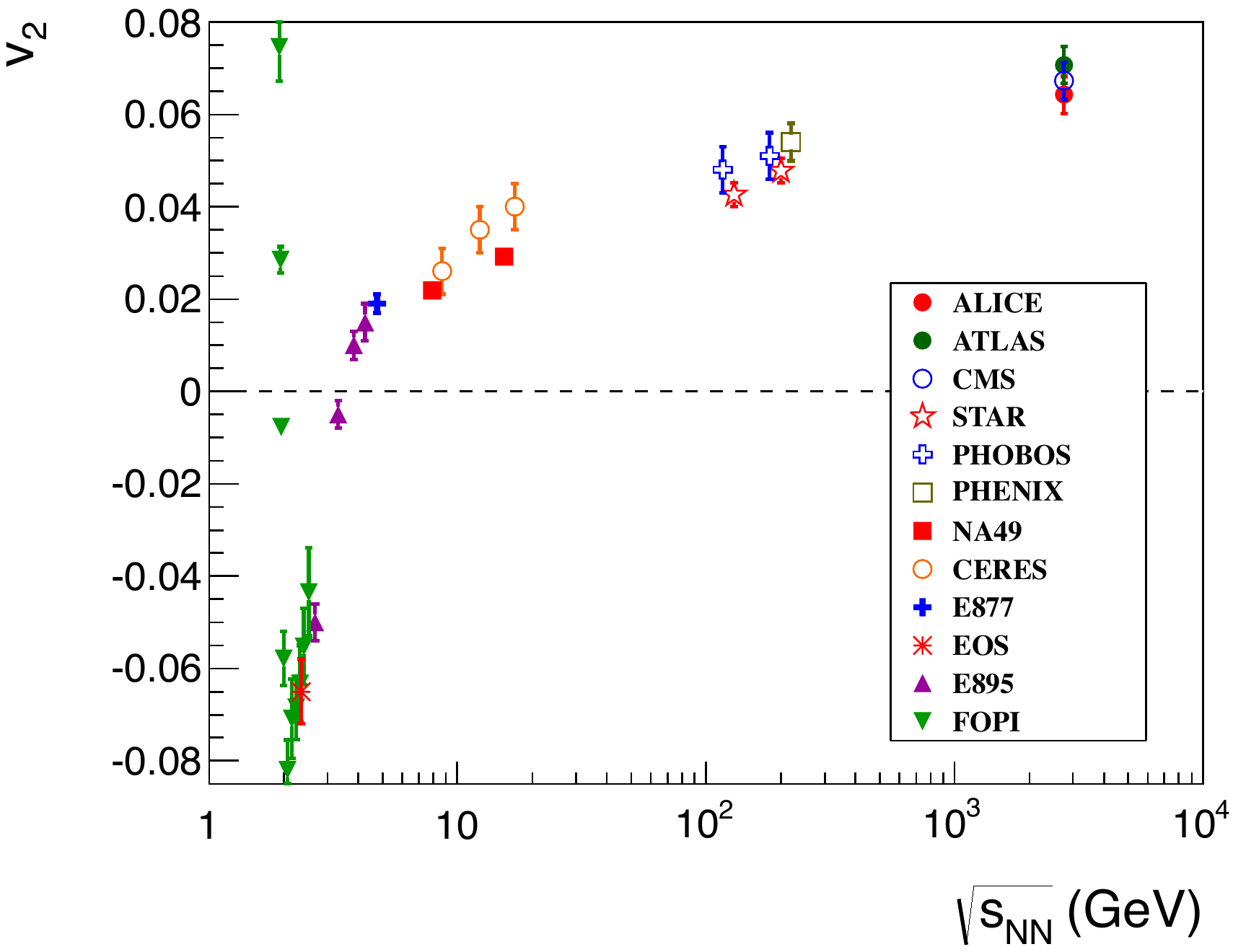}
\includegraphics[width=.48\textwidth, height=0.36\textwidth]{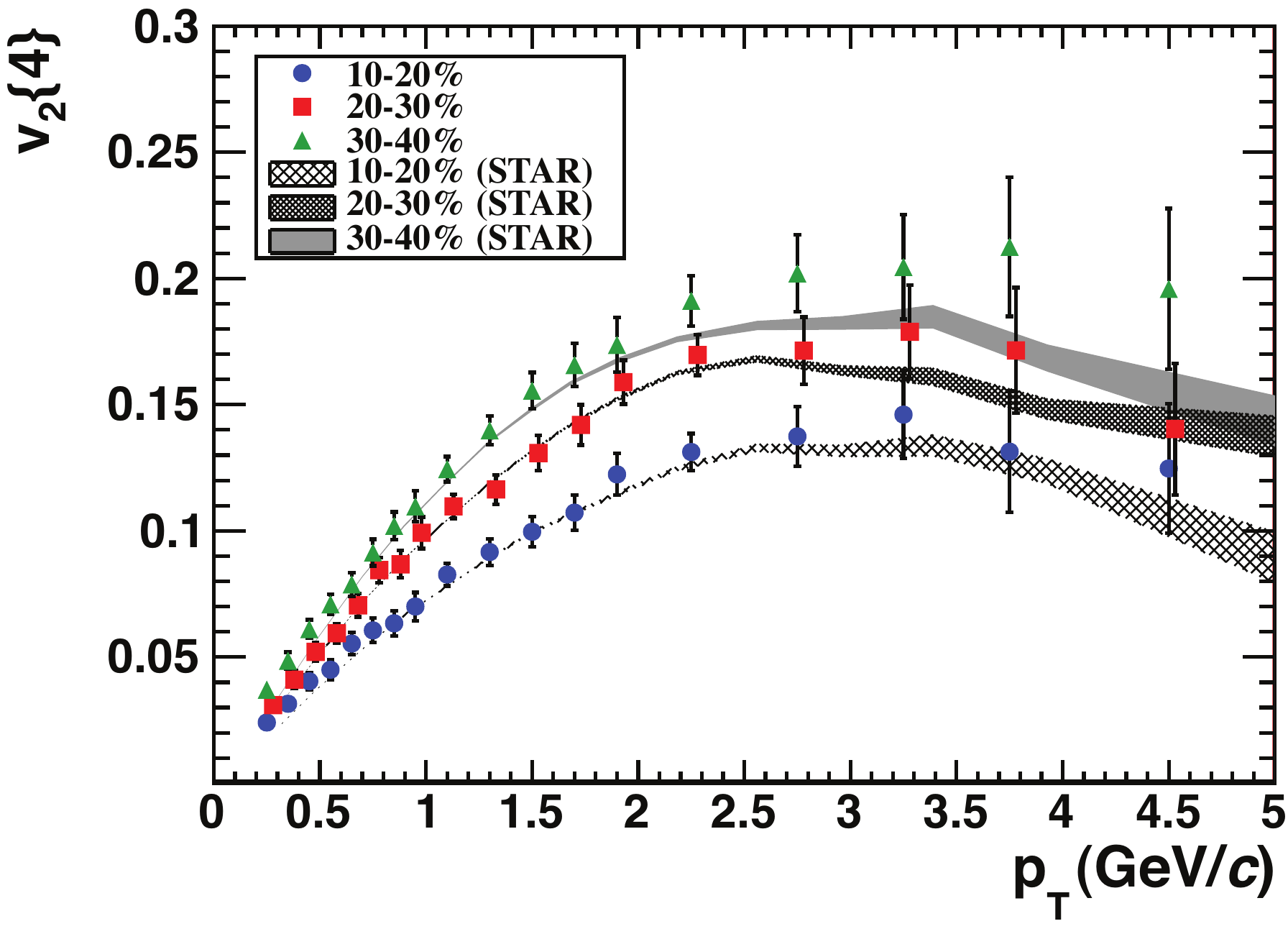} 
\caption{\textbf{Left}: A compilation of data on the dependence of the integrated elliptic flow, v$_2$, on the beam energy. 
The data correspond to ($\sim$20--30\%) \auau or \pbpb most central collisions. Reproduced from Ref.~\cite{Heinz:2013th};
\textbf{Right}: The differential elliptic flow of charged particles. The \pbpb collisions at \sqsn = 2.76 TeV (colored symbols) are compared to \auau 
collisions at \sqsn = 200~GeV (grey lines). Reproduced from Ref.~\cite{Aamodt:2010pa}.
}
\label{v2_exp}   
\end{figure}

The left panel of Figure~\ref{v2_exp} shows the measured energy dependence of the integrated elliptic flow coefficient $v_2$  in one centrality bin. Starting from 
$\sqrt{s_{NN}}$ $\approx$ 5 GeV, there is a continuous increase of $v_2$.  Below this energy, two phenomena occur. At very low energies, due to the rotation 
of the compound system generated in the collision, the emission is in-plane ($v_2>0$). At the laboratory kinetic energy around 100 MeV/nucleon, the preferred 
emission turns into out-of-plane, and $v_2$ becomes negative. The slowly moving spectator matter prevents the in-plane emission of participating nucleons 
or produced pions, which appear to be sqeezed-out of the reaction zone \cite{Venema:1993zz}. As the spectators move faster, their shadowing disappears, 
changing the pattern back to the in-plane emission.

Let us note that at RHIC, for the first time, the magnitude of the elliptic flow (Figure~\ref{v2_exp}) was found to be consistent with the EoS expected from 
the QGP \cite{Adams:2005dq, Heinz:2009xj}. The integrated value of $v_{2}$ for the produced particles increases by 70\% from the top SPS energy to 
the top RHIC energy (see left panel of Figure~\ref{v2_exp}), and it appears to do so smoothly. In comparison to the elliptic flow measurements in \auau collisions 
at $\sqrt{s_{NN}} = 200$ GeV, at the LHC,~$v_2$ increases by about 30\% at $\sqrt{s_{NN}} = 2.76$ TeV.  However, this~increase is not seen in the differential 
elliptic flow of charged particles shown on the right panel of Figure~\ref{v2_exp}. Thus, the bulk medium produced at RHIC and LHC has similar properties, and 
the 30\% increase of $v_2$  between the two energies is due to an enlarged available phase space, resulting in the same increase of the average transverse momentum 
of particles <\pt> between the RHIC and LHC energies.
 
As was first noted in Ref.~\cite{Ollitrault:1992bk}, at high energies, only the interactions among the constituents of matter formed in the initially spatially deformed 
overlap can produce $v_2 > 0$. A transfer of this spatial deformation into momentum space provides a unique signature for re-interactions in the fireball, and 
proves that the matter has undergone significant nontrivial dynamics between its creation and its freeze-out \cite{Heinz:2009xj}. The rapid degradation of 
the initial spatial deformation due to re-scattering causes the ``self-quenching'' of elliptic flow: if the elliptic flow does not develop early (when the collision fireball 
was still spatially deformed), it does not develop at all \cite{Heinz:2009xj}. In particular, the transformation of 
~the rapidly expanding ideal gas of non-interacting quarks 
and gluons into strongly interacting hadrons is unable to produce a sufficient elliptic flow. The elliptic flow thus reflects the pressure due to re-scattering---the induced 
expansion and stiffness of the EoS during the earliest collision stages. Its continuous rise with the energy up to its highest value at the LHC indicates that the early 
pressure also increases.

\begin{figure}[H]
\centering 
 \includegraphics[width=.47\textwidth, height=0.36\textwidth]{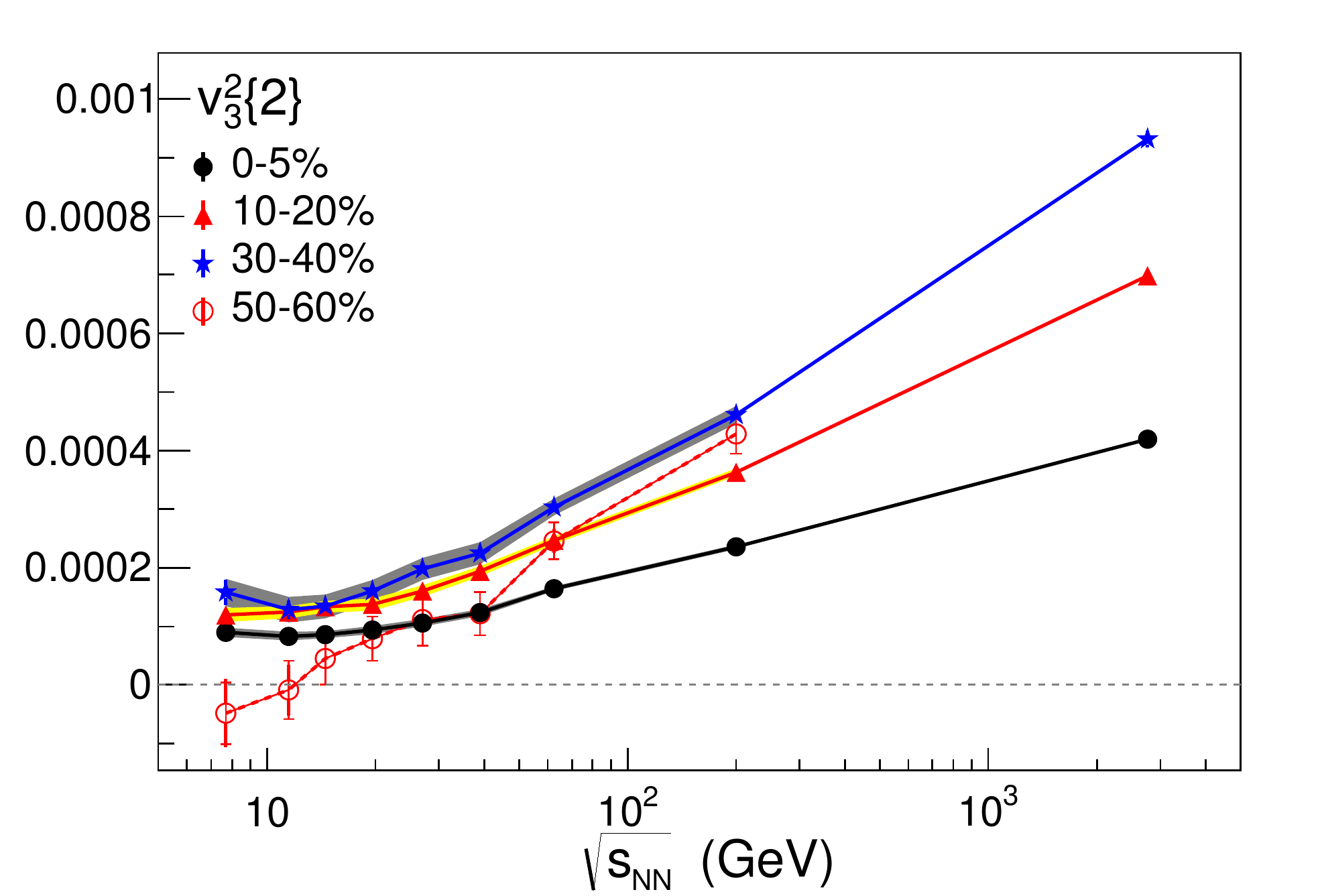} 
\includegraphics[width=.47\textwidth, height=0.36\textwidth]{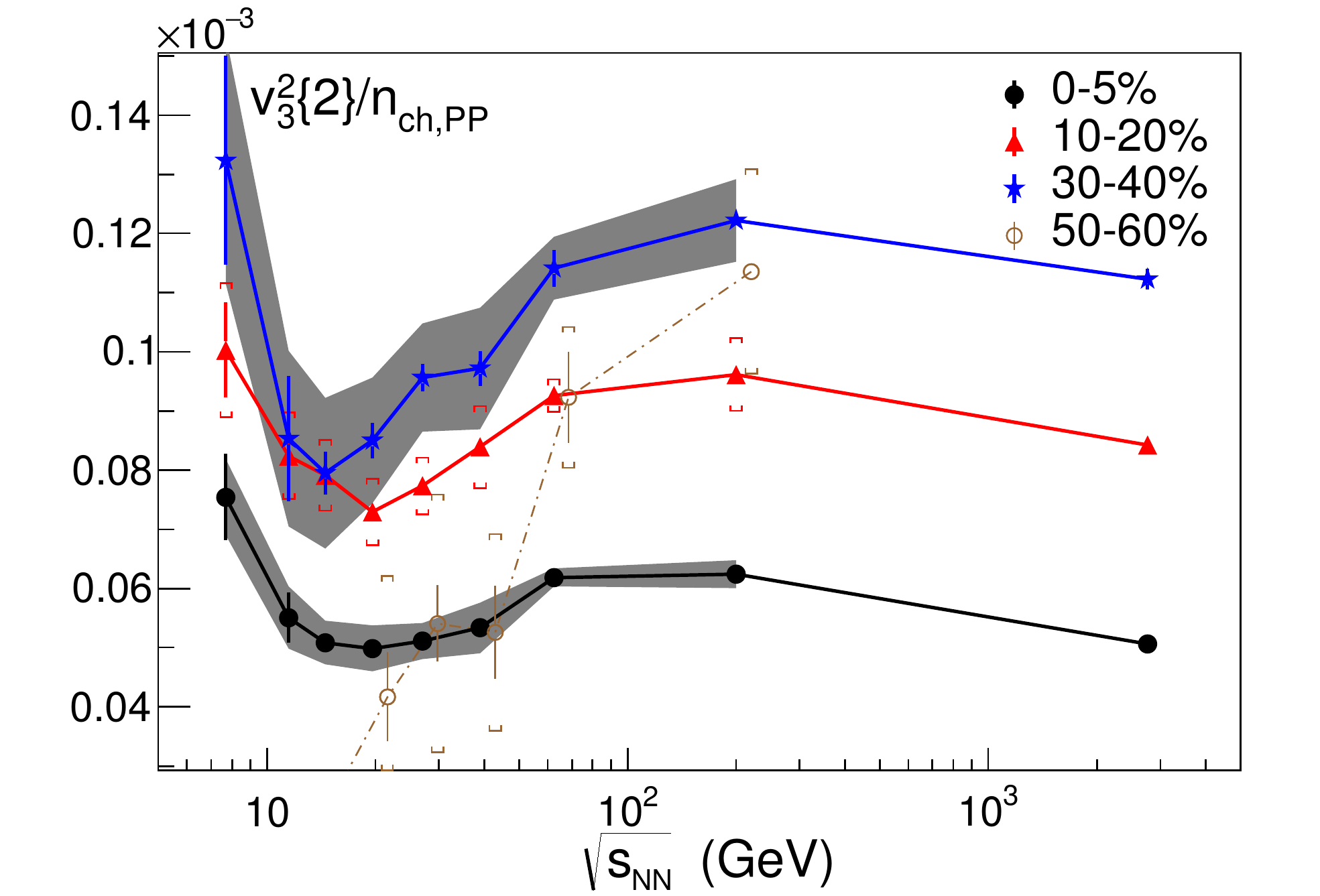} 
\caption{ \textbf{Left}: The energy dependence of \vthree~for four centrality bins. Points at 2.76 TeV corresponding to \pbpb~\cite{ALICE:2011ab}; 
\textbf{Right}: {\vthree} divided by $\rm n_{ch,PP}$ pair. Reproduced from Ref.~\cite{Adamczyk:2016exq}.}
\label{v3_exp}   
\end{figure}

The energy dependence of the integrated triangular flow coefficient {\vthree} of charged hadrons is shown on the left panel of Figure~\ref{v3_exp} 
in four bins of centrality, 0--5\%, 10--20\%, 30--40\%, and 50--60\%. As {\vthree} is sensitive to the fluctuations in the initial matter distribution, 
it is interesting to observe that at \sqsn = 7.7 and 11.5 GeV, values of {\vthree} for 50--60\% centrality become consistent with zero. 
For~more central collisions, however, {\vthree} is finite---even at the lowest energies---and changes very little from 7.7 GeV to 19.6 GeV. Above that, 
it begins to increase more quickly, and roughly linearly with $\log(\sqsn)$. Generally, one would expect that higher energy collisions producing more 
particles should be more effective at converting the initial state geometry fluctuations into {\vthree}. Deviations from that expectation could indicate 
interesting physics, such as a softening of the EoS~\cite{Shuryak:1986nk, Hung:1994eq} discussed already in Section~\ref{subsec:muzero}. This can be 
investigated by scaling {\vthree} by the charged particle rapidity density per participating NN pair, $\rm n_{\mathrm{ch,PP}}$ = \dNch/(0.5\Npart), 
see the left panel of Figure~\ref{v3_exp}. A local minimum of \vthree/$\rm n_{\mathrm{ch,PP}}$ in the region near 15--20 GeV observed in the centrality range 
0--50\% and absent in the more peripheral events could indicate an interesting trend in the pressure developed inside the~system. 

As was already briefly mentioned in Section~\ref{Oism}, recent years have witnessed a growth of interest in studying collective phenomena on event-by-event basis. 
In addition to the key publications~\mbox{\cite{Broniowski:2008vp,Petersen:2010zt,Heinz:2011mh,Noronha-Hostler:2016eow,Lin:2004en,Pang:2012he,Bhalerao:2015iya, 
Xu:2016hmp,Paatelainen:2013eea,Niemi:2015qia}}, it is worth mentioning the study \cite{Gale:2012rq} where the anisotropic flow coefficients $v_1$--$v_5$ 
were computed by combining the IP-Glasma flow with the subsequent relativistic viscous hydrodynamic evolution of matter through the quark--gluon plasma 
and hadron gas phases. The~event-by-event geometric fluctuations in nucleon positions and intrinsic color charge fluctuations at the sub-nucleon scale 
are expected to result in experimentally measurable event-by-event anisotropic flow coefficients \cite{Aad:2013xma}. Let us note that fluctuating initial profiles observed 
in over-many-events integrated triangular and higher odd flow coefficients are also revealed~in difference between various participant plane angles $\Psi_n$ 
introduced in Equation~(\ref{vN}). Correlations between the event plane angles $\Psi_n$ of different harmonic order can not only yield valuable additional insights 
into the initial conditions~\cite{Qiu:2012uy}, but are also experimentally measurable~\cite{Aad:2014fla}. The same is also true for the correlations between 
different flow harmonics~\cite{Qian:2016pau,Giacalone:2016afq,Aad:2015lwa,ALICE:2016kpq}.

\begin{figure}[H]
\includegraphics[width=10.825cm, height=0.4\textwidth]{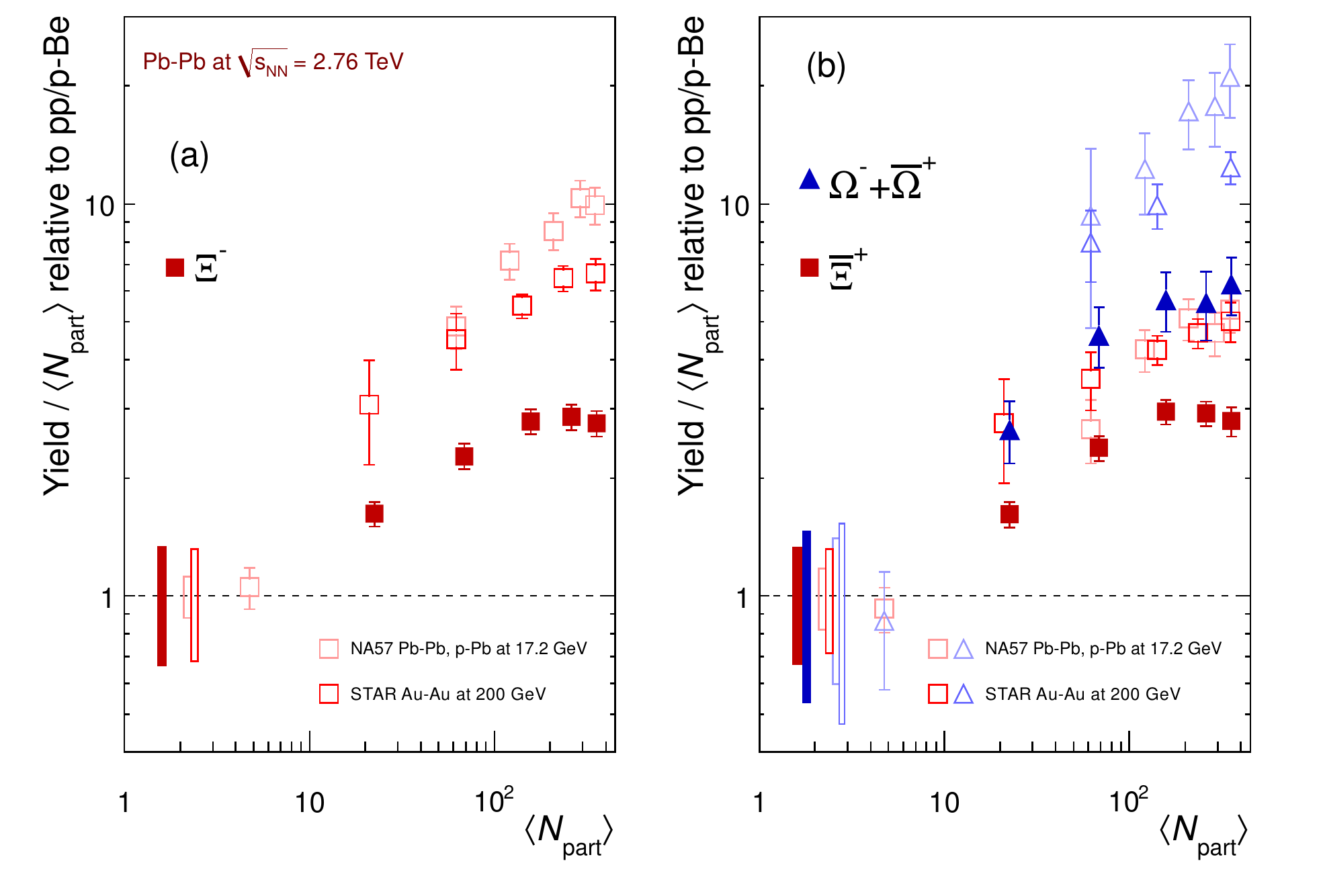}
\hspace{-4mm}
\includegraphics[width=5.3825cm,height=0.4\textwidth]{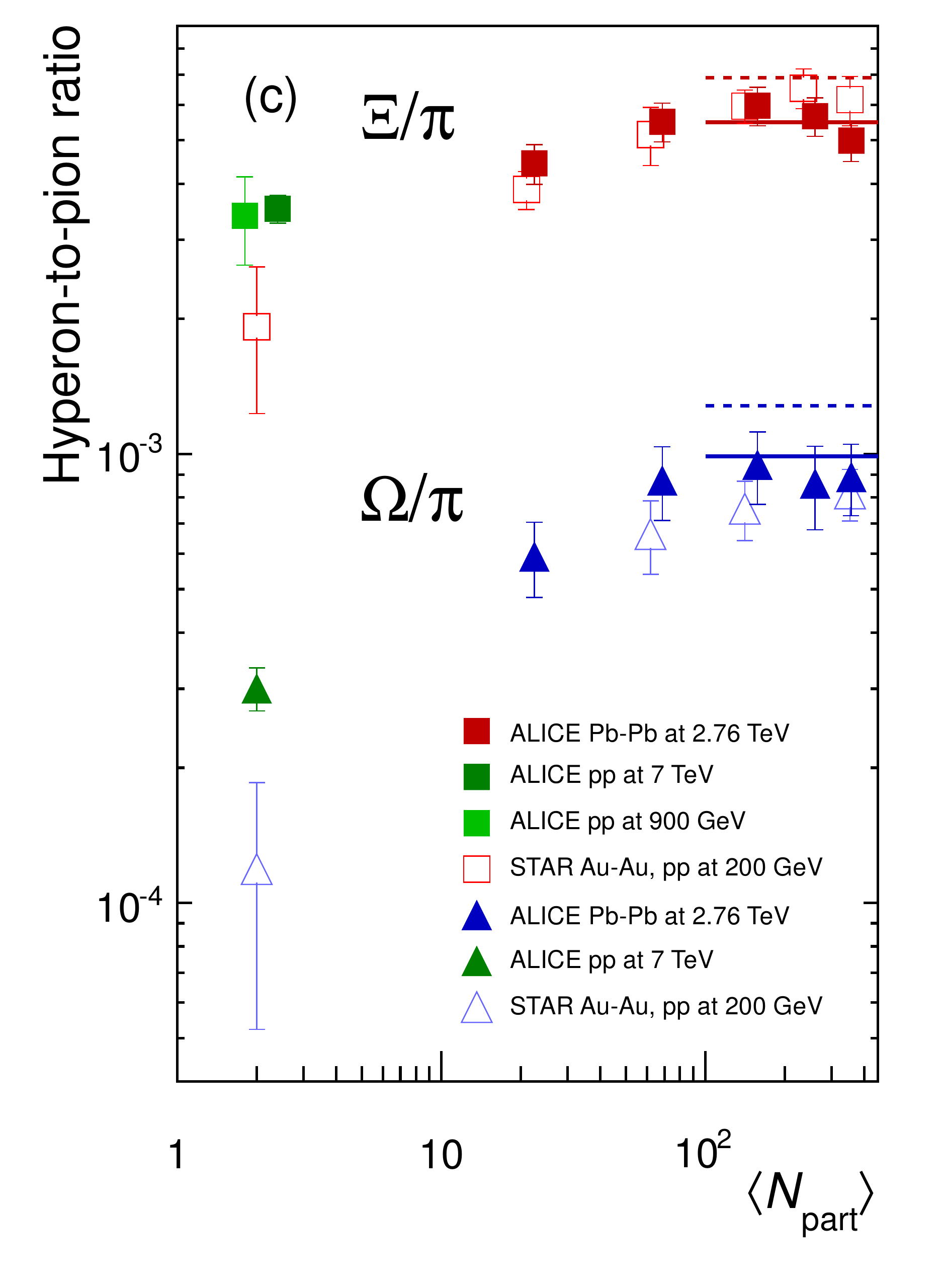}
\vspace{-4mm}
\includegraphics[width=0.495\textwidth]{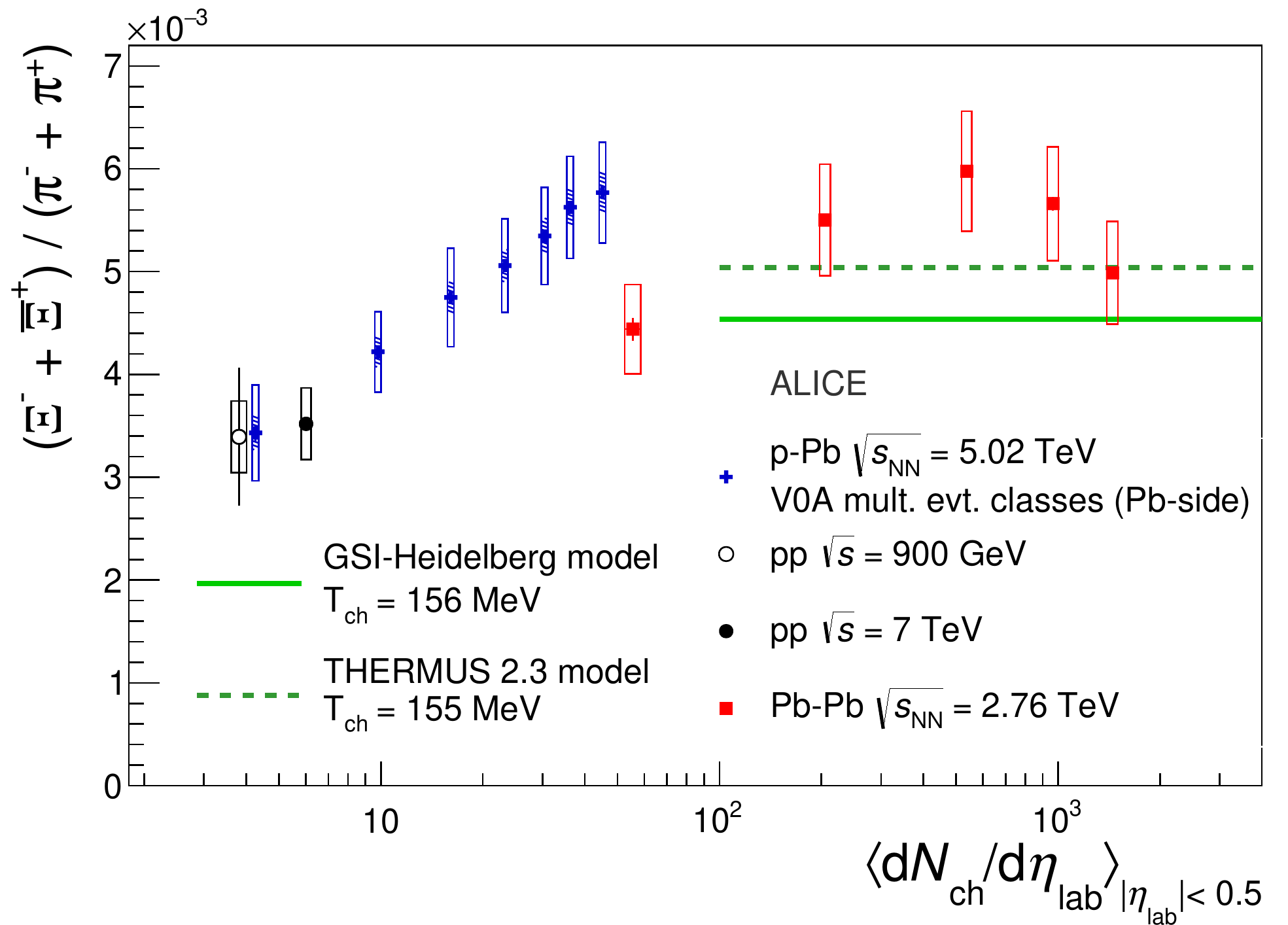}
  \includegraphics[width=0.495\textwidth]{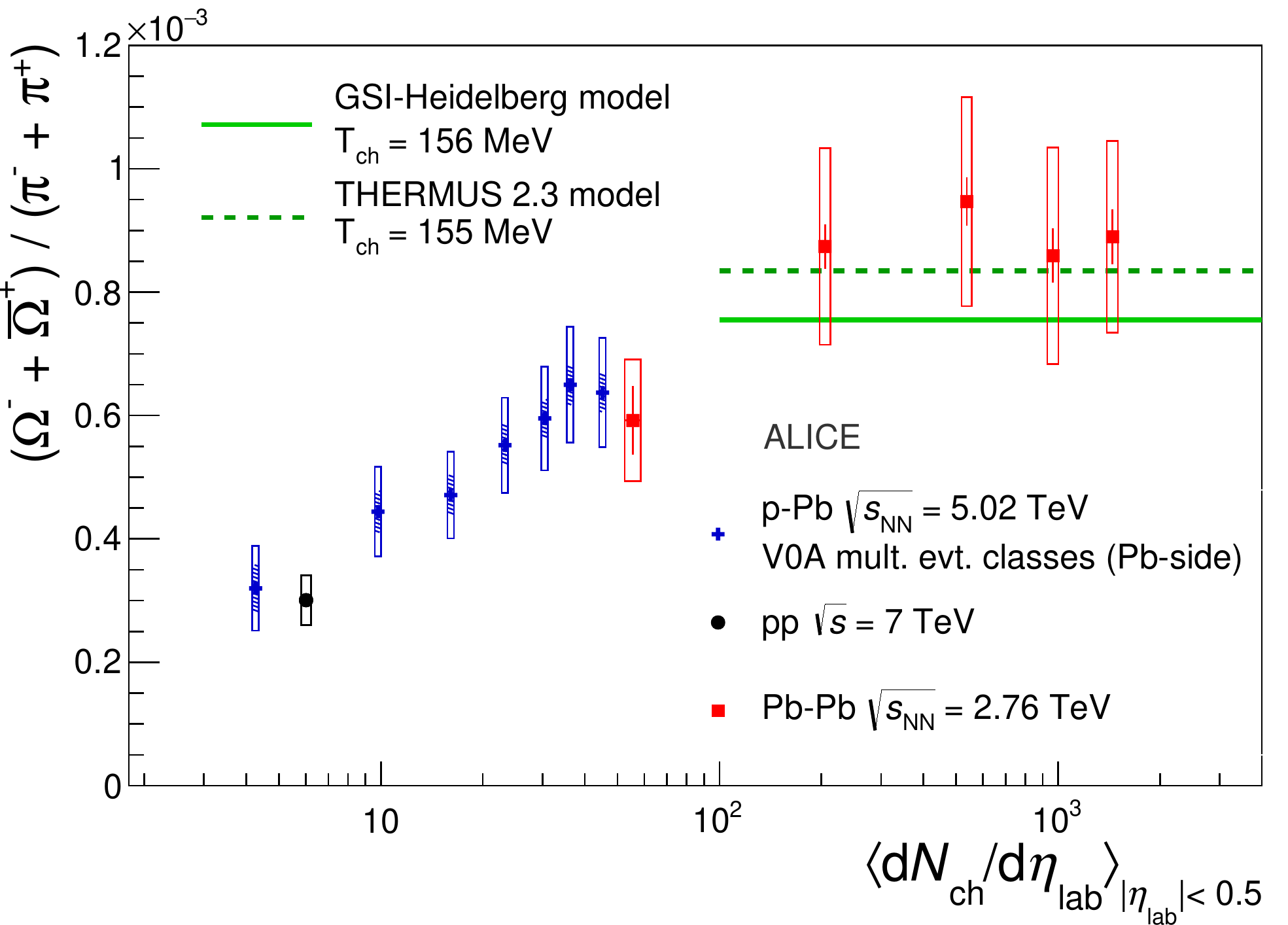}\vspace{12pt}
  
\caption{\textbf{Top}: (\textbf{a},\textbf{b}) Normalized (to \pp or \pbe) yields of multistrange baryons per participants at midrapidity as a function of 
$\left<\Npart\right>$ for the LHC (full symbols), RHIC, and SPS (open symbols) data. Boxes on the dashed line at unity indicate 
the statistical and systematic uncertainties on the \pp\ or \pbe\ reference; (\textbf{c}) Hyperon-to-pion ratios as a function of $\left<\Npart\right>$ 
for \pbpb, \auau, and \pp\ collisions at LHC and RHIC energies. The lines are the thermal model \cite{Andronic:2008gu} (full line) 
and ~\cite{Wheaton:2004qb} (dashed line) predictions. Reproduced from Ref.~\cite{ABELEV:2013zaa}; \textbf{Bottom}: 
(\textbf{left}) ($\Xi^-$+ ${\overline{\Xi}^+}$)/($\pi^+$+$\pi^-$)  and (\textbf{right})  ($\Omega^-$+${\overline{\Omega}^+}$)/($\pi^+$+$\pi^-$) ratios 
as functions of $\left< \dNch \right>$ for \pp, \ppb, and \pbpb collisions at the LHC. The \pbpb  data points \cite{ABELEV:2013zaa}
represent, from left to right, the 60--80\%, 40--60\%, 20--40\%, 10--20\%, and 0--10\% centrality classes. The chemical 
equilibrium predictions by the GSI-Heidelberg \cite{Andronic:2008gu} and the THERMUS~2.3 \cite{Wheaton:2004qb} models are represented 
by the horizontal lines. Reproduced from Ref.~\cite{Adam:2015vsf}.
}
\label{msPbPb_enhancement}
\end{figure}

Enhanced production of hadrons with the quantum numbers not present in colliding matter is one of the oldest signals of the deconfined QGP medium 
\cite{Rafelski:1982pu, Koch:1986ud}. Measurements of the yields of multistrange baryons were carried out at CERN SPS by WA85, and later on 
by WA97/NA57 collaborations since the mid-eighties. After 2000, more data came from RHIC, and starting from 2010 also from the LHC. 
The current status is summarized in the five panels of Figure~\ref{msPbPb_enhancement}. In the top left and middle panels (Figure~\ref{msPbPb_enhancement}a,b), a compilation of 
the results from SPS, RHIC, and the LHC in terms of {\it strangeness enhancement} defined as normalized (to \pp or \pbe) yield per participants is presented. 
On the top right (Figure~\ref{msPbPb_enhancement}c), the hyperon-to-pion ratios as functions of $\left<\Npart\right>$ for \pbpb, \auau, and \pp\ collisions at the LHC and RHIC energies are displayed. 
The normalized yields are larger than unity for all the particles, and increase with their strangeness content. This behaviour is consistent with the picture of 
enhanced $s \bar{s}$ pair production in a hot and dense QGP medium \cite{Rafelski:1982pu, Koch:1986ud}. 

The two bottom plots represent a comparison between the hyperon-to-pion ratios from \pp, \ppb, and \pbpb collisions. Interestingly, the ratios in \ppb collisions 
increase with multiplicity from the values measured in \pp to those observed in \pbpb. The rate of increase is more pronounced for particles with higher 
strangeness content. Let us note that the Grand canonical statistical description of \pbpb data shown as full and dashed lines in Figure~\ref{msPbPb_enhancement} 
may not be appropriate in small multiplicity environments such as those produced in the \ppb case. It appears that for the latter case, the evolution of hyperon-to-pion 
ratios with the event multiplicity is qualitatively well described by the Strangeness Canonical model implemented in \textsc{THERMUS~2.3}~\cite{Wheaton:2004qb}. 
In this case, a local conservation law is applied to the strangeness quantum number within a correlation volume $V_{c}$, while treating the baryon and charge 
quantum numbers grand-canonically within the whole fireball volume $V$ \cite{Adam:2015vsf}. 


\subsection{Hard Probes}
\label{hardpr}

Heavy quarks, quarkonia, and jets---commonly referred to as {\it hard probes}---are created in the first moments after the collision, and are therefore considered 
as key probes of the deconfined QCD medium. Production of these high transverse momentum ($\pt \gg \Lambda_{QCD}$) objects occurs over very short 
time scales ($\tau$$\approx$$1/p_{\text T}$$\approx$0.1 $fm/c$), and can thus probe the evolution of the medium. Since the production cross-sections of 
these energetic particles are calculable using pQCD, they have been long recognised as particularly useful ``tomographic'' probes of 
the QGP~\cite{Bjorken:1982tu, Appel:1985dq, Gyulassy:1990ye}.

Let us start our discussion with the results on the inclusive production of high-\pt hadrons. The latter are interesting on their own because it was there 
where for the first time the suppression pattern was observed \cite{Arsene:2004fa, Back:2004je, Adams:2005dq, Adcox:2004mh}. In an inclusive 
regime, the comparison between $d^2N$/$d\pt d\eta$ (the differential yield of high-\pt hadrons or jets per event in A+B collisions) to that in \pp collisions 
is usually quantified by introducing the nuclear modification factor 
\begin{equation}
\label{R_AB}
\rab(\pt, \eta)=\frac{dN^2_{AB}(\pt)/d{\pt}d\eta}{\left< \Ncoll \right> dN^2_{pp}(\pt)/d{\pt}d\eta}\; .
\end{equation}

\begin{figure}[H]
\centering
\includegraphics[scale=0.28]{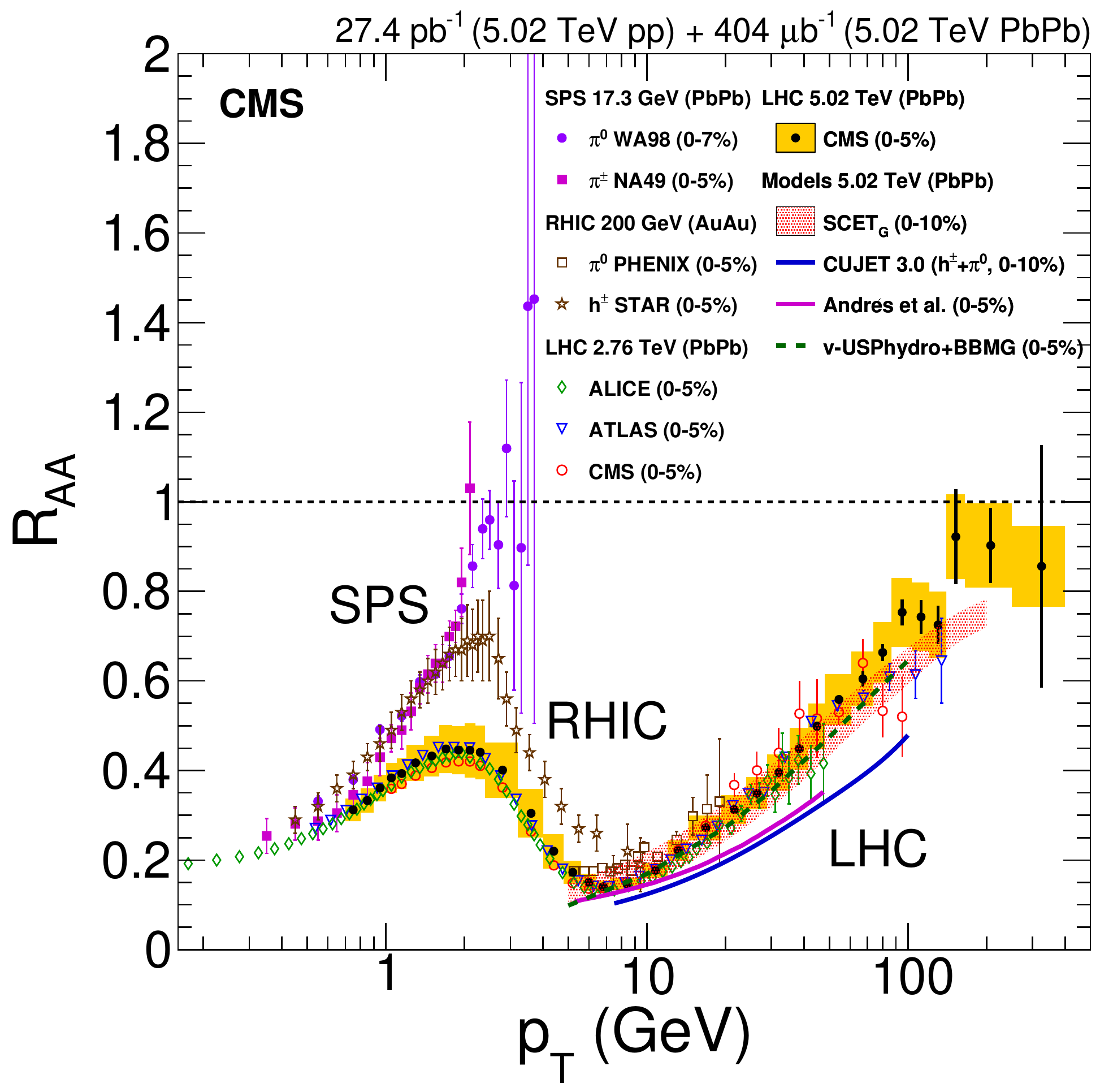}
\includegraphics[scale=0.35]{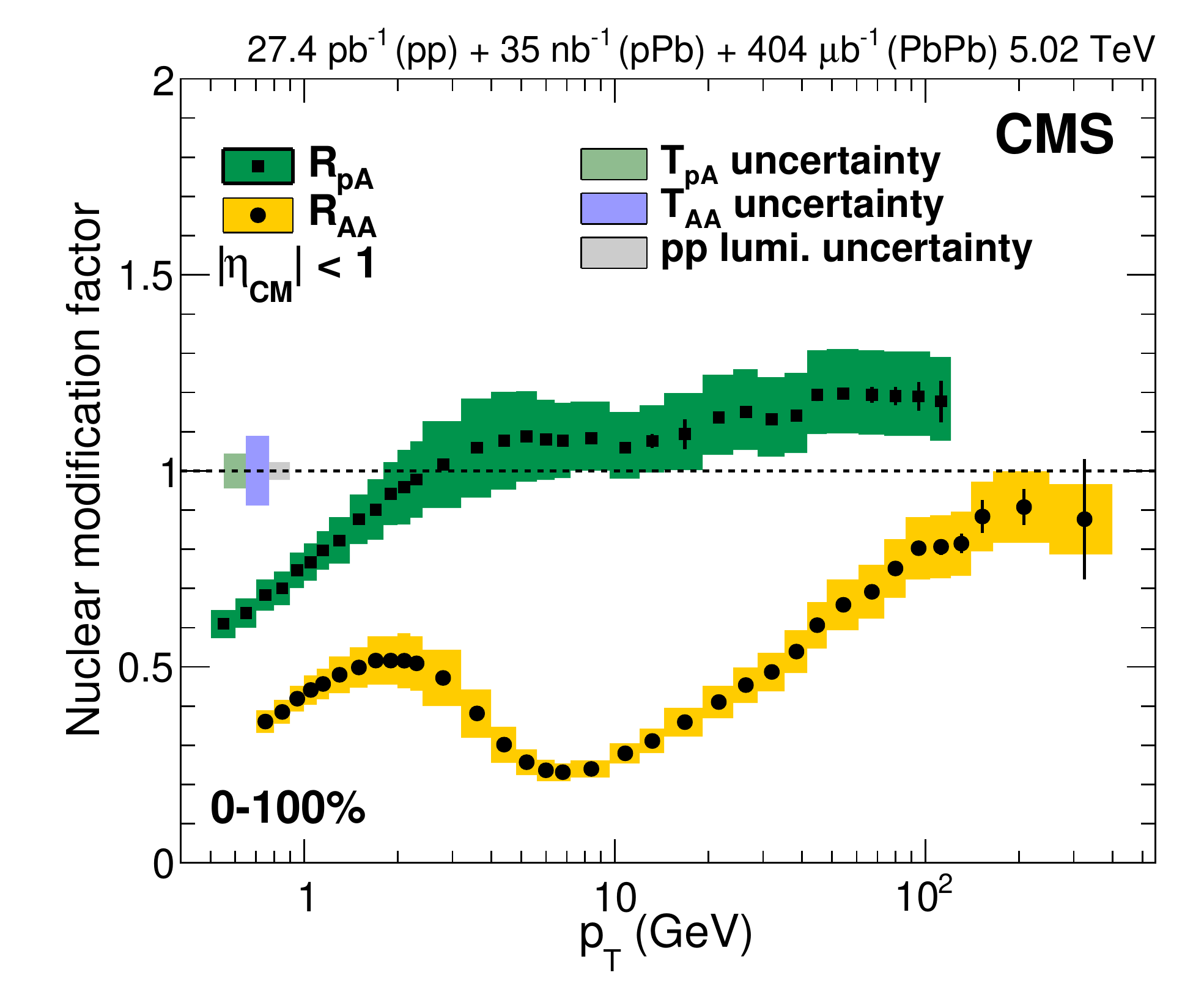} 
\caption{\textbf{Left}: A compilation of the measurements of \raa(\pt) for neutral pions ($\pi^0$), charged hadrons ($h^{\pm}$), 
and charged particles in central heavy-ion collisions at SPS, RHIC, and the LHC. Reproduced from Ref.~\cite{Khachatryan:2016odn};
\textbf{Right}: The nuclear modification factor \rpb(\pt) in \ppb compared to \raa from \pbpb collisions 
at \sqsn = 5.02 TeV at the LHC. Reproduced from Ref.~\cite{Khachatryan:2016odn}.}
\label{raa}
\end{figure}

For collisions of two nuclei behaving as a simple superposition of \Ncoll nucleon--nucleon collisions, the nuclear modification factor 
would be \rab = 1. The data of Figure~\ref{raa} reveal a very different behaviour. The left panel shows a compilation of \raa from \auau and \pbpb collisions, 
the right panel the result of \rpb from three LHC experiments at the same energy \sqsn= 5.02 TeV. In the \raa case, the suppression pattern of high-\pt (>2--3 GeV/c) 
hadrons in the deconfined medium---predicted many years ago \cite{Bjorken:1982tu, Appel:1985dq, Gyulassy:1990ye} as a {\it jet quenching effect}---is clearly visible 
at RHIC and the LHC. However, for proton--nucleus collisions (Figure~\ref{raa}, right panel), no suppression is seen, even at the highest LHC energy.  Moreover, \raa 
in the 5\% most central \pbpb collisions at the LHC shows a maximal suppression by a factor of 7--8 in the \pt region of 6--9 GeV. This dip is followed by 
an increase, which continues up to the highest \pt measured at \sqsn = 5.02 TeV, and approaches unity in the vicinity of \pt = 200 GeV~\cite{Khachatryan:2016odn}.

\subsubsection{High-\pt Hadrons and Jets}
\label{highpt}

\begin{figure}[H]
\centering
\includegraphics[width=.45\textwidth, height=.38\textwidth]{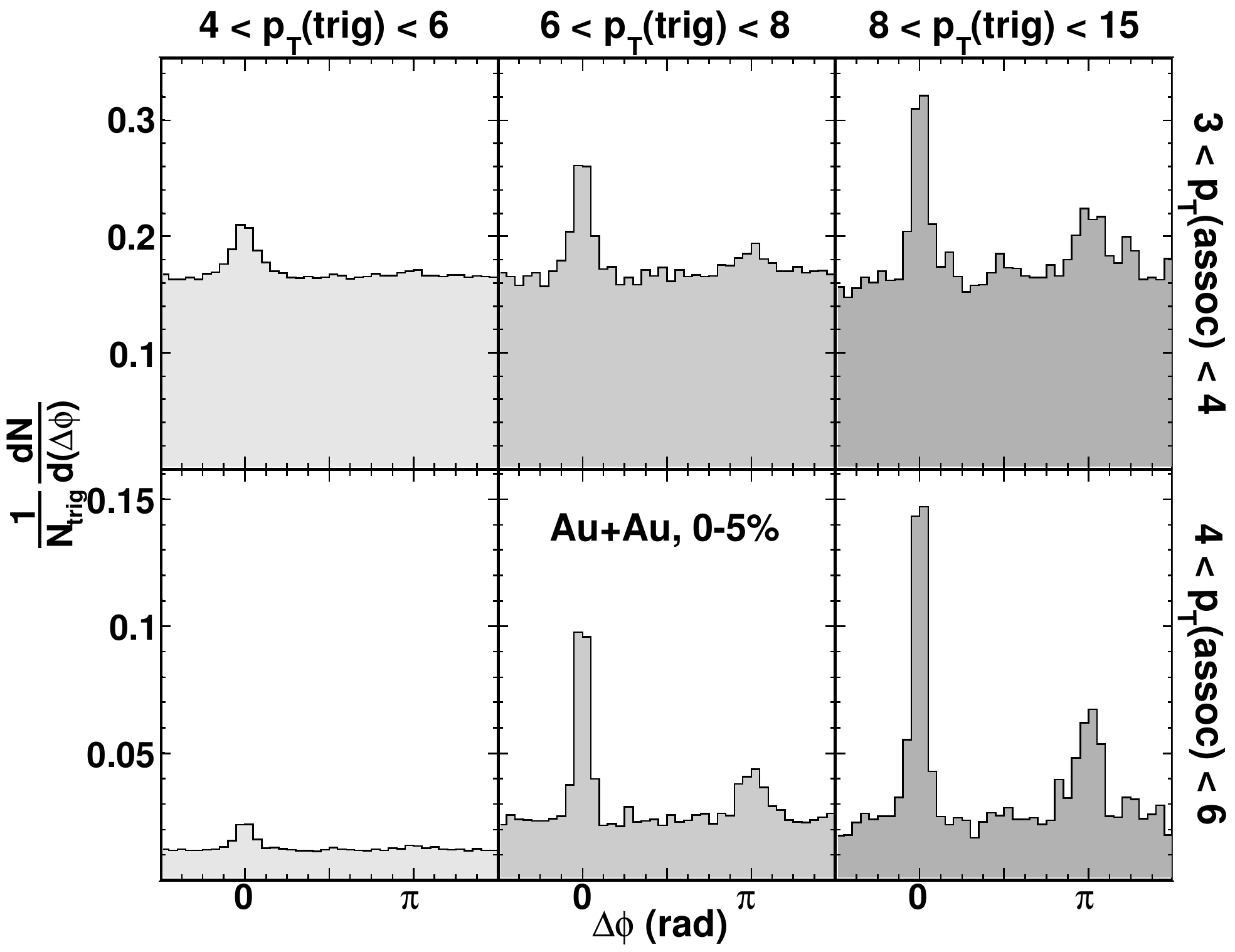} \qquad
\includegraphics[width=.45\textwidth, , height=.38\textwidth]{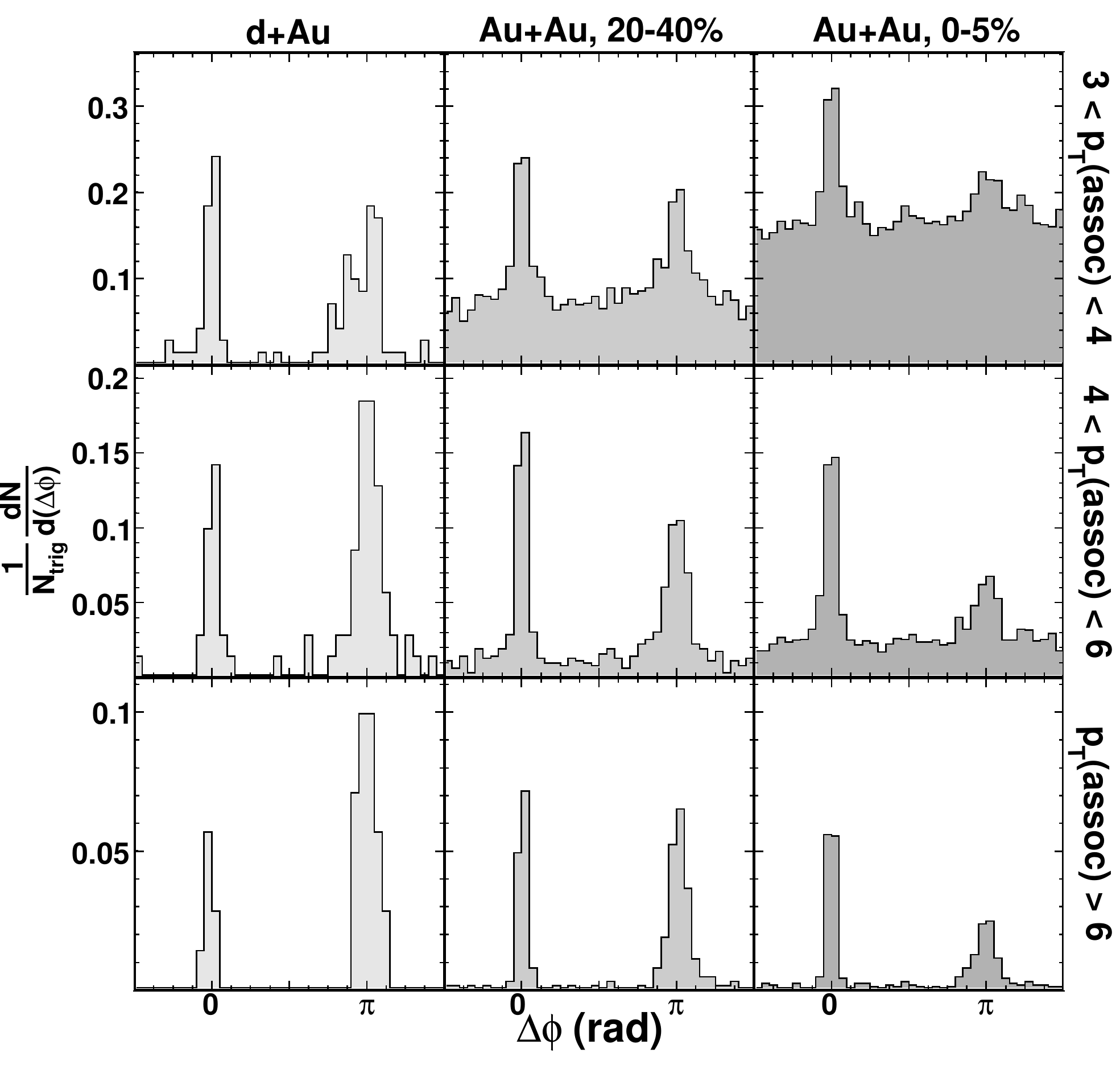} 
\caption{
\textbf{Left}: Azimuthal correlation histograms of high-\pt charged hadron pairs normalized per trigger particle for 0--5\% \auau events 
for various \pTtrig and \pTassoc ranges. The yield in the lower left panel is suppressed due to the constraint $\pTassoc<\pTtrig$; 
\textbf{Right}: The same but for $8 < \pTtrig < 15$ GeV/{\it c}, for \dau, 20--40\% \auau and 0--5\% \auau events. Reproduced from Ref.~\cite{Adams:2006yt}. 
}
\label{dirjets}
\end{figure}

The suppression of high-\pt hadrons in the deconfined medium was thoroughly studied at RHIC using azimuthal correlations between the trigger particle 
and associated particle; see Figure~\ref{dirjets}. Near-side peaks in central (0--5\%) \auau collisions present in all panels of Figure~\ref{dirjets} (left) indicate 
that the correlation is dominated by jet fragmentation. An away-side peak emerges as \pTtrig is increased. The~narrow back-to-back peaks are indicative 
of the azimuthally back-to-back nature of dijets observed in an elementary parton--parton collision. Contrary to the latter, the transverse-momentum imbalance 
of particles from the jet fragmentation due to different path lengths of two hard partons in the medium is apparent. The azimuthal angle difference $\Delta \phi$ 
for the highest \pTtrig range  (8 < \pTtrig < 15 GeV/c)  for mid-central (20--40\%) and central Au+Au collisions---as well as for d+Au collisions---is presented in 
Figure~\ref{dirjets} (right panel). We observe narrow correlation peaks in all three \pTassoc ranges. For each \pTassoc, the nearside peak shows a similar correlation 
strength above background for the three systems, while the away-side correlation strength decreases from \dau to central \auau. For the \dau case, the yield of particles 
on the opposite side $\Delta \phi$ = $\pi$ prevails over the same side. Moreover, for \auau collisions, the nearside yields obtained after subtraction
~of the background 
contribution due to the elliptic flow show a little centrality dependence, while the away-side yields decrease with increasing centrality \cite{Adams:2006yt}.

\begin{figure}[H]
\centering
\includegraphics[trim={ 0 0 0 0}, clip,width=.3\textwidth, height=0.27\textwidth]{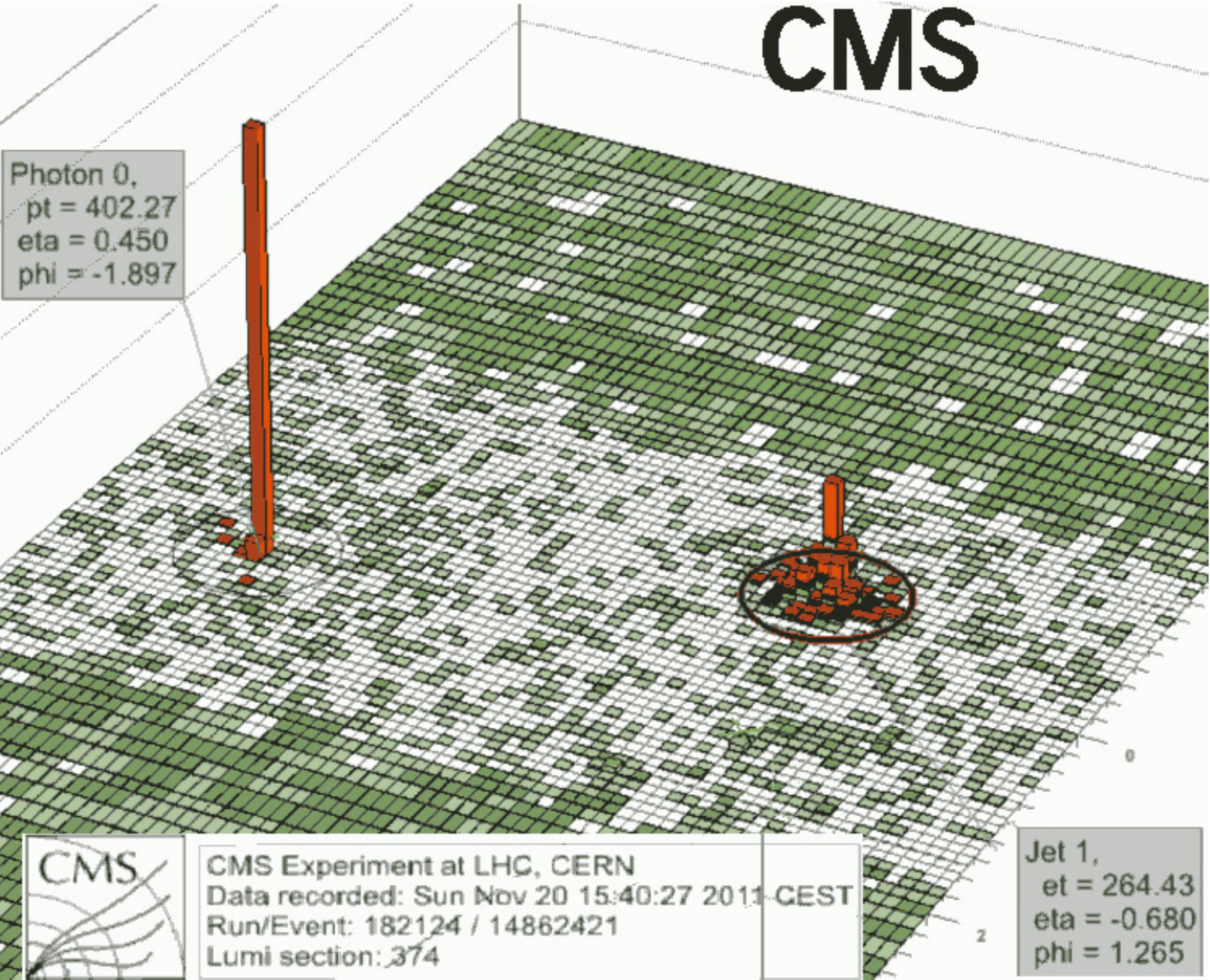} ~
\includegraphics[trim={ 0 .6cm 0 .75cm},clip, width=.289\textwidth, height=0.27\textwidth]{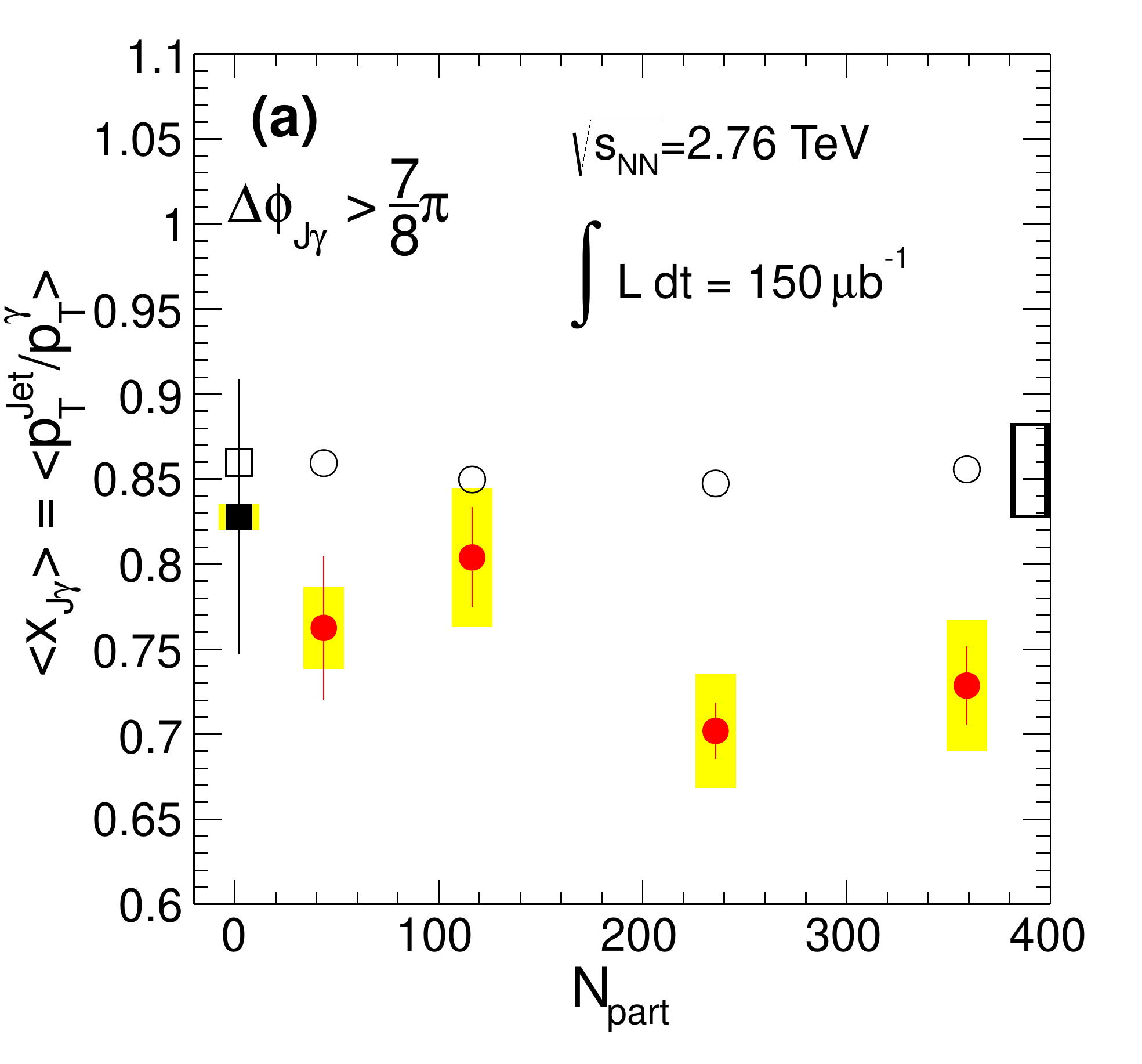} 
\includegraphics[trim={ 0 .6cm 0 .75cm},clip, width=.289\textwidth, height=0.27\textwidth]{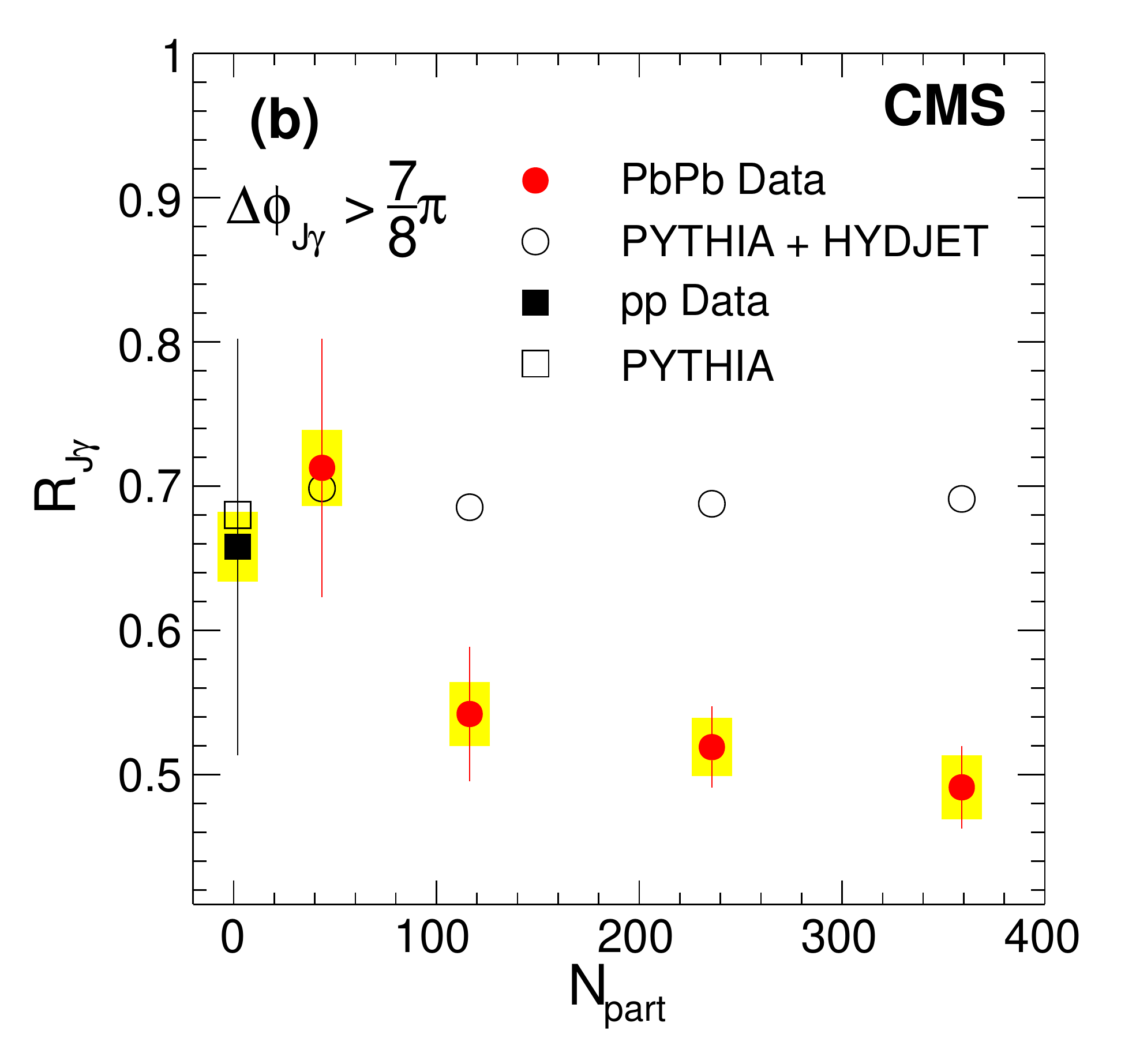} 
\caption{\textbf{Left}:
The energy distribution in the $\eta \times \phi$ plane in a single \pbpb event recorder by the CMS detector at LHC (reproduced from 
Ref.~\cite{d'Enterria:2012iw}); \textbf{Middle} and \textbf{Right}: The average ratio of jet transverse momentum to photon transverse momentum $\left<x_{\text{J}\gamma}\right>$ 
and the average fraction R$_{\text{J}\gamma}$ of isolated photons with an associated jet with energy above 30 GeV as functions of number of participants 
$N_{\text{part}}$. Photons and jets are emitted almost back-to-back in azimuth $\phi_{\text{J}\gamma}$. Reproduced from Ref.~\cite{Chatrchyan:2012gt}.}
\label{gamma_jet}
\end{figure}

Unfortunately, the advantage of the large yield of dijets is offset by a loss of information about the initial properties of the probes (i.e., prior to their interactions with 
the medium). Correlating two probes that both undergo an energy loss also induces a selection bias towards scatterings occurring at---and oriented tangential to---the surface 
of the medium. It is thus interesting to study correlations when one of the particles does not interact strongly with the medium. Triggering on the high-\pt isolated photon 
(i.e.,~not from $\pi^0 \rightarrow 2\gamma$ decays) would do the job. While in \pp collisions an emerging quark jet should balance its transverse momentum with 
the photon, in the heavy-ion collisions, much of its momentum is thermalized while the quark traverses the plasma. This is illustrated in Figure~\ref{gamma_jet} (left panel), 
where a single hard photon with \pt = 402 GeV emerges unhindered from the de-confined medium produced in Pb$+$Pb collisions at the LHC. The accompanying quark jet 
produced via the QCD Compton scattering $qg \rightarrow q\gamma$ loses 1/3 of its energy ($\approx$140 GeV!) inside the hot and dense matter. 

The measurement presented on the middle and right panels of Figure~\ref{gamma_jet} shows that for more central \pbpb collisions, a significant decrease in the ratio 
of jet transverse momentum to photon transverse momentum---$\left<x_{\text{J}\gamma}\right>$---relative to the \textsc{pythia} reference \cite{Sjostrand:2007gs}
is observed. Furthermore, significantly more photons with \pt > 60 GeV/c in \pbpb are observed  to not have an associated jet with \pt > 30 GeV/c jet, compared 
to the reference. However, no significant broadening of the photon + jet azimuthal correlation has been observed.

An important progress in the theoretical understanding of the suppression of energetic partons traversing a deconfined matter was the introduction of the diffusion coefficient 
$\hat{q}$ relevant for the transverse momentum broadening and collisional energy loss of partons (jets) \cite{Baier:1996sk}. This quantity, which is commonly referred to 
as the jet quenching parameter, can be determined either via weak coupling techniques \cite{CaronHuot:2008ni,Laine:2012ht,Blaizot:2014bha}, a combination of lattice 
simulations and dimensionally-reduced effective theory~\cite{Panero:2013pla}, or from the gauge/gravity duality \cite{Liu:2006ug}. Typical estimates for this quantity 
at RHIC and LHC energies range between 5 and 10 GeV$^2$/fm, demonstrating the currently still sizable uncertainties in these calculations.


\subsubsection{Quarkonia}
\label{quarkonia}

Melting of the quarkonia---bound states of heavy quark and anti-quark $q\bar{q}$ where $q$ = $c,b$---due to a colour screening in the deconfined 
hot and dense medium was proposed thirty years ago as a clear and unambiguous signature of the deconfinement \cite{Matsui:1986dk}. 
However, shortly after that, it was noticed that not only diffusion of the heavy quarks from melted quarkonium, but also the drag which charm 
quarks experience when propagating through the plasma is {important} \cite{Svetitsky:1987gq}. 

The latter might lead to an enhancement instead 
of a suppression. This is in variance with the original proposal that the heavy quarks, once screened, simply fly apart. With the advent of the strongly-interacting QGP, 
the Langevin equation model of quarkonium production was formulated, where the charm quark--antiquark pairs evolve on top of a hydrodynamically expanding 
fireball \cite{Young:2008he}. A heavy quark and anti-quark interact with each other according to the screened Cornell potential and interact, independently, 
with the surrounding medium, experiencing both drag and rapidly decorrelating random forces. An extension of this approach to bottomonium production 
\cite{Petreczky:2016etz} shows that a large fraction of $b \bar{b}$ pairs that were located sufficiently close together during the initial hard production 
will remain correlated in the hot medium for a typical lifetime of the system created in heavy-ion collisions. The distribution of the correlated $b \bar{b}$ pair 
in relative distance is such that it will dominantly form 1S bottomonium. A study of quarkonia production in heavy-ion collisions thus provides an interesting 
window not only into static, but also into dynamical properties of the hot, dense, and rapidly expanding medium \cite{Brambilla:2010cs, Andronic:2015wma}. 

\begin{figure}[H]
\centering
\begin{minipage}[b]{0.46\linewidth}
\includegraphics[width=\linewidth, height=.7\linewidth]{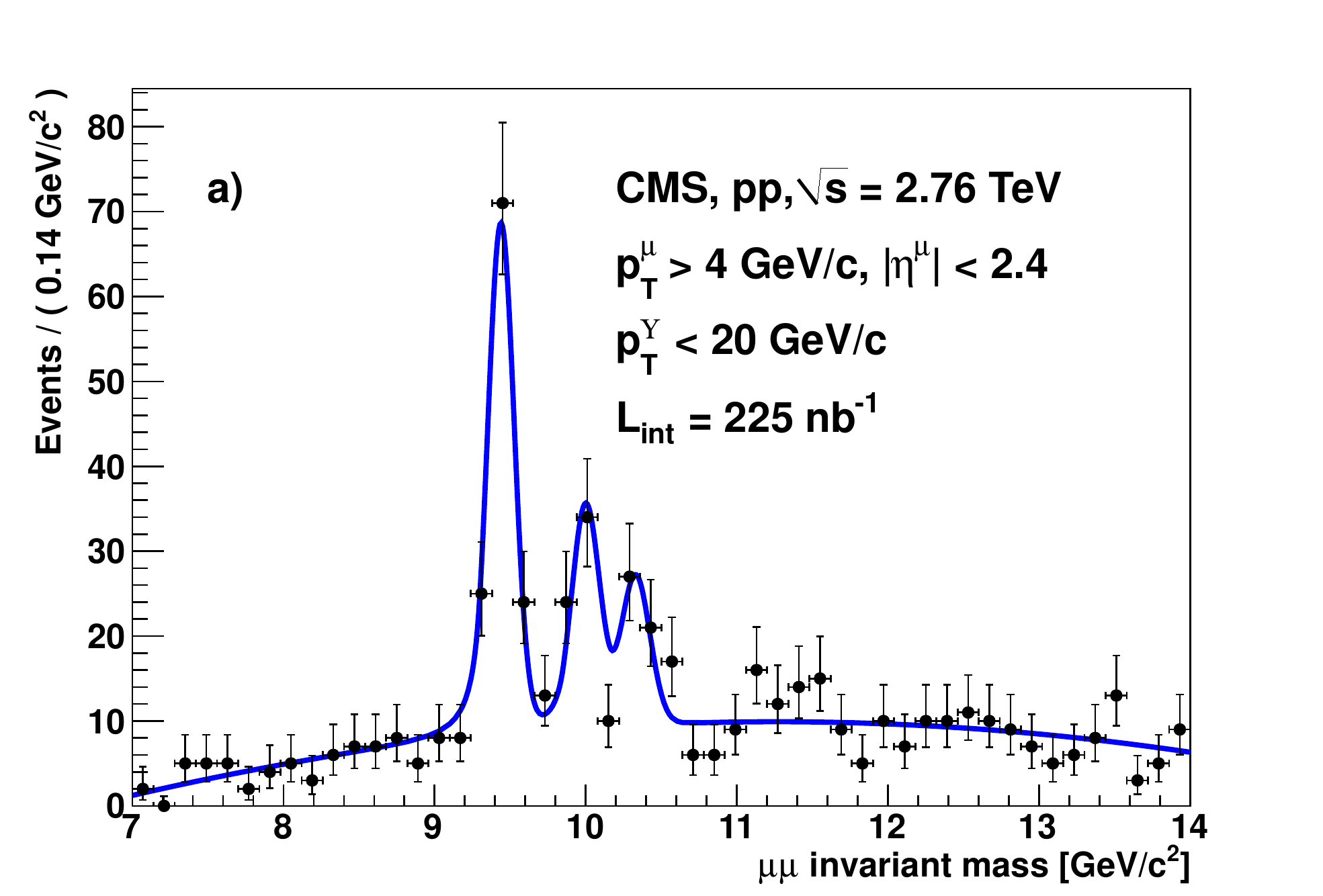} 
\includegraphics[trim={ 0 0 0 1cm}, clip, width=\linewidth, height=.7\linewidth]{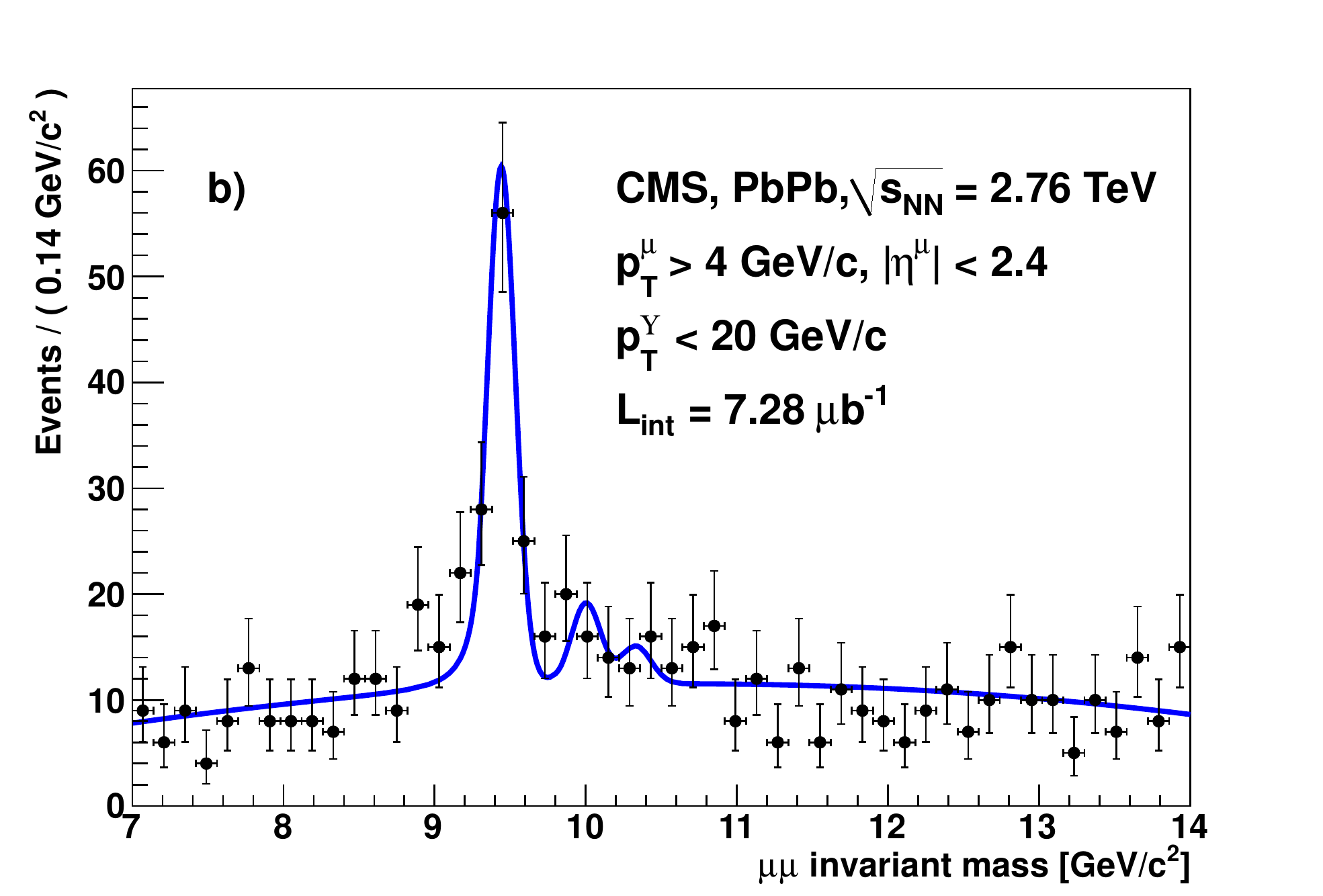} 
\end{minipage}
\begin{minipage}[b]{0.475\linewidth}
\includegraphics[width=\linewidth, height=.66\linewidth]{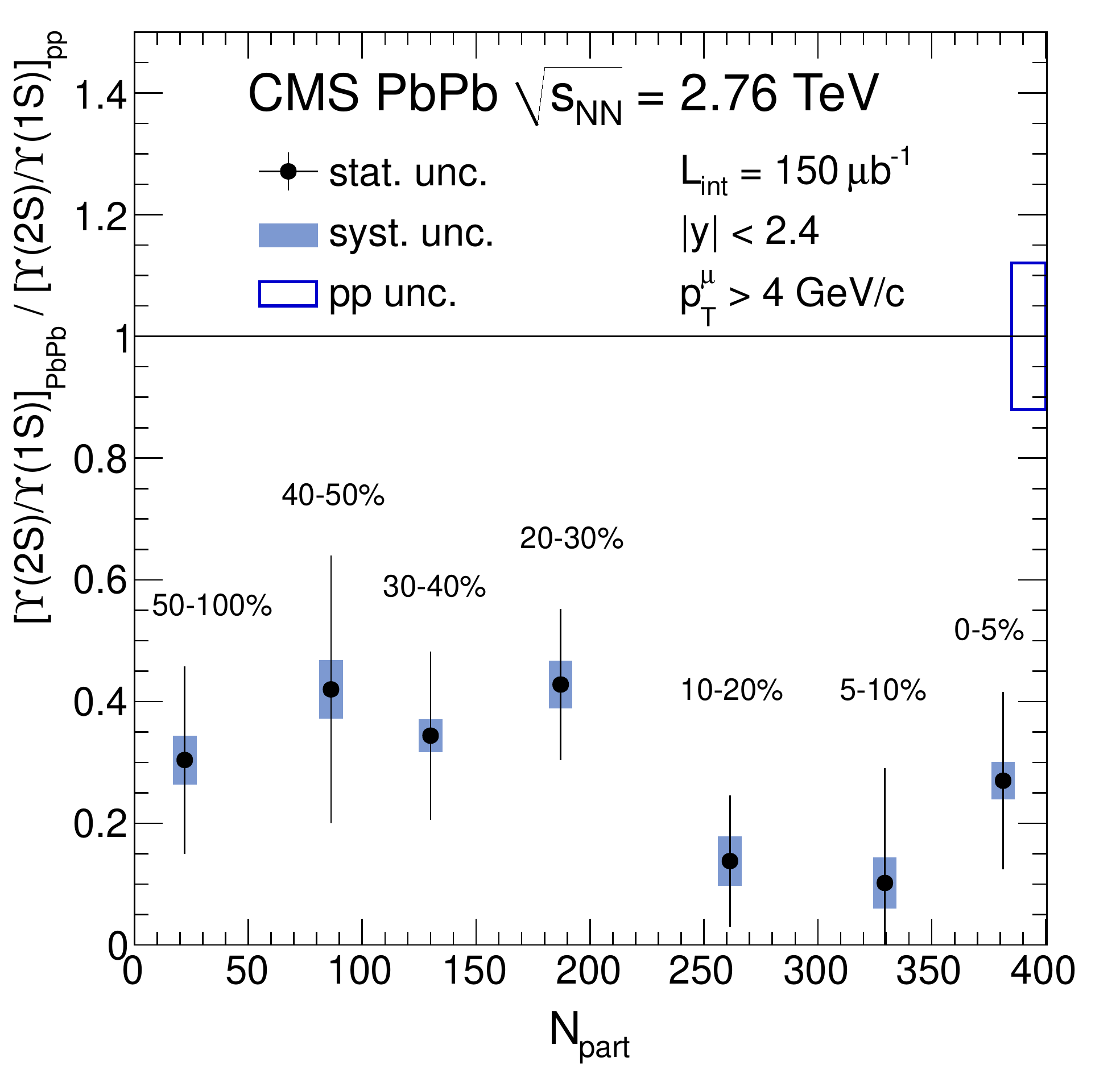} 
\includegraphics[trim={ 0 0.2cm 0 0}, clip,width=\linewidth, height=.66\linewidth]{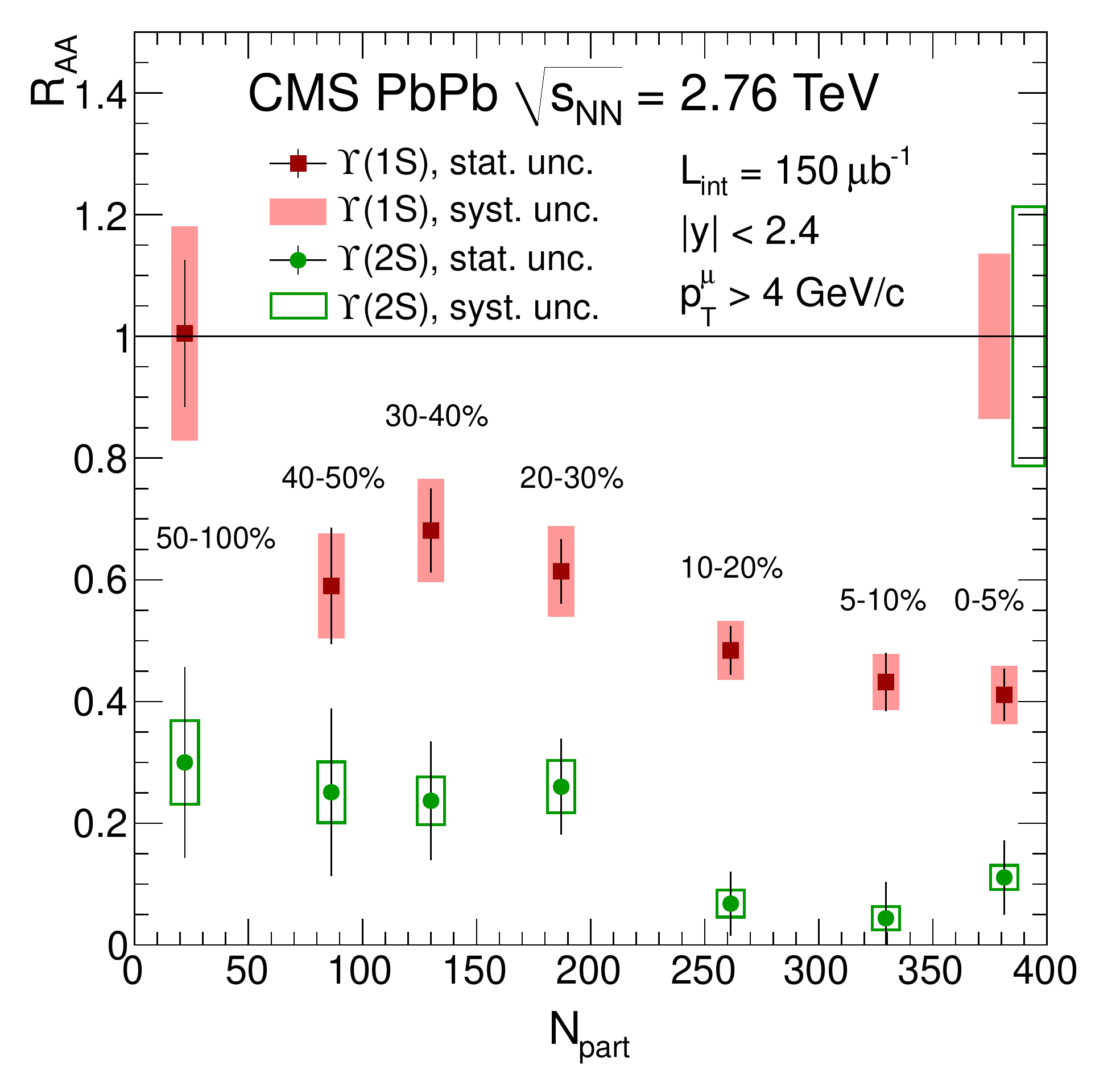} 
\end{minipage}
\caption{
\textbf{Left}:  The di-muon invariant-mass distributions from the (\textbf{a}) \pp  and (\textbf{b}) \pbpb  data at \sqsn= 2.76~TeV;
\textbf{Right}: The centrality dependence (\textbf{top})  of the double ratio  and (\textbf{bottom}) of the nuclear modification factors \raa 
for the \PgUa and \PgUb states. The event centrality bins are indicated by percentage intervals. Reproduced from 
Ref.~\cite{Chatrchyan:2012lxa}.
}
\label{upsilons}
\end{figure}

On the left panels of Figure~\ref{upsilons}, the invariant-mass distributions of $\mu^+\mu^-$ pairs (di-muons) produced in the \pp (a) and \pbpb (b) collisions at 
the LHC are presented. A prominent peak due to the production of the heavy quarkonium state, the bottomonium \PgUa, can be clearly seen in both \pp and \pbpb data. 
Peaks from the higher excited states of \PgU, \PgUb, and \PgUc, although discernible in the \pp case, are barely visible in the \pbpb data. More quantitative 
information on this effect can be found in the right panels of Figure~\ref{upsilons}, where the centrality dependence of the double ratio  $\left[\PgUb/\PgUa\right]_{\rm PbPb}/\left[\PgUb/\PgUa\right]_{\rm pp}$ (top) and of the nuclear modification factors \raa  of \PgUa  and \PgUb (bottom) are displayed. Let us note that 
the observed suppression of the relative yield is in agreement with the expectations that different quarkonium states will dissociate at different temperatures 
with a suppression pattern ordered sequentially with the binding energy; i.e.,~the difference between the mass of a given quarkonium and twice the mass of 
the lightest meson containing the corresponding heavy quark \cite{Digal:2001ue}. Moreover, the observed pattern is now also confirmed in \pbpb collisions 
at \sqsn = 5.02 TeV \cite{CMS:2016ayg}. The double ratio is significantly below unity at all centralities, and no variation with kinematics is observed, confirming 
a strong \PgU suppression in heavy-ion collisions at the LHC.

\subsection{Penetrating Probes}

Electromagnetic probes like photons \cite{Feinberg:1976ua, Shuryak:1978ij} and di-leptons \cite{Domokos:1980ba} (for recent developments, see 
Refs.~\cite{Basso:2016ulb,Basso:2015pba,Goncalves:2016qku}) 
have long been
~expected to provide crucial information on the properties of QGP. 
The absence of strong final-state interactions makes them an ideal {\it penetrating probe} of strongly-interacting matter \cite{Stankus:2005eq}. In collisions 
of ultra-relativistic nuclei, the photons and leptons can be produced either in the initial hard collisions between partons of the incident nuclei (e.g.,~$qg \rightarrow q\gamma$,  
$q\bar{q}\rightarrow\gamma g$, or $q\bar{q}\rightarrow\ell \bar{\ell}$), or radiated from the thermally equilibrated partons and hadrons or via hadronic decays. 
The direct photons are defined to be all produced photons, except those from hadron decays in the last stage of the collision. The high-\pt isolated photon can be 
used to estimate the momentum of the associated parton, allowing a characterisation of the in-medium parton energy loss (see Figure~\ref{gamma_jet}). The prompt 
photons also carry information about the initial state and its possible modifications in nuclei, and should thus be one of the best probes of the gluon saturation. 
The thermal photons emitted from the produced matter in nuclear collisions carry information on the temperature of QGP. 

\begin{figure}[H]
\centering
\includegraphics[trim={ 0 .35cm 0 .8cm}, clip,width=.45\textwidth, height=.47\textwidth]{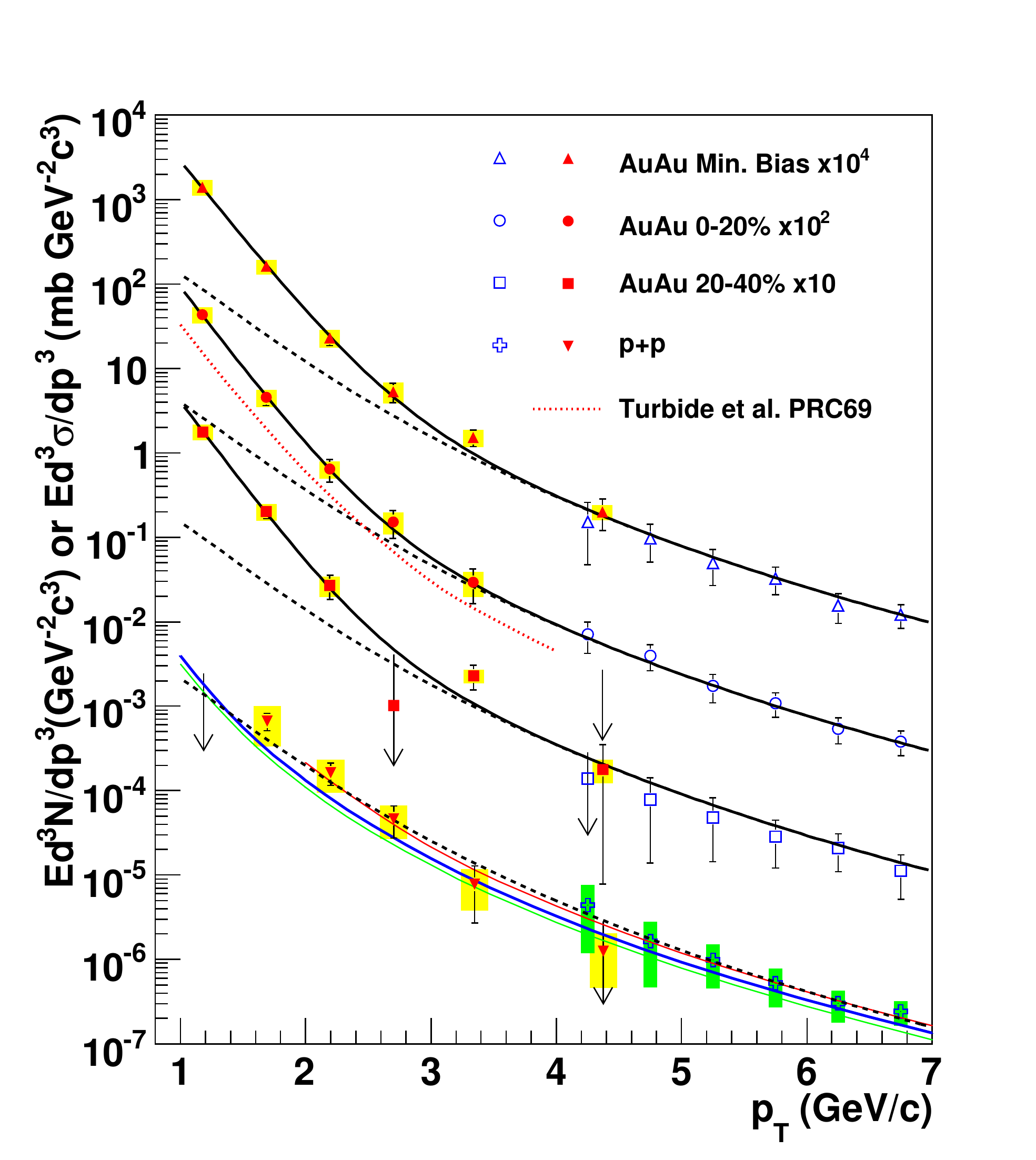} \;
\includegraphics[trim={ 0  0 0 0}, clip, width=.45\textwidth, height=.44\textwidth]{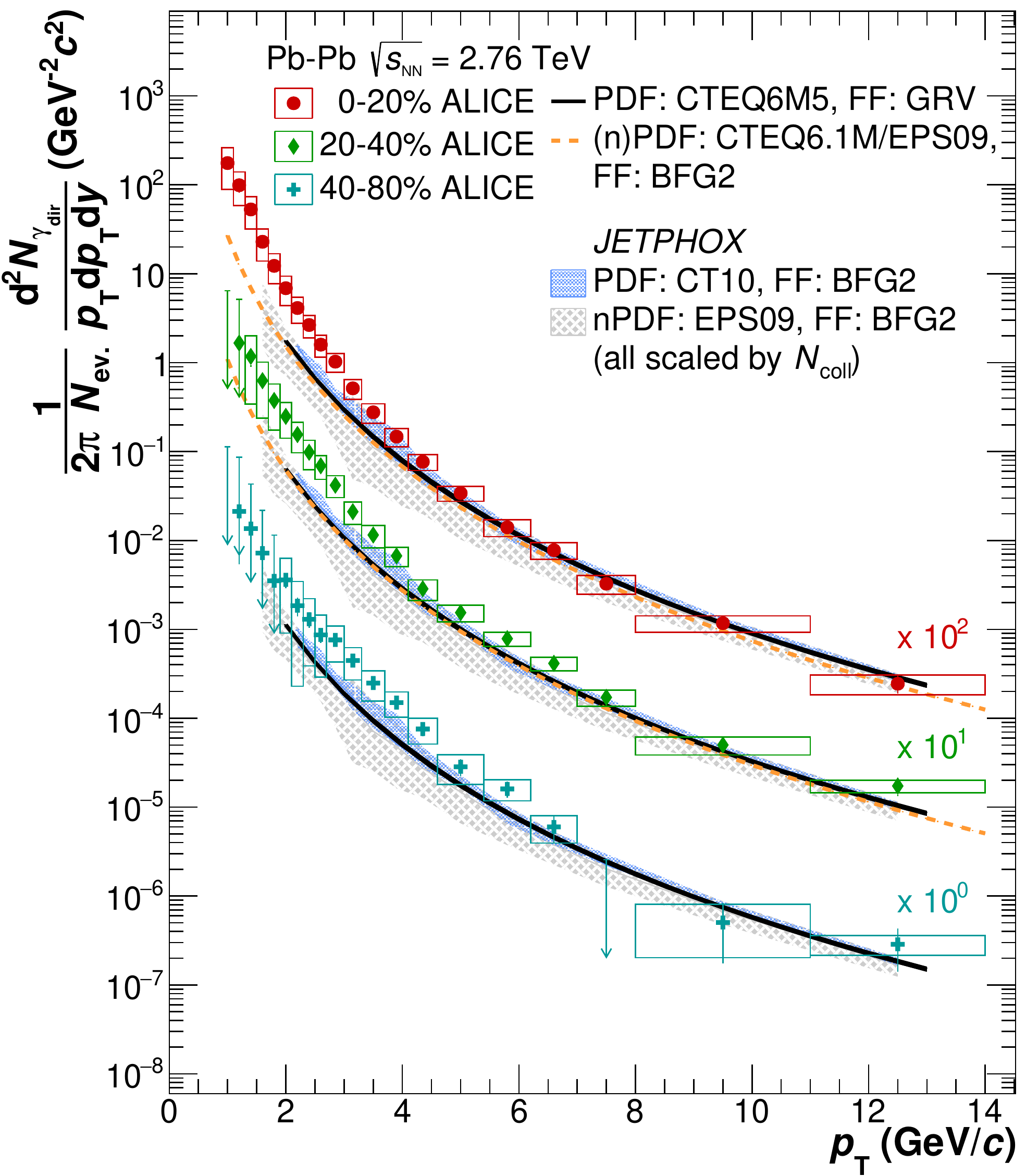}
\caption{\textbf{Left}: The invariant cross-section (\pp) and invariant yield (\auau) of direct photons as functions of \pt. The three curves on the \pp data 
represent the NLO pQCD~calculations, and the dashed curves show a modified power-law fit to the \pp data, scaled by $T_{\rm AA} = \left<\Ncoll\right>/\sigma^{in}_{NN}$. 
The~dashed (\textbf{black}) curves are the same, but with the exponential plus scaled \pp data. The dotted (red) curve near the 0--20\% centrality data is a theory calculation. 
Reproduced from Ref.~\cite{Adare:2008ab}; \textbf{Right}: The direct photon spectra in \pbpb collisions at \sqsn = 2.76 TeV for the 0--20\% (scaled by a factor of 100), 
the 20--40\% (scaled by a factor of 10), and the 40--80\% centrality classes compared to the NLO pQCD predictions for the direct photon yield in $pp$ collisions 
at the same energy, scaled by a number of binary nucleon collisions for each centrality class. Reproduced from Ref.~\cite{Adam:2015lda}.}
\label{dir_gamma}
\end{figure}

The first observation of direct photons in ultra-relativistic heavy-ion collisions has been made by the CERN SPS experiment WA98 \cite{Aggarwal:2000th}. 
In 10\% most central \pbpb collisions at \sqsn= 17.2 GeV, they observed a clear excess of direct photons in the range of \pt > 1.5 GeV/c which was not present 
in more peripheral collisions. The extraction of the direct photon signal (which is extremely difficult) was described in-depth in Ref.~\cite{Aggarwal:2000ps}. 
The data from RHIC and the LHC are presented in Figure~\ref{dir_gamma} on the left and right panels, respectively.

\section{New Developments}
\label{sec:developments}

\subsection{Search for the Critical Point of QCD Phase Diagram}
\label{CEP}

The search for the (tri)critical point (CEP) in the $T - \mu_B$ phase diagram---where the phase transition between the QGP and hadron matter changes 
from the first- to the second-order one---represents one of the most active fields of contemporary ultra-relativistic heavy-ion physics, both experimentally~\mbox{\cite{Aggarwal:2010cw,  Sumbera:2012qb, Sumbera:2013kd, Adare:2015aqk}} and theoretically
~\cite{Stephanov:1998dy, Buballa:2005, Braun-Munzinger:2015hba,  Asakawa:2015ybt}. In order to gain 
more insight into the CEP location, quite advanced techniques from condensed matter physics, such as Finite-Size Scaling (FSS) analysis of data 
\cite{Lacey:2014wqa} or thermal fluctuations characterized by the appropriate cumulants of the partition function~\cite{Stephanov:1998dy, Gavai:2010zn, Asakawa:2015ybt} 
are being exploited. 

The search for the CEP exploiting the potential of the RHIC accelerator complex was mounted by the STAR and PHENIX collaborations in 2010 within the 
{\it Beam Energy Scan} (BES) program \cite{Aggarwal:2010cw}. Going down from the RHIC maximum energy \sqsn = 200 GeV, they have scanned the available 
phase space down to \sqsn = 7.7 GeV. Some of the results from that scan were already mentioned in previous sections, and can be found in Figures~\ref{fig:accelerators}, 
\ref{fig:freeze-out}, \ref{fig:MultMid}, \ref{v2_exp}, \ref{v3_exp}. In the following, we therefore restrict ourselves to the measurements which provide a direct link 
to the lattice results discussed in Section~\ref{subsec:munonzero}. As~already mentioned there, the study of ratios of the Taylor expansion coefficients given by 
Equation~(\ref{fluc})---which are also known also as susceptibilities---seems to be very attractive, since both the temperature and volume dependences drop out. In particular, 
the ratio $\chi_4^B/ \chi_2^B$ calculated from the moments of the net-baryon multiplicity $N_B$ has different values for the hadronic and partonic phases 
\cite{Gavai:2010zn}. For HRG, it equals unity, but is expected to deviate from unity near the CEP. Other interesting ratios of the net-baryon charge moments 
which can be expressed using mean ($M_B$), variance ($\sigma_B^2$), skewness ($S_B$), and kurtosis ($k_B$) of the net baryon number distributions read
\begin{equation}
R_{ij}^B \equiv \frac{\chi_i^B} {\chi_j^B} ~~,~~    R_{12}^B = \frac{M_B}{\sigma_B^2}~~,~~R_{31}^B=\frac{S_B \sigma_B^3}{M_B}~~,~~R_{42}^B = k_B \sigma_B^2 \,.
\label{chi_ratios}
\end{equation}

\begin{figure}[H]
\centering
\includegraphics[width=0.38\textwidth]{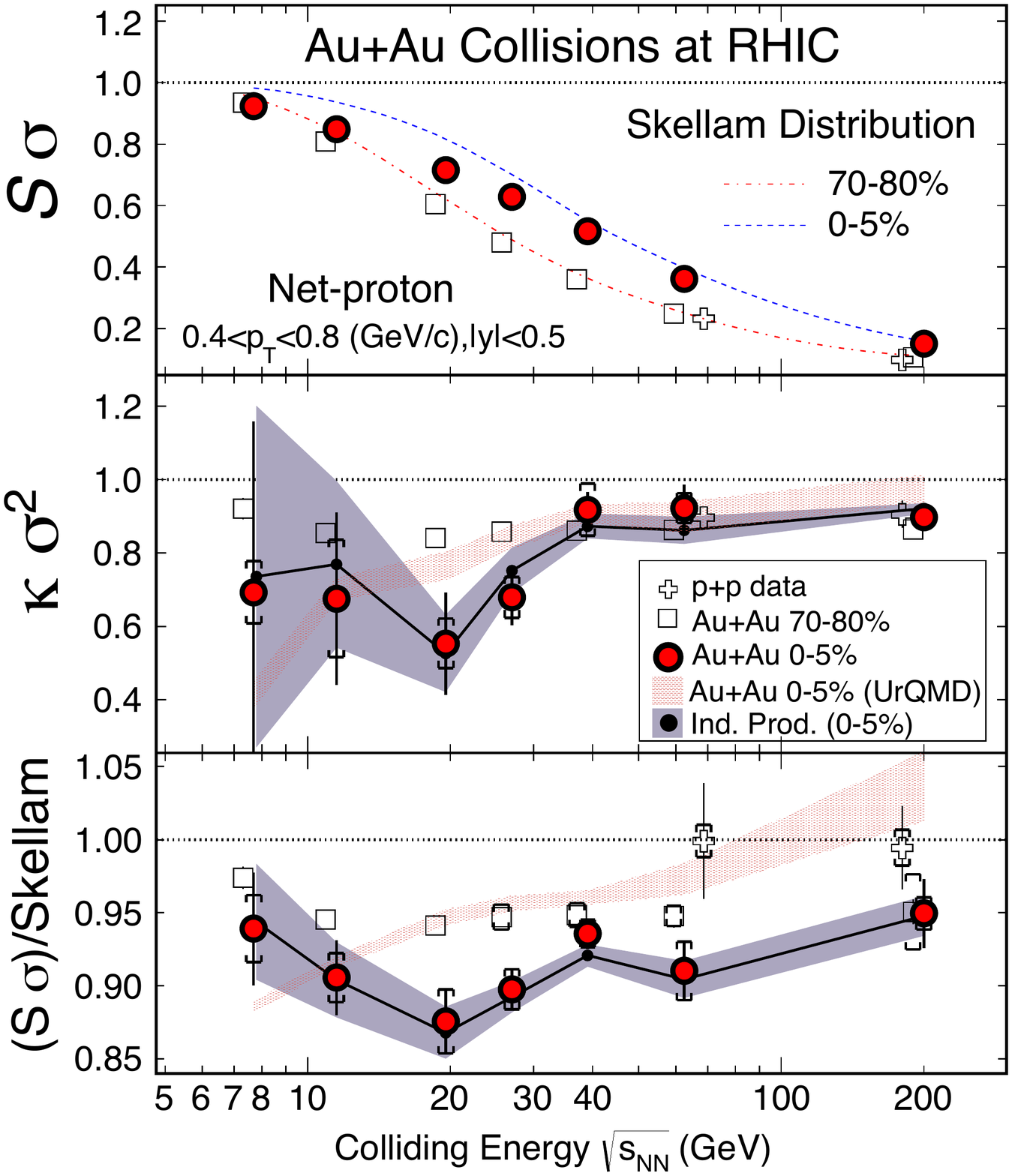} 
\includegraphics[width=0.38\textwidth]{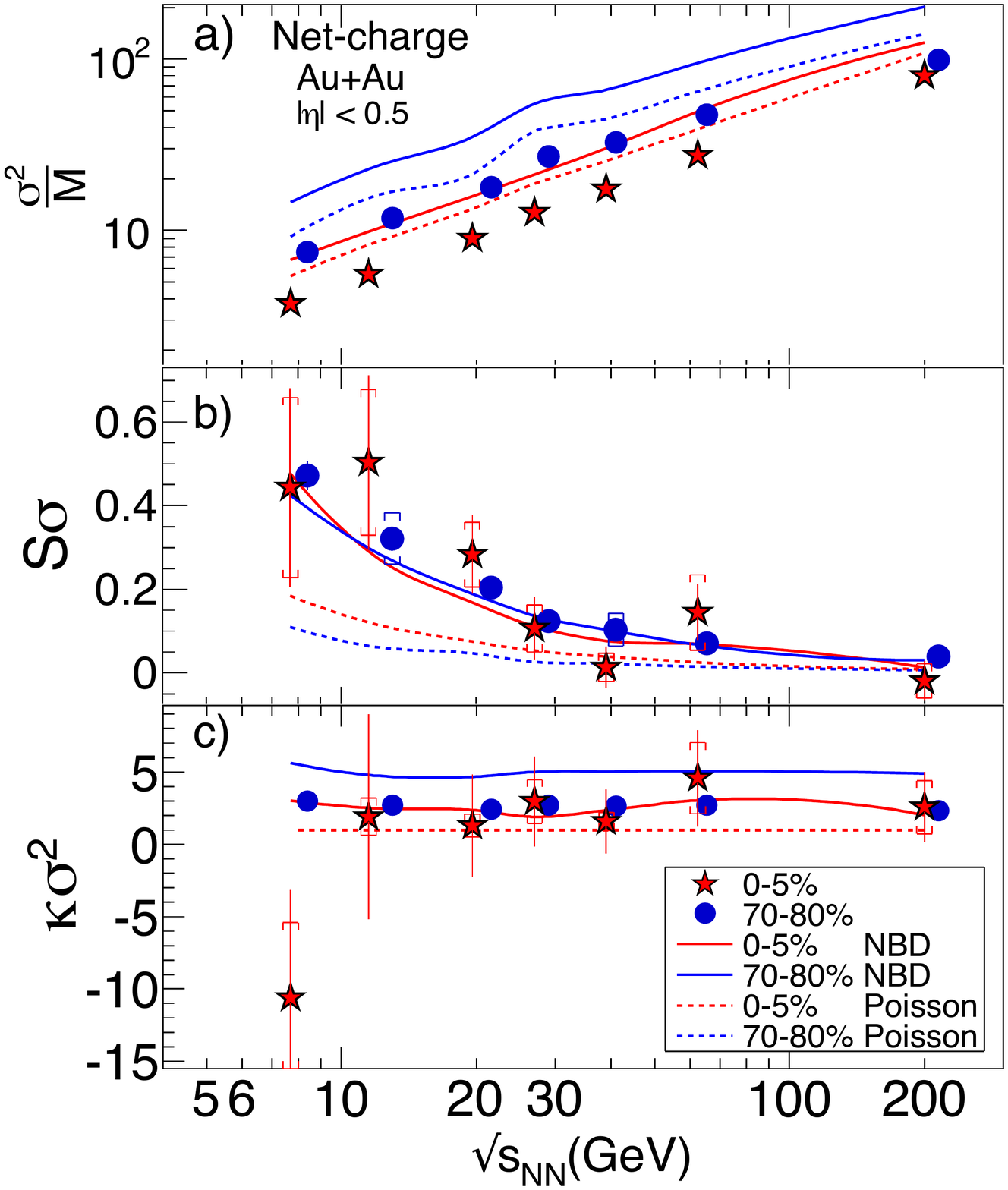}
\vspace*{-.7cm}
\caption{The energy dependence of appropriate combinations of the moments---mean ($M$), variance ($\sigma^{2}$), skewness ($S$), and kurtosis ($\kappa$)---of the multiplicity of conserved charges at mid-rapidity in \auau collisions for seven energies ranging from \sNN~=~7.7 to 200~GeV, and for two centralities, 
the most central (0--5\%) and peripheral (70--80\%) bins. \textbf{Left}: The collision energy and centrality dependence of the net-proton ${\it{S}}\sigma$ and 
$\kappa\sigma^2$ from Au $+$ Au and $p$ $+$ $p$ collisions at RHIC (adapted from Ref.~\cite{Adamczyk:2013dal}); \textbf{Right}: The beam-energy dependence of 
the net-charge multiplicity distributions moments (\textbf{a}) \MV, (\textbf{b}) \Ss, and (\textbf{c}) \KV, after all corrections. Reproduced from Ref.~\cite{Adamczyk:2014fia}. 
The results from the Poisson and the Negative Binomial Distribution (NBD) baselines are superimposed. The values of \KV~ for the Poisson baseline are always unity.}
\label{STAR_fluct}
\end{figure}

Experimentally, the net-baryon number $N_B$ fluctuations and their cumulants are not accessible, and so one has to resort to measurements of the cumulants 
of the net-proton number $N_P$ fluctuations~\cite{Aggarwal:2010wy, Adamczyk:2013dal}. On the other hand, the electric charge fluctuations are experimentally 
accessible~\cite{Adamczyk:2014fia}. This is illustrated in Figure~\ref{STAR_fluct}, where the measurements from the STAR experiment at RHIC are shown. Generally, 
study of the $p_T$ and rapidity acceptance dependence for the moments of the net-proton distributions shows that the larger the acceptance is, the larger 
are the deviations from unity. Preliminary results from the STAR collaboration \cite{Luo:2015ewa} reveal that increasing the transverse momentum acceptance 
of protons and anti-protons from 0.4 < $p_T$ < 0.8 GeV/c to 0.4 < $p_T$ < 2.0 GeV/c leads to a pronounced non-monotonic structure in the energy 
dependence of \KV~of net-proton distributions from 5\% most central \auau collisions. At energies above 39 GeV, the values of \KV are close to unity, while 
for energies below 39 GeV, it shows a significant deviation below unity around 19.6 and 27 GeV and a large increase above unity observed at 7.7 GeV.\vspace{-12pt}

\subsection{Collectivity in Small Systems}
\label{CEP}

Recent years have witnessed a surprising development in multiparticle dynamics of high multiplicity \pp \cite{Khachatryan:2010gv, Khachatryan:2015lva, Aad:2015gqa} 
and \pa  \cite{Aad:2012gla, CMS:2012qk, Abelev:2012ola, Khachatryan:2015waa, Koop:2015wea, Adare:2015ctn} collisions. It all started in 2010 with 
the observation of {\it ridge-like} structures in \pp collisions by the CMS experiment at the LHC \cite{Khachatryan:2010gv}. The surprise was due to the fact 
that a very similar effect was found just a few years before in heavy-ion collisions: first in \auau collisions at \sqsn= 200 GeV at RHIC \cite{Alver:2009id, Abelev:2009af}, 
and later on also in \pbpb collisions at the LHC \cite{Aamodt:2011by} and in \cucu collisions at \sqsn= 62.4 GeV and 200 GeV at RHIC \cite{Agakishiev:2011st}.

In heavy-ion collisions, it was found that pairs of particles are preferentially emitted with small relative azimuthal angles ($\dphi = \phi_1-\phi_2 \approx 0$). 
Surprisingly, this preference persists even when the particles are separated by large pseudo-rapidity ($\eta$) gaps ($-4 < \left|\deta\right| < 2$).
These long-range correlations---known as the ridge---have been traced to the conversion of density anisotropies in the initial overlap of the two nuclei 
into momentum space correlations through subsequent interactions in the expansion~\cite{Voloshin:2004th}.

\begin{figure}[H]
\centerline{
  \mbox{\includegraphics[width=0.7\linewidth]{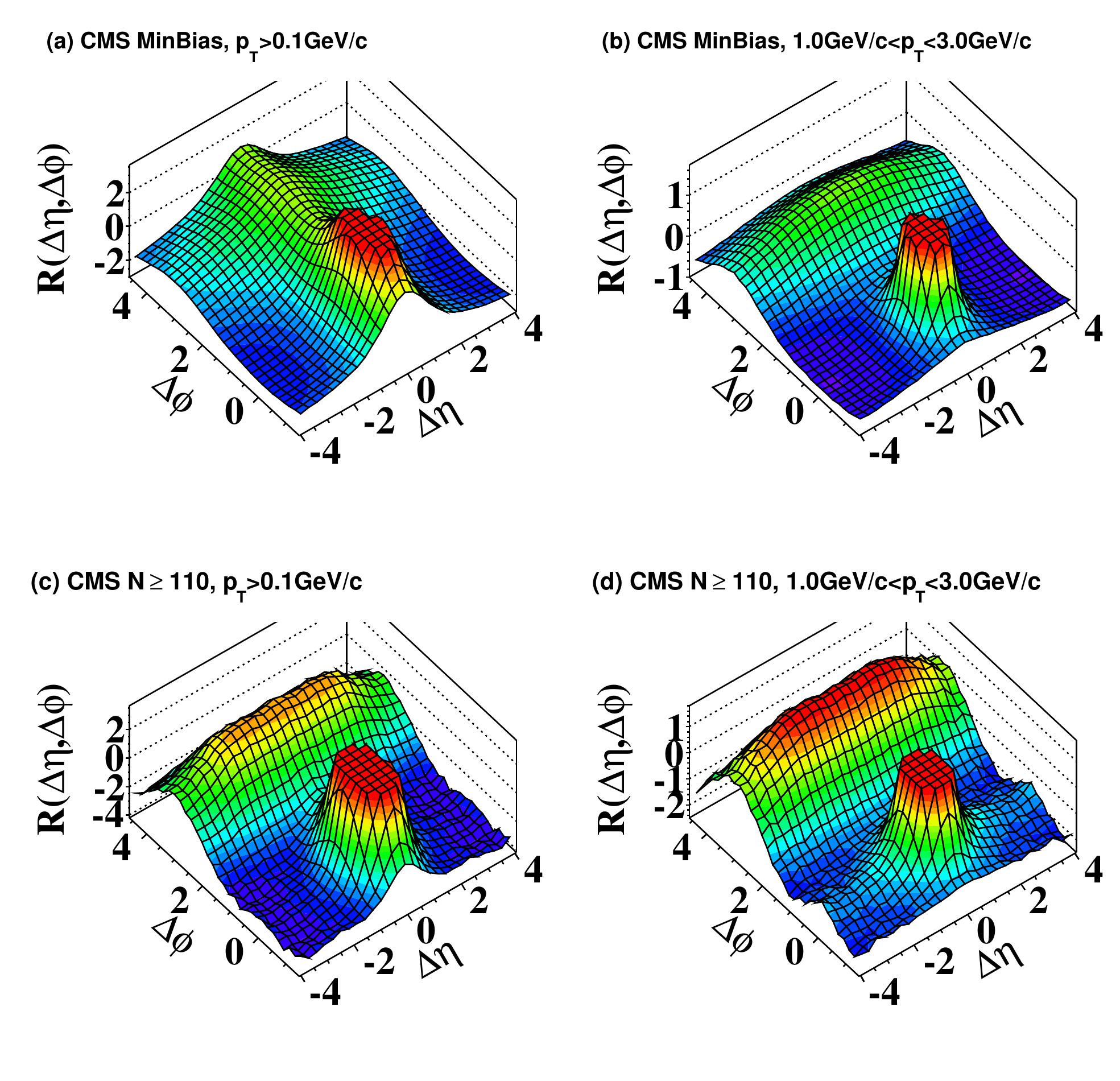}}}
\vspace{-0.3cm}
  \caption{ \label{corr_2D_finaldata_bothenergy}
The 2-D two-particle correlation functions for \pp at \sqs = 7~TeV: (\textbf{a}) minimum bias events with \pt > 0.1~GeV/c; 
(\textbf{b}) minimum bias events with 1 < \pt < 3~GeV/c; (\textbf{c}) high multiplicity ($N_\mathrm{trk}^\mathrm{offline} \geq 110$) events 
with \pt > 0.1~GeV/c; and (\textbf{d}) high multiplicity ($N_\mathrm{trk}^\mathrm{offline} \geq 110$) events with \mbox{1< \pt < 3 GeV/c}.
The sharp near-side peak from jet correlations is cut off in order to better illustrate the structure outside that region. 
Reproduced from Ref.~\cite{Khachatryan:2010gv}.}
\label{2pcorr_incl}
\end{figure}

In \pp minimum bias collisions at the LHC, the peak in the correlation function of particles with \pt > 0.1 GeV/c observed at small angular differences 
(\deta,~\dphi $\approx$ $0$, see Figure~\ref{2pcorr_incl}a) is due to several effects: resonance decays, Bose--Einstein correlations, and near-side jet fragmentation. 
The~fragmentation due to back-to-back jets is visible as a broad elongated ridge around \dphi $\approx$ $\pi$. The~pattern does not change much even when selecting 
the events with very high multiplicity $N \ge 110$ (see Figure~\ref{2pcorr_incl}c). The cut on the multiplicity enhances the relative contribution of high \pt jets, which 
fragment into a large number of particles, and therefore, has a qualitatively similar effect on the shape as the particle cut 1 < \pt < 3 GeV/c on minimum bias events 
(see Figure~\ref{2pcorr_incl}b). However, using now the particle cut 1 < \pt < 3 GeV/c in conjunction with a high multiplicity cut changes the picture dramatically (see Figure~\ref{2pcorr_incl}d). A novel feature never seen before in  \pp collisions at lower energies shows up---a clear and significant ridge-like structure at 
\dphi $\approx$ $0$ extending to $\left| \deta\right|$ of at least four units~\cite{Khachatryan:2010gv}. 

Let us note that for two particles with approximately the same energy $E_1$ $\approx E_2$ $\approx$ $E$, the correlations at \deta $\approx$ $\Delta y$ 
are by the uncertainty relation $\Delta x \approx 1/\Delta p =1/(E \Delta y)$ connected to the correlations in coordinate space. While for pions 
with \deta $\approx$ 1 and  \pt $\approx$ 0.1~GeV/c, we have $\Delta x$ $\approx$ 1 fm, for \deta $\approx$ 4 and \mbox{$\pt$ $\approx$ 1~GeV/c}, 
one gets $\Delta x$ $\approx$ 0.02 fm. It is obvious that at such small inter-parton distances, one enters the realm of the initial state description 
of nuclear collisions when the density of matter even inside the proton fluctuates---see Section~\ref{CGC}. The CGC scenario was recently 
exploited in Ref.~\cite{Rezaeian:2016szi} to predict long-range photon-jet correlations in \pp and \pa collisions at near-side for low transverse 
momenta of the produced photon and jet in high-multiplicity events.

\begin{figure}[H]
\centering
\includegraphics[width=.9\textwidth, height=.5\textwidth]{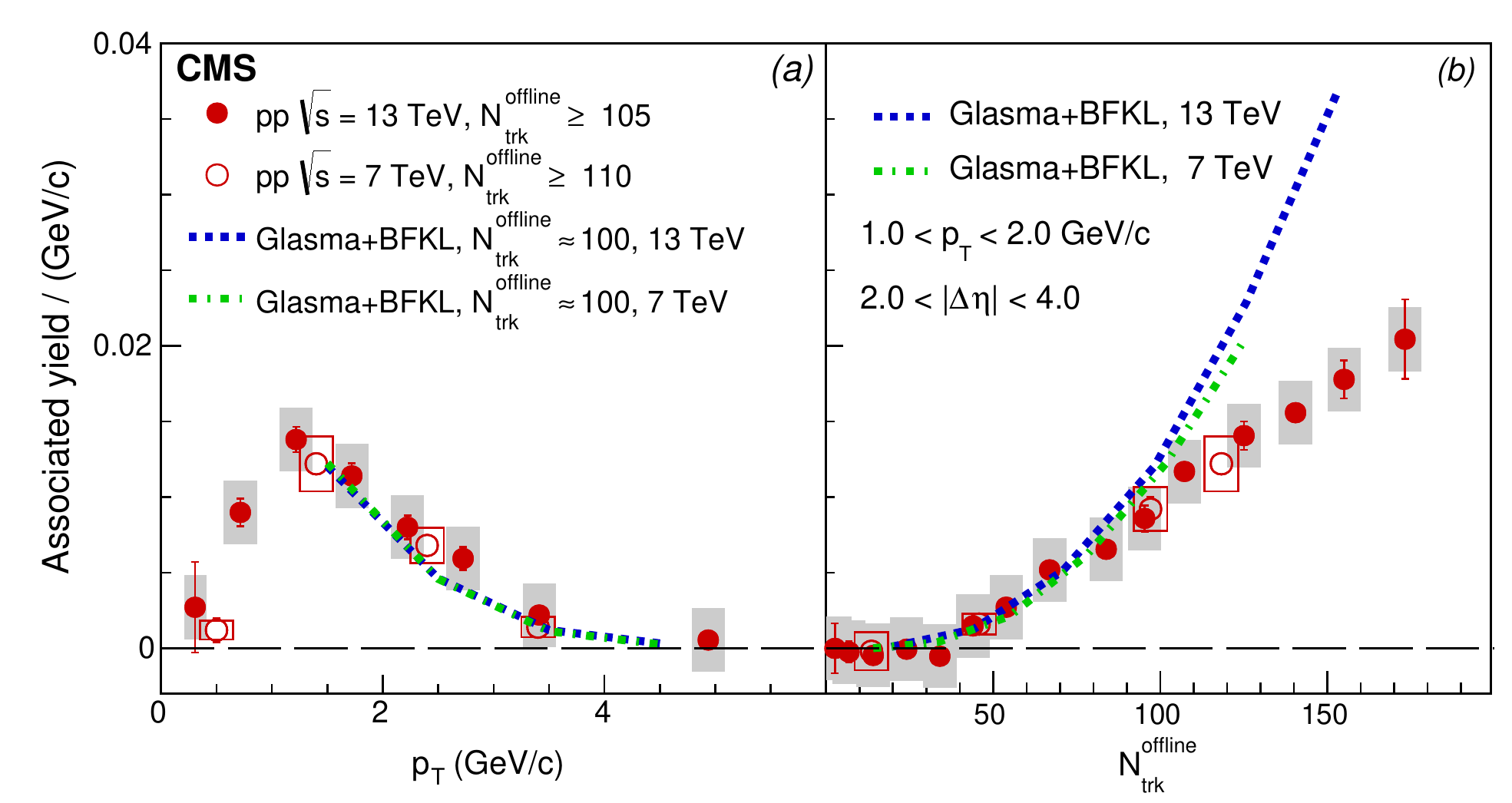} 
    \caption{The associated yield for the near-side of the correlation function averaged over $2<\left| \deta \right| <4$ 
for \pp data at \sqs = 13~TeV (filled circles) and 7~TeV (open circles) \cite{Khachatryan:2015lva}.
Panel (\textbf{a}) shows the associated yield as a function of \pt for events with $\rm N^{\rm offline}_{\rm trk} \geq 105$. 
In panel (\textbf{b}), the associated yield for 1 < \pt < 2 GeV/c is shown as a function of multiplicity---$\rm N^{\rm offline}_{\rm trk}$.
The \pt selection applies to both particles in each pair. Curves represent the predictions of the gluon saturation model~\cite{Dusling:2015rja}.}
\label{Ridge}
  \end{figure}

The relevance of the saturation approach is further supported by the observation of the same ridge-like structure in \pp collisions at \sqs=13 TeV 
\cite{Khachatryan:2015lva, Aad:2015gqa}, see Figure~\ref{Ridge}. For the associated yield of long-range near-side correlations for high-multiplicity events (N >110) peaks in the region  $1 < \pt < 2$~GeV/c, see Figure~\ref{Ridge}a---
the yield reaches a maximum around $\pt$ $\approx$1~GeV/c 
and decreases with increasing \pt. No center-of-mass energy dependence is visible. The multiplicity dependence of the associated yield for $1 < \pt < 2$~GeV/c 
particle pairs is shown in Figure~\ref{Ridge}b. For low-multiplicity events, the associated yield is consistent with zero. At higher multiplicity, the ridge-like 
correlation emerges, with an approximately linear rise of the associated yield with multiplicity for ${\rm N} \geq 40$. Let us note that within the CGC models, 
the observation that the integrated near-side yield as a function of multiplicity is independent of collision energy is a natural consequence of the fact that 
multiparticle production is driven by a single semi-hard saturation scale \cite{Dusling:2015rja}.

\vspace{-12pt}

Another interesting phenomenon observed during recent years is a flow-like pattern in super-central \pp and \ppb collisions at the LHC \cite{CMS:2012qk, Aad:2013fja, ABELEV:2013wsa, Abelev:2014mda, Khachatryan:2014jra, Khachatryan:2015waa} and in 
\dau, \heau, and \pdau collisions at RHIC \cite{Koop:2015wea, Adare:2015ctn}. 
These collisions not only reveal a similar elliptic flow $v_2$, but in some cases, 
also $v_3$ anisotropy previously observed only in collisions of large nuclei. The measurement of higher-order cumulants of the azimuthal distributions has strengthened 
the collective nature interpretation of the anisotropy seen in \ppb collisions. Moreover, the collective radial flow analysis of \pp data enabled authors of 
Ref.~\cite{Hirono:2014dda} to claim that in this case the fireball explosions start with a very small initial size, well below 1 fm. This raises questions about 
whether the perfect liquid sQGP is also formed in these much smaller systems \cite{Bozek:2012gr, Bozek:2013ska, Werner:2013ipa, Romatschke:2015gxa, Niemi:2015ile, Koblesky:2015uba}. Are these flow-like 
structures only similar in appearance to what one observes in heavy-ion collisions, or do they have the same physical origin \cite{Schenke:2014zha, Bzdak:2014dia, Bozek:2015swa}? Obviously, answering these questions 
may help us to understand the emergence of collective phenomena in strongly-interacting systems in general. To make further progress on these fundamental 
questions, more analyses---both experimental and theoretical---are really needed \cite{Antinori:2016zxe}.


\section{Conclusions}
\label{sec:conclusions}

In this review, we have tried to present an up-to-date phenomenological summary of a relatively new and rapidly developing field of 
contemporary physics---the physics of Quark Gluon Plasma. We have also explored its broader ramifications when discussing matter under extreme conditions, 
strongly interacting plasmas, physics of strong electromagnetic fields, history of ultra-relativistic heavy-ion collisions, relativistic hydrodynamics, the role of the QCD 
ground state, and QCD saturation phenomena. In the experimental part, we were overwhelmed by a huge amount of results, and so, in order to keep this review of 
tolerable length, we had to skip several quite important topics. To name just a few---flow and suppression of identified particles \cite{Nouicer:2015jrf, Bala:2016hlf}, 
femtoscopy \cite{Sinyukov:2013zna, Lisa:2016buz}, or identified particle yields and chemical freeze-out conditions \cite{Andronic:2008gu, Roland:2014jsa, Nouicer:2015jrf}. 
We hope that the interested reader will find in the given references enough information to pursue a deeper study of these important subjects. However, in skipping these topics, 
we hope that we have not given up our main goal---to give the reader a possibility to see the QGP landscape at large. At last, we would like to say that it is never enough to stress how important this field is also for other branches of physics, and so we finish with yet another argument. The study of ultra-relativistic heavy-ion collisions 
appears so far to be our only way of studying the phase transitions in non-Abelian gauge theories (most likely taken place in the early universe) under laboratory conditions.


\vspace{6pt}
\acknowledgments{This work was supported in part by the Swedish Research Council, contract number 621-2013-428, the grant LG 13031 
of the Ministry of Education of the Czech Republic and by the grant 13-20841S of the Czech Science Foundation (GACR).}


\renewcommand\bibname{References}


\begin{thebibliography}{999}

\bibitem{Collins:1974ky} 
  Collins, J.C.;  Perry, M.J. Superdense Matter: Neutrons or Asymptotically Free Quarks? \emph{Phys.  Rev. Lett.} {\bf 1975}, \emph{34}, 1353.
  
\bibitem{Cabibbo:1975ig}
Cabibbo, N.; Parisi, G. Exponential hadronic spectrum and quark liberation.
  \emph{Phys. Lett. B} {\bf 1975}, \emph{59}, 67--69. 
  
\bibitem{GellMann:1964nj} 
 Gell-Mann, M.
  A Schematic Model of Baryons and Mesons.
  \emph{Phys. Lett.}  {\bf 1964}, \emph{8}, 214--215. 
  
\bibitem{Zweig:1981pd} 
  Zweig, G.
  An SU(3) Model for Strong Interaction Symmetry and Its Breaking. Version 1.
  CERN-TH-401. Available online: \myurl{http://inspirehep.net/record/11881/files/CM-P00042883.pdf} (accessed on 17~January~2017).

\bibitem{Fritzsch:1973pi} 
 Fritzsch, H.; Gell-Mann, M.; Leutwyler, H.
Advantages of the Color Octet Gluon Picture.
  \emph{Phys. Lett. B} {\bf 1973}, \emph{47}, 365--368.
  
\bibitem{Gross:1973id} 
 Gross, D.J.; Wilczek, F.
Ultraviolet Behavior of Nonabelian Gauge Theories.
 \emph{Phys. Rev. Lett.}  {\bf 1973}, \emph{30}, 1343--1346.
  
\bibitem{Politzer:1973fx} 
 Politzer, H.D. 
Reliable Perturbative Results for Strong Interactions?
  \emph{Phys. Rev. Lett.}  {\bf 1973}, \emph{30}, 1346--1349.

\bibitem{Nambu:1976ay} 
  Nambu, Y.
The Confinement of Quarks.
  \emph{Sci. Am.}  {\bf 1976}, \emph{235}, 48--70.

\bibitem{Yagi:2005yb} 
 Yagi, K.; Hatsuda, T.; Miake, Y.
 {\it Quark-Gluon Plasma: From Big Bang to Little Bang (Cambridge Monographs on Particle Physics, Nuclear Physics and Cosmology)}; Cambridge University Press: Cambridge, UK, 2005.

\bibitem{Schmidt:1992ge} 
  Schmidt, H.R.; Schukraft, J.
The Physics of ultrarelativistic heavy ion collisions.
  \emph{J. Phys. G} {\bf 1993}, \emph{19}, 1705--1795.

\bibitem{Stock:2008ru}
 Stock, R.
Nucleus-Nucleus Collisions and the QCD Matter Phase Diagram. In\emph{ Landolt-B\"{o}rnstein---Group I Elementary Particles, Nuclei
  and Atoms, Vol. 21A, Chapter 7}; Springer: Berlin/Heidelberg, Germany, 2009.

\bibitem{Schukraft:2015dna} 
 Schukraft, J.; Stock, R.
\emph{Toward the Limits of Matter: Ultra-Relativistic Nuclear Collisions at CERN}; World Scientific: Singapore, 2015.
  
\bibitem{Heinz:2000bk} 
  Heinz, U.W.; Jacob, M.
 \emph{Evidence for a New State of Matter: An Assessment of the Results from the CERN Lead Beam Program}; Theoretical Physics Division: Geneva, Switzerland, 2000.

\bibitem{Arsene:2004fa} 
 Arsene, I.; Bearden, I.G.; Beavis, D.; Besliu, C.; Budick, B.; Bøggild, H.; Chasman, C.; Christensen, C.H.; Christiansen, P.; Cibor, J.; et al.
 Quark gluon plasma and color glass condensate at RHIC? The Perspective from the BRAHMS experiment.
  \emph{Nucl. Phys. A} {\bf 2005}, \emph{757}, 1--27.
  
\bibitem{Back:2004je} 
 Back, B.B.; Baker, M.D.; Ballintijn, M.; Barton, D.S.; Becker, B.; Betts, R.R.; Bickley, A.A.; Bindel, R.; Budzanowski, A.; Busza, W.; et al. 
The PHOBOS perspective on discoveries at RHIC.
  \emph{Nucl. Phys. A} {\bf 2005}, \emph{757}, 28--101.
  
\bibitem{Adams:2005dq} 
 Adams, J.; Aggarwal, M.M.; Ahammed, Z.; Amonett, J.; Anderson, B.D.; Arkhipkin, D.; Averichev, G.S.; Badyal, S.K.; Bai, Y.; Balewski, J.; et al. Experimental and theoretical challenges in the search for the quark gluon plasma: The STAR Collaboration's critical assessment of the evidence from RHIC collisions.
  \emph{Nucl.~Phys.~A} {\bf 2005}, \emph{757}, 102--183.
 
\bibitem{Adcox:2004mh} 
  Adcox, K.; Adler, S.S.; Afanasiev, S.; Aidala, C.; Ajitanand, N.N.; Akiba, Y.; Al-Jamel, A.; Alexander, J.; Amirikas, R.; Aoki, K.; et al. 
  Formation of dense partonic matter in relativistic nucleus-nucleus collisions at RHIC: Experimental evaluation by the PHENIX collaboration.
  \emph{Nucl. Phys. A} {\bf 2005}, \emph{757}, 184--283.

\bibitem{Gyulassy:2004zy} 
Gyulassy, M.; McLerran, L.
New forms of QCD matter discovered at RHIC.
  \emph{Nucl. Phys. A} {\bf 2005}, \emph{750}, 30--63.

\bibitem{Rafelski:2003zz} 
  Kapusta, J.; M\"{u}ller, B.; Rafelski, J. {\it Quark-Gluon Plasma: Theoretical Foundations. An Annotated Reprint Collection}; Elsevier: Amsterdam, The Netherlands, 2003.

\bibitem{Csernai:1994xw} 
 Csernai, L.P.
  {\it Introduction to Relativistic Heavy Ion Collisions}; Wiley: Chichester, UK, 1994. 
 
\bibitem{Letessier:2002gp} 
  Letessier, J.; Rafelski, J.
  {\it Hadrons and Quark-Gluon Plasma};
   Cambridge Monographs on Particle Physics, Nuclear Physics and Cosmology; Cambridge University Press: Cambridge, UK, 2002.
  
\bibitem{Vogt:2007zz} 
  Vogt, R.
  {\it Ultrarelativistic Heavy-Ion Collisions};
  Elsevier Science: Amsterdam, The Netherlands, 2007.
 
\bibitem{Florkowski:2010zz} 
  Florkowski, W.
  {\it Phenomenology of Ultra-Relativistic Heavy-Ion Collisions};
 World Scientific: Singapore, 2010. 

\bibitem{Rak:2013yta} 
  Rak, J.; Tannenbaum, M.J.
  {\it High $p_T$ Physics in the Heavy Ion Era}; 
   Cambridge Monographs on Particle Physics, Nuclear Physics and Cosmology; Cambridge University Press: Cambridge, UK, 2013.

\bibitem{BraunMunzinger:2007zz} 
 Braun-Munzinger, P.; Stachel, J.
The quest for the quark-gluon plasma.
  \emph{Nature }{\bf 2007}, \emph{448}, 302--309.

\bibitem{Shuryak:2008eq} 
 Shuryak, E.
Physics of Strongly coupled Quark-Gluon Plasma.
  \emph{Prog. Part. Nucl. Phys.}  {\bf 2009}, \emph{62}, 48--101.
  
\bibitem{Sarkar:2010zza} 
 Sarkar, S.; Satz, H.; Sinha, B.
  The physics of the quark-gluon plasma.
  \emph{Lect. Notes Phys.}  {\bf 2010}, \emph{785}, 1--369.
 
\bibitem{Florkowski:2014yza} 
  Florkowski, W.
Basic phenomenology for relativistic heavy-ion collisions. \emph{Acta Phys. Pol. B} {\bf 2014}, \emph{45}, 2329--2354.
   
\bibitem{Roland:2014jsa} 
  Roland, G.; Safarik, K.; Steinberg, P.
 Heavy-ion collisions at the LHC.
  \emph{Prog. Part. Nucl. Phys.}  {\bf 2014}, \emph{77}, 70--127.

\bibitem{Nouicer:2015jrf} 
  Nouicer, R.
  New State of Nuclear Matter: Nearly Perfect Fluid of Quarks and Gluons in Heavy Ion Collisions at RHIC Energies.
  \emph{Eur. Phys. J. Plus} {\bf 2016}, \emph{131}, 70. 
  
\bibitem{Zeldovich:1966}
 Zeldovich,  Y.B.; Raizer, Y.P. 
  {\it Physics of Shock Waves and High-Temperature Hydrodynamics Phenomena}; Academic Press: New York, NY, USA; London, UK, 1966.

\bibitem{Fortov:2011}
 Fortov, V.I.
 {\it Extreme States of Matter on Earth and in the Cosmos};
 Springer: Berlin/Heidelberg, Germany, 2011.

\bibitem{Oppenheimer:1939ne} 
 Oppenheimer, J.R.; Volkoff, G.M. On Massive Neutron Cores. \emph{Phys. Rev.} {\bf 1939}, \emph{55}, 374--381.

\bibitem{Landau1932}
Landau, L.D. On the theory of stars.
\emph{Physik. Zeits. Sowjetunion} \textbf{1932}, \emph{1}, 285--288.

\bibitem{Zeldovich:1962}
 Zeldovich, Y.B.
  The equation of state at ultrahigh densities and its relativistic limitations.
  \emph{J. Exp. Theor. Phys.}  {\bf 1962}, \emph{14}, 1143--1147.

\bibitem{Baym:1976yu} 
  Baym, G.; Chin, S.A. Can a neutron star be a giant MIT bag?
  \emph{Phys. Lett. B} {\bf 1976}, \emph{62}, 241--244.
  
\bibitem{Gamow:1946eb} 
  Gamow, G.
  Expanding universe and the origin of elements.
  \emph{Phys. Rev.}  {\bf 1946}, \emph{70}, 572--573.

\bibitem{Penzias:1965wn} 
  Penzias, A.A.; Wilson, R.W.
  A Measurement of excess antenna temperature at 4080-Mc/s.
  \emph{Astrophys. J.}  {\bf 1965}, \emph{142}, 419--421.

\bibitem{Sakharov:1966fva} 
  Sakharov, A.D.
 Maximum temperature of thermal radiation.
  \emph{JETP Lett.} {\bf 1966}, \emph{3}, 288--289.
 
\bibitem{Huang:1970iq} 
  Huang, K.; Weinberg, S. Ultimate temperature and the early universe.
  \emph{Phys. Rev. Lett.}  {\bf 1970}, \emph{25}, 895--897.
  
\bibitem{Hagedorn:1965st} 
  Hagedorn, R.
 Statistical thermodynamics of strong interactions at high-energies.
  \emph{Nuovo Cim. Suppl.} {\bf 1965}, \emph{3}, 147--186.
  
\bibitem{Rafelski:2016hnq} 
 Rafelski, J.
 {\it  Melting Hadrons, Boiling Quarks---From Hagedorn Temperature to Ultra-Relativistic Heavy-Ion Collisions at CERN: 
 With a Tribute to Rolf Hagedorn}; Springer: Cham, Switzerland, 2016.

\bibitem{Majumder:2010ik} 
  Majumder, A.; Muller, B.
 Hadron mass spectrum from lattice QCD.
  \emph{Phys. Rev. Lett.} {\bf 2010}, \emph{105}, 252002.

\bibitem{Zeldovich:1967rg} 
  Zeldovich, Y.B.
 The hot model of the universe and the elementary particles.
  \emph{Acta Phys. Hung.} {\bf 1967}, \emph{22}, 51--58.

\bibitem{Zeldovich:1967}
 Zeldovich, Y.B.
  Hot model of Universe.
  \emph{Sov. Phys. Uspechi}  {\bf 1967}, \emph{9}, 602--617.
 
\bibitem{Linde:1978px} 
  Linde, A.D.
  Phase Transitions in Gauge Theories and Cosmology.
 \emph{ Rep. Prog. Phys.} {\bf 1979}, \emph{42}, 389--437.

\bibitem{Bailin:2004zd} 
  Bailin, D.; Love, A.~{\it Cosmology in Gauge Field Theory and String Theory};
 IOP: Bristol, UK, 2004.

\bibitem{Boyanovsky:2006bf} 
  Boyanovsky, D.; de Vega, H.J.; Schwarz, D.J.
  Phase transitions in the early and the present universe.
  \emph{Ann.~Rev. Nucl. Part. Sci.}  {\bf 2006}, \emph{56}, 441--500.

\bibitem{Trodden:1998ym} 
  Trodden, M.
  Electroweak baryogenesis.
  \emph{Rev. Mod. Phys.}  {\bf 1999}, \emph{71}, 1463--1500.
 
\bibitem{Kajantie:1996mn} 
  Kajantie, K.; Laine, M.; Rummukainen, K.; Shaposhnikov, M.E.
  Is there a hot electroweak phase transition at m(H) larger or equal to m(W)?
  \emph{Phys. Rev. Lett.} {\bf 1996}, \emph{77}, 2887--2890.
  
\bibitem{Susskind:1979up} 
  Susskind, L.
Lattice Models of Quark Confinement at High Temperature.
  \emph{Phys. Rev. D} {\bf 1979}, \emph{20}, 2610--2618.

\bibitem{Petreczky:2012rq} 
  Petreczky, P.
  Lattice QCD at non-zero temperature.
  \emph{J. Phys. G} {\bf 2012}, \emph{39}, 093002.
     
\bibitem{Aoki:2006we}
  Aoki, Y.; Endrodi, G.; Fodor, Z.; Katz, S.D.; Szabo, K.K.
  The order of the quantum chromodynamics transition predicted by the
  standard model of particle physics.
  \emph{Nature} {\bf 2006}, \emph{443}, 675--678.

\bibitem{Bhattacharya:2014ara} 
  Bhattacharya, T.; Buchoff, M.I.; Christ, N.H.; Ding, H.-T.; Gupta, R.; Jung, C.; Karsch, F.; Lin, Z.; Mawhinney, R.; McGlynn, G.; et al.
  QCD Phase Transition with Chiral Quarks and Physical Quark Masses.
  \emph{\mbox{Phys. Rev. Lett.}} {\bf 2014}, \emph{113}, 082001.

\bibitem{Cheng:2008} 
Cheng, M.; Christ, N.H.; Datta, S.; van der Heide, J.; Jung, C.; Karsch, F.; Kaczmarek, O.; Laermann, E.; Mawhinney, R.D.; Miao, C.; et al. QCD equation of state with almost physical quark masses.
  \emph{Phys. Rev. D} \textbf{2008}, \emph{77}, 014511.

\bibitem{Rischke:2004}
  Rischke, D. The Quark-Gluon Plasma in Equilibrium.
  \emph{Prog. Part. Nucl. Phys. }\textbf{2004}, \emph{52}, 197--296.

\bibitem{McLerran:2007}
  McLerran, L.; Pisarski, R.D. Phases of Dense Quarks at Large N\_c.
  \emph{Nucl. Phys. A} \textbf{2007}, \emph{796}, 83--100.

\bibitem{Buballa:2005}
  Buballa, M. NJL-model analysis of dense quark matter.
  \emph{Phys. Rept.} \textbf{2005}, \emph{407}, 205--376.

\bibitem{Aggarwal:2010cw} 
  Aggarwal, M.M.; Ahammed, Z.; Alakhverdyants, A.V.; Alekseev, I.; Anderson, B.D.; Arkhipkin, D.; Averichev, G.S.; Balewski, J.; Barnby, L.S.; Baumgart, S.; et al. An Experimental Exploration of the QCD Phase Diagram: The Search for the Critical Point and the Onset of De-confinement. {\em arXiv} {\bf 2010}, arXiv:1007.2613.
  
\bibitem{Itoh:1970}
  Itoh, N.
  Concept of quark matter was mentioned as early as 1970 by Itoh in the context of neutron stars.
  \emph{Prog.~Theor. Phys.} {\bf 1970}, \emph{44}, 291--292.

\bibitem{Shuryak:1977ut} 
  Shuryak, E.V.
  Theory of Hadronic Plasma.
  \emph{Sov. Phys. JETP} {\bf 1978}, \emph{47}, 212--219.

\bibitem{Ichimaru:1982zz} 
  Ichimaru, S.
 Strongly coupled plasmas: High-density classical plasmas and degenerate electron liquids.
  \emph{Rev.~Mod. Phys.}  {\bf 1982}, 54, 1017--1059.

\bibitem{Braun-Munzinger:2015hba} 
  Braun-Munzinger, P.;  Koch, V.;  Sch\"{a}fer, T.;  Stachel, J.
  Properties of hot and dense matter from relativistic heavy ion collisions.
  \emph{Phys. Rep.}  {\bf 2016}, \emph{621}, 76--126.

\bibitem{BraunMunzinger:2008tz} 
  Braun-Munzinger, P.; Wambach, J.
 The Phase Diagram of Strongly-Interacting Matter.
  \emph{Rev. Mod. Phys.}  {\bf 2009}, \emph{81}, 1031--1050.
  
\bibitem{Fukushima:2010bq} 
  Fukushima, K.; Hatsuda, T.
  The phase diagram of dense QCD.
  \emph{Rept. Prog. Phys.}  {\bf 2011}, \emph{74}, 014001.

\bibitem{Bailin:1983bm} 
  Bailin, D.; Love, A.
  Superfluidity and Superconductivity in Relativistic Fermion Systems.
  \emph{Phys. Rep.}  {\bf 1984}, \emph{107}, 325--385.

\bibitem{Buballa:2002wy}
 Buballa, M.; Hosek, J.; Oertel, M.
 Anisotropic admixture in color superconducting quark matter.
 \emph{\mbox{Phys. Rev. Lett.}} {\bf 2003}, \emph{90}, 182002.

\bibitem{Alford:1998mk} 
  Alford, M.G.; Rajagopal, K.; Wilczek, F.
 Color flavor locking and chiral symmetry breaking in high density QCD.
  \emph{Nucl. Phys. B} {\bf 1999}, \emph{537}, 443--458.
  
\bibitem{Becker:2008}
  Becker, W.
  \emph{Neutron Stars and Pulsars};
  Springer: Berlin, Germany, 2009.

\bibitem{Migdal:1978az} 
  Migdal, A.B.
  Pion Fields in Nuclear Matter.
  \emph{Rev. Mod. Phys.}  {\bf 1978}, \emph{50}, 107--172.

\bibitem{D'Enterria:2007xr} 
D'Enterria, D.; Ballintijn, M.; Bedjidian, M.; Hofman, D.; Kodolova, O.; Loizides, C.; Lokthin, I.P.; Lourenço,~C.; Mironov, C.;  et al.
 CMS physics technical design report: Addendum on high density QCD with heavy ions.
  \emph{J. Phys. G} {\bf 2007}, \emph{34}, 2307.

\bibitem{Chaplin}
  Water Structure and Science. Available online: \myurl{http://www1.lsbu.ac.uk/water/water_phase_diagram.html} (accessed on 17 January 2017).

\bibitem{Shifman:1978bx}
 Shifman, M.A.;  Vainshtein, A.I.;  Zakharov, V.I. 
   QCD and Resonance Physics. Theoretical Foundations.
  \emph{Nucl.~Phys. B} {\bf 1979}, \emph{147}, 385--447.  
  \bibitem{Shifman:1978bx-2}
  Shifman, M.A.;  Vainshtein, A.I.;  Zakharov, V.I.  QCD and Resonance Physics: Applications.
  \emph{Nucl.~Phys. B} {\bf 1979}, \emph{147}, 448--518.

\bibitem{Schafer:1996wv} 
 Sch\"afer, T.;~Shuryak, E.V.
   Instantons in QCD.
  \emph{Rev. Mod. Phys.}  {\bf 1998}, \emph{70}, 323--426.
  
\bibitem{Diakonov:2002fq} 
  Diakonov, D.
   Instantons at work.
  \emph{Prog. Part. Nucl. Phys.}  {\bf 2003}, \emph{51}, 173--222.

\bibitem{Diakonov:2009jq} 
  Diakonov, D.
 Topology and confinement.
  \emph{Nucl. Phys. Proc. Suppl.}  {\bf 2009}, \emph{195}, 5--45.

\bibitem{Belavin} 
  Belavin, A.A.;  Polyakov, A.M.;  Schwartz, A.S.;  Tyupkin, Y.S.
  Pseudoparticle Solutions of the Yang-Mills Equations.
  \emph{Phys. Lett. B} {\bf 1975}, \emph{59}, 85--87.

\bibitem{Vainshtein:1981wh} 
  Vainshtein, A.I.;  Zakharov, V.I.;  Novikov, V.A.;  Shifman, M.A.
  ABC's of Instantons.
  \emph{Sov. Phys. Usp.}  {\bf 1982}, \emph{25}, 195--215.

\bibitem{Savvidy}
 {Matinyan, S.G.; Savvidy, G.K.} Vacuum Polarization Induced by the Intense Gauge Field.
 \emph{Nucl. Phys. B} {\bf 1978}, \emph{134}, 539--545.

\bibitem{Pasechnik:2016sbh} 
Pasechnik, R. Quantum Yang–Mills Dark Energy. Universe. \textbf{2016}, \emph{2}, 4.

\bibitem{Pasechnik:2013poa} 
  Pasechnik, R.; Beylin, V.; Vereshkov, G.
  Dark Energy from graviton-mediated interactions  in the QCD vacuum.
  \emph{J. Cosmol. Astropart. Phys.} {\bf 2013}, \emph{1306}, 11.

\bibitem{Pasechnik:2013sga}
  Pasechnik, R.; Beylin, V.; Vereshkov, G.
  Possible compensation of the QCD vacuum contribution to the dark energy.
  \emph{Phys. Rev. D} {\bf 2013}, \emph{88}, 023509.

\bibitem{Pasechnik:2016twe} 
  Pasechnik, R.; Prokhorov, G.; Teryaev, O. Mirror QCD and Cosmological Constant. {\em arXiv} {\bf 2016}, arXiv:1609.09249.

\bibitem{Pagels} 
 Pagels, H.; Tomboulis, E.
   Vacuum of the Quantum Yang-Mills Theory and Magnetostatics.
 \emph{Nucl. Phys. B} {\bf 1978}, {\em 143}, 485--502.

\bibitem{Prokhorov:2013xba} 
  Prokhorov, G.; Pasechnik, R.; Vereshkov, G.
  Dynamics of wave fluctuations in the homogeneous Yang-Mills condensate.
  \emph{J. High Energy Phys.} {\bf 2014}, \emph{2014}, 3.

\bibitem{Abelev:2013vea} 
 Abelev, B.; Adam, J.; Adamová, D.; Adare, A.M.; Aggarwal, M.M.; Rinella, G.A.; Agnello, M.; Agocs, A.G.; Agostinelli, A.; Ahammed, Z.; et al. 
 Centrality dependence of $\pi$, K, p production in Pb-Pb collisions at \mbox{$\sqrt{s_{NN}}$ = 2.76 TeV}.
  \emph{Phys. Rev. C} {\bf 2013}, \emph{88}, 044910.

\bibitem{Abelev:2013pqa} 
Abelev, B.; Adam, J.; Adamová, D.; Adare, A.M.; Aggarwal, M.M.; Rinella, G.A.; Agnello, M.; Agocs, A.G.; Agostinelli, A.; Ahammed, Z.; et al.
Two- and three-pion quantum statistics correlations in Pb-Pb collisions at $\sqrt{{s}_{NN}} =$ 2.76 TeV at the CERN Large Hadron Collider.
  \emph{Phys. Rev. C} {\bf 2014}, \emph{89}, 024911.

\bibitem{Adam:2015pbc} 
Adam, J.; Adamová, D.; Aggarwal, M.M.; Rinella, G.A.; Agnello, M.; Agrawal, N.; Ahammed, Z.; Ahmad, S.; Ahn, S.U.; Aiola, S.; et al.
Multipion Bose-Einstein correlations in pp, p-Pb, and Pb-Pb collisions at energies available at the CERN Large Hadron Collider.
  \emph{Phys. Rev. C} {\bf 2016}, \emph{93}, 054908.

\bibitem{Petran:2013lja} 
  Petr\'an, M.; Letessier, J.; Petr\'a\v{c}ek, V.; Rafelski, J.
 Hadron production and quark-gluon plasma hadronization in Pb-Pb collisions at $\sqrt{s_{NN}}=2.76$ TeV.
  \emph{Phys. Rev. C} {\bf 2013}, \emph{88}, 034907.

\bibitem{Csorgo:1994dd} 
  Csorgo, T.; Csernai, L.P.
 Quark-gluon plasma freezeout from a supercooled state?
  \emph{Phys. Lett. B} {\bf 1994}, \emph{333}, 494--499.

\bibitem{Shuryak:2014zxa} 
  Shuryak, E. Heavy Ion Collisions: Achievements and Challenges. \emph{arXiv} {\bf 2014}, arXiv:1412.8393.

\bibitem{Blaizot:2011xf} 
 Blaizot, J.P.; Gelis, F.; Liao, J.F.; McLerran, L.; Venugopalan, R.
Bose-Einstein Condensation and Thermalization of the Quark Gluon Plasma.
  \emph{Nucl. Phys. A} {\bf 2012}, \emph{873}, 68--80.

\bibitem{Begun:2013nga} 
  Begun, V.; Florkowski, W.; Rybczynski, M.
Explanation of hadron transverse-momentum spectra in heavy-ion collisions at $\sqrt s_{NN} =$ 2.76 TeV within chemical non-equilibrium statistical hadronization model.
  \emph{Phys. Rev.~C} {\bf 2014}, \emph{90}, 014906.

\bibitem{Begun:2014rsa} 
  Begun, V.; Florkowski, W.; Rybczynski, M.
Transverse-momentum spectra of strange particles produced in Pb+Pb collisions at $\sqrt{s_{\rm NN}}=2.76$ TeV in the chemical non-equilibrium model.
  \emph{Phys. Rev. C }{\bf 2014}, \emph{90}, 054912.
  
\bibitem{Begun:2015ifa} 
  Begun, V.; Florkowski, W.
  Bose-Einstein condensation of pions in heavy-ion collisions at the CERN Large Hadron Collider (LHC) energies.
  \emph{Phys. Rev. C} {\bf 2015}, \emph{91}, 054909.

\bibitem{Begun:2016cva} 
  Begun, V.
  Fluctuations as a test of chemical non-equilibrium at the LHC.
  \emph{Phys. Rev. C} {\bf 2016}, \emph{94}, 054904.

\bibitem{Plumer:1985tq} 
 Plumer, M.; Raha, S.; Weiner, R.M.
How Free Is The Quark-Gluon-Plasma? \emph{Nucl. Phys. A} {\bf 1984}, \emph{418}, 549C--557C.

\bibitem{Rischke:1992uv} 
 Rischke, D.H.; Gorenstein, M.I.; Schafer, A.; Stoecker, H.; Greiner, W.
Nonperturbative effects in the SU(3) gluon plasma.
  \emph{Phys. Lett. B} {\bf 1992}, \emph{278}, 19--23.

\bibitem{Bannur:1995np} 
  Bannur, V.M.
Equation of state for a nonideal quark gluon plasma.
  \emph{Phys. Lett. B} {\bf 1995}, \emph{362}, 7--10.

\bibitem{Thoma:2004sp} 
 { Thoma, M.H.} The quark–gluon plasma liquid.
  \emph{J. Phys. G} {\bf 2005}, \emph{31}, L7--L11.

\bibitem{Bannur:2005yx} 
  Bannur, V.M.
Strongly coupled quark gluon plasma (SCQGP).
 \emph{ J. Phys. G} {\bf 2006}, \emph{32}, 993--1002.

\bibitem{Schafer:2009dj} 
  Schafer, T.; Teaney, D.
 Nearly Perfect Fluidity: From Cold Atomic Gases to Hot Quark Gluon Plasmas.
  \emph{Rep. Prog. Phys.}  {\bf 2009}, \emph{72}, 126001.

\bibitem{Bonitz:2010}
  Bonitz, M.; Henning, C.; Block, D.
  Complex plasmas: A laboratory for strong correlations.
  \emph{Rep. Prog. Phys.} {\bf 2010}, \emph{73}, 066501.

\bibitem{Giorgini:2008zz} 
  Giorgini, S.; Pitaevskii, L.P.; Stringari, S.
  Theory of ultracold atomic Fermi gases.
  \emph{Rev. Mod. Phys.}  {\bf 2008}, \emph{80},~1215--1274.

\bibitem{Cao:2010wa} 
Cao, C.; Elliott, E.; Joseph, J.; Wu, H.; Petricka, J.; Sch\"{a}fer, T.; Thomas, J.E.
Universal Quantum Viscosity in a Unitary Fermi Gas.
  \emph{Science} {\bf 2011}, \emph{331}, 58--61.

\bibitem{Johnson:2010zz}
  Johnson, C.V.; Steinberg, P.
  What black holes teach about strongly coupled particles.
  \emph{Phys. Today} {\bf 2010}, \emph{63}, 29--33.

\bibitem{Kovtun:2004de}
  Kovtun, P.; Son, D.T.; Starinets, A.O.
   Viscosity in strongly interacting quantum field theories from black hole physics.
  \emph{Phys. Rev. Lett.}  {\bf 2005}, \emph{94}, 111601.

\bibitem{Thomas:2010zz}
Thomas, J.E. The nearly perfect Fermi gas. \emph{Phys. Today} {\bf 2010}, \emph{63}, 34--37.

\bibitem{Gupta:2003ji} 
  Gupta, S.
  Analyticity and the phase diagram of QCD.
  \emph{Phys. Lett. B} {\bf 2004}, \emph{588}, 136--144.

\bibitem{Bazavov:2014pvz} 
Bazavov, A.; Bhattacharya, T.; DeTar, C.; Ding, H.-T.; Gottlieb, S.; Gupta, R.; Hegde, P.; Heller, U.M.; Karsch,~F.; Laermann, E.; et al.
Equation of state in ( 2+1 )-flavor QCD.
  \emph{Phys. Rev. D} {\bf 2014}, \emph{90}, 094503.

\bibitem{Landau:1980mil} 
Landau, L.D.; Lifshitz, E.M.
  {\it Statistical Physics, Part 1}; Pergamon Press: London, UK, 1980.

\bibitem{Shuryak:1986nk} 
Shuryak, E.V.; Zhirov, O.V. Is the explosion of a quark-gluon plasma found?
  \emph{Phys. Lett. B} {\bf 1986}, \emph{171}, 99--102.

\bibitem{Hung:1994eq} 
Hung, C.M.; Shuryak, E.V. Hydrodynamics near the QCD phase transition: Looking for the longest-lived fireball.
  \emph{Phys. Rev. Lett. } {\bf 1995}, \emph{75}, 4003.

\bibitem{Matsui:1986dk} 
  Matsui, T.; Satz, H.
 $J/\psi$ Suppression by Quark-Gluon Plasma Formation.
  \emph{Phys. Lett. B} {\bf 1986}, \emph{178}, 416--422.

\bibitem{Brambilla:2010cs} 
  Brambilla, N.; Eidelman, S.; Heltsley, B.K.; Vogt, R.; Bodwin, G.T.; Eichten, E.; Frawley, A.D.; Meyer, A.B.; Mitchell, R.E.; Papadimitriou, V.; et al.
Heavy quarkonium: Progress, puzzles, and opportunities.
  \emph{Eur. Phys. J. C} {\bf 2011}, \emph{71}, 1534.

\bibitem{Borsanyi:2015yka} 
 Bors\'{a}nyi, S.; Fodor, Z.; Katz, S.D.; P\'{a}sztor, A.; Szab\'{o}, K.K.; T\"{o}r\"{o}k, C. Static $\overline {\mathrm {Q}}\mathrm {Q}$ pair free energy and screening masses from correlators of Polyakov loops: continuum extrapolated lattice results at the QCD physical point.
 \emph{J. High Energy Phys.} {\bf 2015}, \emph{1504}, 138.

\bibitem{Bazavov:2009us} 
  Bazavov, A.; Petreczky, P.; Velytsky, A. Quarkonium at Finite Temperature. In \emph{Quark-gluon Plasma 4}; World Scientific Publishing Company: Singapore, 2009.

\bibitem{Bazavov:2011nk}   
  Bazavov, A.; Bhattacharya, T.; Cheng, M.; DeTar, C.; Ding, H.-T.; Gottlieb, S.; Gupta, R.; Hegde, P.; Heller,~U.M.; Karsch, F.; et al.
 The chiral and deconfinement aspects of the QCD transition.
  \emph{Phys. Rev.~D} {\bf 2012}, \emph{85}, 054503.

\bibitem{Ding:2015ona} 
  Ding, H.T.; Karsch, F.; Mukherjee, S.
Thermodynamics of strong-interaction matter from Lattice QCD.
  \emph{Int. J. Mod. Phys. E} {\bf 2015}, \emph{24}, 1530007.

\bibitem{Scheid:1974zz} 
  Scheid, W.; Muller, H.; Greiner, W. Nuclear Shock Waves in Heavy-Ion Collisions.
  \emph{Phys. Rev. Lett.} {\bf 1974}, \emph{32}, 741.

\bibitem{Sobel:1975bq} 
Sobel, M.I.; Bethe, H.A.; Siemens, P.J.; Bondorf, J.P. Shock waves in colliding nuclei. \emph{ Nucl. Phys. A} {\bf 1975}, \emph{251}, 502--529.

\bibitem{Chapline:1974zf} Chapline, G.F.; Johnson, M.H.; Teller, E.; Weiss, M.S. Highly excited nuclear matter.
  \emph{Phys. Rev. D} {\bf 1973}, \emph{8}, 4302.

\bibitem{Nagamiya:1982kn} 
  Nagamiya, S.; Gyulassy, M.
 High-energy Nuclear Collisions.
  \emph{Adv. Nucl. Phys.}  {\bf 1984}, \emph{13}, 201--315.

\bibitem{Stock:1985xe}
  Stock, R.
  Particle Production In High-Energy Nucleus Nucleus Collisions.
  \emph{Phys. Rep.}  {\bf 1986}, \emph{135}, 259--315.

\bibitem{Lee:1974ma} 
  Lee, T.D.; Wick, G.C.
Vacuum Stability and Vacuum Excitation in a Spin 0 Field Theory.
  \emph{Phys. Rev. D} {\bf 1974}, \emph{9}, 2291--2316.

\bibitem{Harrison:2002es} 
  Harrison, M.; Peggs, S.G.; Roser, T.
  The RHIC accelerator.
  \emph{Ann. Rev. Nucl. Part. Sci.}  {\bf 2002}, \emph{52}, 425--469.

\bibitem{Evans:2008zzb} 
  Evans, L.; Bryant, P.
 LHC Machine.
  \emph{J. Instrum.} {\bf 2008}, \emph{3}, S08001.
  
\bibitem{Bjorken:1982qr} 
  Bjorken, J.D.
 Highly Relativistic Nucleus-Nucleus Collisions: The Central Rapidity Region.
  \emph{Phys. Rev. D} {\bf 1983}, \emph{27}, 140--151.

\bibitem{Adare:2015bua} 
Adare, A.; Afanasiev, S.; Aidala, C.; Ajitanand, N.N.; Akiba, Y.; Akimoto, R.; Al-Bataineh, H.; Alexander, J.; Alfred, M.; Al-Jamel, A.; et al.
 Transverse energy production and charged-particle multiplicity at midrapidity in various systems from $\sqrt{s_{NN}}=7.7$ to 200 GeV.
 \emph{ Phys. Rev. C} {\bf 2016}, \emph{93}, 024901.

\bibitem{Chatrchyan:2012mb} 
Chatrchyan, S.; Khachatryan, V.; Sirunyan, A.M.; Tumasyan, A.; Adam, W.; Bergauer, T.; Dragicevic, M.; Erö,~J.; Fabjan, C.; et al.
Measurement of the pseudorapidity and centrality dependence of the transverse energy density in PbPb collisions at $\sqrt{s_{NN}}=2.76$ TeV.
  \emph{Phys. Rev. Lett.}  {\bf 2012}, \emph{109}, 152303.

\bibitem{Mitchell:2016fqp} 
Mitchelll, J.T. 
Transverse Energy Measurements from the Beam Energy Scan in PHENIX.
  \emph{Nucl. Phys. A} {\bf 2016}, \emph{956}, 842--845.

\bibitem{Fischer:2013uwj} 
  Fischer, W.; Baltz, A.J.; Blaskiewicz, M.; Gassner, D.; Drees, K.A.; Luo, Y.; Minty, M.; Thieberger, P.; Wilinski,~M.; Pshenichnov, I.A.
Measurement of the total cross section of uranium-uranium collisions at $\sqrt{s_{NN}}=192.8$ GeV.
  \emph{Phys. Rev. C} {\bf 2014}, \emph{89}, 014906.

\bibitem{Fermi:1924tc} 
  Fermi, E.
  On the Theory of the impact between atoms and electrically charged particles.
  \emph{Z. Phys.}  {\bf 1924}, \emph{29}, 315--327.
  
\bibitem{vonWeizsacker:1934nji} 
  Von Weizsacker, C.F.
Radiation emitted in collisions of very fast electrons.
  \emph{Z. Phys.} {\bf 1934}, \emph{88}, 612--625.

\bibitem{Williams:1934ad} 
  Williams, E.J.
Nature of the high-energy particles of penetrating radiation and status of ionization and radiation formulae.
 \emph{Phys. Rev.} {\bf 1934}, \emph{45}, 729--730.

\bibitem{Jackson:1998nia} 
  Jackson, J.D. {\it Classical Electrodynamics};
Wiley: Hoboken, NJ, USA, 1998.

\bibitem{Baur:2001jj} 
  Baur, G.; Hencken, K.; Trautmann, D.; Sadovsky, S.; Kharlov, Y.
Coherent gamma gamma and gamma-A interactions in very peripheral collisions at relativistic ion colliders.
  \emph{Phys. Rep.}  {\bf 2002}, \emph{364}, 359--450.
 
 \bibitem{Bertulani:2005ru} 
 Bertulani, C.A.; Klein, S.R.; Nystrand, J.
 Physics of ultra-peripheral nuclear collisions.
  \emph{Ann. Rev. Nucl. Part.~Sci.}  {\bf 2005}, \emph{55}, 271--310.
 
\bibitem{Baltz:2007kq} 
  Baltz, A.J.; Baur, G.; d'Enterria, D.; Frankfurt, L.; Gelis, F.; Guzey, V.; Hencken, K.; Kharlov, Y.; Klasen, M.; Klein, S.R. The Physics of Ultraperipheral Collisions at the LHC.  
  \emph{Phys. Rep.}  {\bf 2008}, \emph{458}, 1--171. 

\bibitem{Klein:2015qna} 
  Klein, S.R.
Ultra-peripheral Collisions at RHIC: An Experimental Overview. In Proceedings of the Initial States 2014 Conference, Napa, CA, USA, 3--7 December 2014.

\bibitem{Tuchin:2013ie} 
  Tuchin, K.
  Particle production in strong electromagnetic fields in relativistic heavy-ion collisions.
  \emph{Adv. High Energy Phys.}  {\bf 2013}, \emph{2013}, 490495.

\bibitem{Kouveliotou:2003tb} 
  Kouveliotou, C.; Duncan, R.C.; Thompson, C.
  Magnetars.
  \emph{Sci. Am.}  {\bf 2003}, \emph{288N2}, 24.
 
 \bibitem{Kharzeev:2007jp}
  Kharzeev, D.E.; McLerran, L.D.; Warringa, H.J.
The effects of topological charge change in heavy ion collisions: 'Event by
   event P and CP violation.
  \emph{Nucl. Phys. A} {\bf 2008}, \emph{803}, 227--253.
 
 \bibitem{Ding:2010ga} 
  Ding, H.-T.;  Francis, A.; Kaczmarek, O.; Karsch, F.; Laermann, E.; Soeldner, W.
Thermal dilepton rate and electrical conductivity: An analysis of vector current correlation functions in quenched lattice QCD.
  \emph{Phys.~Rev.~D} {\bf 2011}, \emph{83}, 034504.

\bibitem{Amato:2013naa} 
  Aarts, G.; Allton, C.; Amato, A.; Giudice, P.; Hands, S.; Skullerud, J.I.
Electrical conductivity of the quark-gluon plasma across the deconfinement transition.
  \emph{Phys. Rev. Lett.}  {\bf 2013}, \emph{111}, 172001.

\bibitem{Aarts:2014nba} 
  Aarts, G.; Allton, C.; Amato, A.; Giudice, P.; Hands, S.; Skullerud, J.I.
  Electrical conductivity and charge diffusion in thermal QCD from the lattice.
  \emph{J. High Energy Phys.} {\bf 2015}, \emph{1502}, 186.
  
\bibitem{Filip:2015mca} 
  Filip, P.
Decay of Resonaces in Strong Magnetic Field.
  \emph{J. Phys. Conf. Ser.}  {\bf 2015}, \emph{636}, 012013.

\bibitem{Kharzeev:2015znc} 
  Kharzeev, D.E.; Liao, J.; Voloshin, S.A.; Wang, G.
 Chiral magnetic and vortical effects in high-energy nuclear collisions---A status report.
  \emph{Prog. Part. Nucl. Phys.}  {\bf 2016}, \emph{88}, 1--28.
  
\bibitem{Gursoy:2014aka} 
  Gursoy, U.; Kharzeev, D.; Rajagopal, K.
  Magnetohydrodynamics, charged currents and directed flow in heavy ion collisions.
  \emph{Phys. Rev. C} {\bf 2014}, \emph{89}, 054905.

\bibitem{Kharzeev:1998kz} 
  Kharzeev, D.; Pisarski, R.D.; Tytgat, M.H.G.
  Possibility of spontaneous parity violation in hot QCD.
  \emph{Phys.~Rev.~Lett.}  {\bf 1998}, \emph{81}, 512--515.

\bibitem{Abelev:2009ac} 
  Abelev, B.I.; Aggarwal, M.M.; Ahammed, Z.; Alakhverdyants, A.V.; Anderson, B.D.; Arkhipkin, D.; Averichev,~G.S.; Balewski, J.; Barannikova, O.; Barnby, L.S.; et al.
Azimuthal Charged-Particle Correlations and Possible Local Strong Parity Violation.
 \emph{ Phys. Rev. Lett.} {\bf 2009}, \emph{103}, 251601.

\bibitem{Adamczyk:2015eqo} 
 Adamczyk, L.; Adkins, J.K.; Agakishiev, G.; Aggarwal, M.M.; Ahammed, Z.; Alekseev, I.; Alford, J.; Aparin,~A.; Arkhipkin, D.; Aschenauer, E.C.; et al. 
Observation of charge asymmetry dependence of pion elliptic flow and the possible chiral magnetic wave in heavy-ion collisions.
  \emph{Phys. Rev. Lett.}  {\bf 2015}, \emph{114}, 252302.
  
\bibitem{Adamczyk:2014mzf} 
  Adamczyk, L.; Adkins, J.K.; Agakishiev, G.; Aggarwal, M.M.; Ahammed, Z.; Alekseev, I.; Alford, J.; Anson,~C.D.; Aparin, A.; Arkhipkin, D.; et al. 
Beam-energy dependence of charge separation along the magnetic field in Au+Au collisions at RHIC.
  \emph{Phys. Rev. Lett.}  {\bf 2014}, \emph{113}, 052302.

\bibitem{Abelev:2012pa} 
  Abelev, B.; Adam, J.; Adamova, D.; Adare, A.M.; Aggarwal, M.M.; Rinella, G.A.; Agocs, A.G.; Agostinelli,~A.; Salazar, S.A.; Ahammed, Z.; et al.
 Charge separation relative to the reaction plane in Pb-Pb collisions at $\sqrt{s_{NN}}= 2.76$ TeV.
  \emph{Phys. Rev. Lett.}  {\bf 2013}, \emph{110}, 012301.
  
\bibitem{Khachatryan:2016got} 
  Khachatryan, V.; Sirunyan, A.M.; Tumasyan, A.; Adam, W.; Asilar, E.; Bergauer, T.; Brandstetter, J.; Brondolin,~E.; Dragicevic, M.; Ero, J.; et al.
Observation of charge-dependent azimuthal correlations in pPb collisions and its implication for the search for the chiral magnetic effect. {\em arXiv} {\bf 2016}, arXiv:1610.00263.

\bibitem{Schlichting:2010qia} 
  Schlichting, S.; Pratt, S.
Charge conservation at energies available at the BNL Relativistic Heavy Ion Collider and contributions to local parity violation observables.
  \emph{Phys. Rev. C} {\bf 2011}, \emph{83}, 014913.

\bibitem{Skokov:2016yrj} 
  Skokov, V.; Sorensen, P.; Koch, V.; Schlichting, S.; Thomas, J.; Voloshin, S.; Wang, G.; Yee, H.U. Chiral Magnetic Effect Task Force Report. {\em arXiv} {\bf 2016}, arXiv:1608.00982.

\bibitem{Becattini:2016gvu} 
 Becattini, F.; Karpenko, I.; Lisa, M.; Upsal, I.; Voloshin, S. Global Hyperon Polarization at Local Thermodynamic Equilibrium with Vorticity, Magnetic Field and Feed-dow. \emph {arXiv} {\bf 2016}, arXiv:1608.00982.

\bibitem{Aichelin:1991xy}
  Aichelin, J.
'Quantum' molecular dynamics: A Dynamical microscopic n body approach to
   investigate fragment formation and the nuclear equation of state in heavy ion
   collisions.
  \emph{Phys. Rep.}  {\bf 1991}, \emph{202}, 233--360.

\bibitem{Werner:1993uh} 
  Werner, K.
  Strings, pomerons, and the venus model of hadronic interactions at ultrarelativistic energies.
  \emph{Phys. Rep.}  {\bf 1993}, \emph{232}, 87--299.

\bibitem{Gyulassy:1994ew} 
  Gyulassy, M.; Wang, X.N.
HIJING 1.0: A Monte Carlo program for parton and particle production in high-energy hadronic and nuclear collisions.
  \emph{Comput. Phys. Commun.}  {\bf 1994}, \emph{83}, 307--331.

\bibitem{Bass:1998ca}
 Bass, S.A.; Belkacem, M.; Bleicher, M.; Brandstetter, M.; Bravina, L.; Ernst, C.; Gerland, L.; Hofmann, M.; Hofmann, S.; Konopka, J.; et al.
Microscopic models for ultrarelativistic heavy ion collisions.
  \emph{Prog. Part. Nucl.~Phys.}  {\bf 1998}, \emph{41}, 255--369.
  
\bibitem{Humanic:1998ji} 
  Humanic, T.J.
Constraining a simple hadronization model of relativistic heavy ion collisions using hadronic observables.
  \emph{Phys. Rev. C} {\bf 1998}, \emph{57}, 866--876.

\bibitem{Lin:2004en} 
  Lin, Z.W.; Ko, C.M.; Li, B.A.; Zhang, B.; Pal, S.
A Multi-phase transport model for relativistic heavy ion collisions.
  \emph{Phys. Rev. C} {\bf 2005}, \emph{72}, 064901.

\bibitem{Xu:2016hmp} 
 Xu, H.J.; Li, Z.; Song, H.
High-order flow harmonics of identified hadrons in 2.76A TeV Pb + Pb collisions.
  \emph{Phys. Rev. C} {\bf 2016}, \emph{93}, 064905.
  
\bibitem{Roesler:2000he} 
  Roesler, S.; Engel, R.; Ranft, J.
The Monte Carlo Event Generator DPMJET-III. In Proceedings of the Monte Carlo 2000, Lisbon, Portugal, 23--26 October 2000.
  
\bibitem{Pierog:2013ria} 
  Pierog, T.; Karpenko, I.; Katzy, J.M.; Yatsenko, E.; Werner, K.
EPOS LHC: Test of collective hadronization with data measured at the CERN Large Hadron Collider.
  \emph{Phys. Rev. C} {\bf 2015}, \emph{92}, 034906.

\bibitem{Akamatsu:2013wyk} 
  Akamatsu, Y.; Inutsuka, S.I.; Nonaka, C.; Takamoto, M.
A new scheme of causal viscous hydrodynamics for relativistic heavy-ion collisions: A Riemann solver for quark-gluon plasma.
  \emph{J. Comput. Phys. } {\bf 2014}, \emph{256}, 34--54.

\bibitem{Bozek:2012qs} 
  Bozek, P.; Wyskiel-Piekarska, I.
 Particle spectra in Pb-Pb collisions at $\sqrt{s_{NN}}$ = 2.76 TeV. \emph{Phys. Rev. C} {\bf 2012}, \emph{85}, 064915.

\bibitem{deSouza:2015ena} 
  De Souza, R.D.; Koide, T.; Kodama, T.
Hydrodynamic Approaches in Relativistic Heavy Ion Reactions.
  \emph{Prog.~Part. Nucl. Phys.}  {\bf 2016}, \emph{86}, 35--85.

\bibitem{Kolb:2000sd} 
  Kolb, P.F.; Sollfrank, J.; Heinz, U.W.
Anisotropic transverse flow and the quark hadron phase transition.
  \emph{Phys.~Rev. C} {\bf 2000}, \emph{62}, 054909.

\bibitem{Lokhtin:2008xi} 
 Lokhtin, I.P.; Malinina, L.V.; Petrushanko, S.V.; Snigirev, A.M.; Arsene, I.; Tywoniuk, K.
Heavy ion event generator HYDJET++ (HYDrodynamics plus JETs).
  \emph{Comput. Phys. Commun.}  {\bf 2009}, \emph{180}, 779--799.

\bibitem{Hirano:2012kj} 
 Hirano, T.; Huovinen, P.; Murase, K.; Nara, Y.
Integrated Dynamical Approach to Relativistic Heavy Ion Collisions.
  \emph{Prog. Part. Nucl. Phys.}  {\bf 2013}, \emph{70}, 108--158.

\bibitem{Shen:2014vra} 
  Shen, C.; Qiu, Z.; Song, H.; Bernhard, J.; Bass, S.; Heinz, U.
The iEBE-VISHNU code package for relativistic heavy-ion collisions.
  \emph{Comput. Phys. Commun.}  {\bf 2016}, \emph{199}, 61--85.

\bibitem{Abelev:2013haa} 
  Abelev, B.B.; Adam, J.; Adamova, D.; Adare, A.M.; Aggarwal, M.M.; Rinella, G.A.; Agnello, M.; Agocs, A.G.; Agostinelli, A.; Ahammed, Z.;  et al.
Multiplicity Dependence of Pion, Kaon, Proton and Lambda Production in p-Pb Collisions at $\sqrt{s_{NN}}$ = 5.02 TeV.
  \emph{Phys. Lett. B} {\bf 2014}, \emph{728}, 25--38.
 
\bibitem{Adler:2003cb} 
 Adler, S.S.; Afanasiev, S.; Aidala, C.; Ajitanand, N.N.; Akiba, Y.; Alexander, J.; Amirikas, R.; Aphecetche,~L.; Aronson, S.H.; Averbeck, R.; et al.
 Identified charged particle spectra and yields in Au+Au collisions at S(NN)**1/2 = 200-GeV.
  \emph{Phys. Rev. C} {\bf 2004}, \emph{69}, 034909.

\bibitem{Nonaka:2006yn} 
  Nonaka, C.; Bass, S.A.
Space-time evolution of bulk QCD matter.
  \emph{Phys. Rev. C} {\bf 2007}, \emph{75}, 014902.

\bibitem{Muller:1999ys}
  Muller, B.
 Quark matter '99---Theoretical summary: What next?
  \emph{Nucl. Phys. A} {\bf 1999}, \emph{661}, 272--281.

\bibitem{Kittel:2005} 
  Kittel, W.; De Wolf, E.A.~\emph{Soft Multihadron Dynamics}; World Scientific: Singapore, 2005.

\bibitem{Landau:1953}
Landau,  L.D. On the multiparticle production in high-energy collisions.
 \emph{Izv. Akad. Nauk Ser. Fiz.} \textbf{1953}, \emph{17}, 51--46.

\bibitem{Weiner:2005gp}
Weiner, R.M. Experiment and Theory in Computations of the He Atom Ground State. \emph{Int. J. Mod. Phys. E} {\bf 2006}, \emph{15}, 37--56.
  
\bibitem{Heinz:2009xj} 
  Heinz, U.W.
Early collective expansion: Relativistic hydrodynamics and the transport properties of QCD matter.
  In \emph{Landolt-Bornstein}; Springer: Berlin, Germany, 2010; Volume 23, pp. 240--292.

\bibitem{Stoecker:1986ci}
  St\"{o}cker, H.; Greiner, W.
High-Energy Heavy Ion Collisions: Probing The Equation Of State Of Highly
   Excited Hadronic Matter.
  \emph{Phys. Rep.} {\bf 1986}, \emph{137}, 277--392.

\bibitem{Hirano:2008hy} 
  Hirano, T.; van der Kolk, N.; Bilandzic, A.
Hydrodynamics and Flow.
  \emph{Lect. Notes Phys.} {\bf 2010}, \emph{785}, 139--178.

\bibitem{Gale:2013da} 
  Gale, C.; Jeon, S.; Schenke, B.
Hydrodynamic Modeling of Heavy-Ion Collisions.
  \emph{Int. J. Mod. Phys. A} {\bf 2013}, \emph{28}, 1340011.

\bibitem{Jaiswal:2016hex} 
  Jaiswal, A.; Roy, V.
Relativistic hydrodynamics in heavy-ion collisions: General aspects and recent developments.
  \emph{Adv. High Energy Phys.}  {\bf 2016}, \emph{2016}, 9623034.

\bibitem{Eckart:1940te} 
  Eckart, C.
The Thermodynamics of irreversible processes. 3. Relativistic theory of the simple fluid.
  \emph{Phys. Rev.} {\bf 1940}, \emph{58}, 919--924.
  
\bibitem{Nakamura:2004sy} 
  Nakamura, A.; Sakai, S.
Transport coefficients of gluon plasma.
  \emph{Phys. Rev. Lett.}  {\bf 2005}, \emph{94}, 072305.

\bibitem{Karsch:2007jc} 
  Karsch, F.; Kharzeev, D.; Tuchin, K.
Universal properties of bulk viscosity near the QCD phase transition.
 \emph{ Phys. Lett. B} {\bf 2008}, \emph{663}, 217--221.

\bibitem{Muronga:2003ta} 
 Muronga, A.
Causal theories of dissipative relativistic fluid dynamics for nuclear collisions.
  \emph{Phys. Rev. C} {\bf 2004}, \emph{69}, 034903.

\bibitem{Hiscock:1983zz} 
  Hiscock, W.A.; Lindblom, L.
 Stability and causality in dissipative relativistic fluids.
  \emph{Ann. Phys.}  {\bf 1983}, \emph{151}, 466--496.

\bibitem{Heinz:2013th} 
  Heinz, U.; Snellings, R.
 Collective flow and viscosity in relativistic heavy-ion collisions.
  \emph{Ann. Rev. Nucl. Part.~Sci.}  {\bf 2013}, \emph{63}, 123--151. 
    

\bibitem{Abelev:2008ab} 
 Abelev, B.I.; Aggarwal, M.M.; Ahammed, Z.; Anderson, B.D.; Arkhipkin, D.; Averichev, G.S.; Bai, Y.; Balewski,~J.; Barannikova, O.; Barnby, L.S.; et al.
Systematic Measurements of Identified Particle Spectra in $p p, d^+$ Au and Au+Au Collisions from STAR.
  \emph{Phys. Rev. C} {\bf 2009}, \emph{79}, 034909.

\bibitem{Siemens:1978pb} 
 Siemens, P.J.; Rasmussen, J.O.
Evidence for a blast wave from compress nuclear matter.
  \emph{Phys. Rev. Lett.}  {\bf 1979}, \emph{42}, 880--887.
 
\bibitem{Schnedermann:1993ws}
 Schnedermann, E.; Sollfrank, J.; Heinz, U.W.
Thermal phenomenology of hadrons from 200-A/GeV S+S collisions.
  \emph{Phys. Rev. C} {\bf 1993}, \emph{48}, 2462--2475.

\bibitem{Retiere:2003kf} 
  Retiere, F.; Lisa, M.A.
Observable implications of geometrical and dynamical aspects of freeze out in heavy ion collisions.
  \emph{Phys. Rev. C} {\bf 2004}, \emph{70}, 044907.
 
\bibitem{Cooper:1974mv} 
  Cooper, F.; Frye, G.
Comment on the Single Particle Distribution in the Hydrodynamic and Statistical Thermodynamic Models of Multiparticle Production.
  \emph{Phys. Rev. D} {\bf 1974}, \emph{10}, 186--189.

\bibitem{Kumar:2014tca} 
  Kumar, L.
 Systematics of Kinetic Freeze-out Properties in High Energy Collisions from STAR.
  \emph{Nucl. Phys. A} {\bf 2014}, \emph{931}, 1114--1119.

\bibitem{Sjostrand:2007gs} 
 Sjostrand, T.; Mrenna, S.; Skands, P.Z.
A Brief Introduction to PYTHIA 8.1.
 \emph{ Comput. Phys. Commun.}  {\bf 2008}, \emph{178}, 852--867.

\bibitem{Baier:2000sb} 
 Baier, R.; Mueller, A.H.; Schiff, D.; Son, D.T.
 'Bottom up' thermalization in heavy ion collisions.
  \emph{Phys. Lett. B} {\bf 2001}, \emph{502}, 51--58.
 
\bibitem{Xu:2004mz} 
  Xu, Z.; Greiner, C.
Thermalization of gluons in ultrarelativistic heavy ion collisions by including three-body interactions in a parton cascade.
  \emph{Phys. Rev. C} {\bf 2005}, \emph{71}, 064901.
 
\bibitem{Kurkela:2011ti} 
  Kurkela, A.; Moore, G.D.
Thermalization in Weakly Coupled Nonabelian Plasmas.
  \emph{J. High Energy Phys.} {\bf 2011}, \emph{1112}, 44.
  
\bibitem{Romatschke:2016hle} 
  Romatschke, P. Do Nuclear Collisions Create a Locally Equilibrated Quark-Gluon Plasma? {\em Eur. Phys. J. C}  {\bf 2016}, {\em 77}, 21.
  
\bibitem{Broniowski:2008vp} 
  Broniowski, W.; Chojnacki, M.; Florkowski, W.; Kisiel, A.
Uniform Description of Soft Observables in Heavy-Ion Collisions at $\sqrt{s_{\rm NN}}=200$ GeV.
  \emph{Phys. Rev. Lett.}  {\bf 2008}, \emph{101}, 022301.

\bibitem{Petersen:2010zt} 
  Petersen, H.; Coleman-Smith, C.; Bass, S.A.; Wolpert, R.
    Constraining the initial state granularity with bulk observables in Au+Au collisions at $\sqrt{s_{\rm NN}}=200$ GeV. 
  \emph{J. Phys. G} {\bf 2011}, \emph{38}, 045102.
 
 
\bibitem{Heinz:2011mh} 
  Heinz, U.; Moreland, J.S.
    Energy dependent growth of the nucleon and hydrodynamic initial conditions, 
  \emph{Phys. Rev. C} {\bf 2011}, \emph{84}, 054905.

\bibitem{Noronha-Hostler:2016eow} 
 Noronha-Hostler, J.; Betz, B.; Noronha, J.; Gyulassy, M.
    Event-by-event hydrodynamics $+$ jet energy loss: A solution to the $R_{AA} \otimes v_2$ puzzle, 
  \emph{Phys. Rev. Lett.}  {\bf 2016}, \emph{116}, 252301.

 \bibitem{Glauber:1955qq} 
  Glauber, R.J.
    Cross-sections in deuterium at high-energies. 
  \emph{Phys. Rev.}  {\bf 1955}, \emph{100}, 242--248.
  
 \bibitem{Czyz:1969jg} 
  Czyz, W.; Maximon, L.C.
    High-energy, small angle elastic scattering of strongly interacting composite particles. 
  \emph{Ann. Phys.}  {\bf 1969}, \emph{52}, 59--121.
  
 \bibitem{Glauber:1970jm} 
  Glauber, R.J.; Matthiae, G.
    High-energy scattering of protons by nuclei. 
  \emph{Nucl. Phys. B} {\bf 1970}, \emph{21}, 135--157.
 
\bibitem{Glauber:2006gd} 
  Glauber, R.J.
    Quantum Optics and Heavy Ion Physics. 
  \emph{Nucl. Phys. A} {\bf 2006}, \emph{774}, 3--13.
 
\bibitem{Joachain:1973yv} 
  Joachain, C.J.; Quigg, C.
    Multiple scattering expansions in several particle dynamics. 
  \emph{Rev. Mod. Phys.}  {\bf 1974}, \emph{46}, 279--324.
  
\bibitem{Franco:1978yw} 
  Franco, V.; Varma, G.K.
    Collisions Between Composite Particles at Medium and High-Energies. 
  \emph{Phys. Rev. C} {\bf 1978}, \emph{18}, 349--370.
 
 \bibitem{Miller:2007ri} 
  Miller, M.L.; Reygers, K.; Sanders, S.J.; Steinberg, P.
    Glauber modeling in high energy nuclear collisions.
  \emph{Ann.~Rev. Nucl. Part. Sci.}  {\bf 2007}, \emph{57}, 205--243.


\bibitem{Broniowski:2007nz} 
  Broniowski, W.; Rybczynski, M.; Bozek, P.
    GLISSANDO: Glauber initial-state simulation and more. 
  \emph{Comput.~Phys. Commun.}  {\bf 2009}, \emph{180}, 69--83.

\bibitem{Loizides:2014vua} 
  Loizides, C.; Nagle, J.; Steinberg, P.
    Improved version of the PHOBOS Glauber Monte Carlo. 
  \emph{SoftwareX} {\bf 2015}, \emph{1}, 13--18.
 
\bibitem{Abreu:2001kd} 
  Abreu, M.C.;  Alessandro, B.; Alexa, C.;  Arnaldi, R.;  Atayan, M.;  Baglin, C.;  Baldit, A.;  Bedjidian, M.;  Beole, S.; Boldea, V. 
    The Dependence of the anomalous J/psi suppression on the number of participant nucleons. 
  \emph{Phys. Lett. B} {\bf 2001}, \emph{521}, 195--203.

\bibitem{Aad:2012ew} 
  Aad, G.;  Abajyan, T.; Abbott, B.; Abdallah, J.; Khalek, S.A.; Abdelalim, A.A.; Abdinov, O.; Aben, R.; Abi, B.; Abolins, M.; et al. 
    Measurement of $Z$ boson Production in Pb+Pb Collisions at $\sqrt{s_{NN}}=2.76$ TeV with the ATLAS Detector. 
  \emph{Phys. Rev. Lett.}  {\bf 2013}, \emph{110}, 022301.  

\bibitem{Gribov:1968fc} 
  Gribov, V.N.
    A Reggeon Diagram Technique. 
  \emph{Sov. Phys. JETP} {\bf 1968}, \emph{26}, 414--422.
  
\bibitem{Gribov:1968jf} 
  Gribov, V.N.
    Glauber corrections and the interaction between high-energy hadrons and nuclei. 
  \emph{\mbox{Sov. Phys. JETP}} {\bf 1969}, \emph{29}, 483--487.

\bibitem{Abramovsky:1973fm} 
  Abramovsky, V.A.; Gribov, V.N.; Kancheli, O.V.
    Character of Inclusive Spectra and Fluctuations Produced in Inelastic Processes by Multi - Pomeron Exchange. 
  \emph{Yad. Fiz.}  {\bf 1973}, \emph{18}, 595--616.
  
\bibitem{Braun:1988pk} 
  Braun, V.M.; Shabelski, Y.M.
    Multiple Scattering Theory for Inelastic Processes. 
  \emph{Int. J. Mod. Phys. A} {\bf 1988}, \emph{3}, 2417--2501.
  
\bibitem{Drescher:2000ha} 
 Drescher, H.J.; Hladik, M.; Ostapchenko, S.; Pierog, T.; Werner, K.
    Parton based Gribov-Regge theory. 
  \emph{Phys.~Rep.}  {\bf 2001}, \emph{350}, 93--289.

\bibitem{Ostapchenko:2010vb} 
  Ostapchenko, S.
    Monte Carlo treatment of hadronic interactions in enhanced Pomeron scheme: I. QGSJET-II model. 
  \emph{Phys. Rev. D} {\bf 2011}, \emph{83}, 014018.

\bibitem{Pasechnik:2010zs} 
  Pasechnik, R.; Enberg, R.; Ingelman, G.
    QCD rescattering mechanism for diffractive deep inelastic scattering. 
  \emph{Phys. Rev. D} {\bf 2010}, \emph{82}, 054036.

\bibitem{Pasechnik:2010cm} 
  Pasechnik, R.; Enberg, R.; Ingelman, G.
    Diffractive deep inelastic scattering from multiple soft gluon exchange in QCD. 
  \emph{Phys. Lett. B} {\bf 2011}, \emph{695}, 189--193.

\bibitem{Nemchik:2008xy} 
  Nemchik, J.; Petracek, V.; Potashnikova, I.K.; Sumbera, M.
    Nuclear suppression at large forward rapidities in d-Au collisions at relativistic and ultrarelativistic energies. 
  \emph{Phys. Rev. C} {\bf 2008}, \emph{78}, 025213.

\bibitem{Loizides:2016djv} 
  Loizides, C.
    Glauber modeling of high-energy nuclear collisions at the subnucleon level. 
  \emph{Phys. Rev. C} {\bf 2016}, \emph{94}, 024914.



\bibitem{nik} 
 Nikolaev, N.N.; Zakharov, B.G. The triple-pomeron regime and structure function of the pomeron in diffractive deep inelastic scattering at very smallx. \emph{Phys. C} {\bf 1994}, \emph{64}, 631--652.

\bibitem{zkl}
  Kopeliovich, B.Z.; Lapidus, L.I.; Zamolodchikov, A.B.
    Dynamics Of Color In Hadron Diffraction On Nuclei. 
  \emph{JETP Lett.} {\bf 1981}, \emph{33}, 595--597.

\bibitem{nik_dif}
 Nikolaev, N.N.; Zakharov, B.G.
    The Pomeron in diffractive deep inelastic scattering. 
  \emph{J. Exp. Theor. Phys.}  {\bf 1994}, \emph{78}, 598--618.

\bibitem{k95} 
 Kopeliovich, B.Z. In {\it Proceedings of the International Workshop XXIII on Gross Properties 
of Nuclei and Nuclear Excitations, Hirschegg, Austria, 1995}; Feldmeyer, H.,
 Norenberg, W., Eds.; Gesellschaft Schwerionenforschung: Darmstadt, Germany, 1995.

\bibitem{bhq97}
Brodsky, S.J.; Hebecker, A.; Quack, E.
    The Drell-Yan process and factorization in impact parameter space.
  \emph{Phys. Rev. D} {\bf 1997}, \emph{55}, 2584--2590.
  
\bibitem{kst99} 
Kopeliovich, B.Z.; Schafer, A.; Tarasov, A.V. Bremsstrahlung of a quark propagating through a nucleus.
\emph{Phys.~Rev. C} {\bf 1999}, \emph{59}, 1609--1619.

\bibitem{krt01} 
Kopeliovich, B.Z.; Schafer, A.; Tarasov, A.V. The color dipole picture of the Drell-Yan process.
\emph{Phys. Lett. B} {\bf 2001}, \emph{503}, 91--98.

\bibitem{npz-1}
Nikolaev, N.N.; Piller, G.; Zakharov, B.G.
  Quantum coherence in heavy flavor production on nuclei. 
 \emph{Z.~Phys.~A} {\bf 1996}, \emph{354}, 99--105.

\bibitem{npz-2}
  Kopeliovich, B.Z.; Tarasov, A.V.
    Gluon shadowing and heavy flavor production off nuclei.
  \emph{Nucl. Phys. A} {\bf 2002}, \emph{710}, 180--217.

\bibitem{BBGG}
  Bertsch, G.; Brodsky, S.J.; Goldhaber, A.S.; Gunion, J.F. Diffractive Excitation in Quantum Chromodynamics. \emph{Phys. Rev. Lett.} {\bf 1981}, \emph{47}, 297.

\bibitem{BM}
  Brodsky,  S.J.;  Mueller, A. Using nuclei to probe hadronization in QCD. \emph{Phys. Lett. B} {\bf 1988}, \emph{206}, 685--690.

\bibitem{GolecBiernat:1998js}
  Golec-Biernat, K.J.; Wusthoff, M.
    Saturation effects in deep inelastic scattering at low $ Q^2$ and its ... and its implications on diffraction. 
  \emph{Phys. Rev. D} {\bf 1998}, \emph{59}, 014017.

\bibitem{Bartels:2002cj} 
  Bartels,~J.;~Golec-Biernat,~K.;~Kowalski,~H.~Modification of the saturation model: Dokshitzer-Gribov-Lipatov-Altarelli-Parisi evolution. \emph{Phys. Rev. D} {\bf 2002}, \emph{66}, 014001.

\bibitem{kmw-1}
  Kowalski, H.; Motyka, L.; Watt, G. Exclusive diffractive processes at HERA within the dipole picture. 
  \emph{Phys.~Rev. D} {\bf 2006}, \emph{74}, 074016.

\bibitem{kmw-2}
  Watt, G.; Kowalski, H.
    Impact parameter dependent colour glass condensate dipole model. 
  \emph{Phys. Rev. D} {\bf 2008}, \emph{78}, 014016.

\bibitem{ipsatnewfit} 
  Rezaeian, A.H.; Siddikov, M.; van de Klundert, M.; Venugopalan, R. N$\Sigma$c and N$\Sigma$b resonances in the quark-delocalization color-screening model. \emph{Phys. Rev. D} {\bf 2013}, \emph{87}, 034002.

\bibitem{iim} 
  Iancu, E.; Itakura, K.; Munier, S. Saturation and BFKL dynamics in the HERA data at small-x. \emph{Phys. Lett. B} {\bf 2004}, \emph{590}, 199--208.

\bibitem{kkt} 
 Kharzeev, D.; Kovchegov, Y.V.; Tuchin, K. Nuclear modification factor in image collisions: onset of suppression in the color glass condensate. \emph{Phys. Lett. B} {\bf 2004}, \emph{599}, 23--31.

\bibitem{dhj}
  Kharzeev, D.; Kovchegov, Y.V.; Tuchin, K.
    The color glass condensate and hadron production in the forward region. 
  \emph{Nucl. Phys. A} {\bf 2006}, \emph{765}, 464--482.
  
\bibitem{buw} 
  Boer, D.; Utermann, A.; Wessels, E.
    Geometric Scaling at RHIC and LHC. 
  \emph{Phys. Rev. D} {\bf 2008}, \emph{77}, 054014.

\bibitem{Soyez2007}
  Soyez, G. Saturation QCD predictions with heavy quarks at HERA. \emph{Phys. Lett. B} {\bf 2007}, \emph{655}, 32--38.

\bibitem{Kowalski:2003hm}
  Kowalski, H.; Teaney, D. Impact parameter dipole saturation model. \emph{Phys. Rev. D} {\bf 2003}, \emph{68}, 114005.

\bibitem{amirs}
  Rezaeian, A.; Schmidt, I. Impact-parameter dependent color glass condensate dipole model and new combined HERA data. \emph{Phys. Rev. D} {\bf 2013}, \emph{88}, 074016.

\bibitem{Basso:2015pba} 
  Basso, E.; Goncalves, V.P.; Nemchik, J.; Pasechnik, R.; Sumbera, M.
    Drell-Yan phenomenology in the color dipole picture revisited. 
  \emph{Phys. Rev. D} {\bf 2016}, \emph{93}, 034023.

\bibitem{Basso:2016ulb} 
  Basso, E.; Goncalves, V.P.; Krelina, M.; Nemchik, J.; Pasechnik, R.
    Nuclear effects in Drell-Yan pair production in high-energy $pA$ collisions.
  \emph{Phys. Rev. D} {\bf 2016}, \emph{93}, 094027.

\bibitem{Goncalves:2016qku} 
  Basso, E.; Goncalves, V.P.; Krelina, M.; Nemchik, J.; Pasechnik, R.
    Drell-Yan process in $pA$ collisions: The exact treatment of coherence effects. 
  \emph{Phys. Rev. D} {\bf 2016}, \emph{94}, 114009.

\bibitem{kopeliovich-isi-1}
  Kopeliovich, B.Z.; Nemchik, J.; Potashnikova, I.K.; Johnson, M.B.; Schmidt, I. Breakdown of QCD factorization at large Feynman x.
  \emph{Phys. Rev. C} {\bf 2005}, \emph{72}, 054606.

\bibitem{kopeliovich-isi-2}
  Kopeliovich, B.Z.; Nemchik, J.; Potashnikova, I.K.; Schmidt, I. Energy conservation in high-pT nuclear reactions.
  \emph{Int. J. Mod. Phys. E} \textbf{2014}, \emph{23}, 1430006.

\bibitem{phenix-isi-dAu}
Adler, S.S.; Afanasiev, S.; Aidala, C.; Ajitanand, N.N.; Akiba, Y.; Al-Jamel, A.; Alexander, J.; Aoki, K.; Aphecetche, L.; Armendariz, R.; et al.  Centrality Dependence of $\pi$ 0 and $\eta$ Production at Large Transverse Momentum in s N N = 200 GeV d+ Au Collisions.
  \emph{Phys. Rev. Lett.} \textbf{2007}, \emph{98}, 172302.
  
\bibitem{phenix-isi-AuAu-1}
  Afanasiev, S.; Aidala, C.; Ajitanand, N.N.; Akiba, Y.; Al-Jamel, A.; Alexander, J.; Aoki, K.; Aphecetche,~L.; Armendariz, R.; Aronson, S.H.; et al. Measurement of Direct Photons in Au+Au Collisions at \mbox{$\surd$sNN = 200 GeV}.
  \emph{Phys. Rev. Lett.} \textbf{2012}, \emph{109}, 152302.

\bibitem{phenix-isi-AuAu-2}
  Sakaguchi, T. Measurement of electro-magnetic radiation at RHIC-PHENIX. \emph{Nucl. Phys. A} \textbf{2008}, \emph{805}, 355--360.

\bibitem{Landau:1953um} 
   Landau, L.D.; Pomeranchuk, I. 
    Limits of applicability of the theory of bremsstrahlung 
    electrons and pair production at high-energies. 
  \emph{Dokl. Akad. Nauk Ser. Fiz.} {\bf 1953}, \emph{92}, 535--536.

\bibitem{Landau:1953um-2} 
  Landau, L.D.; Pomeranchuk, I. Electron cascade process at very high-energies.
  \emph{Dokl. Akad. Nauk Ser. Fiz.} {\bf 1953}, \emph{92}, 735--738.
  
\bibitem{Migdal:1956tc} 
 Migdal, A.B.
    Bremsstrahlung and pair production in condensed media at high-energies. 
  \emph{Phys. Rev.} {\bf 1956}, \emph{103}, 1811--1820.      

\bibitem{LPM-review}
   Baier, R.; Shiff, D.; Zakharov, B.G. Energy loss in perturbative QCD.
   \emph{Ann. Rev. Nucl. Part. Sci.} {\bf 2000}, \emph{50}, 37--69.

\bibitem{LPM-ion}
  Gyulassy, M.; Wang, X.-N. Multiple collisions and induced gluon Bremsstrahlung in QCD.
  \emph{Nucl. Phys. B} {\bf 1994}, \emph{420}, 583--614.

\bibitem{LPM-ion-1}
  Wang, X.-N.; Gyulassy, M.; Plumer, M. Landau-Pomeranchuk-Migdal effect in QCD and radiative energy loss in a quark-gluon plasma. \emph{Phys. Rev. D} {\bf 1995}, \emph{51}, 3436--3446.

\bibitem{LPM-ion-2}
 Wiedemann, U.A.; Gyulassy, M. Transverse momentum dependence of the Landau-Pomeranchuk-Migdal effect.
  \emph{Nucl. Phys. B} {\bf 1999}, \emph{560}, 345--382.

\bibitem{Zakh}
  {Zakharov, B.G.} Applicability of the eikonal approximation for calculation of the probability of passage of ultrarelativistic positronium through matter. \emph{Sov. J. Nucl. Phys.} {\bf 1987}, \emph{46}, 92--95.

\bibitem{Zakh-1}
  Zakharov, B.G. Radiative energy loss of high-energy quarks in finite-size nuclear matter and quark-gluon plasma.
  \emph{J. Exp. Theor. Phys. Lett.} {\bf 1997}, \emph{65}, 615--620.

\bibitem{Zakh-2}
Zakharov, B.G. Transverse Spectra of Radiation Processes in Medium.
  \emph{J. Exp. Theor. Phys. Lett.} {\bf 1999}, \emph{70}, 176--182.

\bibitem{BDMPS-2}
  Baier, R.; Dokshitzer, Y.L.; Mueller, A.H.; Schiff, D. Angular dependence of the radiative gluon spectrum and the energy loss of hard jets in QCD media. \emph{Phys. Rev. C} {\bf 1999}, \emph{60}, 064902.

\bibitem{BDMPS-1}
  Baier, R.; Dokshitzer, Y.L.; Mueller, A.H.; Schiff, D. Induced Gluon Radiation in a QCD Medium.
  \emph{Phys. Lett.~B} {\bf 1995}, \emph{345}, 277--286.

\bibitem{BDMPS-1-1}
  Baier, R.; Dokshitzer, Y.L.; Mueller, A.H.; Peigne, S.; Schiff, D. The Landau-Pomeranchuk-Migdal effect in QED.
  \emph{Nucl. Phys. B} {\bf 1996}, \emph{478}, 577--597.

\bibitem{BDMPS-1-2}
  Baier, R.; Dokshitzer, Y.L.; Mueller, A.H.; Schiff, D. Medium-induced radiative energy loss; equivalence between the BDMPS and Zakharov formalisms.
  \emph{Nucl. Phys. B} {\bf 1998}, \emph{531}, 403--425.

\bibitem{Nemchik:2014gka} 
  Nemchik, J.; Pasechnik, R.; Potashnikova, I.
    A heuristic description of high- $p_T$ hadron production in heavy ion collisions. 
  \emph{Eur. Phys. J. C} {\bf 2015}, \emph{75}, 95.

\bibitem{kopeliovich-gs}
  Kopeliovich, B.Z.; Schaefer, A.; Tarasov, A.V. Nonperturbative Effects in Gluon Radiation and Photoproduction of Quark Pairs.
  \emph{Phys. Rev. D} {\bf 2000}, \emph{62}, 054022.

\bibitem{kopeliovich-gs-1}
 Kopeliovich, B.Z.; Nemchik, J.; Schaefer, A. Color transparency versus quantum coherence in electroproduction of vector mesons off nuclei.
  \emph{Phys. Rev. C} \textbf{2002}, \emph{65}, 035201.

\bibitem{kopeliovich-gs-2}
  Kopeliovich, B.Z.; Nemchik, J.; Potashnikova, I.K.; Schmidt, I. Gluon Shadowing in DIS off Nuclei.
  \emph{J. Phys. G} {\bf 2008}, \emph{35}, 115010.


\bibitem{Gribov:1984tu}
  Gribov, L.V.; Levin, E.M.; Ryskin, M.G.
    Semihard Processes In QCD. 
  \emph{Phys. Rep.}  {\bf 1983}, \emph{100}, 1--150.
 
\bibitem{Dokshitzer:1977sg} 
  Dokshitzer, Y.L.
    Calculation of the Structure Functions for Deep Inelastic Scattering and e+ e- Annihilation by Perturbation Theory in Quantum Chromodynamics. 
  \emph{Sov. Phys. JETP} {\bf 1977}, \emph{46}, 641--653.

\bibitem{Gribov:1972ri} 
  Gribov, V.N.; Lipatov, L.N.
    Deep inelastic e p scattering in perturbation theory. 
  \emph{Sov. J. Nucl. Phys.}  {\bf 1972}, \emph{15}, 438--450.
  
\bibitem{Altarelli:1977zs} 
  Altarelli, G.; Parisi, G.
    Asymptotic Freedom in Parton Language. 
  \emph{Nucl. Phys. B} {\bf 1977}, \emph{126}, 298--318.

\bibitem{McLerran:1993ka} 
  McLerran, L.D.; Venugopalan, R.
    Gluon distribution functions for very large nuclei at small transverse momentum. 
  \emph{Phys. Rev. D} {\bf 1994}, \emph{49}, 3352--3355.
  
 \bibitem{JalilianMarian:2005jf}
  Jalilian-Marian, J.; Kovchegov, Y.V.
    Saturation physics and deuteron gold collisions at RHIC. 
  \emph{Prog. Part. Nucl.~Phys.}  {\bf 2006}, \emph{56}, 104--231. 
 
 \bibitem{Gelis:2010nm} 
 Gelis, F.; Iancu, E.; Jalilian-Marian, J.; Venugopalan, R.
    The Color Glass Condensate. 
  \emph{Ann. Rev. Nucl. Part. Sci.} {\bf 2010}, \emph{60}, 463--489.

\bibitem{Kharzeev:2004if} 
 Kharzeev, D.; Levin, E.; Nardi, M.
    Color glass condensate at the LHC: Hadron multiplicities in pp, pA and AA collisions. 
  \emph{Nucl. Phys. A} {\bf 2005}, \emph{747}, 609--629.

 \bibitem{JalilianMarian:1997jx} 
  Jalilian-Marian, J.; Kovner, A.; Leonidov, A.; Weigert, H.
    The BFKL equation from the Wilson renormalization group. 
  \emph{Nucl. Phys. B} {\bf 1997}, \emph{504}, 415--431.

\bibitem{JalilianMarian:1997gr} 
  Jalilian-Marian, J.; Kovner, A.; Leonidov, A.; Weigert, H.
    The Wilson renormalization group for low x physics: Towards the high density regime.
  \emph{Phys. Rev. D} {\bf 1998}, \emph{59}, 014014.

\bibitem{Kovner:2000pt} 
 Kovner, A.; Milhano, J.G.; Weigert, H.
    Relating different approaches to nonlinear QCD evolution at finite gluon density. 
  \emph{Phys. Rev. D} {\bf 2000}, \emph{62}, 114005.
  
\bibitem{Weigert:2000gi} 
  Weigert, H.
    Unitarity at small Bjorken x. 
  \emph{Nucl. Phys. A} {\bf 2002}, \emph{703}, 823--860.

\bibitem{Iancu:2000hn} 
 Iancu, E.; Leonidov, A.; McLerran, L.D.
    Nonlinear gluon evolution in the color glass condensate. 1. 
  \emph{Nucl.~Phys.~A} {\bf 2001}, \emph{692}, 583--645.

\bibitem{Ferreiro:2001qy} 
 Ferreiro, E.; Iancu, E.; Leonidov, A.; McLerran, L.
    Nonlinear gluon evolution in the color glass condensate. 2. 
  \emph{Nucl. Phys. A} {\bf 2002}, \emph{703}, 489--538.

\bibitem{GolecBiernat:1999qd} 
  Golec-Biernat, K.J.; Wusthoff, M.
    Saturation in diffractive deep inelastic scattering. 
  \emph{Phys. Rev. D} {\bf 1999}, \emph{60},~114023.

\bibitem{Caola:2008xr}
  Caola, F.; Forte, S.
    Geometric Scaling from GLAP evolution. 
  \emph{Phys. Rev. Lett.}  {\bf 2008}, \emph{101}, 022001.


\bibitem{Kharzeev:2001yq} 
  Kharzeev, D.; Levin, E.; Nardi, M.
    The Onset of classical QCD dynamics in relativistic heavy ion collisions. 
  \emph{Phys. Rev. C} {\bf 2005}, \emph{71}, 054903.
  
\bibitem{Schenke:2012wb} 
 Schenke, B.; Tribedy, P.; Venugopalan, R.
    Fluctuating Glasma initial conditions and flow in heavy ion collisions. 
  \emph{Phys. Rev. Lett.}  {\bf 2012}, \emph{108}, 252301.

\bibitem{Drescher:2006ca} 
  Drescher, H.-J.; Nara, Y.
    Effects of fluctuations on the initial eccentricity from the Color Glass Condensate in heavy ion collisions. 
  \emph{Phys. Rev. C} {\bf 2007}, \emph{75}, 034905. 

\bibitem{Hirano:2010je} 
  Hirano, T.; Huovinen, P.; Nara, Y.
    Elliptic flow in Pb+Pb collisions at $\sqrt{s_{NN}} = 2.76$~TeV: Hybrid model assessment of the first data. 
  \emph{Phys. Rev. C} {\bf 2011}, \emph{84}, 011901.
  
\bibitem{Aamodt:2010pa} 
Aamodt, K.; Abelev, B.; Quintana, A.A.; Adamova, D.; Adare, A.M.; Aggarwal, M.M.; Rinella, G.A.; Agocs, A.G.; Salazar, S.A.; Ahammed, Z.
    Elliptic flow of charged particles in Pb-Pb collisions at 2.76 TeV. 
  \emph{Phys.~Rev.~Lett.}  {\bf 2010}, \emph{105}, 252302.
  

\bibitem{Pang:2012he} 
 Pang, L.; Wang, Q.; Wang, X.N.
    Effects of initial flow velocity fluctuation in event-by-event (3+1)D hydrodynamics.
  \emph{Phys. Rev. C} {\bf 2012}, \emph{86}, 024911.

\bibitem{Bhalerao:2015iya} 
  Bhalerao, R.S.; Jaiswal, A.; Pal, S.
    Collective flow in event-by-event partonic transport plus hydrodynamics hybrid approach. 
  \emph{Phys. Rev. C} {\bf 2015}, \emph{92}, 014903.

\bibitem{Paatelainen:2013eea} 
  Paatelainen, R.; Eskola, K.J.; Niemi, H.; Tuominen, K.
    Fluid dynamics with saturated minijet initial conditions in ultrarelativistic heavy-ion collisions.
  \emph{Phys. Lett. B} {\bf 2014}, \emph{731}, 126--130.

\bibitem{Niemi:2015qia} 
  Niemi, H.; Eskola, K.J.; Paatelainen, R.
    Event-by-event fluctuations in a perturbative QCD + saturation + hydrodynamics model: Determining QCD matter shear viscosity in ultrarelativistic heavy-ion collisions. 
  \emph{Phys. Rev. C} {\bf 2016}, \emph{93}, 024907.

\bibitem{Moreland:2014oya} 
  Moreland, J.S.; Bernhard, J.E.; Bass, S.A.
    Alternative ansatz to wounded nucleon and binary collision scaling in high-energy nuclear collisions. 
  \emph{Phys. Rev. C} {\bf 2015}, \emph{92}, 011901.


\bibitem{Bielcikova:2016lgh} 
  Bielcikova, J. Jets and correlations in heavy-ion collisions. In Proceedings of the 2015 European Physical Society Conference on High Energy Physics, Vienna, Austria, 22--29 July 2015. 

\bibitem{Andronic:2015wma} 
Andronic, A.; Arleo, F.; Arnaldi, R.; Beraudo, A.; Bruna, E.; Caffarri, D.; del Valle, Z.C.; Contreras, J.G.; Dahms, T.; Dainese, A.; et al.
    Heavy-flavour and quarkonium production in the LHC era: From proton-proton to heavy-ion collisions. 
  \emph{Eur. Phys. J. C} {\bf 2016}, \emph{76}, 107.

\bibitem{Rafelski:1982pu} 
  Rafelski, J.; Muller, B.
    Strangeness Production in the Quark-Gluon Plasma. 
  \emph{Phys. Rev. Lett.} {\bf 1982}, \emph{48}, 1066.
  
\bibitem{Koch:1986ud} 
  Koch, P.; Muller, B.; Rafelski, J.
    Strangeness in Relativistic Heavy Ion Collisions. 
  \emph{Phys. Rep.}  {\bf 1986}, \emph{142}, 167--262.

\bibitem{Andronic:2008gu} 
  Andronic, A.; Braun-Munzinger, P.; Stachel, J.
    Thermal hadron production in relativistic nuclear collisions: The Hadron mass spectrum, the horn, and the QCD phase transition. 
  \emph{Phys. Lett. B} {\bf 2009}, \emph{673}, 142--145.

\bibitem{Teaney:2000cw} 
  Teaney, D.; Lauret, J.; Shuryak, E.V.
    Flow at the SPS and RHIC as a quark gluon plasma signature. 
  \emph{Phys.~Rev.~Lett.} {\bf 2001}, \emph{86}, 4783--4786.  
  
\bibitem{Bass_time_evolution}
Modeling Relativistic Heavy-Ion Collisions. Available online: \myurl{https://www.phy.duke.edu/modeling-relativistic-heavy-ion-collisions} (accessed on 17 January 2017).

\bibitem{Tribedy:2011aa} 
  Tribedy, P.; Venugopalan, R.
    QCD saturation at the LHC: Comparisons of models to p + p and A + A data and predictions for p + Pb collisions. 
  \emph{Phys. Lett. B} {\bf 2012}, \emph{710}, 125--133.

\bibitem{ATLAS:2011ag} 
Aad, G.; Abbott, B.; Abdallah, J.; Khalek, S.A.; Abdelalim, A.A.; Abdesselam, A.; Abdinov, O.; Abi,~B.; Abolins, M.; AbouZeid, O.S.; et al.
    Measurement of the centrality dependence of the charged particle pseudorapidity distribution in lead-lead collisions at $\sqrt{s_{NN}}=2.76$ TeV with the ATLAS detector. 
  \emph{Phys.~Lett.~B} {\bf 2012}, \emph{710}, 363--382.

\bibitem{Ollitrault:1992bk} 
  Ollitrault, J.Y.
    Anisotropy as a signature of transverse collective flow. 
  \emph{Phys. Rev. D} {\bf 1992}, \emph{46}, 229--245.
 
\bibitem{Poskanzer:1998yz} 
  Poskanzer, A.M.; Voloshin, S.A.
    Methods for analyzing anisotropic flow in relativistic nuclear collisions. 
  \emph{Phys. Rev. C} {\bf 1998}, \emph{58}, 1671--1678.

\bibitem{Manly:2005zy} 
 Manly, S.
    System size, energy and pseudorapidity dependence of directed and elliptic flow at RHIC. 
  \emph{Nucl.~Phys. A} {\bf 2006}, \emph{774}, 523--526.

\bibitem{ALICE:2011ab} 
 Aamodt, K.; Abelev, B.; Quintana, A.A.; Adamova, D.; Adare, A.M.; Aggarwal, M.M.; Rinella, G.A.; Agocs,~A.G.; Agostinelli, A.; Salazar, S.A.; et al.
    Higher harmonic anisotropic flow measurements of charged particles in Pb-Pb collisions at $\sqrt{s_{NN}}=2.76$ TeV. 
  \emph{Phys. Rev. Lett.} {\bf 2011}, \emph{107}, 032301.

\bibitem{Venema:1993zz} 
Venema, L.B.; Braak, H.; Löhner, H.; Raschke, A.E.; Siemssen, R.H.; Šumbera, M.; Wilschut, H.W.; Berg,~F.D.; Kühn, W.; Metag, V.; et al.  Azimuthal asymmetry of neutral pion emission in Au+Au reactions at 1 GeV/nucleon. 
  \emph{Phys. Rev. Lett.} {\bf 1993}, \emph{71}, 835--838.

\bibitem{Adamczyk:2016exq} 
   Adamczyk, L.; Adkins, J.K.; Agakishiev, G.; Aggarwal, M.M.; Ahammed, Z.; Alekseev, I.; Aparin, A.; Arkhipkin, D.; Aschenauer, E.C.; Attri, A.; et al.
    Beam Energy Dependence of the Third Harmonic of Azimuthal Correlations in Au+Au Collisions at RHIC.
  \emph{Phys. Rev. Lett.} {\bf 2016}, \emph{116}, 112302.

\bibitem{Gale:2012rq} 
 Gale, C.; Jeon, S.; Schenke, B.; Tribedy, P.; Venugopalan, R.
    Event-by-event anisotropic flow in heavy-ion collisions from combined Yang-Mills and viscous fluid dynamics. 
  \emph{Phys. Rev. Lett.} {\bf 2013}, \emph{110}, 012302.

\bibitem{Aad:2013xma} 
Aad, G.; Abajyan, T.; Abbott, B.; Abdallah, J.; Abdel Khalek, S.; Abdelalim, A.A.; Abdinov, O.; Aben, R.; Abi,~B.; Abolins, M.
    Measurement of the distributions of event-by-event flow harmonics in lead-lead collisions at = 2.76 TeV with the ATLAS detector at the LHC.
  \emph{J. High Energy Phys.} {\bf 2013}, \emph{1311}, 183.

\bibitem{Qiu:2012uy} 
  Qiu, Z.; Heinz, U.
    Hydrodynamic event-plane correlations in Pb+Pb collisions at $\sqrt{s}=2.76$A TeV. 
  \emph{Phys.~Lett.~B} {\bf 2012}, \emph{717}, 261--265.

\bibitem{Aad:2014fla} 
 Aad, G.; Abbott, B.; Abdallah, J.; Khalek, S.A.; Abdinov, O.; Aben, R.; Abi, B.; Abolins, M.; Abouzeid, O.; Abramowicz, H.; et al.
    Measurement of event-plane correlations in $\sqrt{s_{NN}}=2.76$ TeV lead-lead collisions with the ATLAS detector. 
  \emph{Phys. Rev. C} {\bf 2014}, \emph{90}, 024905.

\bibitem{Qian:2016pau} 
  Qian, J.; Heinz, U.
    Hydrodynamic flow amplitude correlations in event-by-event fluctuating heavy-ion collisions. 
  \emph{Phys. Rev. C} {\bf 2016}, \emph{94}, 024910.

\bibitem{Giacalone:2016afq} 
  Giacalone, G.; Yan, L.; Noronha-Hostler, J.; Ollitrault, J.Y.
    Symmetric cumulants and event-plane correlations in Pb + Pb collisions. 
  \emph{Phys. Rev. C} {\bf 2016}, \emph{94}, 014906.

\bibitem{Aad:2015lwa} 
Aad, G.; Abbott, B.; Abdallah, J.; Khalek, S.A.; Abdelalim, A.A.; Abdesselam, A.; Abdinov, O.; Abi, B.; Abolins, M.; AbouZeid, O.S.; et al.
    Measurement of the correlation between flow harmonics of different order in lead-lead collisions at $\sqrt{s_{NN}}=2.76$ TeV with the ATLAS detector. 
  \emph{Phys. Rev. C} {\bf 2015}, \emph{92}, 034903.

\bibitem{ALICE:2016kpq} 
Aamodt, K.; Abelev, B.; Quintana, A.A.; Adamova, D.; Adare, A.M.; Aggarwal, M.M.; Rinella, G.A.; Agocs,~A.G.; Agostinelli, A.; Salazar, S.A.; et al.
    Correlated event-by-event fluctuations of flow harmonics in Pb-Pb collisions at $\sqrt{s_{_{\rm NN}}}=2.76$ TeV. 
  \emph{Phys. Rev. Lett.} {\bf 2016}, \emph{117}, 182301.

\bibitem{Wheaton:2004qb} 
  Wheaton, S.; Cleymans, J.
    THERMUS: A Thermal model package for ROOT. 
  \emph{Comput. Phys. Commun.} {\bf 2009}, \emph{180}, 84--106.

\bibitem{ABELEV:2013zaa} 
  Abelev, B.; Adam, J.; Adamová, D.; Adare, A.M.; Aggarwal, M.M.; Rinella, G.A.; Agnello, M.; Agocs, A.G.; Agostinelli, A.; Ahammed, Z.; et al.
    Multi-strange baryon production at mid-rapidity in Pb-Pb collisions at $\sqrt{s_{NN}} = 2.76$ TeV. 
  \emph{Phys. Lett. B} {\bf 2014}, \emph{728}, 216--227.

\bibitem{Adam:2015vsf} 
Adam, J.; Adamová, D.; Aggarwal, M.M.; Rinella, G.A.; Agnello, M.; Agrawal, N.; Ahammed, Z.; Ahmad, S.; Ahn, S.U.; Aiola, S.; et al. 
    Multi-strange baryon production in p-Pb collisions at $\sqrt{s_{NN}}=5.02$ TeV. 
  \emph{Phys.~Lett.~B} {\bf 2016}, \emph{758}, 389--401.

\bibitem{Bjorken:1982tu} 
  Bjorken, J.D. {Energy Loss of Energetic Partons in Quark-Gluon Plasma: Possible Extinction of High p(t) Jets in Hadron-Hadron Collisions}. Avaliable online: http://lss.fnal.gov/archive/1982/pub/Pub-82-059-T.pdf (Accessed on 17 January 2017).
 
\bibitem{Appel:1985dq} 
  Appel, D.A.
    Jets as a Probe of Quark-Gluon Plasmas. 
  \emph{Phys. Rev. D} {\bf 1986}, \emph{33}, 717--722.
  
\bibitem{Gyulassy:1990ye} 
  Gyulassy, M.; Plumer, M.
    Jet Quenching in Dense Matter, The quenching of hard jets in ultrarelativistic nuclear collisions is estimated emphasizing its sensitivity to possible changes in the energy loss mechanism in a quark-gluon plasma. 
  \emph{Phys. Lett. B} {\bf 1990}, \emph{243}, 432--438.

\bibitem{Khachatryan:2016odn} 
Khachatryan, V.; Sirunyan, A.M.;  Tumasyan, A.;  Adam, W.; Asilar, E; Bergauer, T.; Brandstetter, J.;  Brondolin, E.;  Dragicevic, M.;  Erö, J.; et al. Charged-Particle Nuclear Modification Factors in PbPb and pPb Collisions at  $\sqrt{s_\mathbf{NN}}$ = 5.02 TeV. {\em arXiv} {\bf 2016}, arXiv:1611.01664.

\bibitem{Adams:2006yt} 
Adams, J.; Aggarwal, M.M.; Ahammed, Z.; Amonett, J.; Anderson, B.D.; Anderson, M.; Arkhipkin, D.; Averichev, G.S.; Bai, Y.; Balewski, J. Direct Observation of Dijets in Central Au+Au Collisions at $\sqrt{s_\mathbf{NN}}=200$ GeV.
 \emph{\mbox{Phys. Rev. Lett.}} {\bf 2006}, \emph{97}, 162301.

\bibitem{d'Enterria:2012iw} 
  D'Enterria, D.
    Perturbative probes of QCD matter at the Large Hadron Collider. In Proceedings of the 6th International Conference on Quarks and Nuclear Physics, Palaiseau, Paris, 16--20 April 2012.

\bibitem{Chatrchyan:2012gt} 
 Chatrchyan, S.; Khachatryan, V.; Sirunyan, A.M.; Tumasyan, A.; Adam, W.; Bergauer, T.; Dragicevic, M.; Erö, J.; Fabjan, C.; Friedl, M. Studies of jet quenching using isolated-photon+jet correlations in PbPb and $pp$ collisions at $\sqrt{s_{NN}}=2.76$ TeV.  \emph{Phys. Lett. B} {\bf 2013}, \emph{718}, 773--794.

\bibitem{Baier:1996sk} 
  Baier, R.; Dokshitzer, Y.L.; Mueller, A.H.; Peigne, S.; Schiff, D.
    Radiative energy loss and p(T) broadening of high-energy partons in nuclei. 
  \emph{Nucl. Phys. B} {\bf 1997}, \emph{484}, 265--282.

\bibitem{CaronHuot:2008ni} 
  Caron-Huot, S.
    O(g) plasma effects in jet quenching. 
  \emph{Phys. Rev. D} {\bf 2009}, \emph{79}, 065039.

\bibitem{Laine:2012ht} 
 Laine, M.
    A non-perturbative contribution to jet quenching. 
  \emph{Eur. Phys. J. C} {\bf 2012}, \emph{72}, 2233.

\bibitem{Blaizot:2014bha} 
  Blaizot, J.P.; Mehtar-Tani, Y.
    Renormalization of the jet-quenching parameter. 
  \emph{Nucl. Phys. A} {\bf 2014}, \emph{929}, 202--229.

\bibitem{Panero:2013pla} 
  Panero, M.; Rummukainen, K.; Schfer, A.
    Lattice Study of the Jet Quenching Parameter. 
  \emph{Phys. Rev. Lett.} {\bf 2014}, \emph{112}, 162001.

\bibitem{Liu:2006ug} 
  Liu, H.; Rajagopal, K.; Wiedemann, U.A.
    Calculating the jet quenching parameter from AdS/CFT. 
  \emph{Phys.~Rev.~Lett.} {\bf 2006}, \emph{97}, 182301.


\bibitem{Svetitsky:1987gq} 
  Svetitsky, B.
    Diffusion of charmed quarks in the quark-gluon plasma. 
  \emph{Phys. Rev. D} {\bf 1988}, \emph{37}, 2484.

\bibitem{Young:2008he} 
  Young, C.; Shuryak, E.
    Charmonium in strongly coupled quark-gluon plasma. 
  \emph{Phys. Rev. C} {\bf 2009}, \emph{79}, 034907.

\bibitem{Petreczky:2016etz} 
  Petreczky, P.; Young, C. Sequential Bottomonium Production at High Temperatures. {\em arXiv} {\bf 2016}, arXiv:1606.08421.

 \bibitem{Digal:2001ue} 
  Digal, S.; Petreczky, P.; Satz, H.
    Quarkonium feed down and sequential suppression. 
  \emph{Phys. Rev. D} {\bf 2001}, \emph{64},~094015.

\bibitem{CMS:2016ayg} 
  CMS Collaboration.
    \emph{Strong suppression of $\varUpsilon$ excited states in PbPb collisions at  $\sqrt{s_{NN}} = 5.02~\mathrm{TeV}$}; CMS-PAS-HIN-16-008; CMS Collaboration: Cessy, France, 2016.

\bibitem{Chatrchyan:2012lxa} 
Chatrchyan, S.; Khachatryan, V.; Sirunyan, A.M.; Tumasyan, A.; Adam, W.; Aguilo, E.; Bergauer, T.; Dragicevic, M.; Erö, J.; Fabjan, C.; et al.  Observation of sequential Upsilon suppression in PbPb collisions. 
  \emph{Phys. Rev. Lett.} {\bf 2012}, \emph{109}, 222301.

\bibitem{Adare:2008ab} 
Adare, A.; Afanasiev, S.; Aidala, C.; Ajitanand, N.N.; Akiba, Y.; Al-Bataineh, H.; Alexander, J.; Al-Jamel, A.; Aoki, K.; Aphecetche, L.; et al. Enhanced Production of Direct Photons in Au+Au Collisions at $\sqrt{s_{NN}} = 200~\mathrm{GeV}$ and Implications for the Initial Temperature. \emph{Phys. Rev. Lett.} {\bf 2010}, \emph{104}, 132301. 
 
 \bibitem{Adam:2015lda} 
 Adam, J.; Adamova, D.; Aggarwal, M.M.; Aglieri Rinella, G.; Agnello, M.; Agrawal, N.; Ahammed, Z.; Ahn,~S.U.; Aiola, S.; Akindinov, A.
 Direct photon production in Pb-Pb collisions at $\sqrt{s_{NN}} = 2.76~\mathrm{TeV}$.  \emph{Phys.~Lett. B} {\bf 2016}, \emph{754}, 235--248.

\bibitem{Shuryak:1978ij} 
 Shuryak, E.V.
    Quark-Gluon Plasma and Hadronic Production of Leptons, Photons and Psions. 
  \emph{Phys. Lett. B} {\bf 1978}, \emph{78}, 150--153.
  
\bibitem{Feinberg:1976ua} 
  Feinberg, E.L.
    Direct Production of Photons and Dileptons in Thermodynamical Models of Multiple Hadron Production. 
  \emph{Nuovo Cim. A} {\bf 1976}, \emph{34}, 391–412.

\bibitem{Domokos:1980ba} 
Domokos, G.; Goldman, J.I. Quark-matter diagnostics. \emph{Phys. Rev. D} {\bf 1981}, \emph{23}, 203.

\bibitem{Stankus:2005eq} 
  Stankus, P.
    Direct photon production in relativistic heavy-ion collisions. 
  \emph{Ann. Rev. Nucl. Part. Sci.} {\bf 2005}, \emph{55}, 517--554.

\bibitem{Aggarwal:2000th} 
  Aggarwal, M.M.; Agnihotri, A.; Ahammed, Z.; Angelis, A.L.S.; Antonenko, V.; Arefiev, V.; Astakhov, V.; Avdeitchikov, V.; Awes, T.C.; Baba, P.V.K.S.; et al.
    Observation of direct photons in central 158-A-GeV Pb-208 + Pb-208 collisions. 
  \emph{Phys. Rev. Lett.} {\bf 2000}, \emph{85}, 3595.

\bibitem{Aggarwal:2000ps} 
  Aggarwal, M.M.; Aggarwal, M.M.; Agnihotri, A.; Ahammed, Z.; Angelis, A.L.S.; Antonenko, V.; Areev, V.; Astakhov, V.; Avdeitchikov, V.; Awes, T.C.; Baba, P.V.K.S.; et al. Direct Photon Production in 158-A-GeV Pb-208 + Pb-208 Collisions. {\em Phys. Rev. Lett.}  {\bf 2000}, {\em 85}, 3595--3599.

\bibitem{Adamczyk:2013dal} 
  Adamczyk, L.; Adkins, J.K.; Agakishiev, G.; Aggarwal, M.M.; Ahammed, Z.; Alekseev, I.; Alford, J.; Anson, C.D.; Aparin, A.; Arkhipkin, D.; et al.
    Energy Dependence of Moments of Net-proton Multiplicity Distributions at RHIC. 
  \emph{Phys. Rev. Lett.} {\bf 2014}, \emph{112}, 032302. 

\bibitem{Adamczyk:2014fia} 
Adamczyk, L.; Adkins, J.K.; Agakishiev, G.; Aggarwal, M.M.; Ahammed, Z.; Alekseev, I.; Alford, J.; Anson,~C.D.; Aparin, A.; Arkhipkin, D.; et al.
    Beam energy dependence of moments of the net-charge multiplicity distributions in Au+Au collisions at RHIC. 
  \emph{Phys. Rev. Lett.} {\bf 2014}, \emph{113}, 092301.

\bibitem{Sumbera:2012qb} 
  Sumbera, M.
    Soft Physics at RHIC. 
  \emph{EPJ Web. Conf.}  {\bf 2012}, \emph{28}, 03006.
  
\bibitem{Sumbera:2013kd} 
  Sumbera, M.
    Results from STAR Beam Energy Scan Program. 
  \emph{Acta Phys. Pol. Supp.} {\bf 2013}, \emph{6}, 429--436.

\bibitem{Adare:2015aqk} 
Adare, A.; Afanasiev, S.; Aidala, C.; Ajitanand, N.N.; Akiba, Y.; Akimoto, R.; Al-Bataineh, H.; Alexander, J.; Al-Ta'ani, H.; Angerami, A.; et al.
    Measurement of higher cumulants of net-charge multiplicity distributions in Au$+$Au collisions at $\sqrt{s_{_{NN}}}=7.7-200$ GeV. 
  \emph{Phys. Rev. C} {\bf 2016}, \emph{93}, 011901.

\bibitem{Stephanov:1998dy} 
  Stephanov, M.A.; Rajagopal, K.; Shuryak, E.V.
    Signatures of the tricritical point in QCD.
  \emph{Phys. Rev. Lett.} {\bf 1998}, \emph{81}, 4816--4819.

\bibitem{Asakawa:2015ybt} 
  Asakawa, M.; Kitazawa, M.
    Fluctuations of conserved charges in relativistic heavy ion collisions: An~introduction. 
  \emph{Prog. Part. Nucl. Phys.} {\bf 2016}, \emph{90}, 299--342.

\bibitem{Lacey:2014wqa} 
{Lacey, R.A.} Indications for a Critical End Point in the Phase Diagram for Hot and Dense Nuclear Matter.
  \emph{Phys. Rev. Lett.}  {\bf 2015}, \emph{114}, 142301.

\bibitem{Gavai:2010zn} 
  Gavai, R.V.; Gupta, S.
    Lattice QCD predictions for shapes of event distributions along the freezeout curve in heavy-ion collisions. 
  \emph{Phys. Lett. B} {\bf 2011}, \emph{696}, 459--463. 

  
\bibitem{Aggarwal:2010wy} 
  Aggarwal, M.M.; Ahammed, Z.; Alakhverdyants, A.V.; Alekseev, I.; Alford, J.; Anderson, B.D.; Arkhipkin,~D.; Averichev, G.S.; Balewski, J.; Barnby, L.S.; et al.
    Higher Moments of Net-proton Multiplicity Distributions at RHIC.
  \emph{Phys. Rev. Lett.} {\bf 2010}, \emph{105}, 022302.

\bibitem{Luo:2015ewa} 
  Luo, X. 
    Energy Dependence of Moments of Net-Proton and Net-Charge Multiplicity Distributions at STAR. In Proceedings of the 9th International Workshop on Critical Point and Onset of Deconfinement, Bielefeld, Germany, 17--21 November 2014. 

\bibitem{Khachatryan:2010gv} 
Khachatryan, V.; Sirunyan, A.M.; Tumasyan, A.; Adam, W.; Bergauer, T.; Dragicevic, M.; Erö, J.; Fabjan, C.; Friedl, M.; Frühwirth, R.; et al.
    Observation of Long-Range Near-Side Angular Correlations in Proton-Proton Collisions at the LHC. 
  \emph{J. High Energy Phys.} {\bf 2010}, \emph{1009}, 91.

\bibitem{Khachatryan:2015lva} 
Khachatryan, V.; Sirunyan, A.M.; Tumasyan, A.; Adam, W.; Asilar, E.; Bergauer, T.; Brandstetter, J.; Brondolin,~E.; Dragicevic, M.; Erö, J.; et al.
    Measurement of long-range near-side two-particle angular correlations in pp collisions at $\sqrt s =$ 13 TeV. 
  \emph{Phys. Rev. Lett.} {\bf 2016}, \emph{116}, 172302.
 
\bibitem{Aad:2015gqa} 
Aad, G.; Abbott, B.; Abdallah, J.; Abdinov, O.; Aben, R.; Abolins, M.; AbouZeid, O.S.; Abramowicz, H.; Abreu, H.; Abreu, R.; et al.
    Observation of Long-Range Elliptic Azimuthal Anisotropies in $\sqrt{s}=$ 13 and 2.76 TeV $pp$ Collisions with the ATLAS Detector. 
  \emph{Phys. Rev. Lett.} {\bf 2016}, \emph{116}, 172301.

\bibitem{Aad:2012gla} 
   Aad, G.; Abajyan, T.; Abbott, B.; Abdallah, J.; Abdel Khalek, S.; Abdelalim, A.A.; Abdinov, O.; Aben, R.; Abi, B.; Abolins, M.; et al.
    Observation of Associated Near-Side and Away-Side Long-Range Correlations in $\sqrt{s_{NN}}$=5.02??TeV Proton-Lead Collisions with the ATLAS Detector. 
  \emph{Phys. Rev. Lett.} {\bf 2013}, \emph{110}, 182302.

\bibitem{CMS:2012qk} 
 Chatrchyan, S.; Khachatryan, V.; Sirunyan, A.M.; Tumasyan, A.; Adam, W.; Aguilo, E.; Bergauer, T.; Dragicevic, M.; Erö, J.; Fabjan, C.; et al. 
    Observation of long-range near-side angular correlations in proton-lead collisions at the LHC. 
  \emph{Phys. Lett. B} {\bf 2013}, \emph{718}, 795--814.   

\bibitem{Aad:2013fja} 
 Aad, G.; Abajyan, T.; Abbott, B.; Abdallah, J.; Khalek, S.A.; Abdelalim, A.A.; Abdinov, O.; Aben, R.; Abi, B.; Abolins, M.; et al.
    Measurement with the ATLAS detector of multi-particle azimuthal correlations in p+Pb collisions at $\sqrt{s_{NN}} =5.02$ TeV. 
  \emph{Phys. Lett. B} {\bf 2013}, \emph{725}, 60--78.

\bibitem{ABELEV:2013wsa} 
Abelev, B.; Adam, J.; Adamová, D.; Adaredu, A.M.; Aggarwalcc, M.M.; Rinellaag, G.A.; Ci, M.A.; Agocsdt,~A.G.; Agostinelliy, A.; Ahammed, Z.
    Long-range angular correlations of $\rm \pi$, K and p in p-Pb collisions at $\sqrt{s_{\rm NN}}= 5.02$ TeV. 
  \emph{Phys. Lett. B} {\bf 2013}, \emph{726}, 164--177.

\bibitem{Abelev:2014mda} 
Abelev, B.; Adam, J.; Adamová, D.; Aggarwal, M.M.; Rinella, G.A.; Agnello, M.; Agostinelli, A.; Agrawal, N.; Ahammed, Z.; Ahmad, N.; et al.
    Multiparticle azimuthal correlations in p -Pb and Pb-Pb collisions at the CERN Large Hadron Collider. 
  \emph{Phys. Rev. C} {\bf 2014}, \emph{90}, 054901.

\bibitem{Khachatryan:2014jra} 
     Khachatryan, V.; Sirunyan, A.M.; Tumasyan, A.; Adam, W.; Bergauer, T.; Dragicevic, M.; Erö, J.; Fabjan,~C.; Friedl, M.; Frühwirth, R.; et al. Long-range two-particle correlations of strange hadrons with charged particles in pPb and PbPb collisions at LHC energies.
  \emph{Phys. Lett. B} {\bf 2015}, \emph{742}, 200–224.

\bibitem{Khachatryan:2015waa} 
  Khachatryan, V.; Sirunyan, A.M.; Tumasyan, A.; Adam, W.; Bergauer, T.; Dragicevic, M.; Erö, J.;  Friedl, M.; Frühwirth, R.; Ghete, V.M.; et al. Evidence for Collective Multiparticle Correlations in p-Pb Collisions. \emph{Phys.~Rev. Lett.} \textbf{2015}, {\em 115}, 012301.
 
  
\bibitem{Abelev:2012ola} 
Abelev, B.; Adam, J.; Adamova, D.; Adare, A.M.; Aggarwal, M.; Rinella, G.A.; Agnello, M.; Agocs, A.G.; Agostinelli, A.; Ahammed, Z.; et al. Long-range angular correlations on the near and away side in $p$-Pb collisions at $\sqrt{s_{NN}}=5.02$ TeV. \emph{Phys. Lett. B} {\bf 2013}, \emph{719}, 29--41.

\bibitem{Adare:2015ctn} 
Adare, A.; Afanasiev, S.; Aidala, C.; Ajitanand, N.N.; Akiba, Y.; Akimoto, R.; Al-Bataineh, H.; Alexander, J.; Alfred, M.; Al-Ta’ani, H.; Andrews, K.R. Measurements of elliptic and triangular flow in high-multiplicity $^{3}$He$+$Au collisions at $\sqrt{s_{_{NN}}}=200$ GeV.
  \emph{Phys. Rev. Lett.} {\bf 2015}, \emph{115}, 142301.

\bibitem{Koop:2015wea} 
  Orjuela Koop, J.D.; Adare, A.; McGlinchey, D.; Nagle, J.L. Azimuthal anisotropy relative to the participant plane from a multiphase transport model in central p + Au , d + Au , and $^{3}$He + Au collisions at \mbox{$\sqrt{s_{NN}}=200$ GeV}.
  \emph{Phys. Rev. C} {\bf 2015}, \emph{92}, 054903.

\bibitem{Abelev:2009af} 
  Abelev, B.I.; Aggarwal, M.M.; Ahammed, Z.; Alakhverdyants, A.V.; Anderson, B.D.; Arkhipkin, D.; Averichev,~G.S.; Balewski, J.; Barannikova, O.; Barnby, L.S.; Baudot, J.; et al. Long range rapidity correlations and jet production in high energy nuclear collisions.
  \emph{Phys. Rev. C} {\bf 2009}, \emph{80}, 064912.

\bibitem{Alver:2009id} 
  Alver, B.; Back, B.B.; Baker, M.D.; Ballintijn, M.; Barton, D.S.; Betts, R.R.; Bickley, A.A.; Bindel, R.; Busza, W.; Carroll, A.; et al. High transverse momentum triggered correlations over a large pseudorapidity acceptance in Au+Au collisions at s(NN)**1/2 = 200 GeV.
  \emph{Phys. Rev. Lett.} {\bf 2010}, \emph{104}, 062301.
 
\bibitem{Aamodt:2011by} 
 Aamodt, K.; Abelev, B.; Quintana, A.A.; Adamova, D.; Adare, A.M.; Aggarwal, M.M.; Rinella, G.A.; Agocs,~A.G.; Agostinelli, A.; Salazar, S.A.; et al. Harmonic decomposition of two-particle angular correlations in Pb-Pb collisions at $\sqrt{s_{NN}}=$ 2.76 TeV. \emph{Phys. Lett. B} { \bf 2012}, {\em 708}, 249--264.

\bibitem{Agakishiev:2011st} 
 Agakishiev, G.; Aggarwal, M.M.; Ahammed, Z.; Alakhverdyants, A.V.; Alekseev, I.; Alford, J.; Anderson,~B.D.; Anson, C.D.; Arkhipkin, D.; Averichev, G.S.; et al. 
    System size and energy dependence of near-side di-hadron correlations. 
  \emph{Phys. Rev. C} {\bf 2012}, \emph{85}, 014903.
 
\bibitem{Voloshin:2004th} 
  Voloshin, S.A.
    Two particle rapidity, transverse momentum, and azimuthal correlations in relativistic nuclear collisions and transverse radial expansion. 
  \emph{Nucl. Phys. A} {\bf 2005}, \emph{749}, 287--290.

\bibitem{Rezaeian:2016szi} 
  Rezaeian, A.H.
    Photon-jet ridge at RHIC and the LHC. 
  \emph{Phys. Rev. D} {\bf 2016}, \emph{93}, 094030.
 
\bibitem{Dusling:2015rja} 
  Dusling, K.; Tribedy, P.; Venugopalan, R.
    Energy dependence of the ridge in high multiplicity proton-proton collisions. 
  \emph{Phys. Rev. D} {\bf 2016}, \emph{93}, 014034.
 
\bibitem{Hirono:2014dda} 
  Hirono, Y.; Shuryak, E.
    Femtoscopic signature of strong radial flow in high-multiplicity $pp$ collisions. 
  \emph{Phys.~Rev. C} {\bf 2015}, \emph{91}, 054915.

\bibitem{Bozek:2012gr} 
 Bozek,  P.; Broniowski, W.
    Correlations from hydrodynamic flow in p-Pb collisions. 
  \emph{Phys. Lett. B} {\bf 2013}, \emph{718},~1557.

\bibitem{Bozek:2013ska} 
  Bozek, P.; Broniowski, W.; Torrieri, G.
    Mass hierarchy in identified particle distributions in proton-lead collisions. 
  \emph{Phys. Rev. Lett.} {\bf 2013}, \emph{111}, 172303.

\bibitem{Werner:2013ipa} 
  Werner, K.; Bleicher, M.; Guiot, B.; Karpenko, I.; Pierog, T.
    Evidence for Flow from Hydrodynamic Simulations of $p$-Pb Collisions at 5.02 TeV from $\nu_2$ Mass Splitting. 
  \emph{Phys. Rev. Lett.} {\bf 2014}, \emph{112}, 232301.

\bibitem{Romatschke:2015gxa} 
 Romatschke, P.
    Light-Heavy Ion Collisions: A window into pre-equilibrium QCD dynamics? 
  \emph{Eur. Phys. J. C} {\bf 2015}, \emph{75}, 305.

\bibitem{Niemi:2015ile} 
  Niemi, H.
    Recent developments in hydrodynamics and collectivity in small systems. In Proceedings of the European Physical Society Conference on High Energy Physics (EPS-HEP2015), Vienna, Austria, 22--29~July~2015. 
  
\bibitem{Koblesky:2015uba} 
  Koblesky, T.
    Collectivity in Small QCD Systems. In Proceedings of the CIPANP2015 Conference, Vail, CO, USA, 19--24 May 2015.

\bibitem{Schenke:2014zha} 
  Schenke, B.; Venugopalan, R.
    Eccentric protons? Sensitivity of flow to system size and shape in p+p, p+Pb and Pb+Pb collisions. 
  \emph{Phys. Rev. Lett.} {\bf 2014}, \emph{113}, 102301.

\bibitem{Bzdak:2014dia} 
  Bzdak, A.; Ma, G.L.
    Elliptic and triangular flow in $p$+Pb and peripheral Pb+Pb collisions from parton scatterings.
  \emph{Phys. Rev. Lett.} {\bf 2014}, \emph{113}, 252301.

\bibitem{Bozek:2015swa} 
  Bozek, P.; Bzdak, A.; Ma, G.L.
    Rapidity dependence of elliptic and triangular flow in proton?nucleus collisions from collective dynamics. 
  \emph{Phys. Lett. B} {\bf 2015}, \emph{748}, 301--305.

\bibitem{Antinori:2016zxe} 
Antinori,  F.; Becattini,  F.;  Braun-Munzinger, P.;  Chujo, T.;  Hamagaki, H.; Harris, J.;  Heinz, U.; Hippolyte,~B.; Hirano, T.;  Jacak, B.; et al. Thoughts on Heavy-Ion Physics in the High Luminosity Era: The Soft Sector. {\em arXiv} {\bf 2016},	arXiv:1604.03310.

\bibitem{Bala:2016hlf} 
  Bala, R.; Bautista, I.; Bielcikova, J.; Ortiz, A.
    Heavy-ion physics at the LHC: Review of Run I results. 
  \emph{Int. J. Mod. Phys. E} {\bf 2016}, \emph{25}, 1642006.

\bibitem{Sinyukov:2013zna} 
  Sinyukov, Y.M.; Akkelin, S.V.; Karpenko, I.A.; Shapoval, V.M.
 Femtoscopic and Nonfemtoscopic Two-Particle Correlations in $A + A$ and $p + p$ Collisions at RHIC and LHC Energies. 
  \emph{Adv. High Energy Phys.} {\bf 2013}, \emph{2013},~198928.

\bibitem{Lisa:2016buz} 
  Lisa, M.
    Timescales in heavy ion collisions. 
  \emph{Acta Phys. Pol. B} {\bf 2016}, \emph{47}, 1847--1855.
  
\end{thebibliography}
\end{document}